\documentclass[amssymb,floatfix,nofootinbib,nobibnotes, longbibliography,superscriptaddress]{revtex4-1}

\pdfoutput=1
\usepackage[utf8]{inputenc}
\usepackage{amsmath}
\usepackage{amssymb}
\usepackage{color}
\usepackage{lineno}
\usepackage{graphicx}
\usepackage{xfrac}
\usepackage{xspace}
\usepackage{wrapfig}

\usepackage{url}

\usepackage{array,colortbl,xcolor}
\usepackage[caption=false]{subfig}

\definecolor{darkblue}{rgb}{0.0,0,0.5}
\definecolor{darkgreen}{rgb}{0.0,0.3,0.0}
\definecolor{redish}{rgb}{0.675,0,0.2}
\definecolor{red}{rgb}{0.8,0,0}
\definecolor{green}{rgb}{0,0.6,0}
\definecolor{bluish}{rgb}{0.2,0.2,0.675}
\definecolor{mygrey}{rgb}{0.6,0.6,0.6}

\newcommand{\pineappl}{\texttt{PineAPPL}\xspace}
\newcommand{\yadism}{\texttt{yadism}\xspace}
\newcommand{\eko}{\texttt{EKO}\xspace}
\newcommand{\appl}{\texttt{APPLgrid}\xspace}
\newcommand{\fastnlo}{\texttt{fastNLO}\xspace}
\newcommand{\applfast}{\texttt{APPLFast}\xspace}
\newcommand{\lhapdf}{\texttt{LHAPDF}\xspace}
\newcommand{\nnpdf}{NNPDF\xspace}
\newcommand{\madgraph}{\texttt{Madgraph5\_aMC@NLO}\xspace}
\newcommand{\xfitter}{\texttt{xFitter}\xspace}
\newcommand{\validphys}{\texttt{validphys}\xspace}
\newcommand{\reportengine}{\texttt{reportengine}\xspace}

\usepackage[unicode=true, bookmarks=false, linkcolor = darkblue, citecolor = redish, breaklinks=false, colorlinks=true, hyperfootnotes=true]{hyperref}

\begin{document}
\pagenumbering{gobble}
\title{\Large \bf Snowmass 2021 whitepaper: Proton structure at the precision frontier}
\date{October 20, 2022}
\newcommand{\la}{\thanks{Leading author}}

\newcommand{\afECT}{\affiliation{European Centre for Theoretical Studies in Nuclear Physics and Related Areas (ECT*), I-38123 Villazzano, Trento, Italy}}
\newcommand{\afFBK}{\affiliation{Fondazione Bruno Kessler (FBK), I-38123 Povo, Trento, Italy}}
\newcommand{\afTIFPA}{\affiliation{INFN-TIFPA Trento Institute of Fundamental Physics and Applications, I-38123 Povo, Trento, Italy}}
\newcommand{\afFNAL}{\affiliation{Fermi National Accelerator Laboratory, Batavia, IL 60510, USA}}
\newcommand{\afIIT}{\affiliation{Department of Physics, Illinois Institute of Technology, Chicago, IL 60616, USA}}
\newcommand{\afAdiyaman}{\affiliation{Adiyaman University, Faculty of Arts and Sciences, Department of Physics, 02040 Adiyaman, Turkey}}
\newcommand{\afBaruch}{\affiliation{Baruch College, City University of New York, NY, USA}}
\newcommand{\afBrandeis}{\affiliation{Brandeis University, Waltham, MA 02453, USA}}
\newcommand{\afCERN}{\affiliation{CERN, EP Department, CH-1211 Geneva 23, Switzerland}}
\newcommand{\afDAMTP}{\affiliation{DAMTP, University of Cambridge, Cambridge, CB3 0WA, UK}}
\newcommand{\afNorthwestern}{\affiliation{Department  of  Physics  \&  Astronomy,  Northwestern  University,  Evanston,  IL  60208,  USA}}
\newcommand{\afKansas}{\affiliation{Department of Physics and Astronomy,  University of Kansas, Lawrence, KS 66045, USA}}
\newcommand{\afHopkins}{\affiliation{Department of Physics and Astronomy, Johns Hopkins University, Baltimore, Maryland 21218, USA}}
\newcommand{\afMSU}{\affiliation{Department of Physics and Astronomy, Michigan State University, East Lansing, MI 48824, USA}}
\newcommand{\afUCL}{\affiliation{Department of Physics and Astronomy, University College London, London,  WC1E 6BT, UK}}
\newcommand{\afPitt}{\affiliation{Pittsburgh Particle Physics,
  Astrophysics, and Cosmology Center,\\ Department of Physics and Astronomy, University of Pittsburgh, Pittsburgh, PA 15260, USA}}
\newcommand{\afVrije}{\affiliation{Department of Physics and Astronomy, Vrije Universiteit, NL-1081 HV Amsterdam, The Netherlands}}
\newcommand{\afNIKHEF}{\affiliation{Nikhef Theory Group, Science Park 105, 1098 XG Amsterdam, The Netherlands}}
\newcommand{\afSMU}{\affiliation{Department of Physics, Southern Methodist University, Dallas, TX 75275-0175, USA}}
\newcommand{\afDESY}{\affiliation{Deutsches Elektronen-Synchrotron, DESY, Germany}}
\newcommand{\afCagliari}{\affiliation{Dipartimento di Fisica, Universit\`a di Cagliari and INFN, Sezione di Cagliari, I-09042 Monserrato (CA), Italy}}
\newcommand{\afEdinburgh}{\affiliation{Higgs Centre, University of Edinburgh, JCMB, KB, Edinburgh EH9 3JZ, Scotland, UK}}
\newcommand{\afANL}{\affiliation{High Energy Physics Division, Argonne National Laboratory, Argonne, IL 60439, USA}}
\newcommand{\afHamburg}{\affiliation{II. Institute for Theoretical Physics, Hamburg University, Luruper Chaussee 149, D-22761 Hamburg, Germany}}
\newcommand{\afUNAM}{\affiliation{Instituto de F\'isica, Universidad Nacional Aut\'onoma de M\'exico, Ciudad de M\'exico, Mexico}}
\newcommand{\afIGFAE}{\affiliation{Instituto Galego de F\'{\i}sica de Altas Enerx\'{\i}as IGFAE, Universidade de Santiago de Compostela, 15782 Santiago de Compostela, Galicia, Spain}}
\newcommand{\afIRFU}{\affiliation{IRFU, CEA, Universit\'e Paris-Saclay, F-91191 Gif-sur-Yvette, France}}
\newcommand{\afKrakow}{\affiliation{Jagiellonian University, 31-007 Krak\'ow, Poland}}
\newcommand{\afIFJKrakow}{\affiliation{Institute of Nuclear Physics, Polish Academy of Sciences, 31-342, Krak\'ow, Poland}}
\newcommand{\afKSU}{\affiliation{Kennesaw State University, Kennesaw, GA 30144, USA}}
\newcommand{\afLIP}{\affiliation{Laborat{\' o}rio de Instrumenta\c{c}{\~ a}o e F{\' \i}sica Experimental de Part{\' \i}culas (LIP), P-1649-003 Lisboa, Portugal}}
\newcommand{\afLisboa}{\affiliation{Faculdade de Ciencias, Universidade de Lisboa, 1749-016 Lisboa, Portugal}}
\newcommand{\afLPNHE}{\affiliation{LPNHE, Sorbonne Universit\'e, Universit\'e de Paris, CNRS/IN2P3, Paris, France}}
\newcommand{\afMaxPlanck}{\affiliation{Max-Planck-Institut für Physik, München, Germany}}
\newcommand{\afORNL}{\affiliation{ORNL, Physics Division, Oak Ridge, TN, USA}}
\newcommand{\afPeierls}{\affiliation{Rudolf Peierls Centre for Theoretical Physics, University of Oxford, Oxford, OX1 3PU, UK}}
\newcommand{\afSLAC}{\affiliation{SLAC National Accelerator Laboratory, Stanford University, Stanford, CA 94039, USA}}
\newcommand{\afMilano}{\affiliation{ Dipartimento di Fisica, Universit\`a di Milano and INFN, Sezione di Milano, Milano, Italy}}
\newcommand{\afWurzburg}{\affiliation{Universit\"at W\"urzburg, Institut f\"ur Theoretische Physik und Astrophysik, Emil-Hilb-Weg 22, 97074 W\"urzburg, Germany}}
\newcommand{\afLiverpool}{\affiliation{Department of Physics, University of Liverpool, UK}}
\newcommand{\afOxford}{\affiliation{Department of Physics, University of Oxford, UK}}
\newcommand{\afGlasgow}{\affiliation{School of Physics \& Astronomy, University of Glasgow, Glasgow G12 8QQ, Scotland, UK}}
\newcommand{\afOregon}{\affiliation{Department of Physics, University of Oregon, Eugene, OR 97401, USA}}
\newcommand{\afBilbao}{\affiliation{Department of Physics, University of the Basque Country UPV/EHU, 48080 Bilbao, Spain}}
\newcommand{\afIKERBASQUE}{\affiliation{IKERBASQUE, Basque Foundation for Science, 48013 Bilbao, Spain}}
\newcommand{\afSussex}{\affiliation{Department of Physics and Astronomy, University of Sussex, Brighton, UK}}
\newcommand{\afKIT}{\affiliation{KIT, Karlsruhe, Germany}}
\newcommand{\afHumboldt}{\affiliation{Humboldt-Universität zu Berlin, D-12489 Berlin,
Germany}}
\newcommand{\afPuebla}{\affiliation{Universidad de las Américas,  San Andrés Cholula 72820 Puebla, Mexico}}

\author{S.~Amoroso            }    \afDESY                                          
\author{A.~Apyan              }    \afBrandeis                                   
\author{N.~Armesto            }\la \afIGFAE                                 
\author{R.\ D.~Ball           }\la \afEdinburgh                            
\author{V.~Bertone            }\la \afIRFU                                  
\author{C.~Bissolotti         }\la \afANL                                
\author{J.~Bl\"umlein         }    \afDESY                                  
\author{R.~Boughezal          }\la \afANL                                 
\author{G.~Bozzi              }    \afCagliari                                   
\author{D.~Britzger           }\la \afMaxPlanck 

\author{A.~Buckley           }\la \afGlasgow 

\author{A.~Candido            }\la \afMilano

\author{S.~Carrazza            }\la \afMilano       
\author{F.~G.~Celiberto      }\la \afECT\afFBK\afTIFPA                       
\author{S.~Cerci              }    \afAdiyaman                                     
\author{G.~Chachamis          }    \afLIP                                                       
\author{A.\ M.~Cooper-Sarkar  }\la \afOxford                           
\author{A.~Courtoy            }\la \afUNAM                                      
\author{T.~Cridge             }\la \afUCL                                        
\author{J.~M.~Cruz-Martinez  }\la \afMilano                             
\author{F.~Giuli              }\la \afCERN                                        
\author{M.\ G.~Guzzi          }\la \afKSU                                     
\author{C.~Gwenlan            }\la \afOxford                                       
\author{L.\ A.~Harland-Lang   }\la \afPeierls                          
\author{F.~Hekhorn            }\la \afMilano                               

\author{M.~Hentschinski}\la \afPuebla

\author{T.\ J.~Hobbs          }\la \afFNAL\afIIT                              
\author{S.~Hoeche             }    \afFNAL      

\author{A.~Huss          }\la \afCERN

\author{J.~Huston             }\la \afMSU                               

\author{S.~Jadach    }    \afIFJKrakow

\author{J.~Jalilian-Marian    }    \afBaruch                               
\author{M.~Klein              }    \afLiverpool                                   
\author{G.~K.~Krintiras      }    \afKansas            

\author{H.-W. Lin      }    \afMSU

\author{C.~Loizides           }    \afORNL                                        
\author{G.~Magni              }\la \afVrije\afNIKHEF                                 
\author{B.~Malaescu           }\la \afLPNHE                                    
\author{B.~Mistlberger        }\la \afSLAC                                  
\author{S.~Moch               }\la \afHamburg                                      
\author{P.\ M.~Nadolsky       }\la \thanks{Corresponding author}\email{nadolsky@smu.edu} \afSMU  
\author{E.\ R.~Nocera         }\la \afEdinburgh                              
\author{F.~I.~Olness         }\la \afSMU                                    
\author{F.~Petriello          }\la \afNorthwestern\afANL     

\author{J.~Pires               }\la \afLIP \afLisboa  

\author{K.~Rabbertz               }\la \afKIT                             
\author{J.~Rojo               }\la \afVrije\afNIKHEF                               
\author{C.~Royon}\la \afKansas

\author{G.~Schnell             }\la \afBilbao \afIKERBASQUE 

\author{C.~Schwan             }\la \afWurzburg                                      
\author{A.~Si\'odmok            }\la \afKrakow    

\author{D.\ E.~Soper          }\la \afOregon

\author{M.~Sutton          }\la \afSussex

\author{R.\ S.~Thorne         }\la \afUCL                                    
\author{M.~Ubiali             }\la \thanks{Corresponding author}\email{M.Ubiali@damtp.cam.ac.uk} \afDAMTP              
\author{G.~Vita               }\la \afSLAC                                         
\author{J. H.~Weber               }\la \afHumboldt         

\author{J. Whitehead               }\la \afIFJKrakow         
\author{K.~Xie                }\la \afPitt
\author{C.-P.~Yuan            }    \afMSU                                          
\author{B.~Zhou               }\la \afHopkins

\maketitle
\newpage

%
\begin{flushright}
Edinburgh 2022/08, FERMILAB-PUB-22-222-QIS-SCD-T, MPP-2022-32,\\
PITT-PACC-2211, SLAC-PUB-17652, SMU-HEP-22-02, TIF-UNIMI-2022-6
\vspace{1\baselineskip}
\end{flushright}

\begin{center}
{\Large \bf Snowmass 2021 whitepaper: Proton structure at the precision frontier}
\end{center}

\vspace{2\baselineskip}

\begin{center}
{\bf ABSTRACT}
\end{center}
An overwhelming number of theoretical predictions for hadron colliders require parton distribution functions (PDFs), which are an important ingredient of theory infrastructure for the next generation of high-energy experiments. This whitepaper summarizes the status and future prospects for determination of high-precision PDFs applicable in a wide range of energies and experiments, in particular in precision tests of the Standard Model and in new physics searches at the high-luminosity Large Hadron Collider and Electron-Ion Collider. We discuss the envisioned advancements in experimental measurements, QCD theory, global analysis methodology, and computing that are necessary to bring unpolarized PDFs in the nucleon to the N2LO and N3LO accuracy in the QCD coupling strength. Special attention is given to the new tasks that emerge in the era of the precision PDF analysis, such as those focusing on the robust control of systematic factors both in experimental measurements and theoretical computations. Various synergies between experimental and theoretical studies of the hadron structure are explored, including opportunities for studying PDFs for nuclear and meson targets, PDFs with electroweak contributions or dependence on the transverse momentum, for incisive comparisons between phenomenological models for the PDFs and computations on discrete lattice, and for cross-fertilization with machine learning/AI approaches. 

\begin{center}
{\bf Submitted to the US Community Study on the Future of Particle Physics (Snowmass 2021)}
\end{center}

\vspace{2\baselineskip}

\tableofcontents
\newpage
\clearpage \newpage 
\pagenumbering{arabic}

\section{Introduction}
{\it Leading authors: M. Ubiali, P. Nadolsky}

Precision phenomenology at hadron colliders relies upon an accurate estimate of the uncertainty in Standard Model (SM) predictions. An overwhelming number of theoretical predictions for hadron colliders require parton distribution functions (PDFs) \cite{Harland-Lang:2014zoa,Dulat:2015mca,Abramowicz:2015mha,Accardi:2016qay,Alekhin:2017kpj,Ball:2017nwa,Hou:2019efy,Bailey:2019yze,Bailey:2020ooq,Ball:2021leu,ATLAS:2021vod}, the nonperturbative functions quantifying probabilities for finding quarks and gluons in hadrons in high-energy scattering processes. 

In the last decade, we witness a revolution in computing hard scattering cross sections in perturbative quantum chromodynamics (QCD) to a high accuracy. This progress is achieved by including radiative contributions up to the second and third order in the strong coupling constant (N2LO and N3LO, respectively). 
A similar progress in understanding PDFs beyond the current level is critical for realizing the physics programs of the high-energy run (Run III) of the Large Hadron Collider (LHC) and of the high-luminosity runs (HL-LHC).  Limitations in the knowledge of the PDFs constrain the accuracy of measurements of the Higgs boson couplings and electroweak parameters in key channels at the HL-LHC \cite{Cepeda:2019klc,UbialiLOI}. By knowing the PDFs for the gluon and other quark flavors to 1-2\% accuracy, one greatly reduces the total uncertainties on the Higgs couplings in gluon-gluon fusion and electroweak boson fusion. The energy reach in searches for very heavy new particles at the HL-LHC can be extended to higher masses by better knowing the PDFs at the largest momentum fractions, $x>0.1$, and by pinning down the flavor composition of the partonic sea~\cite{Brady:2011hb}. As interest grows in hadron scattering at very small partonic momentum fractions, $x < 10^{-5}$, at hadron colliders (HL-LHC, LHeC, FCC-hh) as well as in the astrophysics experiments, one must include effects of small-$x$ resummation and saturation in QCD theory and, when warranted, in the PDFs \cite{Ball:2017otu}.

PDFs contribute to precise measurements of the QCD coupling constant, heavy-quark masses, weak boson mass, and electroweak flavor-mixing parameters. This requires continuous benchmarking and improvements of the theoretical framework, particularly in the perturbative approach adopted for the computation of observables in a PDF fit~\cite{Butterworth:2015oua,Accardi:2016ndt}.
As lattice QCD techniques advance in computations of PDFs from the first principles, unpolarized phenomenological PDFs serve as important benchmarks for testing the lattice QCD methods \cite{Lin:2017snn,Constantinou:2020hdm,Constantinou:2022yye}. Namely, precisely determined phenomenological parametrizations for PDFs in the nucleon serve as a reference to validate lattice and nonperturbative QCD calculations. Methods of the precision PDF analysis are increasingly applied to explore the nuclear and meson structure, and they inspire related approaches in the studies of 3-dimensional hadron structure, including dependence on transverse momentum and spin. 

\subsection{PDF analyses as a part of HEP theory infrastructure}
\begin{wrapfigure}{r}{0.36\textwidth}
    \centering
    \includegraphics[width=0.36\textwidth]{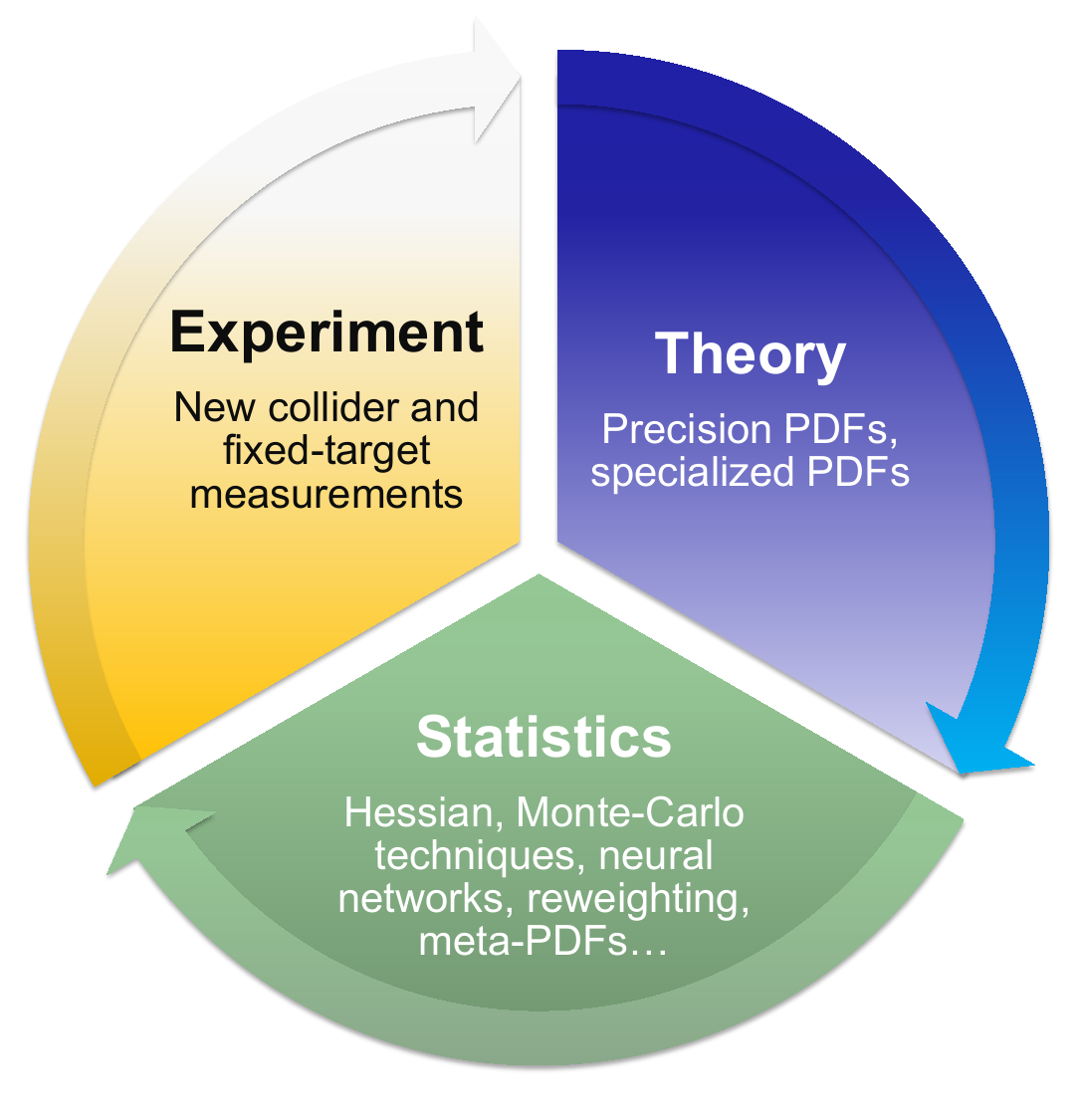}
    \caption{\label{fig:ThreeDomainsOfPDFAnalysis} Three constituent domains of the modern PDF analysis. }
\end{wrapfigure}
In this whitepaper prepared for the Snowmass'2021 planning study \cite{Snowmass2021Website} in the United States, we emphasize that the accurate determination of PDFs constitutes a critical component of theory infrastructure for current and future hadronic experiments, together with the development of Monte-Carlo event generators \cite{Campbell:2022qmc} and multi-loop calculations in QFT. Obtaining such accurate PDFs necessitates continued advancements in the areas of quantum field theory, experimental measurements, and statistical methods.

A typical computation of a cross section for a hadron-scattering process includes two parts, a hard cross section quantifying scattering rates for weakly interacting partons, and several functions quantifying probabilities for either finding partons in the initial-state hadrons or for partons fragmenting into final-state hadrons. While the hard cross section can be computed algorithmically using increasingly sophisticated perturbative techniques, the long-distance nonperturbative functions are found by other methods, most commonly using a large-scale, or global, analysis of hadronic scattering data. The PDFs are quintessential nonperturbative functions of the latter kind. They are ubiquitously used in hadron collider experiments. 

N2LO and N3LO precision of hard cross sections requires equally accurate PDFs. The PDFs generally fall into two classes, {\it general-purpose} (suitable for the majority of applications) and {\it specialized} (suitable for certain applications or obtained using special techniques) ones. The PDF determination at the modern precision level is an exciting research area that incorporates advancements in the three frontiers illustrated in Fig.~\ref{fig:ThreeDomainsOfPDFAnalysis}.
First, new experimental measurements must be performed with consistent control of accuracy at all stages of the analysis. Second, new multi-loop theoretical cross sections must be computed and implemented in an optimal form in the global PDF fit. Third, the PDFs and their uncertainties consistent with the fitted data sets must be determined in a statistically robust way and delivered in a convenient format to a wide range of users.

There is a significant overlap and cross pollination among these three research areas. Progress toward the next generation of PDFs necessary to achieve the physics goals of the planned experiments, including the HL-LHC and EIC, should therefore happen at the intersection of state-of-the-art particle experiments, quantum field theory, multivariate data science and artificial intelligence, as well as high-performance computing. This research field presents ample opportunities for training of students and postdocs, who develop mathematical and theoretical skills applicable in many areas of science and industry.

\subsection{Exploring PDFs in Snowmass community planning studies}

Already at the first Snowmass DPF Summer Studies in 1980's, theoretical issues and practical methods for determination of PDFs were in the focus of the attention of the HEP community, given the pivotal role of PDFs in predicting QCD processes. It was realized that progress in collider studies is impossible without trusted PDF parametrizations, such as the seminal Eichten-Hinchliffe-Lane-Quigg (EHLQ) PDFs \cite{Eichten:1984eu} published in the run-up to the Superconducting SuperCollider. The Snowmass community studies in the 1990's and 2000's have stimulated understanding of the hadron structure through increasingly precise experiments at FNAL, CERN, and DESY. 

The 2021 Snowmass Community Planning Exercise has drawn a large group of participants to explore multi-faceted aspects of PDFs and their applications through collaborative meetings that took place in the Energy Frontier Topical Groups (EF06, as well as EF05, EF07, and other groups) over nearly two years. Many of these aspects are reviewed in this whitepaper, although the full coverage of all involved aspects would be prohibitively extensive. Table~\ref{table:Table1} illustrates the progress that has been made since the previous Snowmass Summer Study completed in 2013. Using the 2013 Working Group Report on Quantum Chromodynamics \cite{Campbell:2013qaa} as a reference, we compare the status of select topics in the 2013 study and in the current study. The main part of the whitepaper details these and many other topics in the order listed in Sec.~\ref{sec:Organization}. We also highlight the challenges and tasks that need to be addressed to advance our understanding. The article is organized in thematic sections written by groups of leading authors to represent the rich tapestry of ideas and approaches in the PDF analysis field. The assessments in individual sections are provided by the leading authors. Quantitative estimates for projected PDF uncertainties reflect the assumptions and methodologies used by the leading authors. The material underwent gentle overall editing primarily for style and consistency of notations.

\begin{table}[t]
    \begin{tabular}{>{\raggedright}p{0.3\textwidth}>{\raggedright}p{0.35\textwidth}>{\raggedright}p{0.35\textwidth}}
\hline 
\textbf{TOPIC} & \textbf{STATUS, Snowmass'2013} & \textbf{STATUS, Snowmass'2021}\tabularnewline
\hline 
\rowcolor{lightgray}Achieved accuracy of PDFs & N2LO for evolution, DIS and vector boson production & N2LO for all key processes; N3LO for some processes\tabularnewline
PDFs with NLO EW contributions & MSTW'04 QED, NNPDF2.3 QED & LuXQED and other photon PDFs from several groups; PDFs with leptons
and massive bosons\tabularnewline
\rowcolor{lightgray} PDFs with resummations & Small x (in progress) & Small-x and threshold resummations implemented in several PDF sets\tabularnewline
Available LHC processes to determine nucleon PDFs & $W/Z$, single-incl. jet, high-$p_{T}$ $Z,$ $t\overline{t}$, $W+c$
production at 7 and 8 TeV & + $t\overline{t}$, single-top, dijet, $\gamma/W/Z+$jet, low-Q Drell
Yan pairs, \ldots{} at 7, 8, 13 TeV\tabularnewline
\rowcolor{lightgray} Current, planned \& proposed experiments to probe PDFs & LHC Run-2\\DIS: LHeC & LHC Run-3, HL-LHC\\DIS: EIC, LHeC, MuIC, \ldots\tabularnewline
Benchmarking of PDFs for the LHC & PDF4LHC'2015 recommendation in preparation & PDF4LHC'21 recommendation issued\tabularnewline
\rowcolor{lightgray} Precision analysis of specialized PDFs &  & Transverse-momentum dependent PDFs, nuclear, meson PDFs \tabularnewline
\hline 
\hline 
\multicolumn{3}{c}{\vspace{6pt} \textbf{NEW TASKS in the HL-LHC ERA} } \tabularnewline
\rowcolor{lightgray} Obtain complete N2LO and N3LO predictions for PDF-sensitive processes & Improve models for correlated systematic errors & Find ways to constrain large-x PDFs without relying on nuclear targets\tabularnewline
Develop and benchmark fast N2LO interfaces & Estimate N2LO/N3LO theory uncertainties & New methods to combine PDF ensembles, estimate PDF uncertainties,
deliver PDFs for applications\tabularnewline
\hline 
\end{tabular}
    \caption{Top part: Some of the PDF-focused topics explored in Snowmass’2013 \cite{Campbell:2013qaa} and '2021 studies. Bottom part: a selection of new critical tasks for the development of a new generation of PDFs that achieve the objectives of the physics program at the high-luminosity LHC.}
    \label{table:Table1}
\end{table}

\subsection{New frontiers in PDF analyses in the HL-LHC era}
The bottom part of Table~\ref{table:Table1} summarizes some of the new tasks for the PDF analysis that emerge in the era of precision QCD.
Several ingredients of the global fits are  essential for robust modeling of the proton structure. Solutions for PDFs must reflect all allowed variations associated with statistical, systematical errors in the experiments, as well as with relevant error correlations. 
Needless to say, the most precise N2LO or even N3LO theoretical cross sections should be preferably used, when possible, as a prerequisite for achieving the highest accuracy. However, accuracy of the theoretical predictions used in the fit also depends on the other factors and must be properly estimated.  At the same time, given the complexity of N2LO/N3LO calculations, their fast approximate implementations (such as {\it fast NNLO} interfaces) must be developed for the PDF analyses. Control of experimental and theoretical uncertainties requires, in particular, to either fit the experiments that are minimally affected by the unknown factors (for example, to include cross sections only on proton, rather than on nuclear targets to minimize the associated uncertainties), or to estimate the associated uncertainty of these unknown factors in the fit.  
Finally, the published range of solutions for the PDFs must account for acceptable variations in methodology, which encompasses such components as the functional forms adopted to parametrize PDFs at some initial energy scale, propagation of experimental uncertainties into the error associated with the fitted functions, the diverse statistical inference techniques, as well as implementation of physical constraints on the PDFs, such as QCD sum rules, positivity of physical observables, and integrability of relevant PDF combinations. Methodological advancements should also include development of practical standards for the delivery of PDFs to a wide range of users. The format of the PDF delivery must optimize for accuracy, versatility, and speed across a broad range of applications -- a non-trivial task, given the ubiquity of the PDF uses by both experimentalists and theorists. The PDF4LHC working group \cite{PDF4LHCWG} leads the development of such standards and delivery formats for the LHC community. In particular, the recently released 2021 recommendation of the PDF4LHC working group (PDF4LHC21 \cite{Ball:2022hsh}) supersedes the previous recommendation issued in 2015 \cite{Butterworth:2015oua}. The PDF4LHC21 recommendation document stipulates guidelines for applications of PDFs and computation of PDF uncertainties at the LHC. With this document, the PDF4LHC working group also distributes combined N2LO PDF4LHC21 error sets (available in the \lhapdf library \cite{LHAPDFWebsite}) to streamline computations with PDFs across typical LHC studies, such as searches for new physics or theoretical simulations. However, comparisons to individual PDF ensembles from the groups, rather than combined ones, remain necessary in the most precise measurements, such as tests of electroweak precision symmetry breaking and Higgs boson physics. 

The rest of the whitepaper discusses all these critical tasks in more detail. We wish to highlight some of the pertinent issues here. 

Recent PDF analyses indicate that the LHC data are increasingly crucial in pinning down the parton densities, and its constraining power will become even more crucial in the HL-LHC run~\cite{AbdulKhalek:2018rok}. At the same time, new experiments on the deeply-inelastic scattering (DIS), in particular, at the Electron-Ion Collider planned at BNL in the USA, may be at least as instrumental as the LHC, and in some important cases more instrumental, in constraining the relevant PDF combinations \cite{AbdulKhalek:2021gbh}. Even more precise measurements of the PDFs in DIS may be obtained at the Large Hadron-Electron Collider (LHeC \cite{LHeC:2020van}) and Muon-Ion Collider (MuIC \cite{Acosta:2022ejc}).  

To elevate the accuracy of PDFs in the next decade, new experiments and theory calculations must implement consistent error control at all stages, from experimental measurements to the distribution of final PDFs. In particular, while there is a reasonable overall agreement between the various experiments in terms of their preferences for the PDFs \cite{Ball:2017nwa,Hou:2019efy,Bailey:2020ooq,Ball:2021leu}, detailed testing with several methods reveals some disagreements (tensions) among the most precise experiments. The strength of these disagreements is about the same in the three recent global fits.\footnote{For example, the $\chi^2/N_{pt}$ values for the LHC data sets and for all data sets tend to be elevated, as compared to the statistics expectation, in a similar fashion in the CT18, MSHT'20, and NNPDF3.1.1 N2LO global fits, cf. Tables 2.1-2.3 in \cite{Ball:2022hsh}.} The disagreements are sometimes stronger than would be expected simply based on random differences between theory and data~\cite{Kovarik:2019xvh,Hou:2019efy,ATLAS:2021vod}. 
Furthermore, the 2021  comparisons of PDF sets by the PDF4LHC working group \cite{Ball:2022hsh} suggest that the differences among the global PDF ensembles have increased in some cases compared to the PDF4LHC15 combination ~\cite{Butterworth:2015oua}. The fitting groups regularly perform thorough benchmark studies~\cite{Ball:2022hsh} to understand the underlying reasons.  In the course of such exercises, the various groups observe good agreement among their theoretical predictions for the most critical data sets when using the same PDF ensemble as the input. At the same time, when fitting the same data using freely varied PDF parametrizations, the groups arrive at mutually consistent, yet not exactly identical best-fit PDF parametrizations and especially PDF uncertainties. These exercises rule out ``trivial" causes for the disagreements among the groups/data sets, such as an incorrect theoretical calculation or improper implementation of an experimental data set. Rather, sometimes the methodological differences, for example due to the fitting techniques, treatment of systematic uncertainties in the data sets, PDF functional forms, or the definition of the PDF errors, can be as large or even larger than the PDF uncertainties from the propagation of experimental errors. Increasing the precision of future PDFs must address such issues. Section~\ref{sec:benchmarking} summarizes the ongoing efforts in this direction. 

Assuming that the possible tensions among the fitted data can be eliminated, as otherwise they may weaken the current HL-LHC projections, there is a hope to arrive at a situation in which, after years of trying to reduce PDF uncertainties, the parton luminosity uncertainties goes down to about 1\% in the central rapidity region and for QCD scales around the $Z$ pole. Nominal uncertainties may go down to 0.3-0.5\% within a decade, provided we obtain consistent constraints from the near-future experiments. Can we really trust PDFs to that level of precision? 

In such a situation, the precision versus accuracy challenge becomes crucial. In some cases, when a new PDF analysis including new data is released by a PDF-fitting collaboration, shifts from the previous to the new PDF set may be larger than the nominal PDF uncertainties. This does not undermine the accuracy of a PDF determination {\it per se}, as long as the origin of such shifts has been identified, and all aspects of the fit are kept under control. In the other cases, the uncertainties provided by the group may already include an estimated contribution from such behind-the-scene factors. This is to say that the span of the uncertainties may vary among the different PDF sets depending on how such situations are handled. 

As far as experimental data are concerned, one of the key challenges has to do with the data sets which, as the luminosity increases, are more and more dominated by correlated systematics. These highly-correlated data sets may destabilize convergence of the fits if small changes in the data covariance matrix lead to dramatic changes in the fully correlated $\chi^2$ to the data. Studies of covariance matrix stabilisation and of the effects of decorrelating the systematics are ongoing and will become increasingly vital. They require a strong synergy between theorists and experimentalists. See Sect.~\ref{sec:MethodologyDeliveryCorrelations} for a detailed discussion. 

As far as the fitting methodology is concerned, several aspects are at play. With the traditional fitting technique based on the minimization of the log-likelihood $\chi^2$, the functional forms of the assumed PDF parametrizations are an important factor that must be carefully handled. The PDF parametrization must be flexible "just enough" to obtain good description of the data without overfitting. Significant progress has been made since the 2013 Snowmass study to understand the dependence of PDFs and their uncertainties on the parametrization form. Some examples of this progress include a more flexible parametrization introduced in the MSHT'20 study \cite{Bailey:2020ooq}, which in particular results in a change in the down quark valence PDFs compared to the previous fits; a cross-validation test proposed to determine the optimal number of parameters for a given PDF parametrization form \cite{Kovarik:2019xvh}, similarly in its spirit to the cross-validation condition that prevents over-training of neural networks; a study of 250+ alternative functional forms for the PDFs to determine the component of the PDF uncertainty due to the parametrization in the CT18 analysis \cite{Hou:2019efy}.

Various components of the fitting methodologies undergo continuous improvements and are subjected to increasingly incisive tests. Statistical closure tests~\cite{DelDebbio:2021whr} may become crucial for the modern PDF sets -- they are already used to test the robustness of the {\tt NNPDF} sets since {\tt NNPDF3.0}~\cite{Ball:2014uwa}.   
The idea of a closure test is that the PDFs determined from pseudodata generated from a known underlying law must correctly reproduce the statistical distributions expected on the basis of the assumed experimental uncertainties. While the closure test validates the performance of the fitting methodology  with the idealized pseudodata, different kinds of tests have been developed for validation with the real-life data sets that are not perfectly consistent. This is the idea behind the strong goodness-of-fit tests that were developed in~\cite{Kovarik:2019xvh} and applied in the CT18 analysis~\cite{Hou:2019efy}. The strong goodness-of-fit criteria demand internal consistency of the probability distribution in a global fit, in addition to requiring an excellent $\chi^2$ describing the overall quality of the fit. 

Another crucial element for the future progress is the theory framework, comprising both the implementation of new calculations in the PDF fits and estimations of theoretical uncertainties on the PDFs. While the theory error introduced by truncation of the perturbative QCD series was believed to be generally less important than the experimental uncertainties, it becomes significant at the present level of precision and must be taken into account. The effort towards the determination of theory uncertainties in fixed-order PDF fits (discussed in Sect.~\ref{sec:thunc}) and the multi-pronged work towards N3LO PDFs (discussed in Sect.~\ref{sec:n3lo}) will be paramount in the next few years. 

Other sources of uncertainties will become crucial in the future. For
example, as even more high-energy data from the LHC are included in PDF fits, the tails of the distributions that are used in PDF determination are potentially affected by new physics effects. To make sure that new physics is not absorbed or “fitted away” in the PDFs, one would either have to exclude these data, thus losing potentially important constraints, or carefully disentangle the Standard Model and new physics effects. More details are provided in Sect.\ref{sec:smeft}. 

If the advancements along the described directions are realized, the HL-LHC projections~\cite{Khalek:2018mdn} will be extremely encouraging, with a foreseen reduction of PDF uncertainties by a factor of 2 to 3.  Given the scope of the outstanding questions, 
this progress will require a broad effort from the HEP community to maintain elevated standards at all stages of the experimental measurements, theoretical computations, and global PDF fits themselves. Accomplishing this goal depends on a dedicated collaboration among experimentalists and theorists.
Clearly the precision physics frontier opens up new fascinating challenges also for the exploration of hadron structure.  

\subsection{Organization of the whitepaper \label{sec:Organization}}
Section~\ref{sec:PDFsAndApplications} compares the latest PDF parametrizations and partonic luminosities from various groups. It also discusses predictions for benchmark LHC measurements and applications of PDFs in studies of electroweak symmetry breaking, searches for new physics, and combined fits of the parameters in the Standard Model and its effective field theory (EFT) extensions. 

Section~\ref{sec:Experiments} summarizes some of the main applications of PDFs in the experimental analyses. It reviews promising scattering processes at the LHC that can provide further constraints on the PDFs. Then, Section~\ref{sec:EICPDF} reviews prospects for obtaining incisive constraints on the unpolarized, spin-dependent, and nuclear PDFs at the planned Electron-Ion Collider at BNL. The potential for determination of PDFs at the Large Hadron-Electron Collider is explored in Sec.~\ref{sec:LHeC}. Sections~\ref{sec:PDFNeutrinoPheno} and \ref{sec:ForwardPDF} explore connections of the PDFs with the neutrino-scattering and forward-scattering experiments. 

Section~\ref{sec:Theory}, dedicated to theoretical aspects of the PDF analyses, begins with a review of the progress toward achieving PDF evolution and computing hard cross sections at N3LO accuracy in Sections~\ref{sec:n3lo} and \ref{sec:N3LOXsec}, followed by a discussion of electroweak radiative contributions for the PDF fits in Section~\ref{sec:EW}. The role of all-order resummations at very small and very large partonic momentum fractions is addressed in Sec.~\ref{sec:Resummations}. It ends with a list of theoretical developments needed beyond fixed-order QCD and EW and with a discussion on the factorization schemes needed for event generators. 

Section~\ref{sec:Methodology} addresses methodological aspects of global fits, starting with the pivotal role of the models for experimental systematic uncertainties for the future PDF fits, and proceeding to the various approaches for the estimate of theoretical uncertainties on the PDFs, machine learning applications in the context of PDF determinations, delivery of PDFs, and the combination of PDF uncertainties without or with data-driven correlations.

Section~\ref{sec:Lattice} presents an overview of the calculations of the QCD coupling strength and PDFs on the lattice -- the rapidly growing field that holds the promise to predict the hadron structure, including the spin-independent and other types of the PDFs, starting from the first QCD principles. This is followed by a summary of prospects for determination of nuclear and meson PDFs in Section~\ref{sec:nuc_PDFs}, and then by an overview of the planned studies of transverse-momentum dependent PDFs in Section~\ref{sec:TMDs}.

Numerical computations constitute the essential part of the PDF analyses. Section~\ref{sec:computing} reviews publicly available computer programs and resources to perform PDF fits and use PDFs in HEP applications. In this section, we discuss the \lhapdf library providing PDF parametrizations, the \xfitter and \texttt{NNPDF}  open-source codes for global fits, as well as \appl, \texttt{Fast(N)NLO}, and \pineappl interfaces for fast computations of QCD and EW radiative contributions. 

Section~\ref{sec:benchmarking} summarizes recent studies by the PDF4LHC working group to benchmark and combine PDF ensembles for LHC applications. This section also reviews the latest recommendation \cite{Ball:2022hsh} from the PDF4LHC working group on using the N2LO PDFs in various LHC contests.

Conclusions for the whitepaper are provided in Section~\ref{sec:Conclusions}.
\section{Modern PDFs and their applications
\label{sec:PDFsAndApplications}}

In this section we start by comparing the most recent PDF determinations presented by several PDF fitting collaborations. We then turn on discussing modern applications of PDFs, particularly focussing on the role of PDFs in Higgs physics, searches for physics beyond the Standard Model (BSM), and global analyses of Standard Model and Effective Field Theory (SMEFT) parameters. 

\subsection{Comparisons of PDFs}

{\it Leading authors: T. Cridge, F. Giuli, J. Huston, M. Ubiali, A. M. Cooper-Sarkar, K. Xie \\ \vspace{6pt}} 

In this section, we compare the recent N2LO PDF sets: NNPDF4.0~\cite{Ball:2021leu}, CT18~\cite{Hou:2019efy}, MSHT20~\cite{Bailey:2020ooq}, the ABMP16 set with $\alpha_s(M_Z)=0.118$~\cite{Alekhin:2017kpj} and the ATLASpdf21 set \cite{ATLAS:2021vod}.

\begin{figure}[htb]
    \centering
    \includegraphics[width=0.95\textwidth]{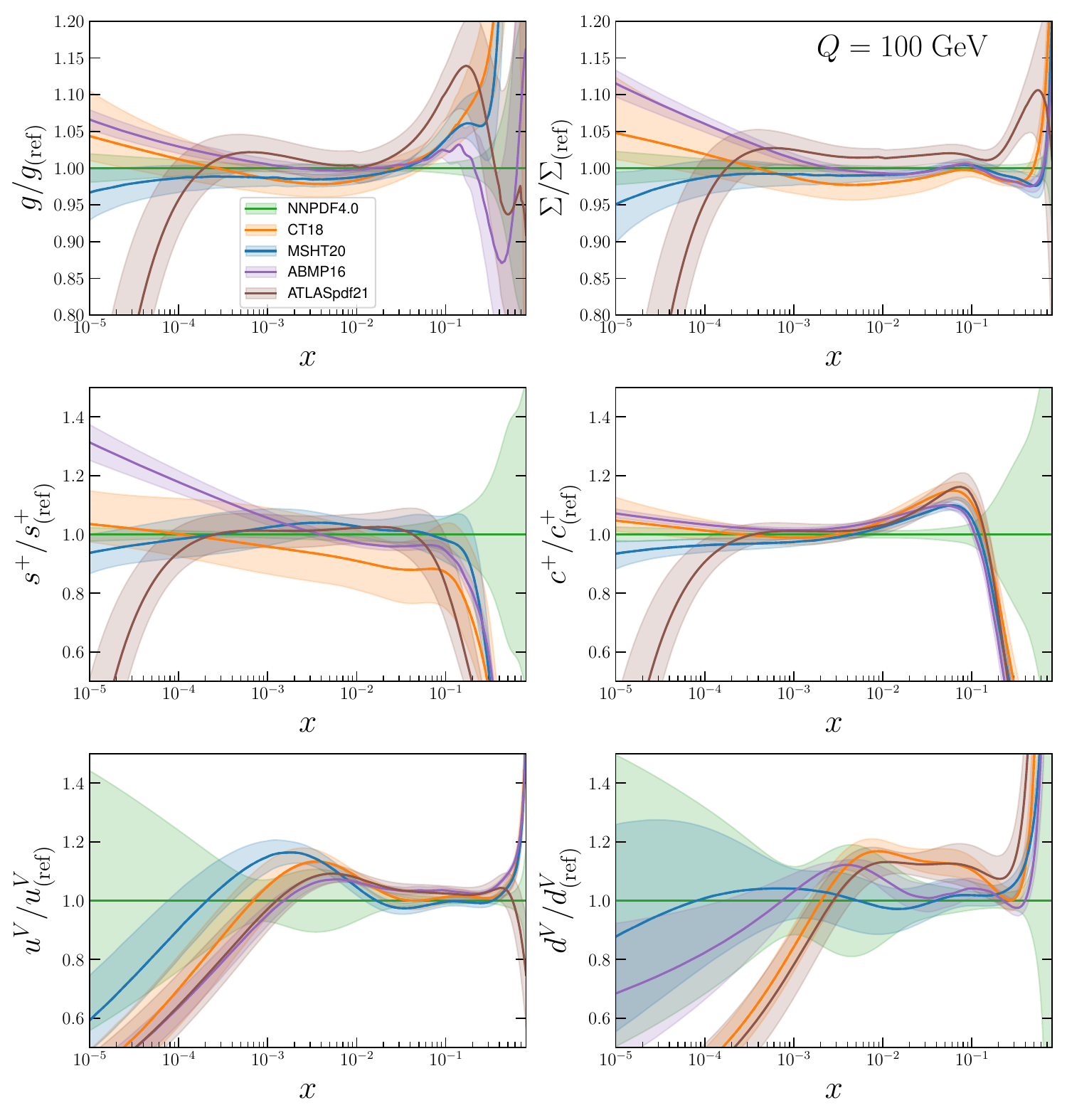}
    \caption{\label{fig:pdfratios} Comparison of the PDFs at $Q=100$ GeV. The PDFs shown are the N2LO sets of NNPDF4.0, CT18, MSHT20, ABMP16 with $\alpha_s(M_Z)=0.118$, and ATLASpdf21. The comparisons are plotted as ratios to the NNPDF4.0 central value  for the gluon $g$, singlet $\Sigma$, total strangeness $s^+=s+\bar{s}$, total charm $c^+ = c+\bar{c}$, up valence $u^V$, and down valence $d^V$ PDFs. The bands indicate the respective 1$\sigma$ uncertainty bands.}
\end{figure}

\begin{figure}[htb]
    \centering
    \includegraphics[width=0.95\textwidth]{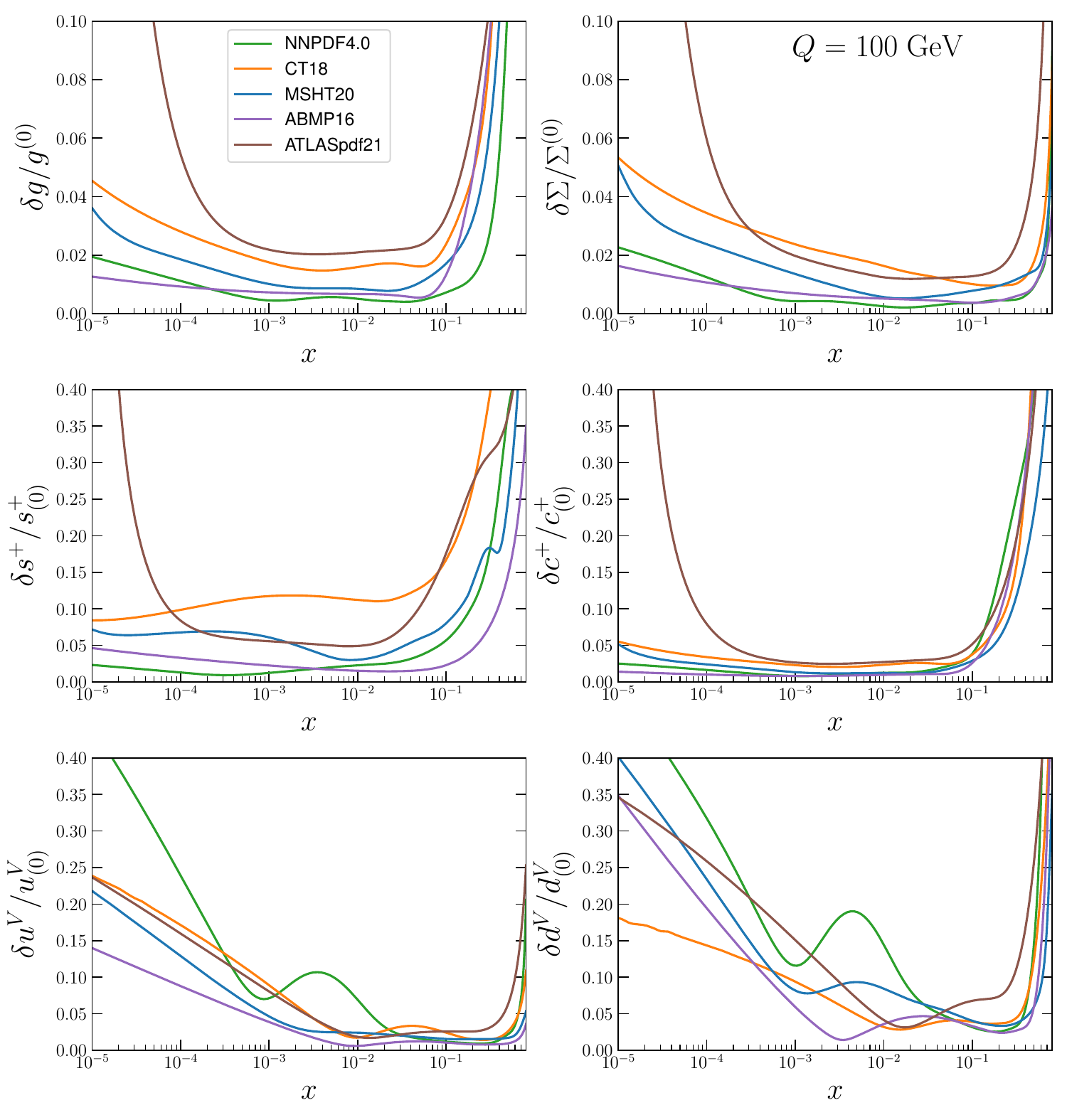}
    \caption{\label{fig:pdferrors} Comparison of the symmetrized $1\sigma$ PDF uncertainties at $Q=100$ GeV for the gluon $g$, singlet $\Sigma$, total strangeness $s^+=s+\bar{s}$, total charm $c^+ = c+\bar{c}$, up valence $u^V$ and down valence $d^V$ PDFs. The PDF sets shown are the N2LO sets of NNPDF4.0, CT18, MSHT20, ABMP16 with $\alpha_s(M_Z)=0.118$ and ATLASpdf21. }
\end{figure}
We start with a comparison at the level of the PDFs themselves in Fig.~\ref{fig:pdfratios}, before looking at parton luminosities and phenomenological predictions.  
Starting with the gluon PDF, we see general agreement between the different groups over the range $10^{-4} \lesssim x \lesssim 10^{-1}$ within $1\sigma$ uncertainties. 
The differences at higher $x$ reflect reflect different selections and treatments of high-$x$ collider data, also influencing the gluon-gluon luminosity at the LHC at high final-state invariant masses.  The singlet $\Sigma$, representing the sum of all the quarks and antiquarks in the proton (up to 5 flavors), is in better agreement, mostly within $1\sigma$ between all the PDF groups until very low $x$, where ABMP16 starts deviating at around $x \sim 10^{-3}$. ATLASpdf21 PDFs are in agreement with the three global fits within $2\sigma$ except at very low and very high $x$. The ATLASpdf21 analysis applies a cut of $Q^2 > 10$ GeV$^2$ on HERA data because of doubts about the adequacy of N2LO DGLAP to describe the HERA data at low $x$ and $Q$ below this cut. These PDFs are designed for use at higher $x$, $x \gtrsim 10 ^{-4}$. Indeed, the deviation of the ATLASpdf21 from the others at the lowest $x$ values demarcates the region where low-$x$ physics effects may need to be considered. 

The total strangeness has undergone notable changes in some sets over the past few years upon the inclusion of new LHC precision Drell-Yan data. The strangeness PDFs from MSHT20 and NNPDF4.0 are in excellent mutual agreement until (very) high $x$ and are raised relatively to CT18 as a result of inclusion of the ATLAS 7~TeV $W,Z$ data \cite{ATLAS:2011qdp}, known to raise the strangeness in the intermediate-to-high $x$ region. The CT group does not include these data in the default CT18 fit, given too high $\chi^2$ of this data set in all recent fits, but these data are included in the complementary CT18Z and CT18A, which have increased strangeness PDF compared to CT18. ABMP16 and ATLASpdf21 show differences at high and low $x$ but also agree in the intermediate $x$ range. ATLASpdf21, in particular, agrees with NNPDF4.0 and MSHT20 very well from $x \sim 10^{-4}$ to $x \sim 0.1$, again reflecting the influence from the high-precision ATLAS $W,Z$ data from LHC Run I.

Meanwhile, the charm PDF shows substantial differences between NNPDF4.0 and the other groups, with the latter largely in agreement with one another. In contrast to the default fits by the other groups, NNPDF4.0 introduces a fitted charm parametrization at the initial scale of DGLAP evolution \cite{Ball:2016neh}, in addition to the perturbative charm treatment that generates the charm PDF purely from perturbative gluon radiation. QCD theory allows introduction of a nonzero initial condition for the evolution of the charm PDF due to nonperturbative (intrinsic) production \cite{Brodsky:1980pb, Brodsky:1981se, Brodsky:2015fna, Blumlein:2015qcn}. In dedicated PDF fits, the fitted charm PDF at the initial evolution scale carries between 0 and 1 percent of the proton's momentum, depending on the fit's settings \cite{Jimenez-Delgado:2014zga,Hou:2017khm, Ball:2022qks}. In NNPDF4.0, this contribution raises the charm at high $x$, as seen in the figure. 

Finally, we compare the valence PDFs. The up valence PDFs show reasonable agreement from $x\sim 0.5$ down to $x \lesssim 10^{-2}$, below which MSHT20 and CT18 (and to a lesser extent ABMP16 and ATLASpdf21) prefer a different shape relatively to NNPDF4.0. At low $x$, constraints on the vanishing valence PDFs remain weak, with the shape differences at low $x$ possibly imposed through the valence sum rules. The down valence shows good consistency among NNPDF4.0, MSHT20 and ABMP16, while CT18 and ATLASpdf21 prefer a substantially distinct shape. For CT18 this is related in part to their lower strangeness PDF in the $10^{-2} \lesssim x \lesssim 10^{-1}$ region, with an increased down PDF compensating to some extent. 

In the comparison of the PDF uncertainties shown in Fig.~\ref{fig:pdferrors}, some appreciable differences in the uncertainty bands reflect a variety of factors, including methodologies and the selections of fitted data sets. For example, ATLASpdf21 PDFs generally have larger uncertainties at low and high $x$ as a result of their reduced data sets relatively to those of the global fitting groups. ABMP16 may have smaller error bands as a result of their application of the $\Delta\chi^2=1$ criterion to define the $1\sigma$ error bands.  These and other sources of the differences and their impact on the LHC phenomenology have been thoroughly investigated in the context of the PDF4LHC21 benchmark studies~\cite{Cridge:2021qjj,Ball:2022hsh}. They are further discussed in Sec.~\ref{sec:benchmarking} as well as in \cite{Ball:2022hsh,Kovarik:2019xvh}. 

\begin{figure}[htb]
\centering
\includegraphics[width=0.49\textwidth]{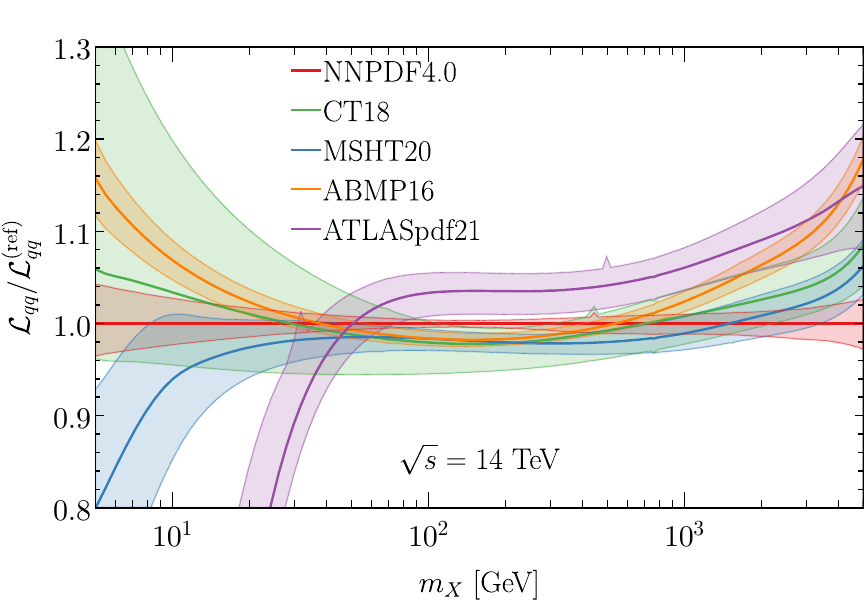}
\includegraphics[width=0.49\textwidth]{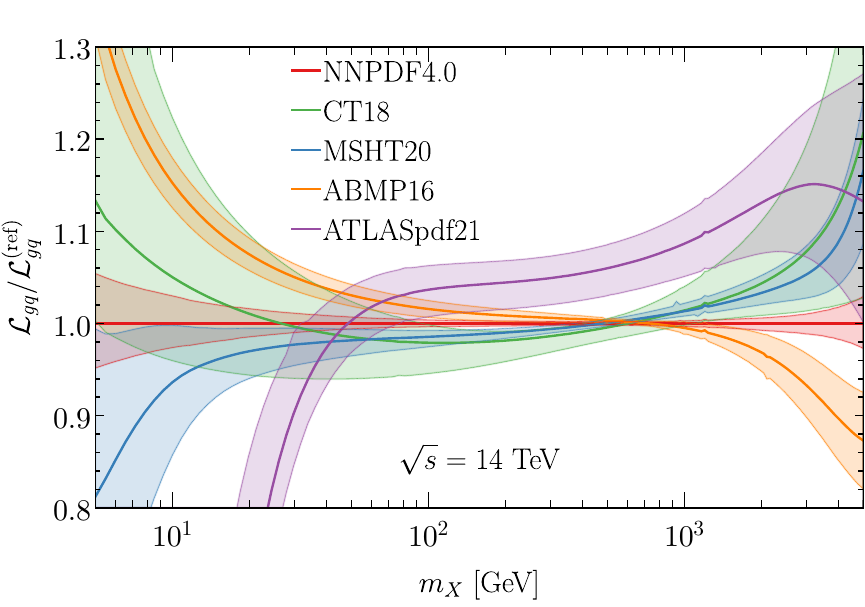}
\includegraphics[width=0.49\textwidth]{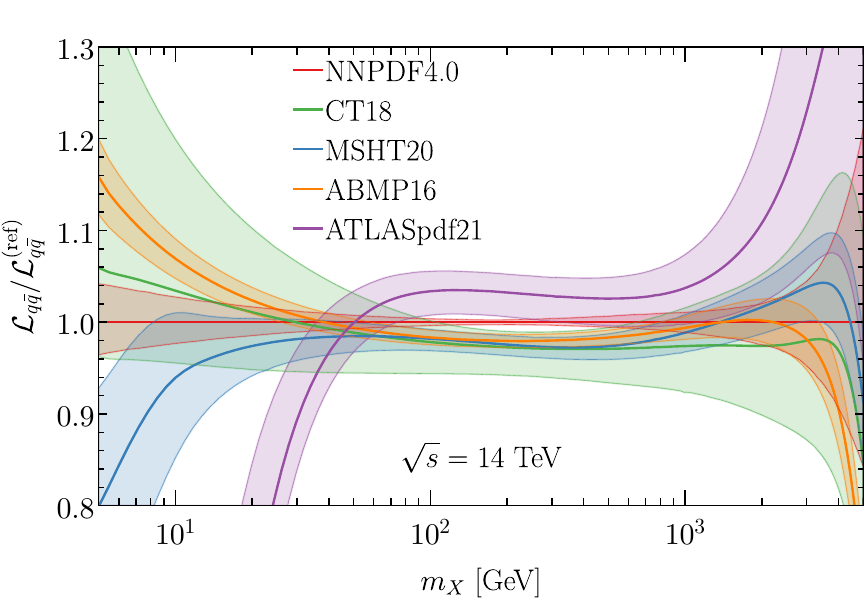}
\includegraphics[width=0.49\textwidth]{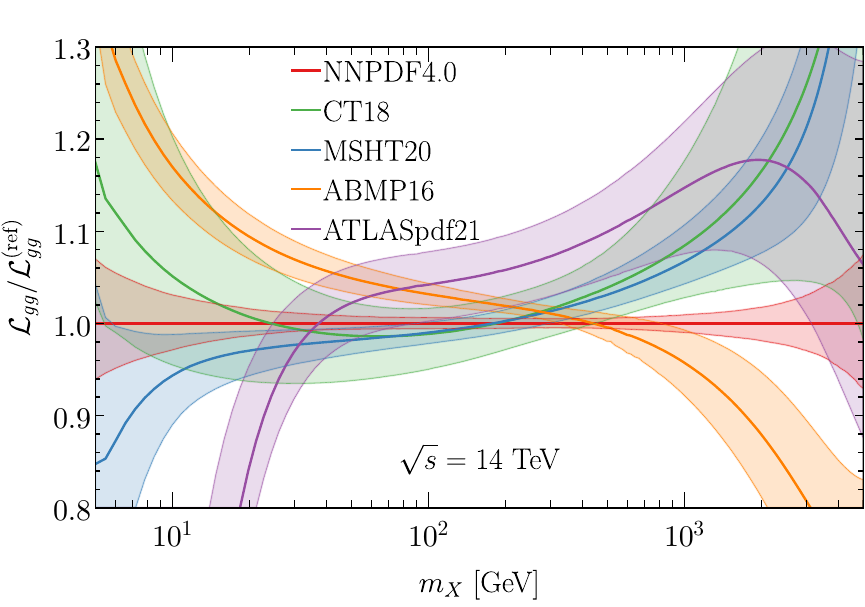}
    \caption{\label{fig:lumi} Comparison, as a function of the invariant mass $m_X$, of the parton luminosities at $\sqrt{s}=14$ TeV, computed using N2LO NNPDF4.0, CT18, MSHT20, ABMP16 with $\alpha_s(M_Z)=0.118$, and ATLASpdf21. 
    The ratio to the NNPDF4.0 central value and the relative 1$\sigma$ uncertainty are shown for each parton combination. 
    }
\end{figure}

While the above differences in the central PDFs and uncertainties have implications for precision measurements, there is an overall agreement among the PDF sets in the $x,Q$ regions with strong data constraints, which also leads to the general accord in the parton luminosities.
In Fig.~\ref{fig:lumi} we compare, as a function of the invariant mass $m_X$, the N2LO parton luminosities at $\sqrt{s}=14$ TeV.  The parton luminosities are defined as~\cite{Campbell:2016lzl}
\begin{equation}
   {\mathcal L}_{ij}(m_X^2,\mu_F^2) = \frac{1}{1+\delta_{ij}}\frac{1}{s}\int_{\tau}^{1}\frac{dx}{x}\left[f_i(x,\mu_F^2)f_j(\tau/x,\mu_F^2)
   +(i\leftrightarrow j)\right] 
\end{equation}
where $\tau=m_X^2/s$, and the factorization scale is chosen as $\mu_F^2=m_X^2$. We have summed over quark flavors $i$ in combinations
\begin{equation}
 \mathcal{L}_{q\bar{q}}=\sum_{i}\mathcal{L}_{q_{i}\bar{q}_i}, ~
 \mathcal{L}_{qq}=\sum_{i}(\mathcal{L}_{q_{i}q_{i}}+\mathcal{L}_{\bar{q}_{i}\bar{q}_{i}}), ~
  \mathcal{L}_{gq}=\sum_{i}(\mathcal{L}_{gq_{i}}+\mathcal{L}_{g\bar{q}_{i}}).
\end{equation}
 For each parton combination, we show the ratio of the central value and the $1\sigma$ uncertainty to the NNPDF4.0 central value.
All luminosities agree within uncertainties in the region around $m_X\sim$ 100 GeV, relevant e.g. for electroweak boson production. 
The ATLASpdf21 luminosities differ at low scale, $m_X \lesssim 40~$GeV, because of the cut on the low-$x,Q^2$ HERA data, as already remarked. 
The quark-quark and quark-antiquark luminosities are otherwise in reasonable agreement --- within $2\sigma$ --- over the most relevant mass range. 
%
For the luminosities of the gluon sector in the high-mass region, $m_X\sim $ 1 TeV, however, the gluon-gluon and gluon-quark luminosities for NNPDF4.0 are rather smaller than MSHT20, CT18
and ATLASpdf21,
while they are larger than ABMP16. These differences are possibly a consequence of both methodology and differences in data included. For instance, NNPDF4.0 include some data that are sensitive to the high-$x$ gluon and are not used by the other groups, such as the dijet cross sections at 7 TeV and the $t\bar{t}$ differential distributions from the LHC Run II. Other differences in data inclusion and treatment are discussed in Sect.~\ref{sec:benchmarking}. 

As for the luminosity uncertainties, NNPDF4.0 generally displays the smallest uncertainty among the groups, with some exceptions, and with the ABMP16 uncertainty being smaller in some cases, such as the gluon-gluon luminosity for a low invariant mass. These reflect the PDF uncertainties seen in Fig.~\ref{fig:pdferrors}, where this general pattern also exists.
%

\begin{figure}[t]
\centering
\includegraphics[width=0.45\textwidth]{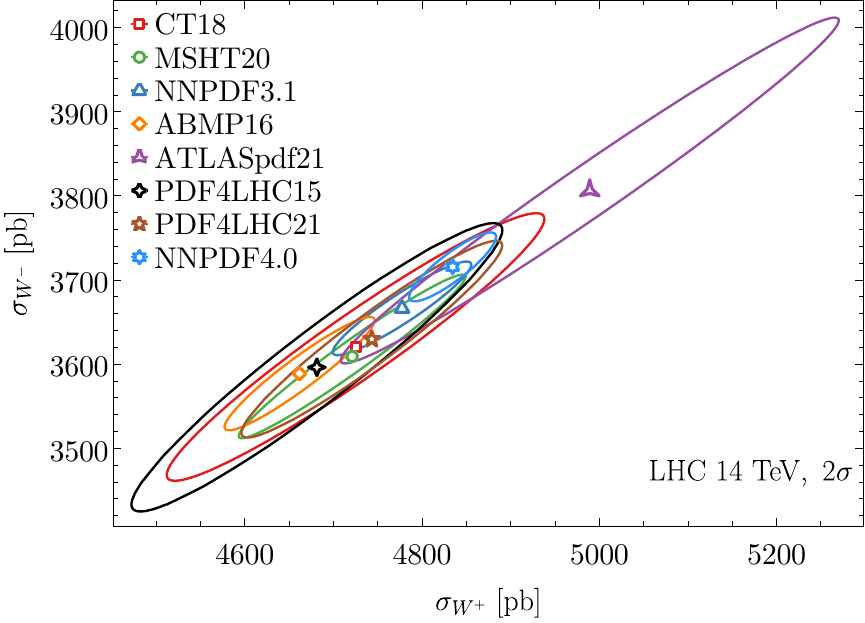}
\includegraphics[width=0.45\textwidth]{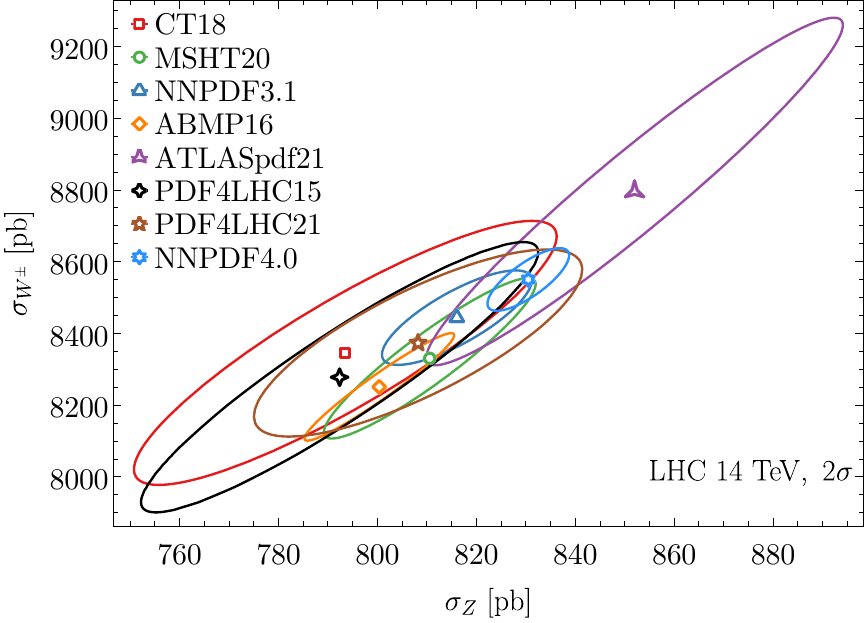}
\includegraphics[width=0.45\textwidth]{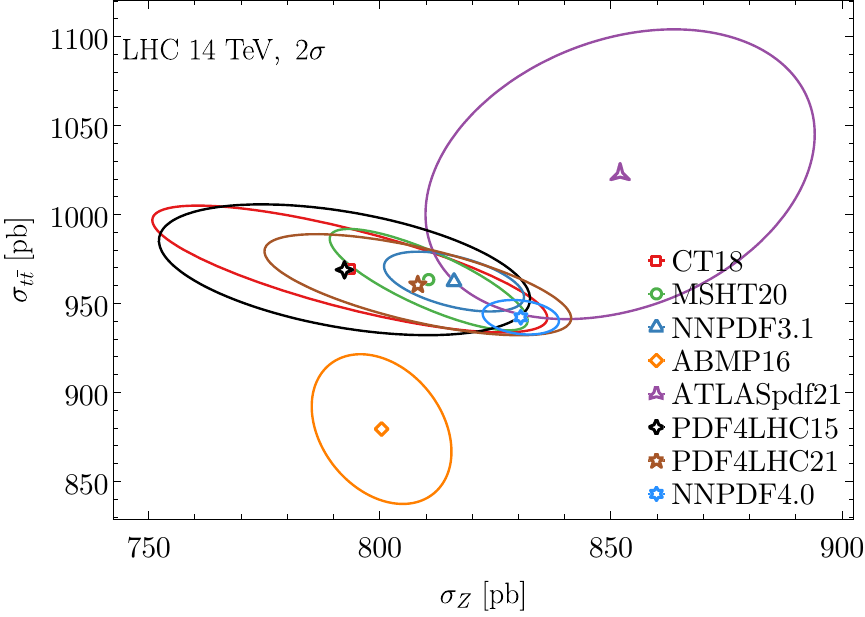}
\includegraphics[width=0.45\textwidth]{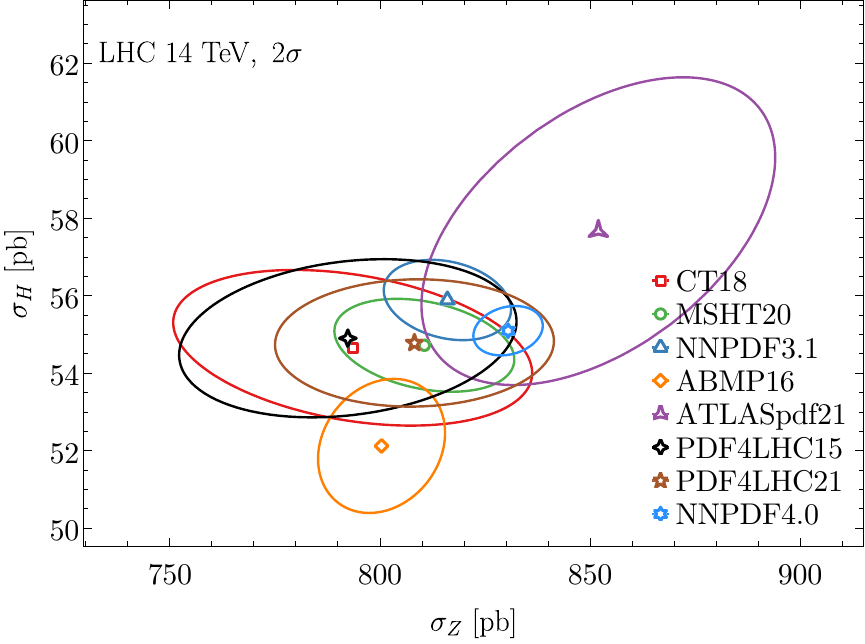}
\includegraphics[width=0.45\textwidth]{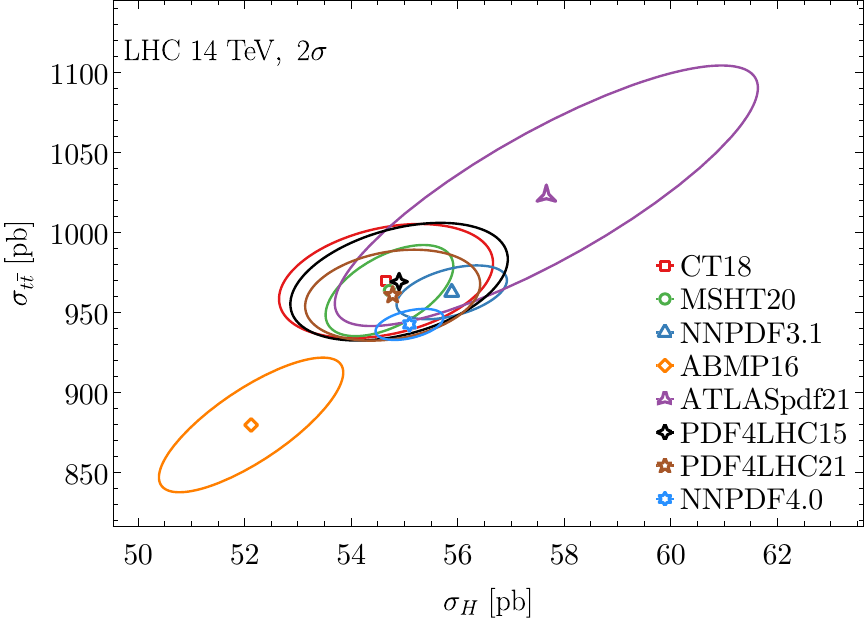}
\includegraphics[width=0.45\textwidth]{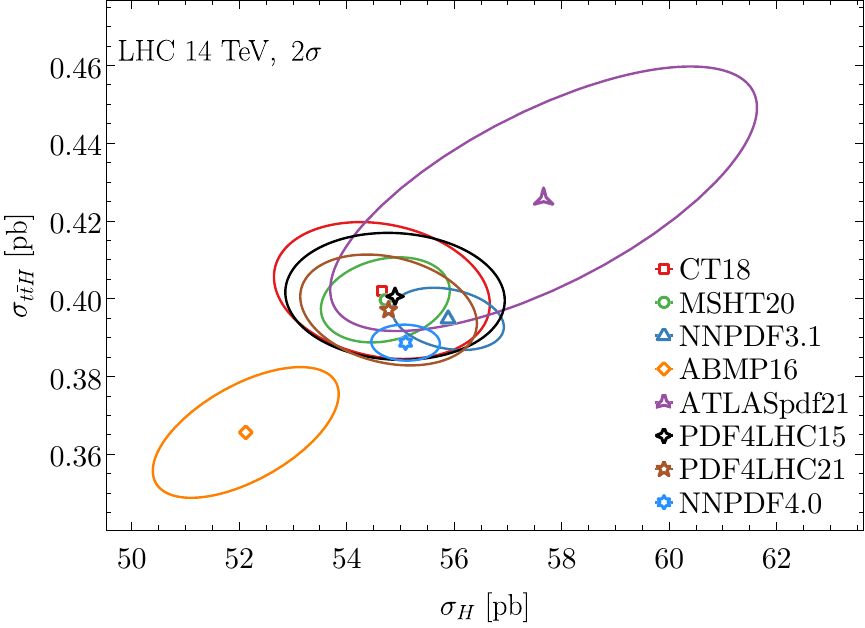}
\caption{Comparison between theoretical predictions 
for the $2\sigma$ correlation ellipses for pairs of inclusive cross sections among the $W^\pm,\ Z,\ t\bar{t},\ H,\ t\bar{t}H$ production processes at the LHC 14 TeV, comparing the predictions
based on PDF4LHC21~\cite{Ball:2022hsh} with those from the previous combination PDF4LHC15~\cite{Butterworth:2015oua} and the individual
NNPDF4.0, CT18, MSHT20, ABMP16 with $\alpha_s(M_Z)=0.118$, and ATLASpdf21 releases. 
\label{fig:LHCpheno}}
\end{figure}

\begin{figure}[tb]
\centering
\includegraphics[width=0.45\textwidth]{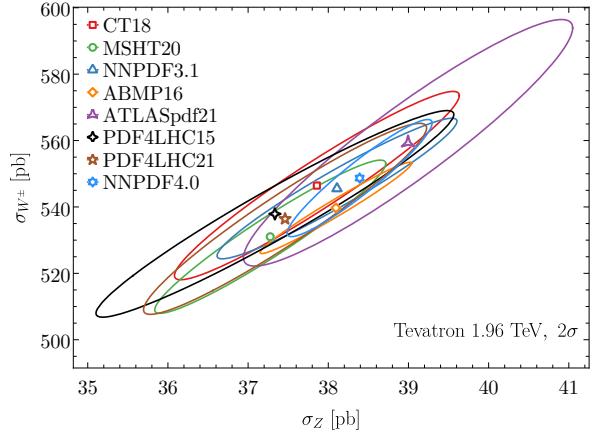}
\caption{Same as Fig.~\ref{fig:LHCpheno}, for inclusive cross sections of $W^\pm$ and $Z$ boson production processes at the Tevatron 1.96 TeV. 
\label{fig:TevatronPheno}}
\end{figure}

Next, we assess how the differences at the level of PDFs and parton luminosities in Figs.~\ref{fig:pdfratios}-\ref{fig:lumi} influence theoretical predictions for LHC total cross sections. These are displayed in Fig.~\ref{fig:LHCpheno}, where we present $2\sigma$ ellipses for pairs of inclusive cross sections among $W^\pm$, $Z$, $t\bar{t}$, $H$, $t\bar{t}H$ production at $\sqrt{s}=14$ TeV. The $W^\pm/Z$ cross sections are defined in the ATLAS 13 TeV fiducial volume~\cite{ATLAS:2016fij}, while others
correspond to the full phase space. 
In addition to the predictions based on the PDF ensembles displayed  in Figs.~\ref{fig:pdfratios}-\ref{fig:lumi} 
(with the PDF uncertainties rescaled to the $2\sigma$ probability), the plots also consider
the recent PDF4LHC21 combined PDF ensemble~\cite{Ball:2022hsh} described in Section~\ref{sec:benchmarking}, and the previous PDF4LHC15 combination~\cite{Butterworth:2015oua}.
There is a general agreement between the correlated predictions, with ATLASpdf21 predictions displaying larger uncertainties compared to the other sets and touching the PDF4LHC21 $2\sigma$ boundaries for the $t\bar{t}$ and $Z$ ellipses, and with ABMP16 giving lower predictions for $H$ and $t\bar{t}H$ cross sections. Generally, NNPDF4.0 predictions are at the boundary of the MSHT20 ellipses, with smaller error bands. MSHT20 are also generally in agreement with CT18, albeit with the latter having notably larger error ellipses. 

Finally, Fig.~\ref{fig:TevatronPheno} presents an analogous comparison of total cross sections for $W^\pm$ and $Z$ boson production in $p\bar p$ collisions in the Tevatron Run-2. 

\subsection{Applications of PDFs to Higgs physics, BSM searches, SMEFT tests}

PDFs are a crucial input at the LHC. Their uncertainty is a key component of theory uncertainties in Higgs physics, a limiting  factor in the mass reach of experimental searches for heavy BSM particles and the treatment of BSM sensitive data in PDF fits makes the interplay between PDFs and SMEFT tests significant. In what follows we briefly discuss each of these applications in turn and refer to a number of studies and new directions within each of these strands. 

\subsubsection{PDFs and Higgs physics}

{\it Author: Maria Ubiali} 

In the SM, once the Higgs mass $M_H$ is measured, all other parameters of the Higgs sector, such as the strength of its coupling to fermions and vector bosons and its branching ratios, are uniquely determined~\cite{LHCHiggsCrossSectionWorkingGroup:2016ypw}. Any deviation of the Higgs couplings with respect to the SM predictions would be a smoking gun for new physics. Crucially, realising this program requires not only high precision experimental measurements of Higgs boson production and its decay in various channels, but also the calculation of the SM cross sections and decay rates with matching theoretical precision. Despite the progress in the precise determination of PDFs, PDF uncertainty is still one of the largest sources of theoretical uncertainty affecting the predictions for Higgs boson  production~\cite{LHCHiggsCrossSectionWorkingGroup:2016ypw,Gao:2017yyd}. 

\begin{figure}[b]
    \centering
    \includegraphics[width=0.49\textwidth]{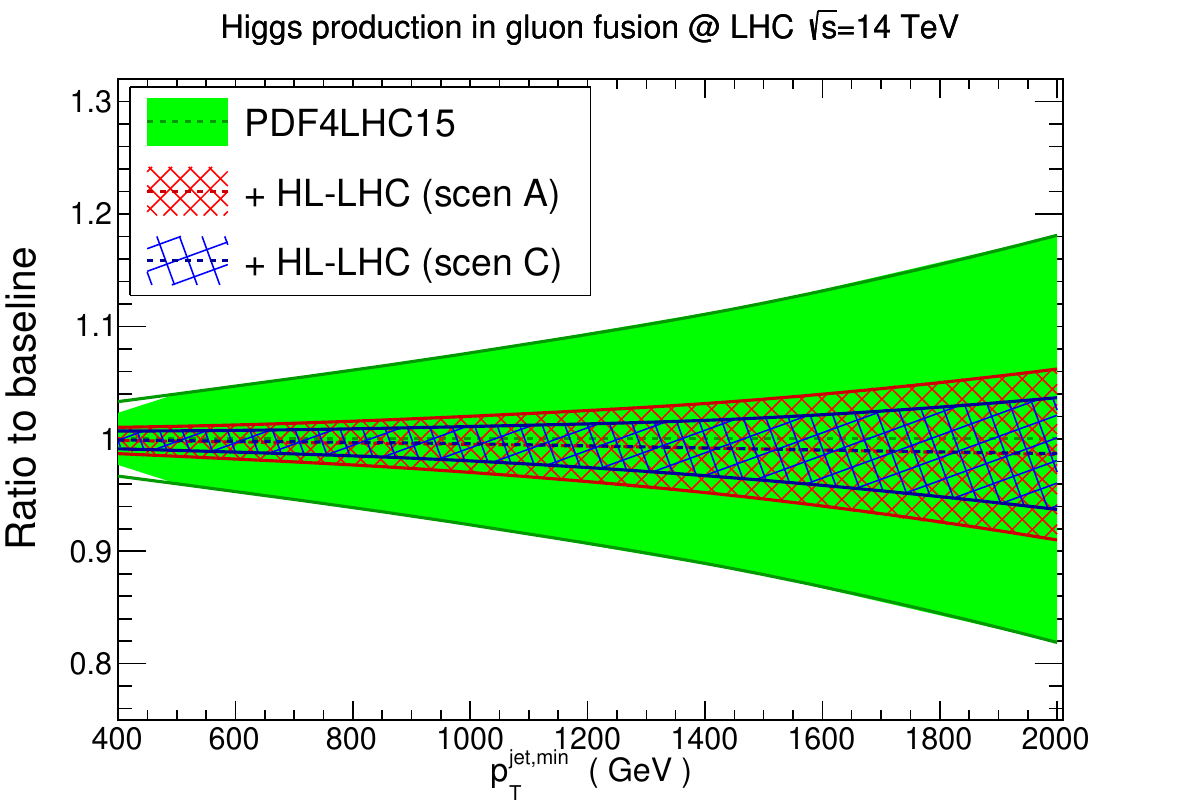}
        \includegraphics[width=0.49\textwidth]{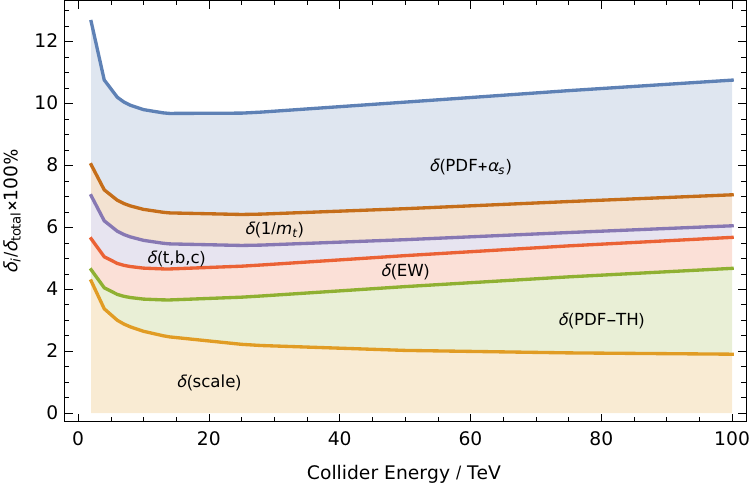}
    \caption{\label{fig:higgs} {\bf Left panel:} Comparison of the predictions for Higgs production via gluon fusion at $\sqrt{s} = 14$ TeV at between the PDF4LHC15 baseline~\cite{Butterworth:2015oua} and the HL–LHC profiled sets in the conservative and optimistic scenarios of Ref.~\cite{AbdulKhalek:2018rok}. Results are shown normalised to the central value of PDF4LHC15. Taken from~\cite{AbdulKhalek:2018rok}, see reference for more details about the calculation. \\
    {\bf Right panel:} linear sum of the different sources of relative uncertainties in the calculation of Higgs production via gluon fusion as a function of the collider energy. Each coloured band represents the size of one particular source of uncertainty. In particular, the component $\delta({\rm PDFs}+\alpha_s)$ corresponds to the uncertainties due to our imprecise knowledge of the strong coupling constant and of PDFs combined in quadrature, while the $\delta({\rm  PDF-th})$ represents the mismatch in the perturbative order of the PDFs, evaluated at N2LO, and the perturbative QCD cross sections evaluated at N3LO, defined as in Eq.~\eqref{eq:mismatch}. Taken from ~\cite{Dulat:2018rbf}.}
\end{figure}

In Ref.~\cite{AbdulKhalek:2018rok} a study of the impact of HL–LHC pseudo–data for a number of PDF–sensitive processes was performed. Different scenarios are considered, from a conservative one with approximately the same systematics as the corresponding baseline measurements from Run I and a factor of 2 reduction for those from Run II, to an optimistic one with a reduction by a factor 2.5 as compared to Run I (II). It was found that the legacy HL–LHC measurements can reduce the uncertainties in the PDF luminosities by a factor between 2 and 5 in comparison to state–of–
the–art fits, depending on the specific flavor combination of the initial state and the invariant mass of the produced final state. As an illustration, on the left panel of Fig.~\ref{fig:higgs} we show a comparison of the PDF uncertainty for Higgs boson production in gluon fusion at $\sqrt{s}\,=\,14$ TeV and its reduction from predictions obtained with the PDF4LHC15~\cite{Butterworth:2015oua}  baseline and the HL–LHC profiled sets in the conservative (scen A) and optimistic (scen C) scenarios. 
%
%

However, the effect of the pure PDF uncertainty is not the end of the story. 
In ~\cite{Cepeda:2019klc}, theoretical predictions for Higgs boson production through gluon fusion at $pp$ collisions are provided as a function of the collider energy $\sqrt{s}$. As it can be observed from the right panel of Fig.~\ref{fig:higgs}, the theoretical uncertainty associated with the predictions is split into various components, including the missing higher order uncertainty $\delta({\rm scale})$ (measured by the usual scale variation procedure) of the N3LO calculation of the $gg\to H$ partonic cross section~\cite{Anastasiou:2016cez,Mistlberger:2018etf}. Electroweak (EW) and approximated mixed QCD-EW corrections as well as effects of finite quark masses are also included in the $\delta({\rm EW})$ component.  Effects due to finite quark masses neglected in the QCD corrections are also accounted for in the $\delta(1/m_t)$ and $\delta(t,b,c)$ components. 
Finally, and most relevant for our discussions are the two components related with PDF uncertainties. On the one hand, the usual component $\delta({\rm PDFs}+\alpha_s)$ corresponding to the uncertainties due to our imprecise knowledge of the strong coupling constant $\alpha_s(M_Z)$ and of the PDFs combined in quadrature. On the other had, the $\delta({\rm  PDF-th})$ components, which represents the mismatch in the perturbative order of the PDFs, evaluated at N2LO, and the perturbative QCD cross sections evaluated at N3LO, defined as
\begin{equation}
    \label{eq:mismatch}
    \delta(\text{PDF-TH})=\frac{1}{2}\Bigg|\frac{\sigma^{\text{N2LO}}_{\text{N2LO-PDF}}-\sigma^{\text{N2LO}}_{\text{NLO-PDF}}}{\sigma^{\text{N2LO}}_{\text{N2LO-PDF}}}\Bigg|.
\end{equation}
As one may observe on the right panel of Fig.~\ref{fig:higgs},  $\delta(\text{PDF-TH})$ leads to a significant uncertainty on N3LO cross section predictions, of the order of several percent in the case of Higgs via gluon fusion as well as in the case of other key LHC observables~\cite{Anastasiou:2016cez,Duhr:2019kwi,Duhr:2021vwj} and is comparable to the regular uncertainty associated with our current understanding of PDF themselves. Of course, the prescription of Eq.~\eqref{eq:mismatch} is a very conservative estimate of the theory uncertainty due to the mismatch between the perturbative order of PDF evolution and partonic cross section. It points to the need of devising a better procedure of estimating theory uncertainties in the now standard N2LO PDF fits (discussed in Sect.~\ref{sec:thunc}) and of moving towards N3LO PDFs (discussed in Sect.~\ref{sec:n3lo}).

\subsubsection{PDFs and BSM searches}
\label{sec:bsm}

{\it Author: Marco Guzzi}  


New physics interactions are currently searched for at the LHC, but are also important for a large variety of 
physics programs at future facilities ({\it e.g.}, HL-LHC, Future Circular Collider (FCC), Super proton proton Collider (SppC), Faser$\nu$~\cite{FASER:2019dxq}).
The interplay between global PDF analyses, precision calculations of matrix elements, and BSM physics would be crucial to accomplish a wide range of physics goals at all these facilities.

As an example, consider the searches for new vector bosons, $Z'$s and $W'$s, from BSM constructions that extend the gauge symmetry group of the SM. The new bosons predicted by different models can have a mass that varies from a fraction of GeV to dozens of TeVs. Their fermion interactions share similar features to those of the $Z$ and $W$ from the SM. Models for $W' / Z'$s in Drell-Yan resonant dilepton production are currently scrutinized at the LHC~\cite{CMS:2018ipm,ATLAS:2019erb}. At high energies, $W' / Z'$s can also be produced in association with another SM vector or scalar boson, or in association with a jet or single heavy quark~\cite{Adelman:2012py,Guzzi:2019ucs}.  
Current LHC bounds on mass disfavor extra vector bosons lighter than approximately 4-5 TeV. 
Production of $W' / Z'$ bosons with larger mass is progressively impacted by PDFs at large $x$ where uncertainties are still large~\cite{Brady:2011hb}.  Constraining PDFs at large $x$ is a very challenging task because there are many effects of comparable size that contribute and affect global PDF analyses in this kinematic region, see a related discussion in Sect.~\ref{sec:TheoryLargeX}. Examples of these are nuclear corrections, higher-twist contributions, presence of intrinsic heavy-quark components, and use of different general mass variable flavor number (GMVFN) schemes. The next run of the LHC, the HL-LHC, and the high-luminosity EIC thus must meet the challenge of constraining the PDFs in the large-$x$ region through a combination of measurements reviewed in Sec.~\ref{sec:Experiments}.

\subsubsection{Interplay of PDF fits and SMEFT fits
\label{sec:smeft}}
{\it Leading authors: R. Boughezal, F. Petriello, M. Ubiali \\ \vspace{6pt}}

If the LHC experiments identify one or more significant deviations from the SM predictions, the most promising way to characterize their possible origin is via Effective Field Theories (EFTs). Even in the absence of any deviations, EFTs can be used to set lower bounds on the scales of a number of new physics scenarios and to steer the future searches~\cite{Manohar:1996cq}. Indeed, for a large class of BSM models, physics at energies well below the mass scale $\Lambda$ of new physics can be parametrized by an EFT, by adding higher dimensional operators to the SM Lagrangian, whose coefficients are suppressed by powers of $\Lambda$.
Such extensions of the SM Lagrangian quantify low-energy interactions that are induced by dynamics at energies far
above the energy scale probed by the LHC experiments.

The analysis of BSM effects via an EFT parametrization is a critical and increasingly active research area. A widely adopted EFT expansion is the Standard Model EFT (SMEFT)~\cite{Brivio:2017vri}, built upon the assumption that all the known particles have the gauge transformation properties predicted by the SM, with their conventional dim-2 and dim-4 interactions being supplemented by new higher-dimensional interactions among all allowed combinations of the SM fields. Such interactions might be generated by massive particles exchanged at the tree level or circulating in loop diagrams.

Although the proton structure parametrized by PDFs is intrinsically a low-energy quantity and, as such, it should in principle be separable from the high-energy new physics imprints, the complexity of the LHC environment might well intertwine them.
Exploiting the full potential of current and future precision measurements at the LHC for indirect BSM searches requires novel data interpretation frameworks to account for hitherto ignored effects, such as the interplay with the PDFs in the high-energy tails of LHC distributions. Indeed, the very same data sets are often used both to determine the PDFs (assuming SM theoretical predictions) and, independently, to constrain the SMEFT Wilson coefficients (assuming SM PDFs). Given that these LHC processes provide significant information for both PDF and SMEFT fits, one must ascertain the extent to which eventual BSM signals can be inadvertently reabsorbed into the PDFs, as well as how current bounds on the EFT coefficients are modified within a consistent simultaneous determination together with the PDFs.

Data sets that may contain information on new physics at high scales, such as inclusive jet production, also typically cover a wide dynamic range in transverse momentum and rapidity. If there is a PDF explanation for any deviation from the SM prediction that is observed at high $p_T$, that explanation has to be universal, {\it i.e.} it also has to explain distributions at similar $x$ values, but at lower transverse momentum, regions where new physics is not expected to produce any notable impacts. In this way, the separate rapidity regions serve as a cross check, both for the PDF determinations themselves and the possible presence of new physics. Care must be taken, however, as tensions between rapidity regions may as well arise from an imperfect knowledge of the rapidity dependence of the experimental systematic errors.

\begin{figure}[tb]
    \centering
    \includegraphics[width=0.49\textwidth]{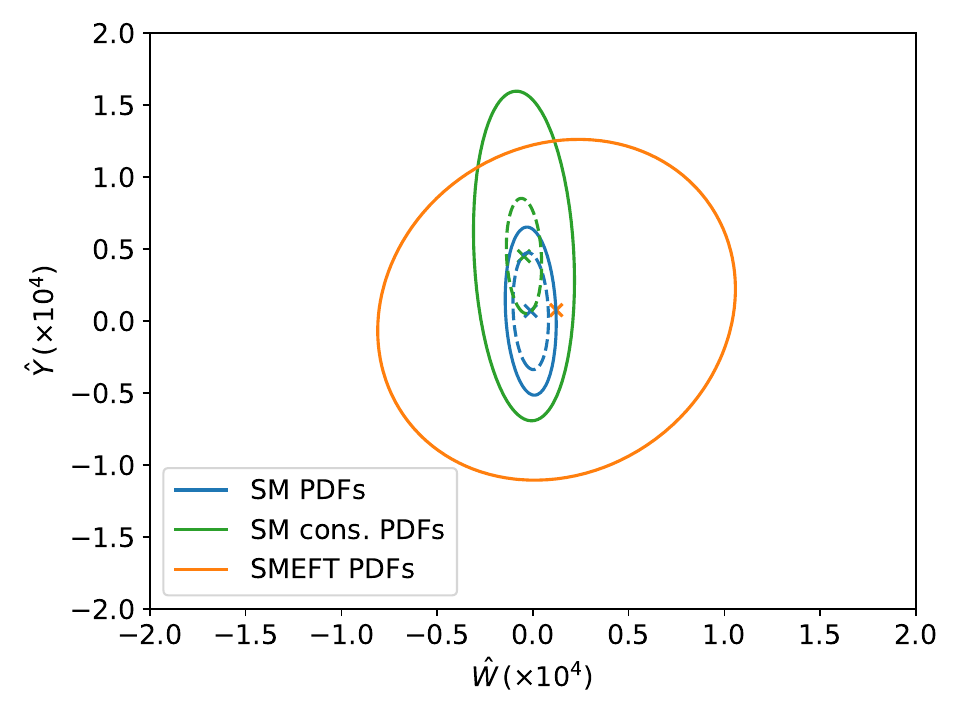}
    \includegraphics[width=0.49\textwidth]{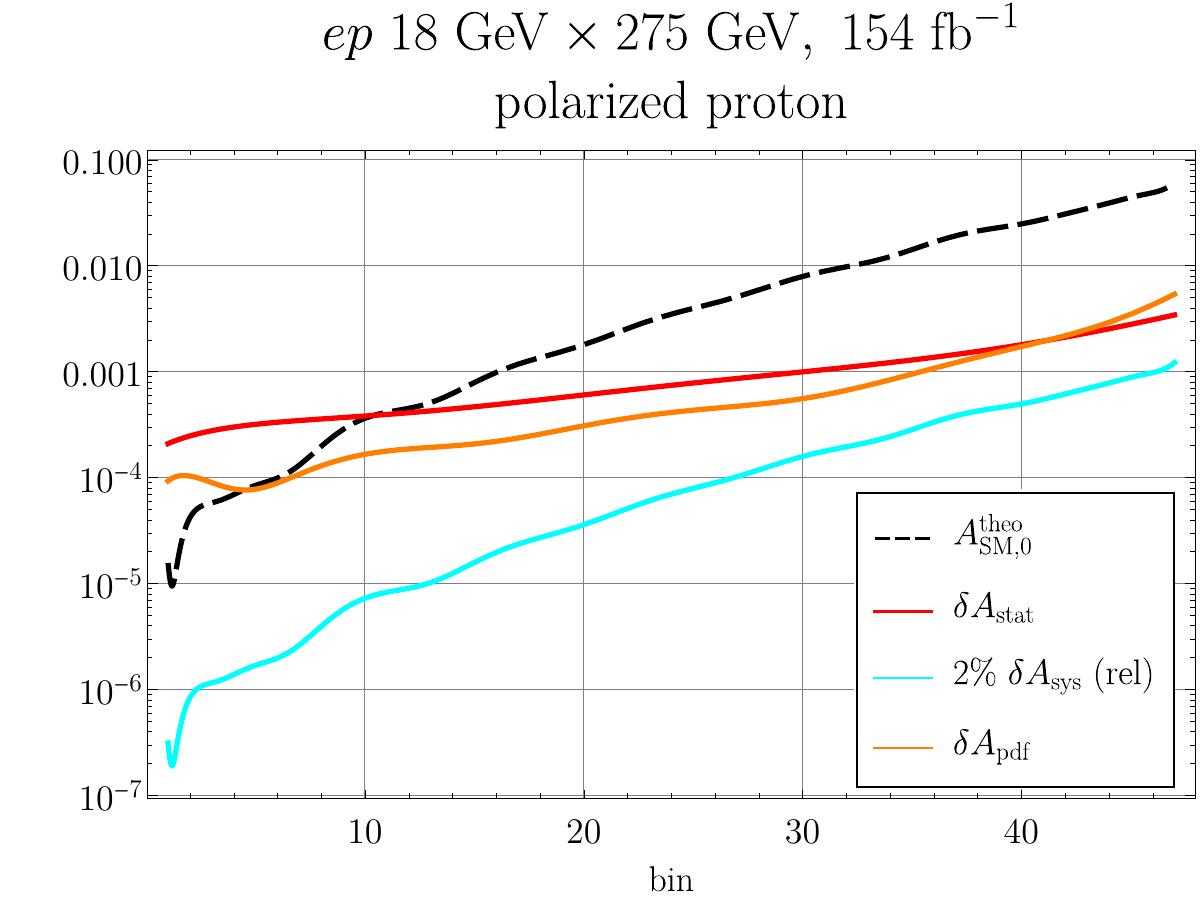}
    \caption{\label{fig:SMEFT_PDFs} {\bf Left panel:} The 95\% confidence level bounds in the plane of the Wilson coefficients obtained in Ref.~\cite{Greljo:2021kvv} using either fixed SM PDFs (blue) or conservative SM PDFs that do not include high-energy data (green). PDF uncertainties are included in the solid lines and not included in the dashed lines. The results are compared to those obtained in a simultaneous fit of SMEFT and PDFs, when the PDFs are allowed to vary when varying the values of the Wilson coefficients (orange).\\ {\bf Right panel:} Error components for the polarization asymmetry at a future EIC as a function of bin number, adapted from Ref.~\cite{Boughezal:2022pmb}. The bins are ordered in the DIS momentum transfer $Q^2$ and Bjorken $x$.}
\end{figure}
Simultaneous determination of the Wilson coefficients of
the SMEFT and of the proton PDFs has been pioneered in several recent studies performed by both theorists and experimentalists ~\cite{Iranipour:2022iak,Liu:2022plj,Greljo:2021kvv,Carrazza:2019sec,ZEUS:2019cou,CMS:2021yzl}. 
These studies reveal that, while with current DIS and Drell-Yan data the interplay is already non-negligible but can
be kept under control, once High Luminosity LHC (HL-LHC) data are considered, neglecting the PDF interplay could potentially miss new physics manifestations or misinterpret them. This is illustrated in the left panel of Fig.~\ref{fig:SMEFT_PDFs}, where it can be observed that including high-mass LHC data both in a fit of PDFs and in a fit of SMEFT coefficients, while neglecting their interplay, could significantly underestimate the uncertainties associated with the EFT parameters. Indeed, the bounds on the Wilson coefficients considered in ~\cite{Greljo:2021kvv} are relaxed when the coefficients and PDFs are varied together. 
The interplay of the old and new experimental constraints can be non-trivial, so the above enlargement of the uncertainties can even
exceed the error estimate when the SMEFT coefficients are fitted with the fixed conservative PDFs that do not include any of the high-mass Drell-Yan sets~\cite{Madigan:2021uho}.

These seminal studies deserve to be further developed to explore a broader number of operators and observables, and at the same time to build a robust methodology for simultaneous determination of the Wilson coefficients of an EFT expansion and the PDF parameters, such as those put forward in~\cite{Liu:2022plj,Iranipour:2022iak}. Other interesting avenues, such as the sensitivity of the LHC data to the presence of dark photons that couple to quarks, have been recently explored~\cite{McCullough:2022hzr}. 

The high polarizations of electron and hadron beams at an EIC would provide unique probes of SMEFT operators complementary to those obtained at the LHC and the HL-LHC~\cite{Boughezal:2020uwq,Boughezal:2021kla}. Maximizing the potential of these measurements requires a precise determination of the polarized PDFs of the nucleons. The effect of PDFs and  systematic errors on SMEFT parameter determinations with polarized deuteron and proton beams at an EIC was recently investigated~\cite{Boughezal:2022pmb}. Particularly in the high-luminosity phase of the EIC, polarized PDF errors are expected to form by far the largest source of systematic error on determinations of SMEFT parameters from polarized proton beams. A summary of the anticipated errors at a high-luminosity EIC with polarized proton beams is shown below in the right panel of  Fig.~\ref{fig:SMEFT_PDFs}. The uncertainty from polarized PDFs on the polarization asymmetry is orders of magnitude larger than the expected beam polarization error and other systematic errors coming from background processes, and it is nearly as large as the expected statistical error over most of the available ($x,Q^2$) parameter space. This provides additional motivation for joint PDF-SMEFT determinations from future polarized deep-inelastic scattering data from the EIC.

\section{Experiments \label{sec:Experiments}}

In this section we first discuss the measurements at the LHC that strongly rely on precise PDFs or are used to constrain the PDFs. We then turn to the opportunities to constrain unpolarized, polarized and nuclear PDFs at the Electron Ion Collider (EIC). Afterwards we describe the opportunities to constrain PDFs at the Large Hadron electron Collider (LHeC). Subsequently, we discuss the importance of PDFs in neutrino phenomenology and the experimental constraints that we expect from new neutrino experimental facilities. Finally, we focus on forward and ultra-high energy scattering processes. 

\subsection{Measurements and applications of PDFs at the LHC \label{sec:PDFLHC}}
{\it Leading authors: A. M. Cooper-Sarkar, J. Huston \\ \vspace{6pt}}

In this section, we first review the measurements from the LHC that are most sensitive to PDFs and are commonly used by the global PDF fitting groups, CT, MSHT and NNPDF. Second, we point out which measurements can be substantially improved at the HL-LHC. Third, we consider measurements which may be most sensitive to new physics and for which PDF uncertainties are the dominant background/uncertainty.

The measurements which are most sensitive to PDFs are:
\begin{itemize}
    \item Inclusive $W$ and $Z/\gamma^*$ boson differential measurements~\cite{ATLAS:2016nqi,ATLAS:2016gic,ATLAS:2017rue,ATLAS:2019fgb,LHCb:2015okr,LHCb:2015mad,LHCb:2015kwa,LHCb:2016fbk,CMS:2016qqr}, as a function of pseudo-rapidity and rapidity respectively. For the $Z/\gamma^*$ production of Drell-Yan lepton pairs, different ranges of dilepton mass are also considered. Furthermore, there is a triple differential $Z/\gamma^*$ measurement in rapidity, mass and the Collins-Soper angle. The experimental precision of the mass peak data is $\sim 0.5\%$. These measurements have impact on the valence PDF distributions. In the LHC kinematic range for ATLAS and CMS, they also have impact on the flavor structure of the sea. It is now the case that N2LO QCD analyses are needed to obtain good fits to these data. NLO-EW predictions are also standardly applied.  When considering the high-mass Drell-Yan process, the photon PDF in the proton is an essential part of the formalism. For the low-mass Drell-Yan and for the higher rapidity ranges probed by LHCb, one may need to move beyond DGLAP to ln(1/x) resummation or non-linear evolution equations~\cite{Ball:2017otu}.
    \item  Inclusive jet and dijet measurements over a range of (di)jet rapidity bins, as a function of the jet's transverse momentum $p_T$ or dijet invariant mass $m_{jj}$~\cite{ATLAS:2014riz,ATLAS:2017kux,ATLAS:2017ble,CMS:2014nvq,CMS:2016lna,CMS:2015jdl}. The data precision ranges from $\sim 5$ to $50\%$. Jet measurements mostly have an impact on the high-$x$ gluon PDF, with minor constraints on the (anti)quark sea composition. The current state of the art is the N2LO QCD and NLO-EW theory, but for jets nonperturbative corrections for hadronisation and underlying event are also applied, and these differ according to the jet radius. Jet angular sizes larger than 0.5 are preferred theoretically. The dijet data are not yet fully exploited in PDF fitting, although several dijet data sets are included in the latest NNPDF release~\cite{Ball:2021leu}. The jet measurements probe the highest scale and hence may require consideration of new physics~\cite{Alte:2017pme}.
    \item $W$-boson + jets and $Z$-boson + jets measurements~\cite{ATLAS:2017irc,ATLAS:2019bsa} extend the kinematic reach of the inclusive $W$ and $Z$ data to higher scale and higher $x$. The measurements are, for example, $p_T^W$ for the $W$ +jets and $y^{\rm jet}$ in bins of $p_T^{\rm jet}$ for the $Z$ +jets. The high-$p_T$ $Z$ boson spectrum~\cite{ATLAS:2015iiu} can also be used instead of $Z$ + jets data. The data precision is $\sim 15\%$. These data have impact on the gluon and on the quark PDFs, both valence and higher-$x$ sea structure. Predictions are made up to N2LO QCD and NLO-EW accuracy, with nonperturbative corrections also needed for the jets. Note that data at low $p_T$ are usually cut out because of the need for nonperturbative modelling of low-$p_T$ resummation, but predictions can still be sensitive to this cut.
    \item $t\bar{t}$ production rates in the form of both the total and differential cross sections~\cite{ATLAS:2015lsn,ATLAS:2016pal,ATLAS:2019hxz,CMS:2017iqf,CMS:2015rld,CMS:2013hon,CMS:2018adi,CMS:2018htd}. The measurements are typically presented as a function of mass $t\bar{t}$, rapidity $t\bar{t}$, average rapidity and average $p_T$,  and they can be doubly differential. The data precision is $\sim 15\%$. These data mostly have an impact on the high-$x$ gluon PDF, although the resulting constraints are not as strong as those from the jet measurements. The measurements can be made in the semileptonic, dilepton, and fully hadronic channels, although the latter have not been used for PDF fitting as yet. Predictions are made at N2LO QCD and NLO-EW, with nonperturbative corrections also needed for the jets in the event.
    \item Direct photon production~\cite{ATLAS:2019drj} is once again being considered as an input to PDF fits. They are measured as a function of $E_T^\gamma$ in bins of the photon pseudorapidity. The data for 8 and 13 TeV have been combined as ratios, with an experimental precision of $\sim 5\%$. These data impact the high-$x$ gluon PDF although less strongly than either the $t\bar{t}$ data and the jet data.  Predictions are made at N2LO QCD and NLO-EW accuracy.
\end{itemize}

Many of the above measurements are already systematics limited, so that improvement is not a matter of accumulating more statistics at the HL-LHC. Nevertheless, it should be noted that high statistics can lead to better systematic uncertainty estimates. Also note that much of the data at 13 TeV from the full statistics runs up to 2018 are not yet included in PDF fits. Processes which may bring improvement are:
\begin{itemize}
    \item Inclusive $W$ and $Z$ production at high rapidity
    \item High-mass Drell-Yan production
    \item Low-mass Drell-Yan production, modulo the small-$x$ modifications mentioned above
    \item For inclusive jet production information on correlations between data sets could allow us to exploit inclusive jet, dijet and even trijet information simultaneously 
    \item $W$ +jet and $Z$ boson +jet data at higher $p_T$
    \item $W/Z$ boson + heavy quark data, particularly $W + c$, which can constrain the strange quark, providing that theoretical calculations can be extended to N2LO. There is some progress in that direction~\cite{Czakon:2020coa} but at present experimental and theoretical jet algorithms are not fully consistent.
    \item More differential information on $t\bar{t}$ production, again assuming that N2LO predictions are available for double or even triple differential distributions.
    \item Single-top distributions have been conisdered by MSHT~\cite{Bailey:2020ooq} and by NNPDF~\cite{Nocera:2019wyk}. The impact is small at present but with better data their use could be extended. 
    \item Isolated photon production and isolated photon-jet correlations at forward rapidity are sensitive to gluon density and saturation effects. Isolated photons originate predominantly in quark–gluon Compton scattering which can be probed over a large range of $x$ and $Q^2$~\cite{ALICE:2020mso}.
    \item Open charm or beauty production at LHCb rapidities has also not been exploited to any extent in the global PDFs, although some PDF studies exist~\cite{Garzelli:2016xmx,Zenaiev:2019ktw}. These have impact on the low-$x$ gluon, and the current NLO theory may need extension to N2LO and plausibly $\ln(1/x)$ resummation. Use of ratios can help to reduce uncertainties.
\end{itemize}

A caution must be raised that data extending to high energy scales may be subject to new physics effects. For example, when looking for new physics in $Z'$ production at very high mass~\cite{ATLAS:2017fih} or in jet production at high-mass~\cite{ATLAS:2015nsi}, we have found that the PDF uncertainty limits our ability to see new-physics signals. Furthermore, we may be 'fitting away' new physics effects in the tails of the distributions of the data that we input. These concerns were also raised in Secs.~\ref{sec:bsm} and \ref{sec:smeft}.
For this reason ATLAS
have considered PDF fits which exclude data at scale $Q > 500$ GeV~\cite{ATLAS:2021vod} (this is mostly the inclusive jet data), and CMS have considered fitting their inclusive jet data using PDF parameters and SMEFT parameters simultaneously~\cite{CMS:2021yzl}. No evidence for new physics has yet been found, but such approaches will have to be pursued in the future, while the interplay between new physics and PDFs will become stronger at the HL-LHC~\cite{Greljo:2021kvv,Iranipour:2022iak}.

New physics can also manifest itself at lower scales by the deviation of Standard Model parameters from their SM values. For example, in recent measurements of the mass of the $W$ boson, $m_W$~\cite{ATLAS:2017rzl} and the weak mixing angle, $\sin^2\theta_W$~\cite{CMS:2018ktx}, the uncertainty due to the PDF used in the extraction is now one of the largest uncertainties. Various strategies have been proposed. Since the PDFs used usually lag behind the new measurements, the PDF can be improved by profiling the same data that are used for the SM parameter measurement. Of course there can be correlations between the SM parameters and the PDF parameters, so ideally a new simultaneous fit should be performed. Another point is that the PDF uncertainty is usually evaluated by comparing the results using different global PDF sets, each of which comes with its own PDF uncertainty. This approach ignores potential correlations between the PDF sets. One may try to reduce this uncertainty by evaluating these correlations, as explored further in Section~\ref{sec:MethodologyDeliveryCorrelations}.

\subsection{PDFs at the Electron-Ion Collider
\label{sec:EICPDF}
}
{\it Leading authors: T. J. Hobbs, E. R. Nocera, and R. S. Thorne\\ \vspace{6pt}}

The construction of an Electron-Ion Collider (EIC)~\cite{Accardi:2012qut,Aschenauer:2017jsk,AbdulKhalek:2021gbh} has been recently approved by the United States
Department of Energy at Brookhaven National Laboratory. The EIC could record the first scattering events as early as 2030. By colliding (polarized) electron, and potentially
positron, beams with proton or ion beams at a center-of-mass energy of up to $\lesssim\! 140$ GeV, the EIC will investigate how partons are distributed in position and momentum spaces within a proton, how the proton spin originates
from the spin and the dynamics of partons, how the nuclear medium modifies parton-level interactions and substructure, and whether gluons saturate within heavy nuclei.
In addition, the EIC will be capable of a range of PDF-related precision measurements in fundamental QCD and electroweak phenomenology. These include new constraints on
standard model inputs like $\alpha_s$ and the heavy-quark masses; novel electroweak probes for BSM physics; precise tests of QCD factorization
theorems; and transition from nonperturbative to perturbative QCD dynamics. Prospects for these EIC studies have been investigated in a recent EIC Yellow
Report~\cite{AbdulKhalek:2021gbh} and form the subject matter of a series of dedicated Snowmass whitepapers \cite{AbdulKhalek:2022hcn,Chekanov:2022sax,Batell:2022ubw}. Here, we discuss the relevance of future EIC measurements for improving the knowledge of the proton PDFs, both unpolarized and longitudinally polarized, and of nuclear PDFs. Below, we discuss each of these in turn based on the cited studies that estimated the EIC potential impact by including EIC pseudodata for an integrated luminosity $\mathcal{L}=10-100~$fb$^{-1}$ and center-of-mass energy $\sqrt{s}=28.6-140$ GeV in the global PDF fits.
We note also that the potential of the EIC to furnish constraining information on the PDFs of other hadrons --- particularly the light mesons --- has been discussed
elsewhere~\cite{Aguilar:2019teb,Arrington:2021biu}; for details, we refer interested readers to these documents.

\subsubsection{Unpolarized Proton PDFs}
\label{subsec:unp_PDFs}

A large quantity of EIC data sensitive to the unpolarized PDFs will be supplied through inclusive neutral-current (NC) and charged-current (CC) DIS cross section measurements involving
electron-DIS collisions with protons and light nuclei, especially the deuteron, $^3$He, $^4$He.
Measurements involving light nuclei could be used to determine either proton PDFs (including a correction~\cite{Bailey:2020ooq,Accardi:2021ysh} or an
uncertainty~\cite{Ball:2020xqw,Ball:2021leu} that accounts for nuclear effects) or nuclear PDFs themselves,
as discussed in Sect.~\ref{subsec:nuc_PDFs}. 
The inclusive NC and CC DIS cross section measurements at the EIC are expected to cover a broad kinematic region that significantly overlaps with the one probed by HERA, and with the EIC instantaneous
luminosities potentially higher by three orders of magnitude. These probes of the $x$-$Q^2$ plane will stretch to much higher values of $x$,
typically up to $x\sim 0.6-0.7$. At sufficiently high center-of-mass energies, this region is expected to be rather insensitive to higher-twist effects, which can be significant at
$W^2 \leq 15-20\, \mathrm{GeV}^2$. These measurements will therefore cleanly constrain PDFs at relatively large $x$. At the same time, the EIC coverage will also extend to softer values of $Q^2$,
allowing a rich phenomenological program to examine power-suppressed contributions like the higher-twist effects. In comparison to previous DIS experiments, systematic uncertainties
will be small, possibly not exceeding 1\%; statistical uncertainties will be even smaller.

The impact of EIC NC and CC inclusive DIS cross section measurements on the proton's unpolarized PDFs were investigated in dedicated studies (see
Sect.~7.1.1 in~\cite{AbdulKhalek:2021gbh} and~\cite{Wang:2018heo,Khalek:2021ulf}), whereby EIC pseudodata were included in a selection of PDF frameworks, namely
CJ~\cite{Accardi:2016qay}, CT~\cite{Hou:2019efy}, JAM~\cite{Moffat:2021dji} and NNPDF~\cite{NNPDF:2017mvq}. The pseudodata were generated for realistic projections of
the energy, luminosity, statistical and systematic uncertainties 
and found to have a potentially strong impact on the (large-$x$) valence PDF
sector, where PDF uncertainties could decrease up to 80\%. The sea PDF sector was predominantly modified in the small-$x$ region, with a decrease
of PDF uncertainties up to 50\%. 

The EIC may also have at its disposal the ability to perform analogous measurements using positron beams --- a possibility explored in Sect.~7.1.1 in~\cite{AbdulKhalek:2021gbh}
as an eventual program upgrade. By exchanging $W$ bosons of a positive charge, positron-initiated CC DIS interactions are capable of probing a combination of
flavor currents that is complementary to electron CC DIS. This potentially constrains the $d$-type PDFs, and, indirectly, the $d/u$ ratio. Beyond this, the use of positron beams may also
allow one to access other parity-violating effects, such as the breaking of the strange--antistrange symmetry or parton-level charge-symmetry violation~\cite{Hobbs:2011vy}.

The EIC will also measure semi-inclusive DIS (SIDIS) processes. Tagged DIS (TDIS) data offer a way to probe the structure of a barely off-shell neutron via semiinclusive tagging
of a slow spectator proton in $e+d\to e^\prime+p+X$ events. The EIC electron DIS data augmented with TDIS data can improve the determination of all
flavors over the whole $x$ range, in particular, for the $d/u$ ratio at large $x$, and to complement JLab experiments by extending their kinematic reach to higher energies. On the other hand, SIDIS data would offer a way to access PDFs for individual quark flavors, given that the valence parton content of the hadron detected in the final state relates
to the fragmenting parton flavor. 
The analysis of SIDIS data requires the simultaneous knowledge of fragmentation functions
(FFs)~\cite{Metz:2016swz}, whose determination will be concurrently improved at the EIC~\cite{Aschenauer:2019kzf,Sato:2019yez,Moffat:2021dji}.
By means of EIC pseudodata 
it was shown~\cite{Aschenauer:2019kzf} that the impact of
pion production SIDIS data on up, down, anti-up and anti-down quark PDFs is moderate, as they are already very well determined. Conversely, the far less known strange PDFs could be
constrained substantially by kaon production SIDIS data, particularly at low $x$.  Final-state tagging of a produced charm quark may also help discriminate among scenarios for the strange sea. Charm-containing jet production would be sensitive to nucleon strangeness and can disentangle patterns of SU(3) symmetry breaking in the light-quark sea \cite{Arratia:2020azl}. Along with direct measurements of the proton's charm structure function, $F^{c\bar{c}}_2$ \cite{GuzziIC:2011,Hobbs:2017fom}, the EIC's charm-tagging ability may
possibly constrain a nonperturbative component of the charm quark PDF.

\subsubsection{Polarized Proton PDFs}
\label{subsec:pol_PDFs}

The EIC will allow for the longitudinal polarization of both the colliding nucleon (and light nuclei) and lepton beams, {\it i.e.}, along their direction of motion. This is a unique feature of the EIC, specifically designed to probe the longitudinal spin structure of the proton. In the key inclusive DIS measurements, besides the parity-conserving
longitudinal double-spin asymmetry, the EIC will access also the parity-violating asymmetry, see, {\it e.g.}, Sect.~18.2 in~\cite{ParticleDataGroup:2020ssz} for a definition.
In the numerator of the latter observable, the parity-conserving contributions from the photon exchange and the vector--vector part of the $Z$-boson exchange cancel exactly, leaving
the dominant contribution from the interference between the photon exchange and the axial-vector part of the $Z$-boson exchange. While the parity-conserving asymmetry probes the sum
of polarized quark and antiquark distributions, the parity-violating asymmetry probes their difference. The combination of the two is one of the cleanest ways to separate quark and
antiquark polarizations.

Parity-conserving and parity-violating polarized DIS asymmetries are expected to expand the kinematic coverage of current DIS measurements significantly, roughly by one order of magnitude
or more, down to $x\sim 10^{-4}$ and up to $Q^2\sim 1000$~GeV$^2$, see, {\it e.g.}, Fig.~1 in~\cite{Ethier:2020way}. In addition to the increased sensitivity to quark, antiquark and gluon
polarized PDFs at small values of $x$, the wide $Q^2$-coverage of the EIC will probe scaling violations in the polarized structure function $g_1$, offering significant additional constraints
on the gluon polarized PDF.

The impact of parity-conserving and parity-violating longitudinal spin asymmetries was investigated in dedicated studies (see Sect.~7.1.2 in~\cite{AbdulKhalek:2021gbh}
and~\cite{Aschenauer:2012ve,Aschenauer:2013iia,Aschenauer:2015ata,Ball:2013tyh,Zhou:2021llj}), whereby EIC pseudodata were included in a selection of polarized PDF frameworks, namely
DSSV~\cite{deFlorian:2014yva,DeFlorian:2019xxt}, JAM~\cite{Ethier:2017zbq}, and NNPDF~\cite{Nocera:2014gqa}. 
If one assumed $\mathrm{SU}(3)$ flavor symmetry for the axial-vector charges, the uncertainty on the first moment of the gluon polarized PDF reduced by up to 80--90\%, and that for the first moment of the sum of all quark and antiquark polarized PDFs was reduced by around 80\%. The uncertainty reduction was more modest if one did not impose the $\mathrm{SU}(3)$ symmetry: the uncertainty on the gluon moment decreased by about 60\%, and no clear reduction in the uncertainty of the quark and antiquark moment was seen. The reductions in the uncertainty of the moments depended on the extrapolation of the PDFs to very small $x$. Be that as it may, these results will test to which extent the small-$x$ dipole formalism~\cite{Kovchegov:2015pbl,Kovchegov:2016zex,Kovchegov:2016weo,Kovchegov:2018znm} holds, and which fraction of the proton spin cannot be
ascribed to the spin of quarks and gluons~\cite{Ji:1996ek}.

Concerning sea quark polarized PDFs,
measurements of SIDIS cross sections with polarized beams can significantly reduce the uncertainties on up, down and strange antiquarks,
see~\cite{Borsa:2020lsz}. Identification of kaons in the final state may, in particular, shed light on the strange sea polarization, whose shape cannot be determined from
parity-conserving DIS asymmetries and is usually constrained by assuming exact $\mathrm{SU}(3)$ flavor symmetry and a relation of its first moment to hyperon beta-decay constants. The
EIC SIDIS data will possibly establish whether there is a non-zero strange polarization at $x>0.5\cdot 10^{-5}$~\cite{AbdulKhalek:2021gbh}, assuming small uncertainties on fragmentation functions~\cite{Ethier:2017zbq}. 
In a similar spirit, the use of DIS and SIDIS longitudinal spin asymmetries, instead of cross sections,
may require the simultaneous determination of unpolarized PDFs~\cite{Zhou:2022wzm}. Concerning the gluon polarized PDF, processes such as photon-gluon fusion in EIC production of
back-to-back partonic jets with large transverse momentum have been shown to be feasible~\cite{Page:2019gbf}. Dijet longitudinal double-spin asymmetries could be measured
with a moderate integrated luminosity; these could be used as a cross-check of the more stringent constraint on the gluon polarized PDF provided by the evolution of the polarized
structure function $g_1$. Finally, additional constraints on the gluon polarized PDFs could come from a measurement of the heavy-quark contribution to the polarized structure function,
in a manner similar to studies at HERA for the unpolarized case~\cite{Hekhorn:2018ywm,Anderle:2021hpa}, though theoretical precision is somewhat limited here.

\subsubsection{Nuclear PDFs}
\label{subsec:nuc_PDFs}

The EIC will be capable of colliding (un)polarized light ion beams and unpolarized heavier ions with beams of electrons, and potentially, positrons. Inclusive NC DIS cross section
measurements are envisioned using $^4$He, C, Ca, Au and Pb nuclei. Their kinematic coverage will roughly double that of currently available data, both at low $x$ and at high $Q^2$,
see, {\it e.g.}, Fig.~7.66 in~\cite{AbdulKhalek:2021gbh}. As with the DIS program involving proton collisions, systematic and statistical uncertainties are projected to be small in
comparison with previous experiments.  These measurements are therefore expected to constrain quark and gluon nuclear PDFs to unprecedented precision. The gluon nuclear PDF could
be further constrained by heavy-flavor cross section measurements, obtained by tagging the decay products of $D$ mesons originating from charm fragmentation.

The impact of NC DIS cross sections in electron--ion collisions on nuclear PDFs was studied in Sect.~7.3.3~of Ref.~\cite{AbdulKhalek:2021gbh} and in Ref.\cite{Khalek:2021ulf}.
In~\cite{AbdulKhalek:2021gbh}, similarly to the case of unpolarized and polarized proton PDFs, pseudodata were included in three different frameworks, namely EPPS~\cite{Eskola:2016oht},
nCTEQ~\cite{Kovarik:2015cma} and nNNPDF2.0~\cite{AbdulKhalek:2020yuc}. 
It was found that EIC NC DIS cross section measurements could reduce the quark and gluon PDF uncertainties for nuclei in a wide range of atomic mass
values both at small and large $x$, by up to a factor of two. The reduction is such that nuclear PDF uncertainties may no longer encompass the difference between predictions obtained with
a free proton or with a proton bound in a nucleus, {\it e.g.}, as currently assumed to be the case when modeling the interactions of ultrahigh energy cosmic neutrinos with matter~\cite{Khalek:2021ulf}. The impact of
heavy-flavor production was studied in~\cite{Aschenauer:2017oxs}, where a similar reduction of the gluon nuclear PDF uncertainty was found at large $x$.

Because the EIC will have the capability to operate with a range of nuclei, from deuterium to lead, the dependence of nuclear PDFs on the atomic mass number $A$ will be investigated. Current parametrizations assume that this dependence is continuous, and determine it by analyzing data for different nuclei at the same time. The abundance of EIC
measurements will make it possible to determine nuclear PDFs independently for each nucleus; the dependence on $A$ could therefore be studied {\it a posteriori}. Finally, because proton
and ion beams will be used in a consistent experimental framework, the level of sophistication of PDF analyses may need to improve, in particular to allow for a combined, simultaneous
determination of proton and nuclear PDFs. This may reduce inaccuracies that follow from using the former as input to the latter and {\it vice versa}: nuclear PDFs in the analysis of
nuclear data included in proton PDF determinations; and proton PDFs as the boundary condition for analyses of nuclear PDFs.


\subsection{The Large Hadron electron Collider (LHeC) \label{sec:LHeC}}
{\it Leading authors: N. Armesto, D. Britzger, C. Gwenlan, M. Klein, F. I. Olness\\ \vspace{6pt}}

The proposed Large Hadron electron Collider
experiment (LHeC)~\cite{Jacob:1984vnf,LHeCStudyGroup:2012zhm,LHeC:2020van,Andre:2022xeh} at CERN
will provide a unique set of electron-proton/nucleus collision 
data. It will afford superior sensitivity to PDFs and related
subjects through highly precise measurements of neutral-current and charged-current
deep-inelastic scattering (NC and CC DIS) cross sections, jet-production cross
sections in DIS, as well as heavy-flavor cross sections in NC and CC
DIS.
The LHeC experiment can (only) be realized in the 2030s at the
HL-LHC, and it is also the \emph{cleanest high-resolution
  microscope} that can be attained in the next decade due to its
unprecedented resolution of the partonic constituents and dynamics in hadronic
matter down to $x$-values as small as $10^{-6}$, and up to $x\sim0.9$.

The LHeC experiment will add to the HL-LHC a new high-energy high-intensity electron
accelerator based on an energy-recovery-linac (ERL)
technology~\cite{Angal-Kalinin:2017iup,Adolphsen:2022ibf,Andre:2022xeh}, which provides an electron beam energy, $E_e$, of 50 to 60\,GeV.
The electron beam will be collided with one of the proton beams from
the HL-LHC, thus resulting in an $ep$ center-of-mass energy of
1.3\,TeV.
Further running modes will provide positron-proton ($e^+p$),
lepton-nucleus ($e^\pm$A), proton-proton or nucleus-nucleus
collision data~\cite{Andre:2022xeh}.
For $ep$ collisions, the luminosity will reach
$10^{34}\,\text{cm}^{-2}\text{s}^{-1}$, so the LHeC could provide
about $50$\,fb$^{-1}$  during an initial 3-year run (which would be equivalent to 50 times the
entire accumulated HERA data set), and will finally reach an integrated
luminosity of about 1\,fb$^{-1}$ after the HL-LHC era.
The data taking of $ep$ collisions with the LHeC
experiment will take place at LHC interaction point 2 (IP2) and
will be performed concurrently to the $pp$ data taking with ATLAS, CMS and
LHCb at the other three IPs.
Recently discussed new considerations on the accelerator, and particularly
on the design of the interaction region, explore a unique three-beam interaction
point, where lepton-hadron and hadron-hadron collisions can be
recorded with a single experiment~\cite{Andre:2022xeh}.
Together with a symmetric detector design, as is commonly used at
hadron colliders, the LHeC physics program could further comprise the
physics of the ALICE3 program, and that would even benefit  
considerably from the improved calibration that can be obtained from the $ep$ collision data.
The following discussion will focus on the LHeC; however,
essentially all the results carry forward to the FCC-eh~\cite{FCC:2018vvp},
which is designed to utilize the same ERL technology, and would further extend the
rich physics program of the LHeC to even higher energies.

\begin{figure}[htbp!]
    \includegraphics[width=0.38\textwidth,trim={0 70 0 40},clip]{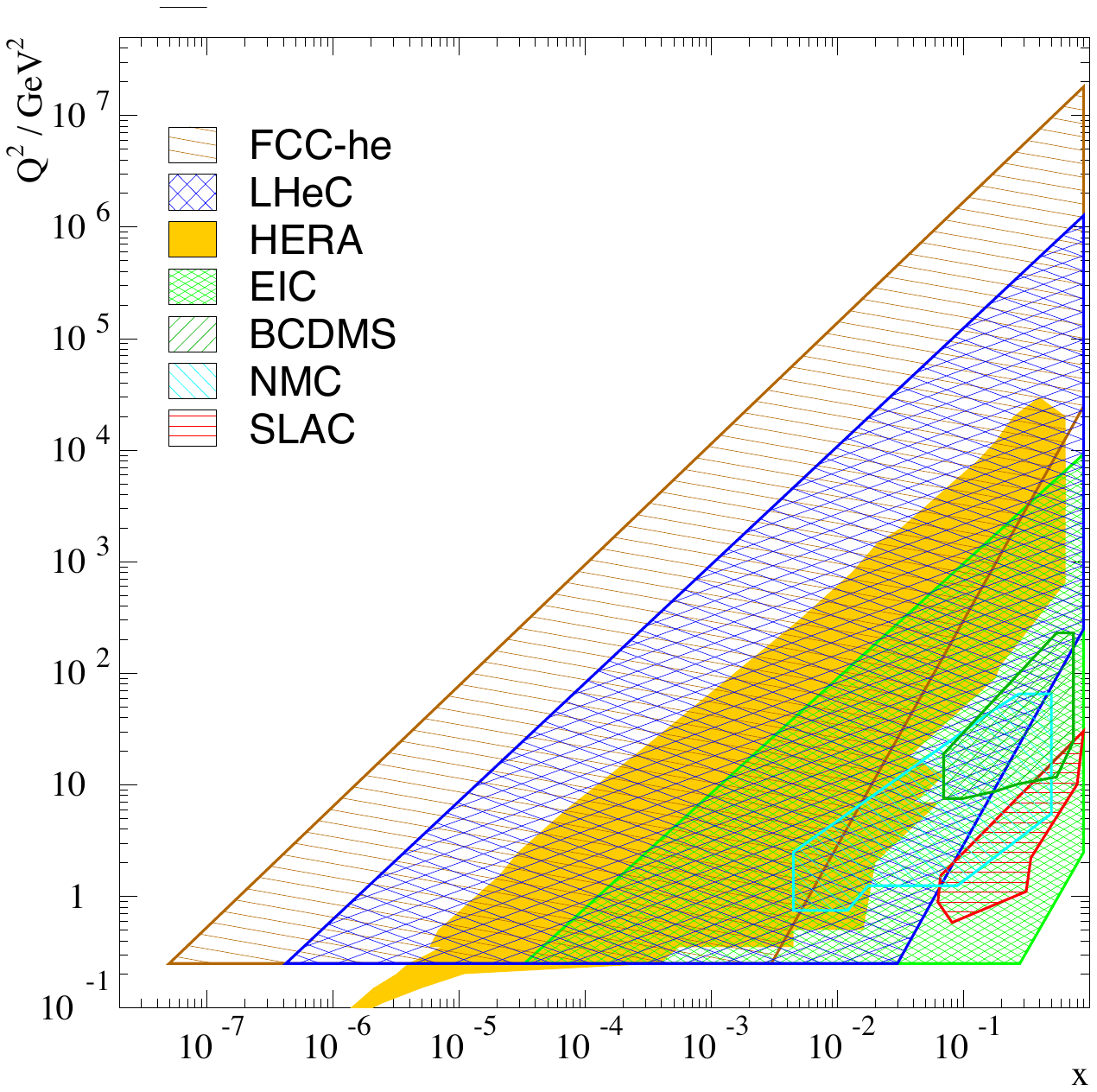}
    \hspace{0.05\textwidth}
    \includegraphics[width=0.52\textwidth]{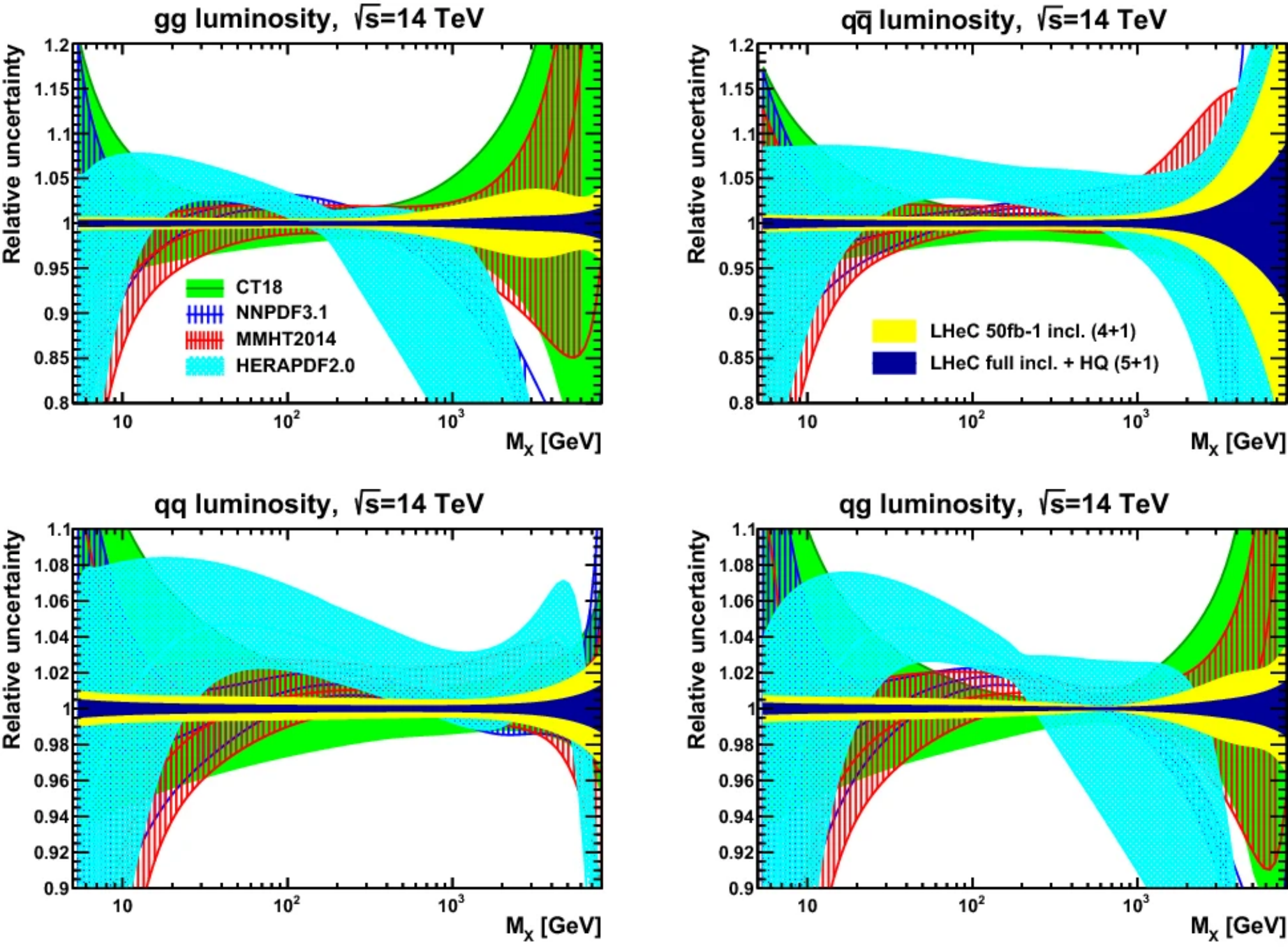}
    \caption{
      Left:
      Coverage of the kinematic plane in deep inelastic lepton-proton scattering
      by some initial fixed target experiments (SLAC,NMS, BCDMS), and
      by the $ep$ colliders:
      the EIC (green), HERA (yellow), the LHeC (blue) and the FCC-eh (brown).
      Figure from Ref.~\cite{LHeC:2020van}.
%
    Right: Expected precision for the parton-parton luminosities as a
      function of $M_X$ in Drell-Yan scattering at the LHC at
      $\sqrt{s}=14\,$TeV for three recent PDFs (shaded areas) and for PDFs
      from LHeC (full areas)
      shown for  an initial 3-year LHeC run (yellow)  
      and the full LHeC data set (dark blue). 
      Figure from Ref.~\cite{Andre:2022xeh}.
    }
    \label{fig:lhec}
\end{figure}

Studies on the expected sensitivity of LHeC data on PDFs were presented in
Refs.~\cite{LHeCStudyGroup:2012zhm,LHeC:2020van,AbdulKhalek:2019mps}, where
simulated NC and CC DIS data, including a full set of statistical and
systematic uncertainties~\cite{Klein:1564929,LHeC:2020van}, were investigated.
The coverage of the $\{x,Q^2\}$  kinematic plane of the LHeC $ep$ data is displayed in
Fig.~\ref{fig:lhec} (left) and compared to HERA, EIC, FCC-eh and
fixed-target experiments.
The data at the LHeC span a considerable kinematic range in $Q^2$ up to 
$10^6\,\rm{GeV}^{2}$, and $x$ in the range of
$10^{-6}\lesssim x\lesssim0.9$.
The measurements of inclusive NC and CC DIS cross sections at the LHeC benefit from the excellent
calibration opportunities in $ep$ collider experiments, from
high-acceptance detectors with modern detector
technologies~\cite{Andre:2022xeh}, from sophisticated data analysis
algorithms~\cite{Arratia:2021tsq}, and, of course, from high
statistical precision. Typical total uncertainties in the bulk
kinematic region will be of the order of 0.8 to 1.4\,\%.
Using such inclusive NC and CC DIS cross section measurements (which are expressed as combinations of the structure functions $F_2$, $xF_3$ and $F_L$)
as well as heavy quark production, the partonic structure of the
proton (and nuclei) can, for the first time be completely resolved in a single experiment.
The high energy collisions allow weak
probes ($W^\pm$, $Z$) to dominate the interaction at larger $Q^2$
values, which permits the up and down sea-quark PDFs and the valence
quark distributions to be resolved in the full range of $x$. Data
with different longitudinal electron-beam polarisation ($P_e=-0.8$, 0,
or +0.8) also enhance the sensitivity.
Obviously, independent data from the $pp$ experiments will further 
improve the PDFs, but also introduce new theoretical challenges.
The possibility to take positron-proton collision data greatly enhances the
precision determination of the down-quark PDF.
Dedicated data at different $ep$ center-of-mass energies give access
to the longitudinal structure functions $F_L$~\cite{LHeC:2020van}.

The gluon PDF is insufficiently known today, while it is of crucial importance for
precision Higgs, electroweak and top-quark physics at the
(HL-)LHC~\cite{Dainese:2019rgk}. The large-$x$ gluon may be
important for new physics searches.
The LHeC will constrain the gluon to percent accuracy for all $x$ values probed, by using a variety
of measurements, primarily from scaling violations ($\partial F_2/\partial\log Q^2$) as well as the longitudinal structure function $F_L$.
The measurement of jet cross sections in the Breit frame provides a further 
constraint on the gluon, since jets are predominantly initiated in the
boson-gluon fusion channel.

The size of the strange quark PDF is a long-standing puzzle.
Measurements ranging from fixed-target to collider experiments have not resolved this
important question, and the $x$ dependence of $xs(x,Q^2)$ is rather unknown, 
and it may differ from that of $x\bar{d}$ or $x(\bar{u}+\bar{d})$.
A direct measurement of $xs(x,Q^2)$ and the resolution of the complete
light-quark struture of the proton over a wide $x$ range is a
fundamental goal of the LHeC.
To cite one example, the precise extraction of the strange PDF can be
performed directly using the charm production process in CC DIS
($Ws\to c$)~\cite{xFitterDevelopersTeam:2019ygc}.


The LHeC will provide unprecedented precision measurements on heavy-quark ($c$ and $b$) production to resolve  a variety of outstanding questions: 
To what extent do the universality and factorization
theorems work in the presence of heavy quarks?
Are the current theoretical tools sufficient to address the 
multi-scale paradigm we encounter when adding new heavy-quark mass scales? 
Are charm quarks radiatively generated, or is there also
an intrinsic charm quark component in the proton?
Using  charm and beauty tagging 
with high precision in NC $ep$ scattering, 
the LHeC can completely resolve  components of the proton by flavor.   
%
%
It will directly access top
quark production in a DIS environment, 
allowing for single top production ($W b \to t$), 
top pair production ($g \to t\bar{t}$), 
and even investigation of the top-quark PDF.
%

In addition to probing the gluon PDF (see above), jet production cross sections in NC DIS in the Breit frame
at the LHeC will have high sensitivity to the
strong coupling constant
$\alpha_s(\mu)$~\cite{H1:2017bml,LHeC:2020van},
since jet cross sections are proportional to $\mathcal{O}(\alpha_s)$
already in leading-order QCD.
At the LHeC, jets with transverse momenta from 3\,GeV up to 500\,GeV
will be recorded.
Due to the over-constrained kinematics in NC DIS, the jet energy
scale can be calibrated with high precision to reduce the uncertainty
below 0.3--0.5\,\%, a value 
significantly smaller than in the present LHC experiments (in part also
because of the absence of pile-up and underlying event).
This translates into an uncertainty of about 1 to 5\,\% on the jet
cross sections in the Breit frame~\cite{LHeC:2020van}.
In a simultaneous PDF+$\alpha_s$ fit, where inclusive DIS and jet
pseudodata were included,  an uncertainty in the strong coupling
constant of  $\delta\alpha_s(m_Z)= \pm0.00018$ was projected~\cite{LHeC:2020van}, 
which is a factor of 6 smaller than the present world average value. It will be a
challenge to match such experimental precision with equally accurate
theoretical predictions (c.f.\ Sect.~\ref{sec:Theory}).
Related measurements of the hadronic final state, like event shapes,
$N$-jettiness observables, jet substructure observables or multi-jet
cross sections, may all be included in PDF determinations and, commonly,
provide predominantly a sensitivity to the gluon distribution, or to the
valence quarks at high $x$.

Interestingly, the large luminosity of the LHeC would provide high
experimental precision at large $x$, where inclusive DIS data are
sensitive to $\alpha_s$ through scaling violations, as well as enable precision
measurements of $F_L$ at high $y$.
Consequently, the strong coupling constant can be determined together
with the PDFs already from inclusive NC and CC DIS data alone, something,
that was not possible with HERA data~\cite{H1:2015ubc}.
A PDF+$\alpha_s$ analysis of inclusive DIS pseudodata yields an
uncertainty in $\alpha_s$ of $\pm0.00022$, which again imposes a real
challenge to provide accurate theoretical predictions at N3LO or
even beyond.
These studies underline the extraordinary high precision of the
inclusive DIS data from LHeC to QCD phenomena, which are otherwise
inaccessible experimentally.

Many low-$x$ and high-$x$ phenomena can be studied at the
LHeC due to the high luminosity of the accelerators, the large acceptance of the LHeC detector, and the high $ep$
center-of-mass energy.
In addition, since LHeC processes include only one initial-state hadron (in contrast to LHC
$pp$ data), the PDF determinations are free from the low-$x$--high-$x$ correlations, and these two extreme regions can be studied separately, with high precision.
The very high luminosity leads to ample
statistics in the large-$x$ region at such a high $Q^2$ that higher
twist effects become negligible. This region is especially
important for constraining BSM signatures with large
mass scales at the LHC.
At small $x$, the gluon and sea quark densities,
as discovered at HERA, rise so much that nonlinear and
possibly saturation effects may become manifest~\cite{LHeC:2020van}. The LHeC
can study them reliably in both $ep$ and $e$A collisions.
With new measurements of diffractive DIS cross sections, the field of diffractive
parton distribution functions will gain new interest~\cite{Armesto:2019gxy,LHeC:2020van}.

Beyond the collinear PDFs, semi-inclusive measurements of
jets and vector mesons, and especially exclusive vector meson production and Deeply Virtual Compton Scattering (DVCS),
the latter a process established at HERA, will shed light on 
the transverse structure of the proton in a new kinematic
range. These measurements allow us to access the Wigner
distribution $W(x, k_\text{T} , b_\text{T} )$; one can think of it as the
``master'' parton distribution. When integrating the Wigner
distribution over the transverse momentum ($k_\text{T}$), one obtains
a Generalized Parton Distribution (GPD) $f_\text{GPD} (x, b_\text{T})$,
while if we integrate over the impact parameter ($b_\text{T}$), one
obtains a Transverse Momentum Dependent (TMD) PDF
$f_\text{TMD}(x, k_\text{T})$.
Due to the considerably higher $ep$ center-of-mass energy, the LHeC
will investigate both TMDs and GPDs down to much lower $x$ and higher $Q^2$ than
the EIC, and thus provide a complementary perspective  (c.f.\ Sect.~\ref{sec:TMDs}), and shed light on their evolution with $x$ and $Q^2$.
Precision measurements of lepton-jet decorrelation
observables, measured in the laboratory rest frame,
may be sensitive probes of TMD dynamics~\cite{Liu:2018trl,H1:2021wkz}, as well as various event shapes.

While HERA inclusive NC and CC DIS data have relevant sensitivity to PDFs, their
sensitivity to further parameters in so-called PDF+$X$ fits is rather
limited. For example, $X$ could be $\alpha_s$~\cite{H1:2015ubc} or 
electroweak parameters~\cite{H1:2018mkk}.
The high luminosity of the LHeC will change that picture significantly,
and, as just discussed in the context of $\alpha_s$, above, the LHeC inclusive DIS
data will have significant sensitivity to parameters other than PDFs.
The sensitivity of inclusive DIS data to electroweak
paramaters was studied in Refs.~\cite{Britzger:2020kgg,LHeC:2020van} using
PDF+$X$ fits, where N2LO QCD and NLO EW corrections were employed.
It is found, that in the on-shell renormalization scheme, the mass of
the $W$-boson can be determined with an uncertainty of $\delta
m_W=\pm6\,$MeV, which is at a level where EW theory uncertainties are
significant.
More interestingly, the leptonic effective weak mixing angle at the
mass of the $Z$-boson can be determined with an uncertainty of
$\delta\sin^2\theta^{\rm eff}_{\text{W},f}=\pm0.00015$~\cite{Britzger:2020kgg}, which is of
comparable size to the LEP+SLD combination~\cite{ALEPH:2005ab} or
the HL-LHC prospects~\cite{Azzi:2019yne}.
Even parameters contributing beyond the leading-order formalism can be tested with LHeC
inclusive DIS data, and, for example, the oblique parameters $S$, $T$
and $U$~\cite{Peskin:1991sw} can be studied, as well as modifications
to the higher-order form factors $\rho_{\text{NC,CC},f}$ or
$\kappa_{\text{NC},f}$~\cite{Britzger:2020kgg}.
The prospects for the FCC-eh are, of course, even more promising due
to increased $\sqrt{s}$ and $\mathcal{L}$~\cite{Britzger:2022abi}.
The sensitivity to other quantities can be considered as well, such as the proton radius,
contact interaction~\cite{LHeCStudyGroup:2012zhm} or EFT parameters, and these will provide rich
physics opportunities with PDF+$X$ studies.
Moreover, with the inclusion of HL-LHC $pp$ data, combined fits of PDFs with
SM/BSM parameters will gain a considerable attention in the 2030s.

Although PDFs in the LHC kinematic range can also be constrained from (HL-)LHC data
themselves~\cite{AbdulKhalek:2018rok} (see also
sect.~\ref{sec:PDFLHC}), the importance of constraints from an
independent experiment should not be underestimated.
Already today, many LHC measurements at the LHC are limited by systematic uncertainties, and those affect the precision of the PDFs.
A valid application of such PDFs for LHC phenomenology is therefore
non-trivial (c.f.\ Sect.~\ref{sec:MethodologyExpSystUncertainties}), 
or requires comprehensive simultaneous PDF+$X$ analyses
(see Sect.~\ref{sec:PDFplusX}). For the FCC-hh, the small-$x$ dynamics will affect production of particles with masses $\mathcal{O}(100)$\,GeV, including Higgs~\cite{LHeC:2020van}.
Consequently, an independent DIS experiment to determine the proton PDFs is
of high importance for achieving the physics goals of the
HL-LHC $pp$ program.
The projected uncertainties of the parton luminosities at the
HL-LHC in $pp$ collisions at $\sqrt{s}=14$\,TeV of LHeC PDFs are displayed in
Fig.~\ref{fig:lhec} (right).
We project that the uncertainties will reduce by an order of
magnitude, compared to modern PDFs, and the improvement is
particularly pronounced at the electroweak scale.
The improvement in PDF uncertainties afforded by the LHeC can also be predicted to exceed
those of the HL-LHC PDFs, in particular for applications in SM phenomenology.
As an example, it was studied in Ref.~\cite{Azzi:2019yne}, that LHeC data
would reduce the PDF uncertainty in the measurement of the $W$ boson by ATLAS to
only $\pm1.6$\,MeV, while it would be $\pm3.7$ to $\pm5.8$\,MeV with HL-LHC data. 

While the LHeC can completely resolve the proton PDF
flavors without using any nuclear data, an option of lepton beam scattering on the 
LHC heavy-ion beam would allow exploration of nuclear PDFs as well. $e^\pm$A collisions at the LHeC~\cite{LHeCStudyGroup:2012zhm,LHeC:2020van} will be performed at $\sqrt{s}\simeq 0.8$\,TeV per nucleon (for Pb) with per nucleon instantaneous luminosities $\sim 7 \cdot 10^{32}\,\text{cm}^{-2}\text{s}^{-1}$. They will allow, as in $ep$, complete unfolding of the PDFs of a single nucleus for the first time, without the use of fixed target or hadron-nucleus data. The corresponding uncertainties will be considerably smaller than those in present global fits due to the use of single nucleus data (therefore with no need of functional initial conditions depending on nuclear size) obtained in a single experiment (thus, large tolerances are not required). The data can also be used for global fits, and the single nucleus PDFs for precision checks of collinear factorization when used for predictions in proton-nucleus and nucleus-nucleus collisions.

Studies of diffraction on nuclei present, as in the case of the EIC, the challenge of forward instrumentation required to distinguish coherent from incoherent diffraction. If such separation can be achieved, diffractive nuclear PDFs will be measured in a large kinematic domain, comparable to that in $ep$ collisions~\cite{Armesto:2019gxy,LHeC:2020van}. Also nuclear GPDs and TMDs will be studied in the nuclear case, using the same observables employed in $ep$. Finally, the eventual discovery and verification of the current explanation of the non-linear saturation regime of QCD as a density effect requires both decreasing $x$ and increasing A, making $e$A collisions essential.

As a final remark on $e$A collisions, precise knowledge of the nuclear partonic structure in the collinear regime and beyond -- in a kinematic region matching that of the corresponding hadronic colliders, and of the QCD dynamics at small $x$ or high energies, is central for heavy-ion collisions at the LHC or FCC-hh. The characterization of the hot dense medium produced in ion-ion collisions, the Quark-Gluon Plasma (QGP), suffers from large uncertainties derived from the present lack of knowledge on these aspects~\cite{LHeC:2020van}. In studies of the nuclear medium, many observables taken as signatures of QGP formation --- the most prominent of them being the ridge ---
are found in smaller colliding systems, both proton-proton and proton-nucleus ones. Measurements at the LHeC and the FCC-eh can clarify these issues. 

Undoubtedly, HERA had an outstanding impact on our present knowledge of
the proton structure.
The LHeC, with its 1000~times larger luminosity (and higher
center-of-mass energy and kinematic reach), will equally advance the field and will
provide the relevant experimental input data for precision PDF physics in
the 2030s.
Furthermore, such independent PDFs are of crucial importance to
achieve the physics goal of the HL-LHC program.


\subsection{PDFs for neutrino phenomenology
\label{sec:PDFNeutrinoPheno}}
{\it Leading authors: T. J. Hobbs, K. Xie, and B. Zhou \\ \vspace{6pt}}

 PDFs play a crucial role in neutrino interactions above a few GeV, a kinematical region which is dominated by DIS between neutrinos and the target nucleus~\cite{Gandhi:1995tf, Gandhi:1998ri, CooperSarkar:2011pa, Connolly:2011vc, Chen:2013dza, Bertone:2018dse}. The precision of theoretical predictions for the relevant cross sections is pivotal for medium-, high-, and ultra-high-energy neutrino physics and astrophysics. The relevant experiments include DUNE~\cite{DUNE:2015lol}, Super-Kamionkande~\cite{SK_web}, Hyper-Kamionkande~\cite{Hyper-Kamiokande:2018ofw}, IceCube~\cite{IceCube_web}, KM3NeT~\cite{Adrian-Martinez:2016fdl}, Baikal-GVD~\cite{Baikal-GVD:2018isr},  IceCube-Gen2~\cite{IceCube-Gen2:2020qha}, ANITA~\cite{ANITA:2010hzc}, ARA~\cite{ARA:2015wxq}, and GRAND~\cite{GRAND:2018iaj}.

Above $\sim\! 100$~GeV and $\sim\! 100$~TeV, charm- and top-quark production are especially important for DIS~\cite{Quigg:1986mb}, but, even at lower energies, the precision of cross section calculations depends in part on a proper accounting for heavy-quark mass effects~\cite{Gao:2021fle}.  At leading order, a crude approximation of mass dependence is realized by slow rescaling of the quark's light-cone momentum fraction in the PDFs~\cite{Phillips:1982bn}.  At higher orders, various formalisms have been developed~\cite{Collins:1998rz, Aivazis:1993pi, Buza:1996wv, Thorne:1997ga, Cacciari:1998it} to include mass effect to arbitrary accuracy. 

At PeV and EeV energies, DIS probes kinematical regions of very small $x$ and large $Q^2$. PDFs have very limited data from colliders and fixed-target experiments~\cite{Hou:2019efy} with direct sensitivity to this region; as a result, there is currently significant dependence on extrapolations to these largely unfitted regions. Moreover, at very small $x$, the perturbative expansion is not stable, such that resummation corrections must be included in the DGLAP formalism~\cite{Bertone:2018dse}.  The color dipole approach~\cite{Gluck:2010rw, Goncalves:2010ay, Albacete:2015zra, Arguelles:2015wba} provides an efficient way to account for resummation and saturation effects. For different formalisms and input PDFs, the predicted cross sections can differ by as much as a factor of a few in the ultra-high energy regime. 

Nuclear effects on PDFs are also important, as DIS mostly happens on the nucleus for neutrino detection (water/ice, argon, lead, {\it etc}.) and via propagation through the Earth (iron, oxygen, silicon, {\it etc}.).  An important effect at higher energies (related to the behavior of nuclear PDFs at low $x$) is nuclear shadowing~\cite{Frankfurt:2011cs}, {\it i.e.}, the relative depletion of nuclear structure functions as compared to their free-nucleon counterparts. Nuclear PDFs, such as \cite{Kovarik:2015cma, Eskola:2016oht, AbdulKhalek:2019mzd, Walt:2019slu}, explore this phenomenology systematically, as discussed in Sec.~\ref{sec:nuc_PDFs}. DIS cross sections calculated using nuclear PDFs show 5-15\% suppression at PeV energies, but the uncertainty of the nuclear corrections is still large~\cite{Bertone:2018dse, Garcia:2020jwr, Klein:2020nuk}.  Another nuclear effect is isospin dependence. In $W$-boson exchanges, neutrinos interact differently with protons and neutrons, while current works treat the Earth as an isoscalar target, which is not true for some nuclear components of the Earth. It would be very interesting to examine the nuclear isospin dependence with various models or assumptions, as discussed in Sec.~\ref{sec:nuc_PDFs} and \ref{subsec:nuc_PDFs}. 

On the other hand, the photon PDF is also important for neutrino phenomenology, as it is the most important input for calculating neutrino-nucleus $W$ boson production~\cite{Seckel:1997kk, Alikhanov:2015kla, Zhou:2019vxt, Zhou:2019frk}, $\nu_\ell + A \rightarrow \ell^- + W^+ + X$, where a neutrino couples to an in-nucleus photon through a charged lepton or $W$ boson split from the neutrino. See Fig.~1 of Ref.~\cite{Zhou:2019vxt} for relevant diagrams. The cross sections of this process are the second largest for high-energy neutrinos, reaching 5--10\% of DIS on water and 10--15\% on iron~\cite{Zhou:2019vxt, Zhou:2019frk}. Therefore, a precise parametrization of the photon PDF in various nuclei is important for accurate determinations of the cross sections for this process.

Dimuon events (involving two energetic muons emanating from one neutrino interaction) at accelerator-based neutrino experiments have been very important to measuring the strange-quark PDF in the region of $x\!\gtrsim\! 0.01$ and $Q\!\lesssim\!10$~GeV.  Recently, Ref.~\cite{Zhou:2021xuh} proposed that high-energy neutrino telescopes like IceCube and IceCube-Gen2 can achieve a higher level of sensitivity in dimuon detection due to the small vertical spacing between the detector's digital optical modules. This work further predicted that IceCube and IceCube-Gen2 can detect $\simeq\! 1000$ dimuon events in 10 years, and that these events can probe the strange-quark PDF in the region of $x\!\gtrsim\! 0.01$ and $10\!\lesssim\! Q\!\lesssim\! 100$ GeV.  These dimuons events can also be used to detect production of $W$ bosons as another means of probing the nuclear photon PDF as discussed above.

The FASER$\nu$~\cite{FASER:2019dxq} experiment --- one of the experiments in the CERN Forward Physics Facility area discussed in Sec.~\ref{sec:ForwardPDF} --- is designed to detect neutrinos produced in proton-proton collisions in the far-forward region of the ATLAS detector. Neutrinos produced from decays of heavy mesons such as the $B, D$ {\it etc}., can provide information on the heavy-flavor and gluon PDFs. Considering the far-forward kinematics, the FASER$\nu$ measurement is able to probe the PDFs down to $x\!\sim\!10^{-8}$~\cite{Anchordoqui:2021ghd,Feng:2022inv}, a region which has not been probed by existing experiments. In this scenario, new data would provide fresh insights into how the PDFs behave at very low $x$, possibly informing a new understanding of QCD, especially with respect to the small-$x$ behavior as discussed in Sec.~\ref{subsec:smallx}.

Complementary to the considerations discussed above for neutrino scattering at the TeV
scale and beyond, PDFs also play an important role at lower energies in GeV-scale experiments like the
upcoming DUNE/LBNF effort~\cite{DUNE:2015lol} at Fermilab.
Long-baseline experiments like DUNE depend upon precise control over the neutrino-nuclear interaction
over a wide range of scattering energies, $E_\nu$, to achieve their target sensitivities to the neutrino-mass
hierarchy and a possible $\mathrm{CP}$-violating phase, $\delta_\mathrm{CP}$, in the neutrino sector.
In the case of DUNE, the anticipated neutrino flux will peak near $E_\nu\!\sim\! 2.5\,\mathrm{GeV}$, with
a substantial tail to higher energies. In this region, the neutrino-nuclear cross section must be
determined from a complicated mix of underlying processes, including quasi-elastic scattering,
resonance excitation, and deeply-inelastic scattering. The latter of these dominates the cross
section at successively higher values of $E_\nu$, but, in the few-GeV region, it has an important
dependence on various nonperturbative effects, including contributions from higher-twist
({\it i.e.}, twist-4) and target-mass corrections. These must be systematically assessed and controlled
in the delicate resonance-to-DIS transition region in a context in which nuclear effects are
also critical. For this reason, nuclear PDF studies and extrapolations to the lower $W$ and $Q^2$
values of greatest relevance to DUNE will be a priority for enhancing understanding of the DUNE neutrino-nuclear
program.

\subsection{Forward (and ultra-high energy) scattering processes at the LHC \label{sec:ForwardPDF}}
{\it Leading authors:  M. Guzzi, L. A. Harland-Lang,  M. Hentchinski, K. Xie, with a contribution from C. Loizides
\vspace{6pt}}

The kinematic regime of scattering processes in the very forward region is outside of the range of genuine validity of DGLAP picture. One may ask up to which values of the momentum fractions or rapidities the DGLAP picture remains valid. Multiple experiments have been proposed to study forward production in the future. Here we focus on LHCb, FPF, and ALICE FoCal as examples of forward physics experiments in which calculations in the collinear factorization picture can provide useful guidance and in-depth tests of QCD.

The LHCb experiment is designed for precision physics in the forward region. Properties of final-state particles in the forward configuration can be used to probe PDFs in regions at both small and large $x$  ($x\sim (Q/\sqrt{s})e^{\pm y}$). Measurements of heavy-flavor charm and bottom production at LHCb provide constraints on the gluon and heavy-flavor PDFs at $x\sim 5\times 10^{-6}$~\cite{PROSA:2015yid}. Such a small value of $x$ is currently not covered by other LHC experiments. The large $x$ region has been recently studied~\cite{LHCb:2021stx} at LHCb
where cross section measurements of $Z$ bosons to in association with a charm quark were used to probe the existence of an intrinsic charm (IC) component of the proton. 
In addition, a recent study from the CTEQ-TEA group~\cite{Hou:2019efy}, has shown high sensitivity
of Drell-Yan dilepton production at LHCb to quark and antiquark PDFs at small $x$, especially strangeness. Future high-luminosity measurements at LHCb will be critical to set stronger constraints on PDFs~\cite{Deng:2020sol} and to explore small-$x$ dynamics~\cite{Celiberto:2018muu}.

The ALICE detector at the LHC, equipped with a dedicated forward calorimeter system (FoCal)~\cite{ALICE:2020mso} will allow us to start a new program to investigate small-$x$ gluon distributions of hadrons and nuclei.
FoCal is designed as a highly granular Si+W electromagnetic calorimeter combined with a conventional sampling hadronic calorimeter covering pseudorapidities of $3.4<\eta<5.8$.
Its performance is optimized to measure isolated-photon spectra at forward rapidity in the range of about $4 < p_{T} < 20$~GeV/$c$ with high precision even at the lowest momenta.
This kinematic reach with photons is equivalent to constraining the gluon distribution in Pb nuclei down to Bjorken-$x$ of about $10^{-5}$ over a large range of $Q^2$~\cite{Hentschinski:2022xnd}.
In addition to the photon measurements, FoCal will allow us to measure photon-jet and jet-jet correlations, as well as J/$\psi$ production in ultra-peripheral collisions. These processes are strongly sensitive to non-linear effects at small-$x$.

The Forward Physics Facility (FPF) is a proposal at CERN to complement the existing experimental program with a range of far--forward detectors that will be in particular be able to collect a significant data sample of neutrinos produced due to particle production in the central ATLAS detector. The FASER, FASER$\nu$ and SND experiments will begin taking data in 2022, while there is a dedicated proposal to extend these by creating space in the far-forward region for a suite of upgraded experiments that would run during the High Luminosity LHC (HL-LHC) era, see~\cite{Anchordoqui:2021ghd,Feng:2022inv}. Along with the range of BSM and neutrino physics studies that this will permit, there is promising potential for information about the proton and nuclear PDFs to be provided by the FPF. 

In more detail, the neutrino flux will be created via very forward production of particles in proton-proton collisions at ATLAS, such as light hadrons or charmed mesons. The flux will be sensitive to the proton PDFs at both rather low and high $x$. Measuring these neutrinos at the FPF thus has the potential to probe the PDFs (e.g. the gluon) in a rather poorly constrained low-$x$ region, where a range of BFKL resummation and nonlinear effects may be present. At high $x$, there is a distinct sensitivity to possible intrinsic charm in the proton, which is theoretically expected to be enhanced in the high-$x$ region. Several studies have investigated the possible existence of this intrinsic charm, including the recent measurements of $Z$+charm production by the LHCb experiment~\cite{LHCb:2021stx}, which hints at its presence. However, the analyses arrive at different conclusions about the allowed bounds on the IC \cite{Jimenez-Delgado:2014zga, Hou:2017khm, Ball:2022qks}, and hence the FPF could shed light on this unresolved question.

In addition, the FPF detector will effectively operate as a neutrino-induced deep-inelastic scattering experiment with TeV-scale neutrino beams. Measurements of the resulting DIS structure functions will provide a valuable handle on the partonic structure of both nucleons and nuclei, in particular concerning quark flavor separation. Of particular note is the potential for measurements of charm-tagged neutrino structure functions, which would provide further information about the possible tensions between such existing data and measurements from the LHC on $W,Z$ production. Moreover, the FPF will host not only emulsion experiments, which allow several kinds of charmed baryons and mesons to be tagged by reconstructing in detail the topology of their decays, but also experiments allowing to tag charm quarks using the dimuon signature.


\section{Theory \label{sec:Theory}}

{\it Leading authors: S.-O. Moch, B. Mistlberger, G. Magni, with a contribution from J. Bl\"umlein}\\

LHC particle physics phenomenology at the percent accuracy allows us to stringently test our understanding of fundamental interactions, and it is experimentally feasible at the LHC.
In particular, the precise measurement of observables involving highly energetic electroweak bosons, top quarks or jets of QCD radiation shed light on some of the most pressing questions of modern particle physics.
This motivates a large effort to improve our theoretical capabilities to predict hadronic scattering cross sections at the enhanced level of precision required to extract the desired information from LHC data.
PDFs are the backbone of such predictions. 
Theoretical developments will play a crucial role in future improvements of PDFs and consequently are of great importance to our aim of maximally utilizing LHC data.

\subsection{PDF evolution at N3LO}
\label{sec:n3lo}
QCD factorization allows to express observables in 
{\it ep} and {\it pp} hard scattering with large momentum transfer schematically as
\begin{equation}
\label{eq:OepOpp}
  O^{\,ep} \,=\, f_{i}^{} \,\otimes\, c_i^{\,\rm o} \, , \qquad
  O^{\,pp} \,=\, f_{i}^{} \,\otimes\, f_{k}^{} \,\otimes\, c_{ik}^{\,\rm o}\, ,
\end{equation}
where the PDFs of the proton with dependence 
on the momentum fraction $x$ are denoted by $f_{i\,}^{}(x,\mu^2)$ 
and the process dependent partonic cross sections (coefficient functions) by $c^{\,\rm o}$.
QCD factorization holds up to power corrections and is performed at the 
(renormalization and factorization) scale $\mu$, which is 
taken to be of the order of a physical hard scale. 

The scale dependence of the PDFs is governed by the well-known evolution equations~\cite{Altarelli:1977zs,Dokshitzer:1977sg,Gribov:1972ri}
\begin{equation}
\label{eq:Evol}
 \frac{\partial}{\partial \ln \mu^2} \, f_i^{}(x,\mu^2) 
 \,=\, 
 \left[ P^{}_{ik}(\alpha_s(\mu^2)) \otimes f_k^{}(\mu^2) \right]\!(x) \, ,
\end{equation}
where $\otimes$ denotes the Mellin convolution with the evolution kernels,
i.e. the splitting functions $P^{}_{ik}$.
The latter are calculable in QCD perturbation theory and, 
together with the coefficient functions $c^{\,\rm o}$, 
can be expanded in powers of the strong coupling constant $a_s \equiv \alpha_s(\mu^2)/(4\pi)$, 
\begin{eqnarray}
\label{eq:Pexp}
P \, &=& 
a_s\, P^{\,(0)} 
+\, a_s^{2}\, P^{\,(1)}
+\, a_s^{3}\, P^{\,(2)} 
+\, a_s^{4}\, P^{\,(3)} 
+\, \ldots\, , 
\\
\label{eq:Cexp}
c^{\,\rm o}_{a} &=& 
a_s^{\:n_{\rm o}} \big[ \, c_{\rm o}^{\,(0)} 
+\, a_s\, c_{\rm o}^{\,(1)} 
+\, a_s^{2}\, c_{\rm o}^{\,(2)} \,
+\, a_s^{3}\, c_{\rm o}^{\,(3)} 
+\, \ldots \,\big] \, .
\end{eqnarray}
N$^n$LO parton distribution functions are extracted from hadronic scattering cross sections by fitting cross section predictions using N$^n$LO hadronic cross sections to data. 
N2LO PDFs represent the state of the art, where the first three terms in Eqs.~(\ref{eq:Pexp}) and (\ref{eq:Cexp}) provide
the N2LO predictions for the observables (\ref{eq:OepOpp}).
Currently, this is standard approximation for many hard processes and for PDF
determinations, see~\cite{Moch:2004pa,Moch:2014sna,Moch:2015usa,Vogt:2004mw,Ablinger:2014nga,Ablinger:2017tan,Blumlein:2021enk,Blumlein:2021ryt}
for the corresponding N2LO splitting functions.

Tackling the next perturbative order - N3LO - requires significant improvements and requires an unified effort from the theoretical particle physics community. 
The benefit to the particle physics phenomenology program is nevertheless clear: a consistent extraction and application of N3LO PDFs will result in more reliable predictions of scattering cross sections and ultimately in a reduction of uncertainties due to our limited knowledge of PDFs. 
In particular, work on the four-loop splitting functions in Eq.~(\ref{eq:Pexp}) to ensure 
QCD evolution equations at N3LO accuracy is ongoing~\cite{Moch:2017uml,Moch:2018wjh,Moch:2021qrk}. 
The  massless and massive Wilson coefficients are known~\cite{Blumlein:2006be,Vermaseren:2005qc,Ablinger:2014vwa}. 
With these results, the flavor non-singlet N3LO contributions to DIS can be implemented already now, since
all ingredients are known to sufficient accuracy in the relevant range of
parton kinematics, following e.g.,~\cite{Blumlein:2006be,Blumlein:2021lmf}. 

Finally there is another piece that has to be taken into account when looking at N3LO PDF evolution: 
the matching conditions for different flavor number schemes. 
In fact, if the number of active, light flavors that are participating in the DGLAP equation changes by one unit,
the distributions do not behave in the same matter above and below the threshold: in particularly
the new quark distributions $q_{n_f+1}(x,\mu) = h(x,\mu)$ did not take part in the evolution below the threshold,
but above they do. The nontrivial contribution of these matching conditions is of order $\mathcal{O}(a_s^2)$~\cite{Buza:1996wv} 
and have been computed almost completely also at N3LO~\cite{Bierenbaum:2009mv,Ablinger:2010ty,Ablinger:2014vwa,Ablinger:2014uka,Behring:2014eya,Blumlein:2017wxd,Ablinger:2014lka,Ablinger:2014nga,Blumlein:2022ndg},
allowing for consistent N3LO PDF evolution of  the heavy quarks.

In this context, a number of programs can solve DGLAP evolution equations at N2LO~\cite{Vogt:2004ns,Salam:2008qg,Bertone:2013vaa}, 
but not yet at N3LO. The recently released \eko~\cite{Candido:2022tld,candido_alessandro_2022_5896965} is able to perform the full evolution 
up to N2LO, and it contains some ingredients needed to N3LO, such as $a_s$ running and the matching conditions;
however the implementation of $\mathcal{O}(\alpha_s^3)$ splitting function is still work in progress.

The work on the determination of all N3LO splitting functions and matching conditions, along with their implementation in public codes, is paramount and will be one of the most important development in the precision physics program of the next decade. An approximated analysis of N3LO PDFs has been recently presented in Ref.~\cite{McGowan:2022nag}. 

\subsection{N3LO Cross Sections And Perturbative Uncertainties \label{sec:N3LOXsec}}

The current frontier in QCD perturbation theory is posed by third order - N3LO - predictions. 
Calculations for partonic scattering cross sections, Eq.~\eqref{eq:Cexp}, for DIS processes at N3LO are already readily available~\cite{Ablinger:2014nga,Ablinger:2014vwa,Ablinger:2017err,Vermaseren:2005qc,Moch:2004xu,Moch:2007gx,Moch:2007rq,Moch:2008fj}.
Predictions for cornerstone LHC process at N3LO are a very active field of development and are available for key inclusive cross sections~\cite{Anastasiou:2015vya,Banfi:2015pju,Mistlberger:2018etf,Dulat:2018bfe,Duhr:2020sdp,Duhr:2020seh,Duhr:2020kzd,Duhr:2019kwi,Duhr:2021vwj,Chen:2019fhs,Dreyer:2016oyx,Dreyer:2018qbw} as well as some fully differential predictions~\cite{Cieri:2018oms,Chen:2021ibm,Camarda:2021jsw,Camarda:2021ict,Billis:2021ecs}. 
The overall picture that emerges from these computations is that corrections at N3LO are of the order of a few percent, and residual uncertainties due to the truncation of the perturbative expansion in the strong coupling constant are comparable to, or subdominant with respect to, other sources of theoretical uncertainties.
In particular, uncertainties on parton distribution functions often represent the largest of the residual theoretical uncertainties.

Currently, N3LO PDFs are not available, and computations of hadronic cross sections at N3LO are consequently performed using N2LO PDFs as inputs. While this procedure is theoretically sound, it naturally leads to the question of the phenomenological impact of N3LO PDFs on such predictions. 
To quantify the answer to this question in terms of an uncertainty due to missing N3LO PDFs, an ad-hoc procedure was introduced in Ref.~\cite{Anastasiou:2016cez}. The authors defined an uncertainty by the relative difference of computing an N2LO cross section once with N2LO and once with NLO PDFs, and reducing the size of this uncertainty by half in order to account for a perturbative reduction of the impact of N3LO over N2LO corrections. The definition is given in Section~II.B.1, Eq.~\eqref{eq:mismatch}.
Albeit ad-hoc, $\delta(\text{PDF-TH})$ results in a significant uncertainty (several percent) on N3LO cross section predictions for key LHC observables~\cite{Anastasiou:2016cez,Duhr:2019kwi,Duhr:2021vwj}. It is consequently of comparable size as the regular uncertainty associated with our current understanding of PDFs themselves.
Moving toward a better procedure of estimating the uncertainty due to the mismatch of the perturbative order of PDFs and partonic cross section calculations is very desirable.

Extracting N3LO parton distribution will require a large range of cross sections computed at this perturbative order. Some of these computations will not be available imminently.
Consequently, it is necessary to develop a scheme to set perturbative uncertainties on PDFs that takes into account the effect of fitting cross sections to predictions based on a mix of N2LO and N3LO calculations.
As the overall precision in the determination grows, it will be paramount to consistently treat uncertainties of the theoretical input cross sections.
Theory uncertainties are important even in N2LO
fits~\cite{Harland-Lang:2018bxd,Ablinger:2014vwa,Ablinger:2014nga,AbdulKhalek:2019bux,AbdulKhalek:2019ihb} and will be even more so, as we progress towards N3LO PDFs.
We refer to Section~\ref{sec:Methodology} for more details. 
Furthermore, fitting PDFs requires flexible and fast frameworks for the computation of N3LO cross sections.
The development of such framework requires the collaboration of multiple research groups and should be supported by our field. 
Another key ingredient is the availability of high-performance computing infrastructure that facilitates the complexity of PDF extractions, the requirement that gets only more complex as we advance toward N3LO PDFs. See Sec.~\ref{sec:computing}. 

\subsection{Electroweak Corrections in PDF fits \label{sec:EW}}
{\it Leading authors: L.A. Harland-Lang, T. J. Hobbs, E. R. Nocera, R. S. Thorne, and K. Xie
\\ \vspace{6pt}}

The level of  precision expected at the LHC and future colliders such as the EIC demands that electroweak (EW) contributions are evaluated at least at the next-to-leading order in the electroweak pertubative series in many theoretical cross sections. 
EW corrections, especially the QED ones, must be implemented both in the partonic cross sections and  the PDFs, including QCD evolution.   
The DGLAP equations for the PDFs can be expanded to include QED parton splittings, automatically resulting in the photon becoming a constituent parton of the proton, with the corresponding photon distribution introduced.  This new distribution leads to photon--initiated (PI) 
subprocesses, which enter as corrections to the purely QCD cross section for processes such as Drell--Yan \cite{ATLAS:2016gic}, 
EW boson--boson scattering \cite{Bierweiler:2012kw}, and Higgs production with an associated EW boson \cite{Denner:2011id}. 
In addition, semi-exclusive \cite{Harland-Lang:2016apc,Harland-Lang:2020veo} and exclusive PI production 
of states with EW couplings has a significant potential as a probe of SM and BSM physics.

The inclusion of QED corrections in the DGLAP evolution equations and the photon PDF goes back about two decades.
MRST provided the first publicly available QED set~\cite{Martin:2004dh}, using splitting kernels at $\mathcal{O}(\alpha)$ 
in QED and a model where the input photon was generated radiatively from the quarks below input. Subsequent sets  
either used similar phenomenological models~\cite{Schmidt:2015zda} or constrained the photon by utilizing the distinctly limited 
sensitivity of  DIS and Drell-Yan data to the PI channel \cite{Ball:2013hta,xFitterDevelopersTeam:2017fxf}. This automatically led to photon PDF uncertainties of 
at least $10\%$ and often considerably more.  Moreover, the distinction between the elastic and inelastic photon emission was rarely considered. 
Refs.~\cite{Martin:2014nqa,Harland-Lang:2016apc,Harland-Lang:2016kog} have shown how a more accurate determination of the input photon distribution could 
be obtained by using the elastic form factors of the proton, which are experimentally 
well determined. However, as discussed long ago in e.g.~\cite{Budnev:1975poe}, in fact the entire contribution to the photon PDF from both elastic and inelastic emissions is directly related to the corresponding  structure functions, $F_{1,2}^{\rm el}$, $F_{1,2}^{\rm inel}$, as was also discussed
in~\cite{Anlauf:1991wr,Blumlein:1993ef,Mukherjee:2003yh,Luszczak:2015aoa}. This basic idea has been realized within 
a precise theoretical framework by the LUXqed group~\cite{Manohar:2017eqh,Manohar:2016nzj}, and they were able to provide a publicly available 
photon PDF with uncertainties which are due overwhelmingly to those from the structure functions used as input. This approach
improves the precision of the photon PDF to the level of a few percent. 
Moreover, QED DGLAP splitting kernels have now been calculated to $\mathcal{O}(\alpha\alpha_S)$ \cite{deFlorian:2015ujt} 
and $\mathcal{O}(\alpha^2)$ \cite{deFlorian:2016gvk}. These are implicit in the LUXqed approach, but also easily implemented 
in DGLAP evolution codes.  

Hence, it is now possible to be far more precise and confident about the photon 
distribution, the related QED modifications to other partons, and the subsequent impact on cross section calculations. The first global PDF set 
including a photon distribution based on the LUXqed approach was produced by the NNPDF group~\cite{Bertone:2017bme}, and this  
was soon followed by QED corrected PDFs based on the MMHT14 PDFs~\cite{Harland-Lang:2019pla}. More recently, the CT group has also produced PDFs 
with QED corrections and a LUXqed-inspired photon distribution~\cite{Xie:2021equ,Xie:2021ajm}. A set based on the MSHT20 PDFs, 
using an extremely similar approach to that in~\cite{Harland-Lang:2019pla}, has appeared very recently~\cite{Cridge:2021pxm}.
The photon distributions in these sets are all based on the same underlying principle, but have differences in the details of their methodology. They each  
now have uncertainties of a few percent, and are all broadly consistent with each other, despite the differences in their approach. This represents a huge improvement in 
the knowledge of the photon content of the proton. However, care must be taken when claiming equivalently high 
precision in the corresponding PI cross sections. Studies of this type are so far often based on calculations at LO in $\alpha$, in which case they will have significantly larger 
scale variation uncertainties than the percent level uncertainty due to the photon PDF. In practice, of the processes entering global 
PDF fits, the PI contributions to off--peak lepton pair production are by far the dominant ones. For these it is most accurate to follow the approach 
of~\cite{Harland-Lang:2016lhw,Harland-Lang:2021zvr}, which applies the structure function (SF) approach to directly calculate the dominant PI contribution 
to lepton pair production away from the $Z$ peak. This provides percent level precision in the cross section prediction here, bypassing the issue of large LO 
scale variations. For many other processes, a standard EW K--factor approach can be taken (or fast interpolation grids, as presented in~\cite{Carrazza:2020gss}), 
although in the majority of cases the impact of PI production is found to be marginal at the current level of precision.

Besides the PI corrections, other EW contributions are relevant for processes used in PDF fits. 
For inclusive jet production, it is possible to use K--factors evaluated from the calculation of~\cite{Dittmaier:2012kx} (see also~\cite{Frederix:2016ost}). 
These do not include QED corrections, and therefore PI production, 
on the premise that the dominant contribution is from the pure weak corrections (a distinction that can be made in a gauge invariant way 
in this case), due to their Sudakov logarithmic enhancement. The size of the overall EW corrections, which is dominated by this source, can be as large as $\sim 10\%$ at the highest jet $p_T$ values. 
For $Z$ $p_T$ data, there is a calculation of~\cite{Denner:2011vu}. It includes mixed $\gamma q$ PI production, found to enter at the per mille level and be significantly smaller than the 
other EW corrections. The total size of the EW corrections can be as large as $\sim 20\%$ at high $p_T^{ll}$ for current data, though it is generally less than this~\cite{ATLAS:2015iiu}.
For the precision $W, Z$ data, corrections can be derived from, for example, the \texttt{MCSANC} generator~\cite{Bardin:2012jk,Bondarenko:2013nu}. The total size 
of the EW corrections is $\sim 0.5\%$ at intermediate and high masses, 
but $\sim 6\%$ in the lowest-mass region. NLO EW corrections can also be calculated using \texttt{FEWZ}~\cite{Li:2012wna} and \texttt{MG5\_aMC v3}~\cite{Frederix:2018nkq}.
For differential top-quark pair production data, EW corrections are calculated in~\cite{Czakon:2017wor} based on an earlier study in~\cite{Pagani:2016caq}. 
These include a very small contribution from the $\gamma g$-initiated channel, calculated using the LUXqed~\cite{Manohar:2017eqh} and CT18qed~\cite{Xie:2021equ}. 
In $t\bar t$ production, the differential distributions that are more sensitive to EW corrections are given in $m_{t\bar{t}}$, $p_{T,t}$, and $p_{T,t\bar{t}}$, especially at large $p_{T}$. The rapidity distributions $y_{t}$ and $y_{t\bar{t}}$ are less sensitive, although some impact can be observed at large rapidity values. 

For differential $WH$ production, the EW correction is found to be enhanced in the large-invariant-mass tail, mainly due to new channel initiated by $\gamma q$~\cite{Xie:2021equ}.
Hence, it is possible to include all significant EW corrections to processes currently involved in providing good constraints on PDFs. However, as precision rises, further calculations of combined QCD and EW contributions will be necessary. 

We also note that it can be useful to provide both the individual elastic
and inelastic photon PDF components, 
$\gamma^{\rm el}(x,Q^2)$ and  $\gamma^{\rm inel}(x,Q^2)$, with $\gamma(x,Q^2) 
= \gamma^{\rm el}(x,Q^2)+\gamma^{\rm inel}(x,Q^2)$. For example, the separate components are used for predictions for exclusive and semi--exclusive PI production~\cite{Harland-Lang:2016apc,Harland-Lang:2020veo}, 
although in this case care must be taken to also include the survival factor probability of no additional particle production due to multi--particle interactions (MPI). 
At high scales, e.g. $Q^2 = 10^4$ ${\rm GeV^2}$, the inelastic component is dominant until very high $x$, while at lower scales, e.g. $Q^2 = 10^2$ ${\rm GeV^2}$, the relative contribution from the elastic 
component is somewhat larger due to the shorter evolution length for (inelastic) $q\to q \gamma$ splitting. QED corrections for neutron PDFs
are as important as those for the proton to do a consistent fit to deuteron and nuclear fixed target data from neutrino ($\nu$N) 
DIS scattering experiments. The QED-corrected neutron PDFs automatically provide isospin violating partons, 
with $u_{(p)} \neq d_{(n)}$, and these were seen in ~\cite{Martin:2004dh} to automatically
reduce the NuTeV $\sin^2\theta_W$ anomaly \cite{NuTeV:2001whx}. The breaking of isospin symmetry may also have implications for the development of nuclear PDFs, in particular at the EIC.We also note that, in addition to the photon, it is also possible to include leptons~\cite{Bertone:2015lqa} and electroweak bosons~\cite{Bauer:2017isx,Fornal:2018znf,Han:2020uid} as the constituents of the nucleon. The effect of the former is very small for almost all processes, while the latter may have more significance at a future very high energy collider.

\subsection{PDFs and resummations at extreme momentum fractions \label{sec:Resummations}}
 \subsubsection{Large $x$ \label{sec:TheoryLargeX}}
 {\it Leading authors: A. Courtoy, D. Soper, M. Ubiali \\ \vspace{6pt}}

The large-$x$ region is the least known PDF kinematic region, as the number of experimental data that constrain $x\gtrsim 0.1$ are less in number and less precise than those that constrain the small and medium-$x$ regions. The resulting large-$x$ PDF uncertainties hamper the precision of both the signal and background theoretical predictions in the high energy tails, which are the focus of both direct and indirect searches for new physics (see Sect.~\ref{sec:bsm}). It is therefore crucial for hadron collider phenomenology to pin down the large-$x$ region. 

Parton distributions at $x > 0.1$ are also of a special interest for theoretical studies, as they can be increasingly connected to  nonperturbative and lattice QCD approaches. These theoretical techniques can be assessed by comparisons against precisely known unpolarized collinear PDFs found from phenomenological analyses and then expanded to predict less experimentally accessible quantities such as spin-dependent PDFs. The connection to lattice QCD is discussed in Sec.~\ref{sec:Lattice}.

The present data impose few little experimental constraints at $x > 0.5$, where various factors may introduce corrections to the simplest collinear factorization framework. Much of the relevant data lie at low $Q$, close to the lower boundary of the validity region for perturbation theory. Several groups develop frameworks to account for corrections (nuclear, target mass, higher twists) that affect extraction of nucleon collinear PDFs in the large-$x$ and low-$Q$ region, {\it e.g.}~\cite{Accardi:2016qay,Cocuzza:2021rfn}. Interplay between these corrections and determination of PDFs at large $x$ will be increasingly relevant in near-future precision experiments~\cite{Bailey:2020ooq,Ball:2020xqw,Accardi:2021ysh}. 
It should be pointed out that, while these types of corrections are most pronounced at low $Q$ and very large $x$, they propagate to smaller $x$ at electroweak $Q$ via DGLAP evolution and may affect percent-level phenomenology. Large-$x$ contributions might also matter for specific kinematics, see, {\it e.g.},~\cite{Bonvini:2015ira,Ball:2022hsh}.

In the realm of perturbative QCD,  it is well-known that fixed-order perturbative calculations, even when computed at N2LO in $\alpha_s$, display
classes of logarithmic contributions that become large in some kinematic regions, thus
spoiling the perturbative expansion in the strong coupling constant $\alpha_s$. Among these 
enhanced logarithmic contributions, there are the high-energy (or small-$x$) contributions that will be discussed in the next section. Here we will focus on another type of logarithmic enhancement of higher order perturbative contributions that is relevant at large $x$~\cite{Sterman:1986aj}.
This class of logarithms appears close to threshold for the production of the final states: this is the large-$x$ kinematic region, and the resummation of logarithms from this region is known as large-$x$, soft gluon, or threshold resummation. The importance
of these contributions varies significantly with both the type and the kinematic regime of
the processes which enter PDF fits. Therefore, their omission can lead to a significant
distortion of the PDFs, thereby reducing their theoretical accuracy.
Some time ago, in Ref.~\cite{Bonvini:2015ira} a set of PDFs was constructed in which fixed-order NLO and N2LO calculations were supplemented with soft-gluon (threshold) resummation up to next-to-leading-log (NLL) and next-to-next-to-leading-log (NNLL) accuracy respectively. This specialized set of PDFs was produced to be used in conjunction with any QCD calculation in which threshold resummation is included at the level of partonic cross sections. These resummed PDF sets, based on the old NNPDF3.0 analysis~\cite{Ball:2012cx}, were extracted from a restricted set of data, namely DIS, Drell-Yan, and top quark pair production data, for which resummed calculations were available in a usable format. The interesting result was that, close to threshold, the inclusion of resummed PDFs can partially compensate the enhancement in resummed matrix elements, leading to resummed hadronic cross sections closer to the fixed-order calculation. On the other hand, far from threshold, resummed PDFs reduce to their fixed-order counterparts. This pointed to the need for a consistent use of resummed PDFs in resummed calculations. 

Within the context of parton shower event generators, the need to sum threshold logarithms arises from a mismatch between the kinematic limits in the evolution of parton distribution functions and the evolution of the parton shower. In part, this means that one should use different PDFs within the splitting functions of the parton shower than the usual $\overline {\mathrm{MS}}$ PDFs used for fixed order perturbation theory \cite{Nagy:2016pwq, Nagy:2017dxh, Nagy:2020gjv}. The most practical way to do this is to transform the $\overline {\mathrm{MS}}$ PDFs, but a more ambitious solution would be to independently fit the PDFs in the needed scheme. For more details, see the discussion in Section~\ref{subsec:evgen}. 

Once the corrections of perturbative QCD are properly accounted for, the obtained PDFs should be consistent with their field-theoretical definition. 
An interesting question is then to which degree the theoretical expectations, such as positivity, quark counting rules, or quark-hadron duality, must influence the shape of phenomenological PDFs~\cite{Candido:2020yat, Collins:2021vke}. Should the allowed PDF solutions reflect these semi-quantitative constraints? This is a topic of the recent phenomenological work \cite{Ball:2016spl,Courtoy:2020fex} and exploration within global fits \cite{Hou:2019efy,Ball:2021leu}. While first principles of QCD need to be fulfilled, empirical testing of various hypotheses for the hadron structure must be mindful of biases introduced by such prior expectations. On the flip side, without the control of associated uncertainties, agreement between a theoretical model and phenomenological PDFs is not sufficient for validating the model; detailed studies of uncertainties in such tests are crucial both on the theoretical and phenomenological side \cite{Courtoy:2020fex}. Anticipated DIS and other measurements at higher $x$ and $Q$ values will advance our knowledge of large-$x$ dynamics~\cite{ZEUS:2020ddd,Courtoy:2021xpb}.

 \subsubsection{Small $x$}
 \label{subsec:smallx}
 {\it Leading authors: R. D. Ball, M. Hentschinski, C. Royon, K. Xie \vspace{6pt}}
 
  For successful runs at any colliders, such as the LHC at CERN or the incoming EIC at BNL~\cite{AbdulKhalek:2021gbh}, and future projects such as FCC at CERN~\cite{FCC:2018vvp}, it is fundamental to understand fully the complete final states. This obviously includes the central part of the detector that is used in BSM searches and also the forward part, the kinematic region close to the outgoing beam remnants after collision. The detailed understanding of final states with high forward multiplicities, as well as those with the absence of energy in the forward region (the so-called rapidity gap), in elastic, diffractive, and central exclusive processes is of greatest importance.
Some of these configurations originate from purely nonperturbative reactions, while others can be explained in terms of multi-parton chains or other extensions of the perturbative QCD parton picture such as the Balitsky-Fadin-Kuraev-Lipatov (BFKL) formalism~\cite{Kuraev:1976ge,Kuraev:1977fs,Balitsky:1978ic}. Future progress in this fundamental area requires the combination of  experimental measurements and theoretical work.

 When the parton momentum fraction $x$ becomes small, small-$x$ logarithms $\log(1/x)$ become significant, and require all-order resummation to obtain a good convergence of the QCD theory. This can be achieved through the BFKL formalism~ at NLL~\cite{Fadin:1996nw,Fadin:1997hr,Fadin:1997zv,Fadin:1998py}, matched to collinear factorization at NLO or N2LO using either the ABF formalism~\cite{Altarelli:1999vw,Altarelli:2000mh,Altarelli:2001ji,Altarelli:2003hk,Altarelli:2005ni,Altarelli:2008aj}, or the closely related CCS approach~\cite{Salam:1998tj,Ciafaloni:1999yw,Ciafaloni:1999au,Ciafaloni:2000cb,Ciafaloni:2003rd,Ciafaloni:2005cg,Ciafaloni:2007gf}. An efficient numerical implementation of the ABF results \cite{Bonvini:2016wki,Bonvini:2017ogt} made it possible to perform a global PDF determination, based on the NNPDF3.1 data set, but also resumming small-$x$ logarithms in parton evolution and structure functions coefficients~\cite{Ball:2017otu}. This analysis found significant evidence for BFKL resummation in the small $x$ and low $Q^2$ region of the HERA structure function data~\cite{H1:2015ubc}. An analysis using the same ABF implementation and \xfitter reached a similar conclusion~\cite{xFitterDevelopersTeam:2018hym}. Future work on using small-$x$ resummation to improve PDF fits will require the high-energy resummation of the hadronic cross sections~\cite{Ball:2001pq,Ball:2007ra,Marzani:2008uh,Diana:2010ef,Marzani:2011yg,Ball:2013bra,Marzani:2015oyb,Forte:2015gve,Muselli:2015kba,Bonvini:2018iwt} included in global fits, which, while technically challenging, is now perfectly feasible. The effects are most likely to be important in the LHCb data, which can probe $x$ as small as $10^{-6}$. 
 
 Eventually, at small enough $x$ and low enough $Q^2$, we enter into the partonic saturation region~\cite{Mueller:2001fv}. The boundary to delineate the small-$x$ resummation region and the saturation one is ambiguous. In the latest round of the CTEQ-TEA global analysis~\cite{Hou:2019efy}, two alternative ensembles, CT18X and CT18Z, were released, in which an $x$-dependent DIS factorization scale was adopted. It is motivated by a partonic saturation model~\cite{GolecBiernat:1998js}, and improves the QCD description of the HERA DIS data, obtaining a similar $\chi^2$ for the same data set as the small-$x$ resummation treatment adopted in NNPDF~\cite{Ball:2017otu} and \xfitter~\cite{xFitterDevelopersTeam:2018hym}. Both approaches obtain an enhancement of gluon PDF at small $x$ and $Q$~\cite{Guzzi:2021fre}. However, the enhancement of the small-$x$ resummation is noticeably larger than the $x$-dependent scale approach, in which the small-$x$ growth is largely tamed toward $x\to10^{-6}$. Some implications for the small-$x$ dynamics have been explored, such as the DIS structure functions~\cite{Guzzi:2021fre}. As expected, both approaches have obtained similar predictions for the transverse structure functions $F_2$. Intriguingly, the longitudinal one $F_L$ is pulled in different directions in the two approaches when Bjorken $x$ is below $10^{-4}$. Future measurements can discriminate between these two distinct approaches.
 
It is obvious that the PDF fits at small $x$ will benefit from a better understanding of small-$x$ multi-gluon kinematics, such as in the BFKL regime or the saturation phenomena. The small-$x$ modifications may be far more pronounced in heavy-ion collisions. Understanding of diffractive events and their effects on PDFs is also fundamental. A dedicated white paper~\cite{Hentschinski:2022xnd} reviews some recent developments and future prospects in the domain of small-$x$ physics, saturation, and diffraction.  This document discusses first the occurrences of BFKL resummation effects in special final states, such as Mueller-Navelet jets, jet gap jets, and heavy quarkonium production. It further addresses TMD factorization at small $x$ and the manifestation of a semi-hard saturation scale in  (generalized) TMD PDFs. More theoretical aspects of low-$x$ physics, probes of the quark-gluon plasma, the possibility of using photon-hadron collisions at the LHC to constraint the hadronic structure at low $x$, as well as the resulting complementarity between  LHC and the EIC are discussed.  The white paper also  briefly reviews diffraction at colliders as well as the possibility to explore further the electroweak theory in central exclusive events using the LHC as a $\gamma\gamma$ collider.

\subsection{Factorization schemes for event generators}
\newcommand{\dd}{\mathrm{d}}
\newcommand{\dsigma}{\dd\sigma}
\newcommand{\alphas}{\alpha_\mathrm{s}}
\newcommand{\alphaem}{\alpha_\mathrm{em}}
\newcommand{\ms}{\mathrm{MS}}
\newcommand{\msbar}{{\overline{\mathrm{MS}}}}
\newcommand{\rA}{\mathrm{A}}
\newcommand{\rB}{\mathrm{B}}
\newcommand{\rC}{\mathrm{C}}
\newcommand{\rD}{\mathrm{D}}
\newcommand{\rF}{\mathrm{F}}
\newcommand{\rM}{\mathrm{M}}
\newcommand{\rR}{\mathrm{R}}
\newcommand{\rPS}{\mathrm{PS}}
\newcommand{\rS}{\mathrm{S}}
\newcommand{\rT}{\mathrm{T}}
\newcommand{\rV}{\mathrm{V}}
\newcommand{\muf}{\mu_\rF}
\newcommand{\mur}{\mu_\rR}
\newcommand{\krk}{\mathrm{Krk}}
\newcommand{\order}[1]{\mathcal{O}\left({#1}\right)}
\newcommand{\ve}{\varepsilon}
\newcommand{\etal}{{et al}., }
\newcommand{\ie}{{i}.{e}.\ }
\newcommand{\eg}{{e}.{g}.\ }
\label{subsec:evgen}
{\it Leading authors: S. Hoeche, A. Siodmok, J. Whitehead \\ \vspace{6pt}}

Defining a PDF requires the choice of a factorization scheme, which governs the allocation
of finite terms between the PDFs and the hard, partonic cross sections. This choice is generally 
a matter of taste and convenience \cite{CTEQ:1994ydo}. In practice, the
majority of QCD hard process calculations and PDF sets adopt the
\({\overline{\mathrm{MS}}}\)-scheme.

Recently there has been renewed interest in developing alternative factorization
schemes \cite{deOliveira:2012qa, Oliveira:2013aug, Jadach:2011cr, Jadach:2015mza, Jadach:2016qti, Jadach:2020xfl, Candido:2020yat}, including
to investigate the positivity of \(\overline{\mathrm{MS}}\) PDFs \cite{Candido:2020yat}
and to simplify Monte Carlo calculations \cite{Jadach:2011cr, Jadach:2015mza, Jadach:2016qti, Jadach:2020xfl,Jadach:2020xfl}.
PDFs in different factorization schemes are related to each other, and to
those in the \({\overline{\mathrm{MS}}}\) scheme, by a transition operator
that mixes PDFs of different flavors,
$
\mathbf{f}^{\mathrm{FS}}
=
\mathbb{K}^{{\overline{\mathrm{MS}}}\to \mathrm{FS}}
\otimes
\mathbf{f}^{{\overline{\mathrm{MS}}}},
$
so that for each flavor \(a\) we have
\[
f^{\mathrm{FS}}_a (x; \mu_\mathrm{F})
=
\sum_b
\int_x^1
\frac{\mathrm{d}\xi}{\xi}
\quad
\mathbb{K}^{{\overline{\mathrm{MS}}}\to \mathrm{FS}}_{ab} \left(\frac{x}{\xi}; \mu_\mathrm{F}\right)
\;
{f}^{{\overline{\mathrm{MS}}}}_b (\xi; \mu_\mathrm{F})
,
\]
where
\[
\mathbb{K}^{{\overline{\mathrm{MS}}}\to \mathrm{FS}}_{ab} (x; \mu)
\equiv
\delta_{ab} \, \delta(1-x)
+
\frac{\alpha_\mathrm{s}(\mu)}{2\pi}
\,
\mathrm{K}^{\mathrm{FS}}_{ab} (x; \mu)
+
\mathcal{O}\left({\alpha_\mathrm{s}^2}\right).
\]
The transformation kernels are often further constrained
to ensure that the transformed PDFs obey the same sum rules as the
input PDFs ({e}.{g}.\ re-imposing momentum sum rules by modifying an
end-point contribution \(\propto \delta(1-x)\) accordingly.
The required independence of predictions from the choice of
factorization scheme is achieved, to NLO accuracy, by a corresponding inverse transformation
of the partonic cross sections.
The freedom to choose a factorization
scheme therefore corresponds to a freedom to remove a common set of
convolution terms \(\mathrm{K}^{\mathrm{FS}}_{ab}\) from all partonic cross sections.

The \texttt{Krk} (formerly \texttt{MC}) factorization scheme \cite{Jadach:2016acv}
exploits this freedom to significantly simplify  
the matching of the parton shower Monte Carlo event generators to 
NLO calculations for the hard process 
by systematically removing the convolution terms \(\mathrm{K}^{\mathrm{FS}}_{ab}\) from
the hard process.
This may be conveniently done within the modified Catani-Seymour (CS)
dipole subtraction method \cite{Catani:1996vz}. 
Within the \texttt{Krk} scheme the transition operator is therefore derived from the
finite and collinear part of the CS integrated-subtraction
and collinear contributions, given by the \(\mathbf{P}\) and
\(\mathbf{K}\) collinear operators, so that
\(\mathbf{K}\) objects as
\[
\mathrm{K}^{\mathrm{Krk}}_{ab} (x; \mu)
=
\mathbf{K}^{ba} (x) + \mathbf{P}^{ba} (x; \mu).
\]

These transition operators to the \texttt{Krk} factorization scheme
are modified so that the NLO corrections to
heavy colour-neutral boson production in $pp$ collisions (Drell-Yan type processes)
and electron-proton scattering (DIS-type processes) are maximally simplified.
They have been applied to several public
\({\overline{\mathrm{MS}}}\) PDF sets.
These PDFs, uniquely, allow NLO-accurate calculations of any such process
using CS dipole subtraction without requiring an on-the-fly convolution.
The \texttt{Krk} factorization scheme has been employed in the
\texttt{KrkNLO} parton-shower matching method, which has been implemented as a
proof-of-concept for the Drell-Yan and (gluon-fusion) Higgs-production
processes in both \texttt{Sherpa} and \texttt{Herwig} \cite{Jadach:2016qti}.

\subsection{Other theoretical developments for future PDF analyses}

{\it Leading authors: B. Mistlberger, S. Moch, R. Ball, P. Nadolsky, M. Ubiali}\\

This section have reviewed numerous advancements in theoretical calculations necessary for determinations and future applications of PDFs. The whole subject is too expansive to be covered in a short contribution. We conclude this section by highlighting briefly several other aspects that are key for obtaining the next generation of PDFs of high accuracy. 
\begin{itemize}
    \item  Effects due to non-zero quark masses become non-negligible at a certain level of precision, and a consistent framework to take them into account in the extraction of PDFs is desirable. While several General-Mass-Variable-Flavor-Number-Scheme (GM-VFNS) calculations \cite{Collins:1998rz} have been implemented in PDF global fits for DIS observables~\cite{Thorne:1997ga,Aivazis:1993pi,Forte:2010ta}, nearly all  observables for hadron colliders in the PDF fits are computed in a Zero-Mass VFNS, thus ignoring the effects associated with the finite mass of heavy quarks. This is justified by the smallness of the $c$ and $b$ quark masses as compared to typical energy scales in the examined LHC observables. However, as the targeted precision increases and lower energy regions are explored, the implementation of $pp$ GM-VFNS calculations will become necessary in global PDF fits. For example, production of charmed and bottom particles at the LHCb requires to account for mass effects at small $p_T$, which can be done over the whole $p_T$ range using an advanced heavy-quark scheme such as SACOT-MPS \cite{Xie:2019thesis} that can be extended to N2LO. 
    \item As discussed in Sec.~\ref{sec:Resummations}, in the kinematic limits where a parton takes up almost all or almost none of the momentum of its hadron, parton distribution functions may be resummed to all orders in perturbation theory to a given logarithmic accuracy. Consistently including such resummation should be a part of future research and ultimately the determination of PDFs.  At small values of parton momentum fractions $x$, the resummation of  corrections to a given logarithmic accuracy to all orders has been considered.
It has been shown~\cite{Blumlein:1995jp,Blumlein:1997em}, however, 
that the yet unknown subleading small-$x$ terms are larger than the leading-order terms. Resummation of the entire tower of subleading terms is needed, at least to the fourth subleading logarithm, to have quantitative control of small-$x$ cross sections~\cite{Moch:2004pa,Vogt:2004mw}.
    \item Studies of theoretical uncertainties and their propagation through the PDF extraction process should be encouraged. In general, the total PDF uncertainty reflects a complex interplay of experimental, theoretical, methodological, and parametrization sources of uncertainties \cite{Kovarik:2019xvh}. These uncertainties cannot be easily separated. For example, reliable theoretical predictions are necessary to understand unfolding and acceptance effects in observables constraining the PDFs. New complexity issues emerge in the global analyses of large data sets. Representative exploration of all contributions to the uncertainties is necessary for confident and accurate predictions \cite{Courtoy:2022ocu}. 
    \item  Fast interfaces of QCD, electroweak and resummation contributions are crucial for the extraction of PDFs and their development should be facilitated. The distribution of final PDF parameterizations in a convenient form  for applications, such as the \lhapdf format, is important to the usability of PDFs for the community. This is partially discussed in Sec.~\ref{sec:computing}.
\item Beyond the fixed-order perturbation theory and the leading-power expansion for
observables in Eq.~(\ref{eq:OepOpp}), several improvements of the theoretical
description are compulsory, depending on the observable under consideration,
and additional care has to be taken.
For the kinematics range covered by currently available data from DIS {\it ep} scattering, 
higher-twist effects become important. 
In the flavor non-singlet case these have been measured in~\cite{Blumlein:2008kz,Blumlein:2012se}, 
and in the singlet case they were determined in \cite{Alekhin:2012ig}.
\end{itemize}

\section{Methodology}
\label{sec:Methodology}

\subsection{Experimental systematic uncertainties in PDF fits \label{sec:MethodologyExpSystUncertainties}}
{\it Leading authors: A. M. Cooper-Sarkar, T. Cridge, F. Giuli, J. Huston, R. S. Thorne  
\\ \vspace{6pt}}

The LHC has accumulated a large amount of data at 7,8 and 13 TeV, for a variety of processes. The data sets vary from purely inclusive processes, such as the $W/Z$ cross sections, to differential measurements over a variety of kinematic variables, such as the Drell-Yan cross section as a function of the invariant mass, rapidity and transverse momentum of the final-state leptons. Due to the large data samples, many of the measured distributions are limited by systematic uncertainties rather than statistics.
Differential measurements over wide kinematic ranges (and over multiple detector regions) that are systematics-limited require detailed knowledge regarding the correlation of the systematic error components over these regions. Such error correlations are difficult to determine experimentally and their imperfect knowledge often results in tensions between rapidity ranges (for example for the case of the ATLAS inclusive jet cross section) or kinematic variables (for example for the ATLAS $t\bar{t}$ rapidity and $t\bar{t}$ mass distributions). Such tensions may mask or diminish the power of the data to determine PDFs and their uncertainties. 

To be specific, many systematic uncertainties are point-to-point correlated within a kinematic distribution and between distributions of the same analysis. There can also be correlations between different analyses due to systematic uncertainties from the same sources. For example, inclusive jet production data are presented as functions of transverse momentum in several bins of jet rapidity, and with many systematic sources correlated between rapidity bins. Alternatively, $t\bar{t}$ production data are presented in terms of several different variables, such as the mass, $m_{tt}$, or rapidity, $y_{tt}$, of the $t\bar{t}$ pair, and the average transverse momentum, $p_T^t$, or rapidity, $y_t$, of the $t\bar{t}$ pair. There are both statistical and systematic correlations between all of these distributions. Finally, since the  $t\bar{t}$ data are measured in the lepton+jets channel, there are potential correlations between the systematic uncertainties, from sources such as the jet energy scale, between the inclusive jet measurements and the $t\bar{t}$ measurements. 

Experimental correlated systematic uncertainties can be taken into account in PDF fits by using a covariance matrix provided by the experimental collaboration, but it is more informative if this information is given as a list of 1$\sigma$ uncertainties due to each source of systematic uncertainty for each data point. This information can then be applied to the fit using nuisance parameters that are common between the data points for the same source. It is the default to consider the same source of systematic uncertainty to be $100\%$ correlated between the data points, and this is the assumption used when constructing a covariance matrix, but this may not be realistic. The advantage of keeping the information split into separate sources is that one can trace the sources of uncertainty, and one can change the degree of correlation in an informed manner by consulting the experimentalists.

In the context of LHC analyses, problems with the treatment of correlated systematic uncertainties first came to light in fits to the ATLAS 7 TeV jet data~\cite{ATLAS:2014riz}.
Whereas good fits could be found to the separate rapidity bins of the data, the fit to all rapidity bins taken together was very poor. The tensions result in $\chi^2$ values that may be acceptable for individual rapidity regions (for the jet fit), but have vanishing probability when fit together. Such tensions may mask or diminish the power of the data to determine PDFs and their uncertainties. The information provided by the separate individual rapidity interval fits greatly reduce the discriminating power of the full data set, as (1) the $x$-range probed is reduced and (2) the systematic error shifts may differ significantly from distribution to distribution, a situation that does not reflect reality.~\footnote{Fits to a wider rapidity range for jet production also help to distinguish between PDF variations and the possible presence of new physics.}

A data set, such as the ATLAS jet cross section, could be divided into its individual rapidity intervals to determine any tensions that may exist, for example in the determination of the high-$x$ gluon distribution, and how these tensions and the constraining power on the PDFs change as decorrelation models are applied. In this context, a bad data set $\chi^2$ may not necessarily represent a disappointing outcome, if the data set’s constraining power is not reduced. In addition, it can be checked whether a decorrelation model that improves the global $\chi^2$ affects the impact of this data set on the PDF fit.

\begin{figure}[t]
    \centering
    \includegraphics[width=0.37\textwidth]{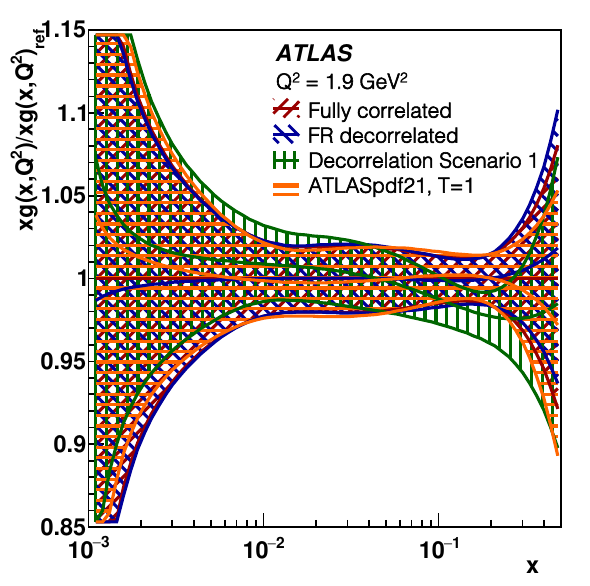}
\quad  \includegraphics[width=0.6\textwidth]{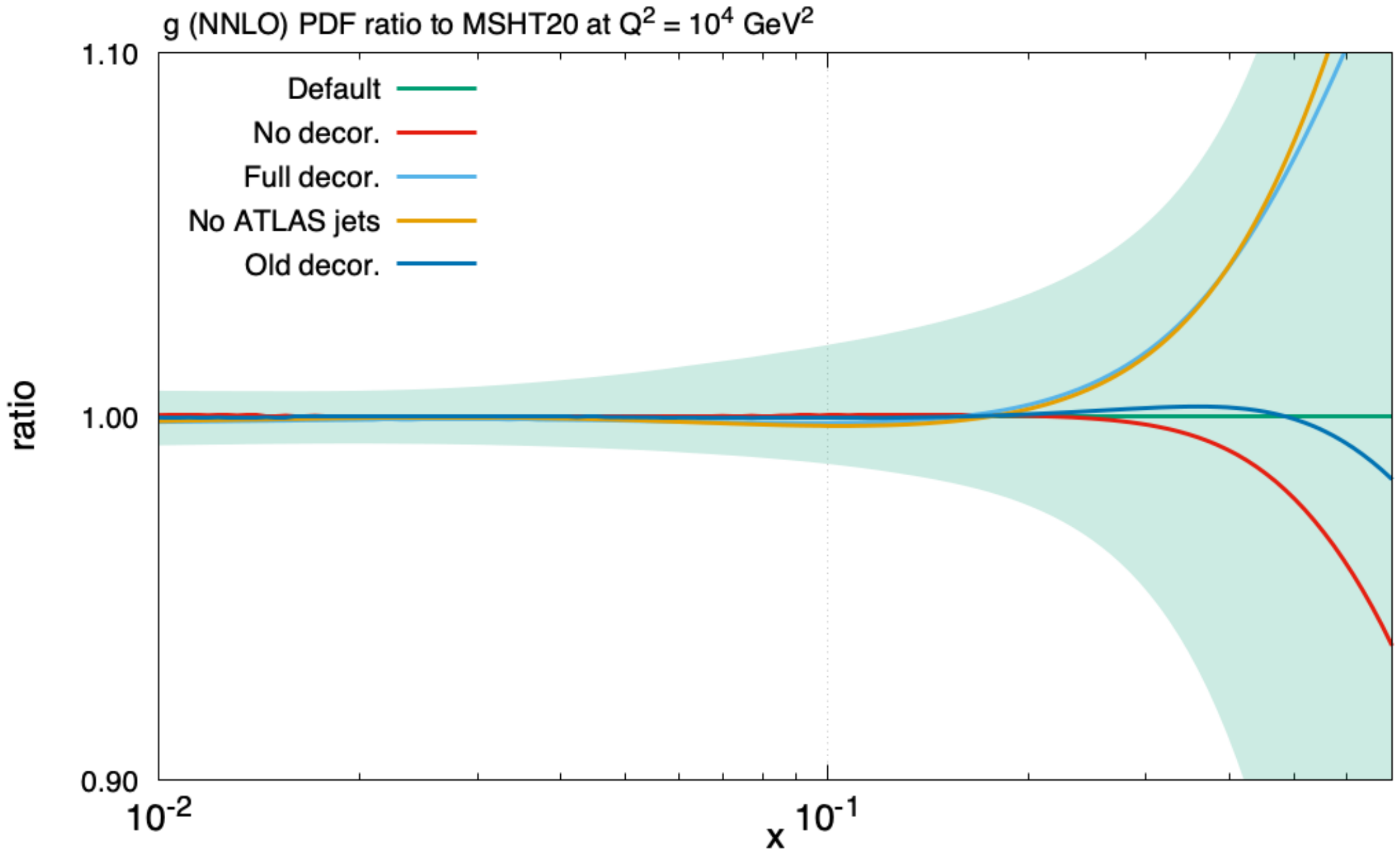}
    \caption{Left: difference in the gluon PDF shown in ratio to the ATLASpdf21 (full correlation) gluon. The default ATLASpdf21 fit uses Decorrelation 2 scenario, and we also show the scenarios with full decorrelation of the flavor response (FR) systematic, and Decorrelation procedure 1. Right: the effect of no decorrelation, the default correlation of \cite{Bailey:2020ooq}, the decorrelation in \cite{Harland-Lang:2017ytb}, and full decorrelation for the MSHT20 gluon.}
    \label{fig:atlasjetdecorr}
\end{figure}

This problem with fitting multiple rapidity intervals has led NNPDF to fit only one rapidity bin~\cite{NNPDF:2017mvq}. Decorrelation models have been developed to reduce these tensions, but they suffer from their somewhat ad hoc nature.  MMHT did an alternative study of the effect of decorrelating some systematic sources between rapidity bins~\cite{Harland-Lang:2017ytb}. However, the most thorough study was carried out by the ATLAS collaboration, who studied the same problem in their ATLAS 8 TeV inclusive jet data~\cite{ATLAS:2017kux}. Some of the systematic uncertainties appertaining to the jet energy scale are evaluated from the difference of two different Monte-Carlo estimates. Such "two-point systematics" are reasonable estimates of uncertainty, but they are far from being Gaussian distributed. These systematic sources are often the largest systematic uncertainties for analyses involving jet production. One may question the convention that these are $100\%$ correlated between data points. It is vital to try the decorrelations solely in collaboration with experimentalists with the knowledge of which sources can be legitimately decorrelated. ATLAS developed some models for the decorrelation of such systematic sources as functions of rapidity and $p_T$~\cite{ATLAS:2017kux}. These models were applied to several of the jet energy scale systematic sources, and some favored combinations of correlation model were suggested. These were then used in a PDF fit using these jet data~\cite{ATLAS:2021vod}. The $\chi^2/NDP$ for fits to these jet data with different levels of decorrelation are summarised in Table~\ref{tab:ATLASjetchi2}. Whereas the $\chi^2$ differ considerably, the difference in the resulting PDFs between the use of full correlation and extreme decorrelation is small, see Fig~\ref{fig:atlasjetdecorr}. 
Another study on ATLAS 7 TeV jet data associated with the MSHT20 global analysis~\cite{Bailey:2020ooq} comes to a similar conclusion when decorrelating two of the jet energy scale systematic between rapidity bins (partial decorrelation) or indeed decorrelating all systematic sources between rapidity bins (full decorrelation), see Fig~\ref{fig:atlasjetdecorr}.
\begin{table}[b!]
\begin{center}
\begin{tabular}{ccccc}
\hline
\hline
 ATLAS 8 TeV Jets R=0.6    &  Fully Correlated   &  FR Decorrelated &  Decorrelation Scenario 1  &  Decorrelation Scenario 2\\
\hline
   $\chi^2/\rm{NDP}$ & 289/171  &  226/171 & 250/171 & 248/171\\
\hline
\hline  
\end{tabular}
\end{center}
\caption{Partial $\chi^2$ for jet data entering the PDF fit, for different levels of decorrelation ranging from fully correlated to an extreme scenario of the jet flavor response (FR) decorrelated between rapidity bins. The Decorrelation Scenarios are described in ref.~\cite{ATLAS:2021vod}. \label{tab:ATLASjetchi2}}
\end{table}

Although studying the impact of various experimental systematic decorrelations can very useful in general, such decorrelation models should be vetted by the original experimental collaboration, i.e., by the measurement's authors  who best understand the origins of those systematics. 

\begin{figure}[t!]
    \centering
    \includegraphics[width=0.44\textwidth]{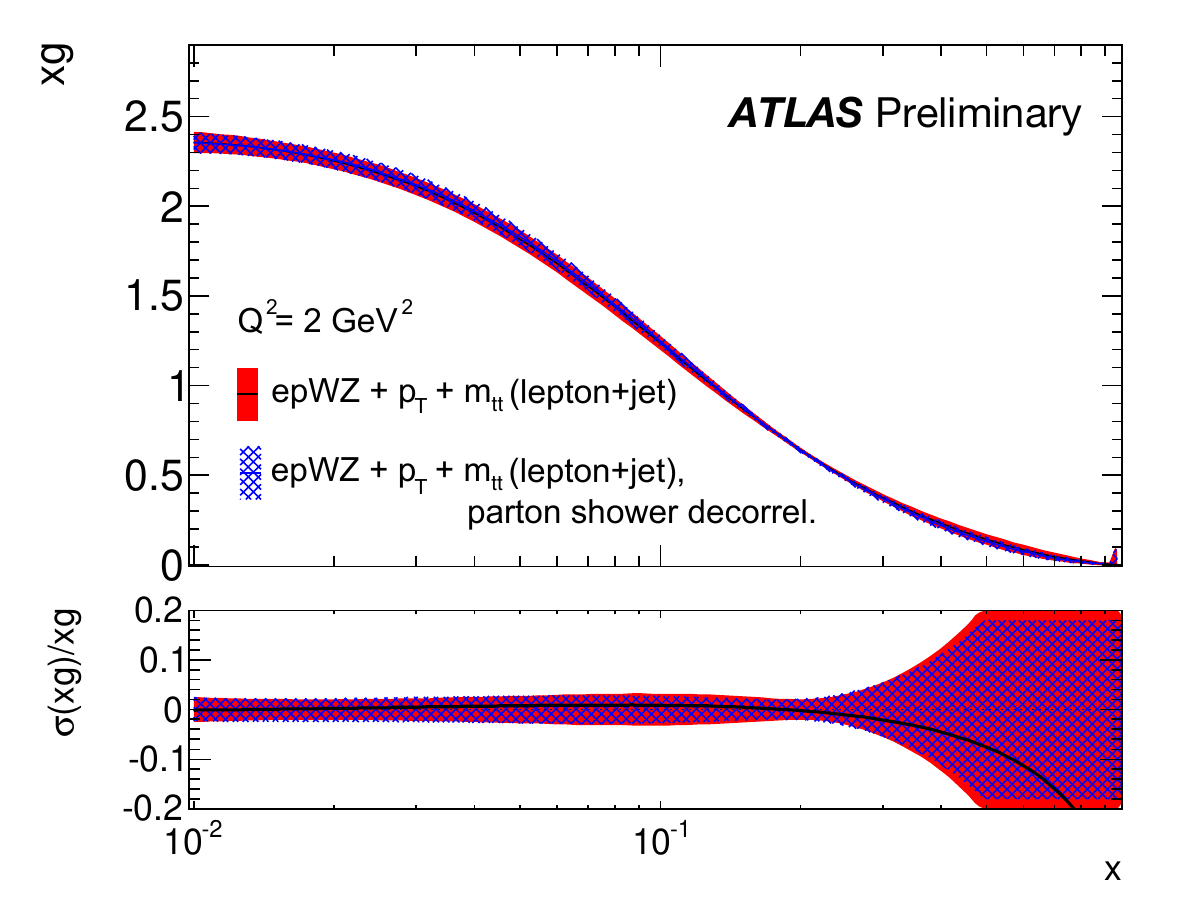}\quad
\includegraphics[width=0.53\textwidth]{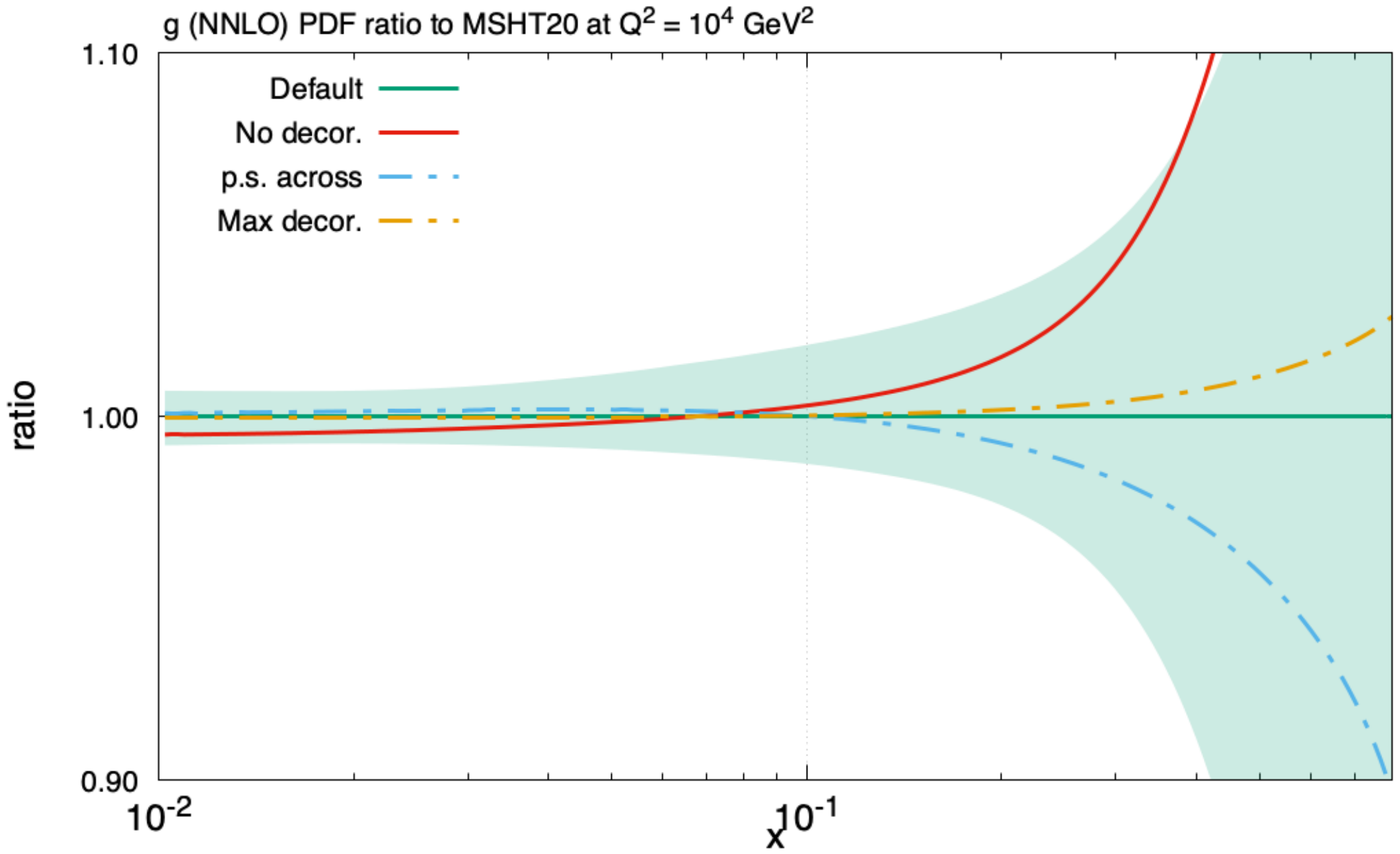}
    \caption{Left: difference in the gluon PDF shown in ratio to the ATLASepWZtop18 gluon (called epWZ+$p^t_T$+$m_{tt}$) . The use of  full correlation of all systematic sources (red) is compared to the result when decorrelating the parton shower systematic between the $p^t_T$ and $m_{tt}$ spectra (blue). Right: Differences between the scenarios with the default systematic error treatment, no decorrelation, decorrelation only across distributions, and full decorrelation for the gluon PDF from \cite{Bailey:2020ooq}.}
    \label{fig:atlastopdecorr}
\end{figure}

The second example of the need to consider some degree of decorrelation of 'two-point systematics' comes up in fits to ATLAS 8 TeV $t\bar{t}$ spectra~\cite{ATLAS:2015lsn}. When these data were first issued, there was no information statistical correlations between the spectra, so that only one spectrum could be fit at once~\cite{NNPDF:2017mvq}. However, such information was provided in Ref.~\cite{ATLAS:2018owm} and is now available in HEPDATA for Ref.~\cite{ATLAS:2015lsn}, and a study of using both systematic and statistical correlations within a PDF fit was made.
The $\chi^2/NDP$ for separate fits to the lepton+jets spectra are given in Table~\ref{tab:ATLAStopchi2}, where it can be seen that the rapidity spectra cannot be fitted well. A further study of fitting the spectra simultaneously was then restricted to the $m_{tt}$ and $p^t_T$ spectra.
Since the $\chi^2$ of the separate fits to $p^t_T$ and $m_{tt}$ adds to 11.3, it was somewhat surprising that a fit with a $100\%$ correlation between all systematic sources yields a joint $\chi^2$ of 45. The answer lies in the correlation of the large two-point systematics related to the models for parton showering, hard scattering and initial/final-state radiation. When the spectra are fitted separately, the nuisance parameters for these sources take very different values for the $p^t_T$ and $m_{tt}$ spectra, see Table~\ref{tab:shifts}, but an assumption of $100\%$ correlation forces them to be the same-- this suits neither spectrum. A fit in which all three of these sources of systematics are decorrelated between the two spectra, or a fit in which just the parton shower model sources are decorrelated produce considerably lower $\chi^2$, see Table~\ref{tab:chi2ptmtt}. The decorrelation of the parton shower systematic has been adopted for the ATLAS PDF fits and for the CT18 PDF fits. Indeed, it has been confirmed recently that, as well as the ATLAS analysis, all of CT, MSHT and NNPDF find problems when fitting all 
distributions simultaneously without some decorrelation~\cite{Cridge:2021qjj}. 
\begin{table}[b!]
\begin{center}
\begin{tabular}{ccccc}
\hline
\hline
  ATLAS 8 TeV $t\bar{t}$ lepton+jets spectrum    &  $m_{tt}$  &  $p^t_T$ &  $y_{tt}$  &  $y_t$\\
\hline
 $\chi^2/\rm{NDP}$ & 3.4/7  &  7.9/8 & 19.7/5 & 18.3/5\\
\hline
\hline  
\end{tabular}
\end{center}
\caption{Partial $\chi^2$ for data sets entering the PDF fit, for each of the top spectra separately. \label{tab:ATLAStopchi2}}
\end{table}
The effect of this decorrelation on the gluon PDF is fortunately small as illustrated with the ATLASepWZtop18 fit~\cite{ATLAS:2018owm} in Fig.~\ref{fig:atlastopdecorr}. 
However, a study by MSHT~\cite{Bailey:2019yze} took the decorrelation further. In order to fit the $y_{tt}$ and $y_t$ rapidity spectra, decorrelation of the parton shower systematic within these spectra is also considered. This decorrelation is done as a trigonometric function of rapidity. This reduces the $\chi^2$ per point for this data set with all for distributions in the MSHT20 fit from 6.84 with no correlation, to 1.69 with correlation between distributions, to 1.04 for the additional 
decorrelation within spectra. 
In this case, although the difference in the gluon PDF between the fully correlated and uncorrelated case is still within uncertainties, it is nevertheless comparable to the difference between and NLO and an N2LO analysis. This approach has been carried into the full MSHT20
analysis~\cite{Bailey:2020ooq}, see Fig.~\ref{fig:atlastopdecorr}. It should also be noted that, the more decorrelation is applied, the less power the data have to constrain PDFs. 
\begin{table}[t!]
\begin{center}
\begin{tabular}{ccccc}
\hline
\hline
   ATLAS 8 TeV $t\bar{t}$ spectra    &  $p^t_T$ & $m_{tt}$  \\
\hline
  hard scattering model& +0.74  &-0.43 \\
  parton shower model &-1.32   &  +0.39\\
  isr/fsr model & -0.47 &  +0.33\\
\hline
\hline  
\end{tabular}
\end{center}
\caption{Shifts of the named nuisance parameters, in units of standard deviations, for the fits to the top spectra separately .\label{tab:shifts}}
\end{table}
\begin{table}
\begin{center}
\begin{tabular}{cccc}
\hline
\hline
  ATLAS 8 TeV $t\bar{t}$ spectra    &  $p^t_T$ and $m_{tt}$  &$p^t_T$ and $m_{tt}$& $p^t_T$ and $m_{tt}$\\
                   & fully correlated & 3 sources decorrelated & parton shower decorrelated\\
\hline
  $\chi^2/\rm{NDP}$ & 45/15 & 11.5/15 & 14.1/15\\
\hline
\hline  
\end{tabular}
\end{center}
\caption{Partial $\chi^2$ for data sets entering the PDF fit for simultaneous 
fits to the $p^t_T$ and $m_{tt}$. \label{tab:chi2ptmtt}}
\end{table}

\begin{figure}[t!]
    \centering
    \includegraphics[width=0.48\textwidth]{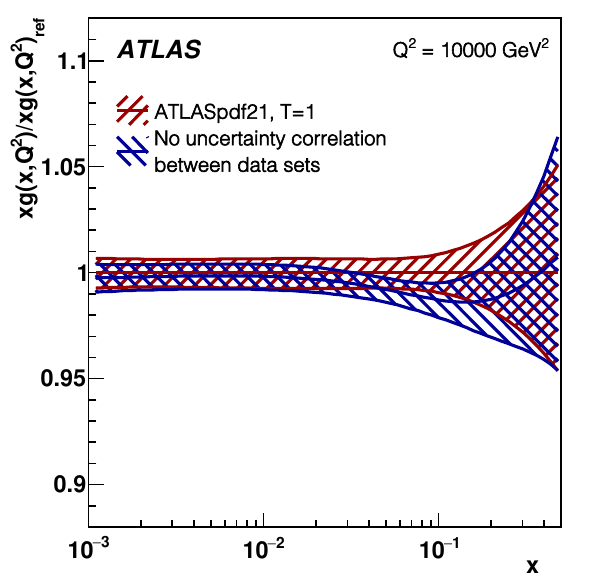}
    \includegraphics[width=0.48\textwidth]{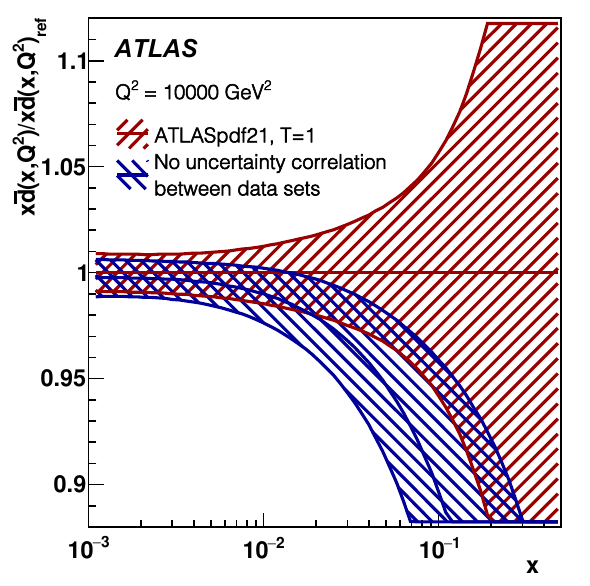}
    \caption{Difference in the $xg$ and $x\bar{d}$ PDFs shown in ratio to the ATLASpdf21 (default) PDFs. The default (red) analysis applies the full correlation of specified systematic sources between the data sets which use jet data, and the alternative (blue) analysis does not apply any correlation of systematics sources between the data sets (apart from the luminosities).}
    \label{fig:atlascorr}
\end{figure}

Finally, although  some sources of systematics can be legitimately decorrelated between spectra of the same analysis, there are other systematic sources for which correlation between different analyses should be considered. This has been studied in a recent ATLAS PDF analysis~\cite{ATLAS:2021vod} ATLASpdf21, where the correlations of various systematic sources have been considered between different analyses which use jet data: inclusive jet data~\cite{ATLAS:2017kux}, $t\bar{t}$ data in the lepton+jets channel~\cite{ATLAS:2015lsn}, $W$ +jets data~\cite{ATLAS:2017irc} and $Z$+jets~\cite{ATLAS:2019bsa} data. The details of the correlated systematic sources considered are given in ref~\cite{ATLAS:2021vod}. Fig.~\ref{fig:atlascorr} shows the difference in the resulting gluon and $x\bar{d}$ PDFs when such correlations between the input data sets are considered, and when they are not. Note that this figure is made for the scale $Q^2=10,000$GeV$^2$ to illustrate that such differences are still visible at LHC scales.

In conclusion, correlations of sources of systematic uncertainty both within and between data sets need to be carefully considered in PDF fits and, although the difference between the resulting PDFs is not large in the best known kinematic region $ 0.01 < x < 0.1$ (corresponding to mass scales $\sim 100$ GeV $\to$ 1 TeV at the LHC), it can nevertheless be large enough to have impact if an ultimate precision of $\sim 1\%$ is sought on PDFs. In the less well-known regions, at smaller and larger mass scales, the impact can be considerably greater.

\newpage
\subsection{Theoretical uncertainties in PDF fits}
\label{sec:thunc}
{\it Leading authors: R.D. Ball, A. M. Cooper-Sarkar  \\ \vspace{6pt}}

Over the last few years, there has been considerable progress in developing new techniques for incorporating theoretical uncertainties into determination of parton distribution functions (PDFs). This work centres on a Bayesian formalism, the “theory covariance matrix”, which can be simply added to the usual experimental covariance matrix used in the PDF fit \cite{Ball:2018lag}.  
\begin{figure}[t!]
    \centering
    \includegraphics[width=0.48\textwidth]{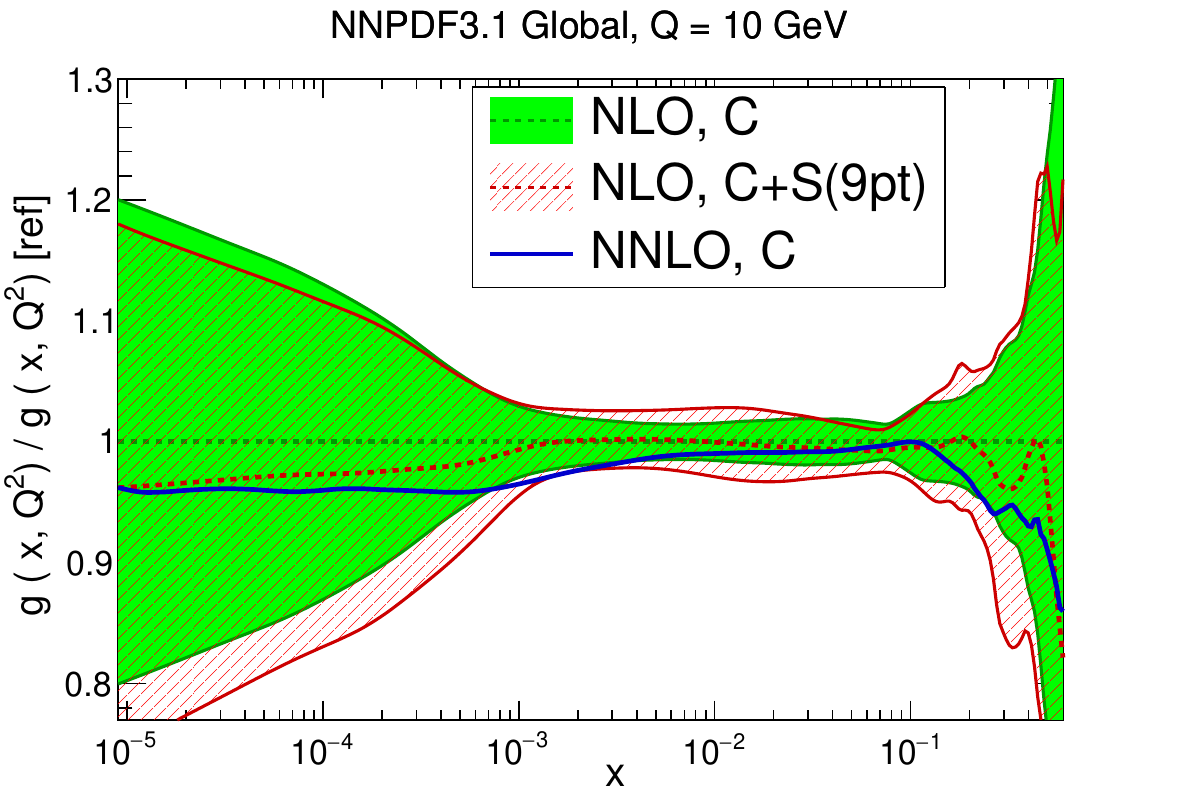}
        \includegraphics[width=0.48\textwidth]{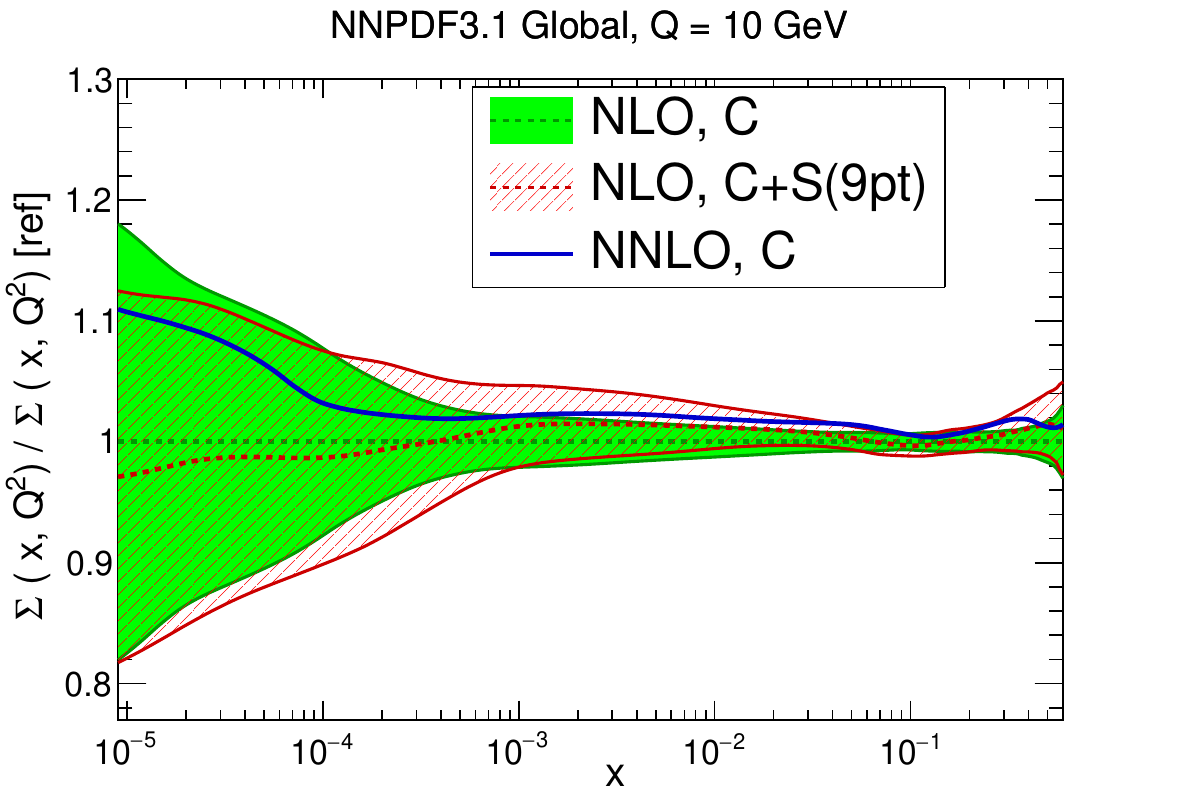}
    \caption{Ratio of the gluon (left) and singlet (right) PDFs for the NLO NNPDF3.1 fit including factorization and renormalization scale uncertainties combined in a 9pt scheme (red), to a fit not including these scale uncertainties (green). Also shown (in blue) is the central value for the N2LO fit (without scale uncertainties). }
    \label{fig:NNPDFscale}
\end{figure}
While the experimental covariance matrix includes the statistical and systematic uncertainties in the measurement of a given cross section, the theoretical covariance matrix incorporates all the various theoretical uncertainties, correlated across different experimental measurements, in the procedure which extracts the PDFs from a global data set. The main assumption is that the theoretical uncertainties are Gaussian and independent of the experimental uncertainties (likewise also generally assumed to be Gaussian). 

\begin{figure}[tb!]
    \centering
    \includegraphics[width=0.45\textwidth]{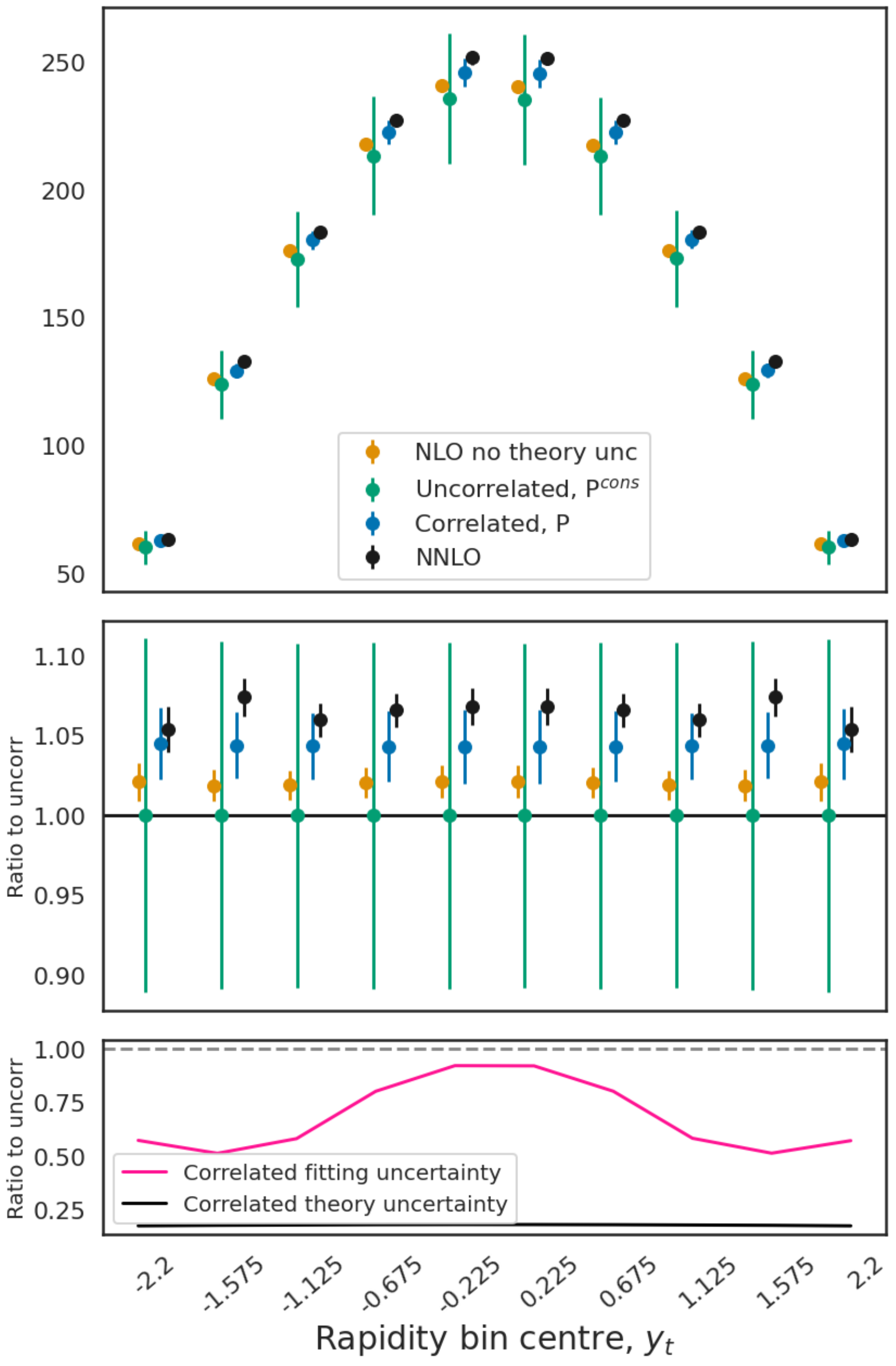}
        \includegraphics[width=0.45\textwidth]{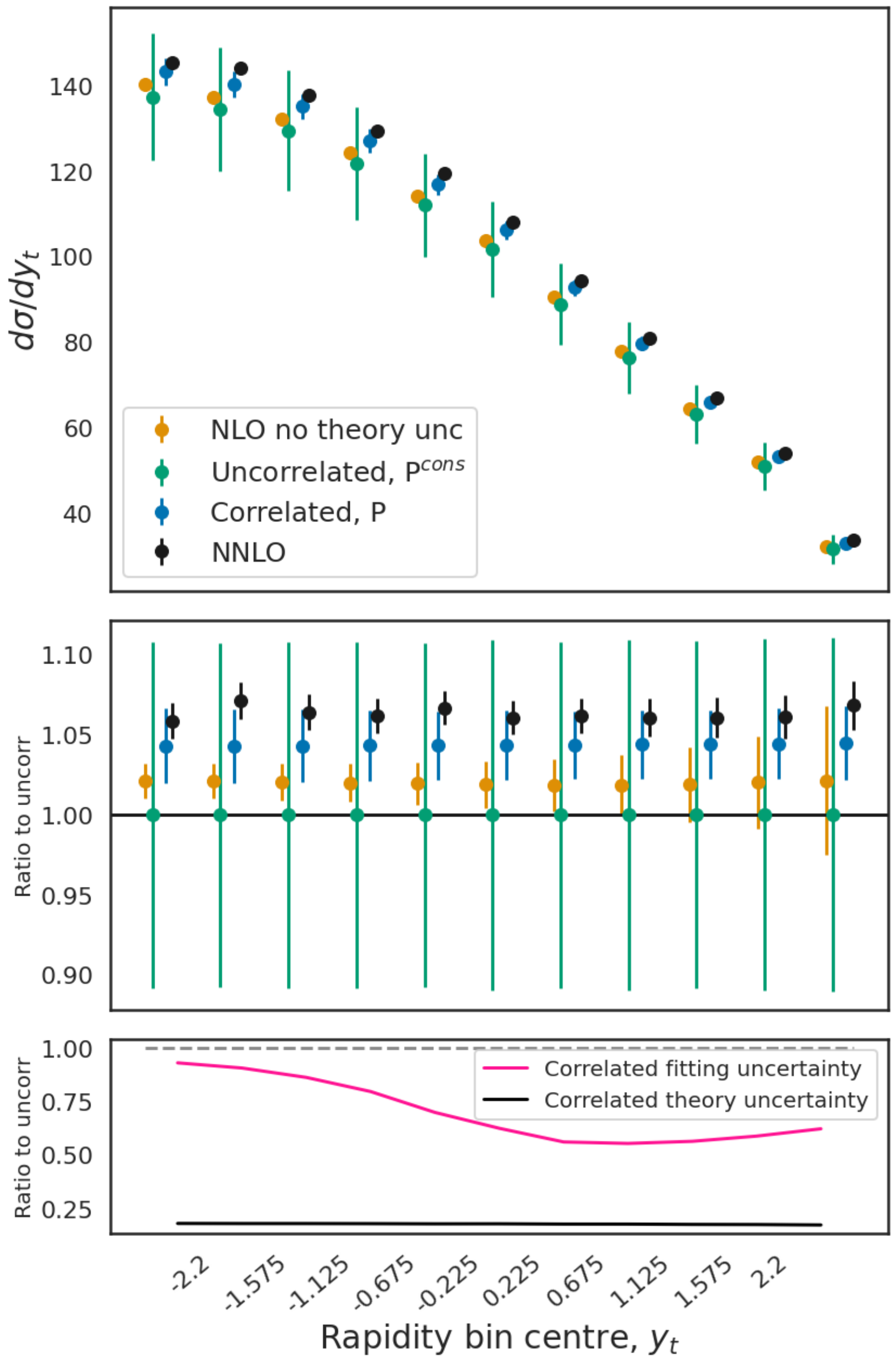}
    \caption{Predictions for $t\bar{t}$ unnormalized rapidity distribution data
taken at 13 TeV by CMS, for the dilepton rapidity distribution (left) and the lepton+jets distribution (right). The four predictions show the NLO NNPDF3.1 fit with no MHOUs, PDF error only; the 
NNPDF3.1 fit with MHOU, and with MHOU in the prediction, but ignoring correlations; the same, but including the correlations between MHOU in fit and prediction; and the N2LO result with no MHOU. In the middle panels,
the same results are shown, but normalized to the uncorrelated result. In the lower panels, we show the fractional
reduction in the PDF uncertainty and the theory in the prediction due to the inclusion of the correlations. Note that these NNPDF3.1 global fits include data for the  $t\bar{t}$ total cross section, but not the rapidity data, so in this example the correlations are particularly strong. }
    \label{fig:DYraps}
\end{figure}
The theory covariance matrix formalism was first applied to incorporate nuclear uncertainties in PDF fits, firstly to data taken on heavy nuclear targets \cite{Ball:2018twp,Pearson:2019upi}, then to data taken on deuteron targets \cite{ Ball:2020xqw, Pearson:2021lmm}. The prior in these examples was determined empirically, through fits to nuclear data. These techniques were then used to incorporate nuclear uncertainties into the NNPDF4.0 fit \cite{Ball:2021leu,NNPDF:2021uiq}. 

More challenging was to apply the new techniques to incorporate missing higher-order uncertainties (MHOU). Here the prior was purely theoretical, determined using scale variations, taking great care to correctly incorporate the correlations between the different processes used in the PDF fit. As a test of principle, the new formalism was applied to the global NLO NNPDF3.1 fit \cite{NNPDF:2019vjt, NNPDF:2019ubu} using a variety of scale variation schemes (5pt, 7pt and 9pt) for variations of the renormalization and factorization scales, and the result was compared to the N2LO fit. The outcome was that, while overall uncertainties in the NLO fit have only increased a little, some shifts were found in central values due to the rebalancing of the data sets in the fit (deweighting the data sets associated with large theoretical uncertainties in the PDF extraction), thereby taking the NLO results closer to the N2LO (see Fig.~\ref{fig:NNPDFscale}).

There has been some debate concerning the appropriate way to use PDFs with MHOUs, since the MHOU in the PDF might be correlated with the MHOU in the matrix elements used in the prediction \cite{Harland-Lang:2018bxd}. This issue can be solved in several ways. One way is to compute the correlations explicitly \cite{Ball:2021icz}, finding that though the correlations can often be ignored (as argued in \cite{NNPDF:2019vjt, NNPDF:2019ubu}), in some circumstances they could lead to significant improvements in the accuracy and precision of predictions 
(see Fig.~\ref{fig:DYraps}). Incidentally, the same correlation machinery could be used to incorporate correlations with PDF uncertainties in extraction of physical parameters, such as $\alpha_s$, $m_W$ or $\theta_W^{\rm eff}$: using PDF sets with a range of fixed physical parameters is not sufficient for parameter extraction, since it does not include the correlation \cite{Forte:2020pyp}.

Another way to incorporate scale uncertainties in PDFs was put forward in ~\cite{Kassabov:2022orn}. There, the idea of a Monte Carlo sampling for propagating experimental uncertainties into the PDF space is extended to the space of factorization and renormalization scales. A prior probability is assigned to each group of QCD scale combinations used to obtain each PDF replica in the Monte Carlo ensemble. A posterior
probability is obtained by selecting replicas that satisfy quality-of-the-fit criteria. Such an approach allows one to exactly match the scale variations in the PDFs with those in the computation
of the partonic cross sections, thus accounting for the full correlations between the two.

\begin{figure}[t!]
    \centering
    \includegraphics[width=0.4\textwidth]{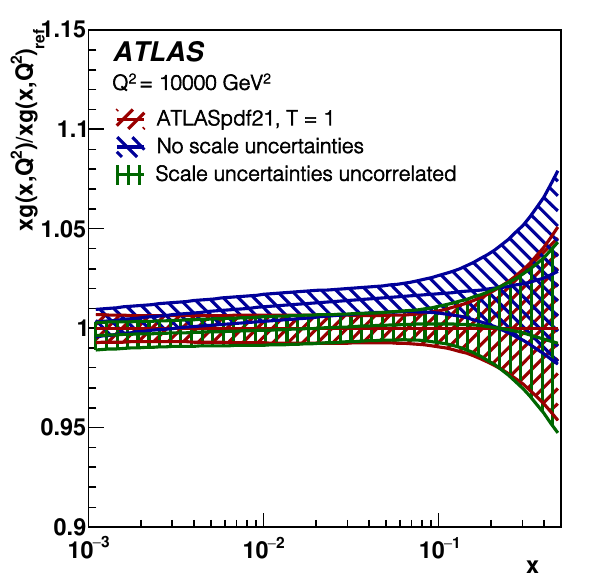}
    \caption{Ratio of the gluon PDF in two fits to the one from the ATLASpdf21 fit  (red) to inclusive $W, Z$ production at 7, 8 TeV, including N2LO scale uncertainties and with scale uncertainties correlated between $W$ and $Z$ and between 7 and 8 TeV data. The blue band is for a fit that does not include these scale uncertainties, and the green band is for the fit that includes them but does not correlate them between 7 and 8 TeV data.}
    \label{fig:atlasscale}
\end{figure}

Theoretical MHOU were recently included in a N2LO PDF determination by the ATLAS collaboration \cite{ATLAS:2021vod}. When calculating $K$ factors for inclusive $W$ and $Z/\gamma^*$ production at 7 and 8 TeV, fully correlated variations of renormalization and factorization scales  were combined in a 5pt scheme and added to the experimental uncertainty in order to estimate the MHOU. The impact is illustrated in Fig.~\ref{fig:atlasscale}. 

In the near future, it is hoped that the next update of the NNPDF4.0 fit \cite{Ball:2021leu,NNPDF:2021uiq},  NNPDF4.1, will include MHOU, correlated across all processes in the global N2LO fit. This will use the newly developed PDF evolution code \eko \cite{Candido:2022tld,candido_alessandro_2022_5896965} that implements an evolution kernel in Mellin space to allow clean variation of the factorization scale and comparison of different truncations of the evolution equations that are equivalent up to MHO corrections. One advantage of fits with MHOU is that they allow inclusion of processes known only at NLO into an N2LO fit, with the MHOU model providing the appropriate deweighting. It will also facilitate the development of N3LO fits, in which the parton evolution will be at (approximate) N3LO, but the processes included in the fit will use a mixture of N2LO and N3LO predictions. These developments would be sufficient for global determination of the strong coupling at N2LO while directly incorporating the MHOU, and then they will allow this determination to be repeated at N3LO.

There has also been progress recently in the development of methods that do not rely on scale variation to estimate MHOU~\cite{Cacciari:2011ze, Bagnaschi:2014wea, Bonvini:2020xeo,Duhr:2021mfd}. These methods generally give rise to priors that are not Gaussian, making it difficult to incorporate them in the theory-covariance matrix formalism (although it should be noted that this problem also exists when estimating experimental systematics, so is perhaps not insurmountable).

\subsection{Machine learning/AI connections}
{\it Leading authors:  S. Carrazza, J. Cruz-Martinez, M. Ubiali \\ \vspace{6pt}}

Machine learning (ML) methods are designed to exploit large data sets in order to reduce complexity and find new features in data. The current most frequently used ML algorithms in HEP are boosted decision trees and neural networks. Machine learning in particle physics is evolving fast, and ML algorithms are already state-of-the-art in many areas of particle physics and will likely be called on to take on a greater role in solving upcoming data analysis and event reconstruction challenges~\cite{Albertsson:2018maf}. 

\subsubsection{PDF determination as a ML problem}
Among various applications, ML techniques have contributed to a better understanding of the proton structure.
PDFs are typically determined by means of a supervised regression model which compares a wide set of experimental data with theoretical predictions computed with a specific PDF parametrization.
A truthful determination of PDFs and of their uncertainties are crucial when producing theoretical predictions for precision studies in high-energy physics.

From a methodological point of view, the choice of a regression model and its uncertainty treatment is a crucial decision, which will impact the quality of PDFs and of theoretical predictions.
The determination of PDFs is a problem very well suited for ML techniques: the functional form is not known, and there is a general consensus on using the log-likelihood $\chi^2$ as the "cost function" to be minimized during the optimization procedure.\footnote{The form of $\chi^2$ may vary e.g. depending on the implementation of nuisance parameters and inclusion of Lagrange Multiplier terms imposing prior constraints.}

The NNPDF collaboration pioneered the usage of Neural Networks (NNs) as universal approximators for model-independent determination of the structure function $F_{2}$ \cite{Forte:2002fg, DelDebbio:2004xtd} and later for full-fledged PDF determinations \cite{Ball:2008by, Ball:2021leu}.
The NNs in these studies have multi-layer feed-forward architectures, also known as multi-layered perceptrons.
In a perceptron, each layer feeds information to the next one in a sequential manner.
Each PDF in the proton is then parametrized at the initial scale  $Q_{0} \simeq 1.6$ GeV of the DGLAP evolution as
\begin{equation}
    f_{i}(x, Q_{0}) = (1-x)^{\beta_{i}}x^{\alpha_{i}}{\rm NN}_{i}(x),
\end{equation}
where the NN plays the same role as an analytical parametrization form used in other PDF determination methods.
The index $i$ represents the parton flavor, while NN$_{i}$ refers to the $i$-th output of the NN. A single network for all parton flavors can also be utilized~\cite{Ball:2021leu}).

\begin{figure}[!t]
 \includegraphics[width=1\textwidth]{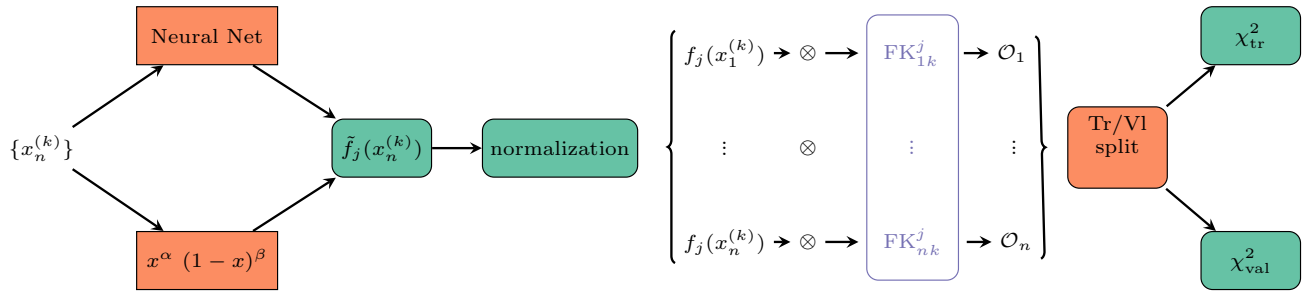}
  \caption{Diagrammatic representation of the calculation of the $\chi^{2}$ in the NNPDF fitting framework as a function of the values of $\{x_n^{(k)}\}$ for the different data sets. Each block indicates an independent component.}
  \label{fig:n3fit}
\end{figure}

The pre-factor $(1-x)^{\beta}x^{\alpha}$, while not strictly necessary \cite{Carrazza:2021yrg}, speeds up the convergence of the network in the extrapolation regions by providing a sensible prior.
The optimization of the parameters of the network has been performed using genetic algorithms until NNPDF3.1~\cite{NNPDF:2017mvq}. Recently, algorithms based on gradient descent have also been implemented within NNPDF framework~\cite{AbdulKhalek:2019mzd, Carrazza:2019mzf}, and the NNPDF4.0 analysis~\cite{Ball:2021leu} is based on them.

The fitting procedure described above requires a number of methodological choices on the exact architecture for the NN, the optimization algorithm and its associated parameters, if any, stopping strategies, etc.
Set of parameters defining the procedure are collectively known as hyperparameters.
In previous versions of NNPDF, these were determined by trial and error. Most recently~\cite{Carrazza:2019mzf,NNPDF:2021uiq} an automatic and systematic hyperparameter scan procedure has been introduced as a fundamental step of the NNPDF methodology.
The faithfulness of the results (of the central value and corresponding uncertainties) are tested by statistical closure tests~\cite{NNPDF:2017mvq,DelDebbio:2021whr}.
These techniques developed by the NNPDF collaboration are not limited to the nucleon PDFs and are also used to determine nuclear PDFs~\cite{Khalek:2022zqe} and fragmentation functions~\cite{Bertone:2017tyb}.

\subsubsection{Simultaneous fits of physics parameters and PDFs}\label{sec:PDFplusX}
Despite the broad consensus on the need for precision, this need  is often reduced to performing better measurements and improving the accuracy of theoretical predictions. However, it is equally important to have a robust framework that is able to globally interpret the LHC data, in particular to spot any subtle deviations from the SM predictions that might arise. While huge progress has been made in determining key ingredients of theoretical predictions from the data, such as the PDFs, $\alpha_s(M_z)$, and EW parameters, it is not yet evident how to combine all these partial fits into a global interpretation of the LHC data that accounts for all relevant correlations.

A very important step in this direction was done in~\cite{H1:2018mkk}, in which a full simultaneous fit of PDFs and EW parameters
was done including full N2LO-QCD and full NLO-EW corrections. Future prospects for such simultaneous fits of PDF and EW precision parameters, using inclusive DIS data, were discussed using projected LHeC data in~\cite{Britzger:2020kgg}, and looking toward a future electron-hadron Circular Collider (FCC-eh) in~\cite{Britzger:2022abi}.

Also very recently, a new methodology dubbed {\tt SIMUnet} was presented in~\cite{Iranipour:2022iak}. It allows for a simultaneous determination of the PDFs alongside any physical parameter that enters theoretical predictions, such as a precision SM parameter or a Wilson coefficient of
some effective field theory (EFT) expansion. The methodology is based on an extension of the {\tt n3fit} methodology, described in the previous section, and the NNPDF4.0 neural network architecture, which treats both the PDFs and the parameters fitted alongside the PDFs on a completely equal footing. 
For this purpose, the NNPDF fitting framework (see Sec.~\ref{sec:nnpdffittingcode}) was extended to incorporate  an extra layer of trainable edges to simultaneously determine the PDFs alongside an arbitrary number of such parameters. The capabilities of the new methodology were illustrated by simultaneously fitting PDFs with a subset of SMEFT Wilson coefficients and showing how the methodology extends naturally to larger subsets of parameters. For example, one could employ the methodology above to yield improved determinations of the PDFs along with the precision parameters, such as the strong coupling constant $\alpha_s(M_z)$, EW couplings of the SM, heavy-quark masses, or a large number of Wilson coefficients in the SMEFT or any other EFT expansion.

Concerning the simultaneous fit of PDFs and new physics parameters, in~\cite{Liu:2022plj}, a joint global fit including both PDFs and a single Wilson coefficient parametrizing a vector-current type lepton-quark contact interaction in the SMEFT was also successfully performed by means of a fast scan in the Wilson coefficient space.  Modifications in the PDFs due to a non-zero Wilson coefficient were studied. Also, the ZEUS and CMS collaborations have performed similar studies of limits on contact interactions in DIS and jet production~\cite{ZEUS:2019cou,CMS:2021yzl}. 

Simultaneous fits of PDFs and other theoretical parameters is certainly a very interesting direction that will receive a big deal of attention in the near future. 

\subsubsection{Other PDF-related ML applications}
Beyond the determination of the PDFs themselves, ML techniques have also been proposed to power up PDF studies. In Ref.~\cite{Carrazza:2015hva}, genetic algorithms are utilized to compress a set of Monte Carlo replicas with the minimal loss of statistical information. In Ref.~\cite{Carrazza:2021hny}, this approach is further expanded with the usage of a generative adversarial model to enhance a PDF set, reducing finite-size artifacts that could be introduced by the compression procedure. In Ref.~\cite{Perez-Salinas:2020nem}, PDFs are approximated using Quantum-ML algorithms in the context of the NNPDF fitting methodology. Another interesting application of ML to the study of PDFs was recently put forward~\cite{Liu:2022plj}. 
Modern PDF analyses require calculations of the log-likelihood functions from thousands of experimental data points, and scans of multidimensional parameter space with tens of degrees of freedom. To overcome lengthy computations of exact log-likelihood functions
in the neighborhood of the best fit and to estimate the PDF uncertainties, the authors put forward NNs and ML techniques to model profile of the log-likelihood functions or cross sections for such a multi-dimensional parameter space. The methodology is applied to the CT18 global analysis and to study the impact of the NOMAD dimuon data on constraining the strange content of the proton. 

\subsection{Delivery of PDFs; PDF ensemble correlations in critical applications \label{sec:MethodologyDeliveryCorrelations}}
{\it Leading authors: R. D. Ball, A. M. Cooper-Sarkar, T. Cridge, B. Malaescu, P. Nadolsky     
\\ 
Contributions from S. Amoroso, A. Apyan, D. Froidevaux, S. Glazov, S.-O. Moch, R. Thorne 
\vspace{6pt}}

An important component of the PDF methodology is the delivery of the PDFs to the users in the form that allows easy yet accurate estimation of a wide range of QCD cross sections and their PDF uncertainties. For this purpose, the PDFs are commonly distributed either as bundles of the central PDF parametrizations and error sets constructed as Hessian eigvenvector sets \cite{Pumplin:2001ct} or as a Monte Carlo (stochastic) ensemble of replicas \cite{Giele:2001mr}. 
All error PDFs propagate the uncertainty from the fitted experimental data. Some PDF ensembles, e.g. \cite{H1:2015ubc,Hou:2019efy,NNPDF:2019ubu,ATLAS:2021vod}, include other sources of uncertainties implicitly or explicitly, such as  
those originating from the choice of input data, the methodology and parametrization, the choice of input theory and the related theoretical uncertainties. 
The error PDFs provide {\it approximations} to the full probability distributions explored in PDF fits. In practice, they reproduce the expectation values and key correlations in the fitted probabilities, while neglecting subleading features to some degree. There is a trade-off between the faithfulness of the reproduction of the full probability and the number of PDF error sets needed for this purpose. Several available methods can be further developed to compress \cite{Gao:2013bia, Carrazza:2015aoa,Carrazza:2015hva, Carrazza:2016htc,Carrazza:2021hny} or diagonalize \cite{Pumplin:2009nm, Dulat:2013kqa} the error PDFs to retain the relevant information with fewer PDF members/replicas. 

As the field advances toward high precision in the LHC Run-3 and at the HL-LHC, more detailed models of PDFs may become necessary in experimental measurements. In particular, in the most precise cases it is observed that measurements performed with different PDF sets can differ by more than the expected PDF uncertainties, without an agreed-upon means to evaluate the degree of compatibility among the different results. A more rigorous and conservative quantitative approach would increase the PDF uncertainty of the measurement (e.g., following Ref.~\cite{Barlow:2002yb}) in the presence of statistically significant tension between results obtained using different PDF sets.
A related question arises about the role of correlations among the PDF ensembles via the fluctuations in their shared fitted data sets.

In an effort taking place within the LHC Standard Model ElectroWeak and the PDF4LHC working groups,
correlations between PDF sets obtained by different groups are being evaluated for the first time~\cite{daniel-PDF4LHC-2018,daniel-LHC-EWWG-2019,bogdan-LHC-EWWG-2019}. 
The study aims to gain precise knowledge of the degree of correlation between different PDF determinations, the essential missing ingredient to evaluate the degree of compatibility between different PDF sets and thereby derive realistic estimates of the overall PDF uncertainties for existing precision SM measurements.
The correlations between different PDF determinations are studied by the means of fits to coherently generated pseudo-experiments, 
first by fitting a reduced ensemble of data sets, and then in a full-scale exercise~\cite{daniel-PDF4LHC-2018} and accounting for fluctuating statistical and systematic correlated experimental uncertainties. The fluctuated data for each generated pseudoexperiment are shared by PDF fits of all partipating groups, and hence the correlations between the fits via their shared data can be studied. A feasibility study of this kind has been already performed \cite{HERAFitterdevelopersTeam:2014fzy} by sharing the fluctuated pseudodata among PDF fits of different perturbative orders. Eventually, a follow-up study may be performed to understand the (de)correlations induced by the use of different parameterizations and fit methodologies, possibly performing comparisons using common theoretical predictions and uncertainties \cite{daniel-PDF4LHC-2018,daniel-LHC-EWWG-2019,bogdan-LHC-EWWG-2019}.

As a related study to these proposals, a determination of the correlations between different PDF sets, using replica ensembles fitted to a common set of data replicas, has now been carried out \cite{Ball:2021dab}. It was found that even when fitted to identical data sets, using common theoretical predictions and parameter settings, different PDF sets are still only partially correlated, since the functional uncertainties arising from different methodologies (in this case, NNPDF3.1 and NNPDF4.0) are still treated as uncorrelated. This suggests it may be challenging to make use of the data or theory correlations to reduce uncertainties when combining different PDF sets or making predictions obtained using different PDF sets, since the methodological correlations are unknown. Exploring this in more details is an interesting focal area for future studies. 

In another recent development, it was pointed out that precisely approximating the full probability with 100-1000 MC replicas is very challenging in models with several tens or hundreds of parameters because of large dimensionality of the associated parameter space \cite{Courtoy:2022ocu}. Representative sampling is needed to adequately explore the space of available PDF solutions, and differences in sampling with independent methodologies are related to the issue of (the absence of) methodological correlations.
\section{PDFs and the strong coupling from lattice QCD \label{sec:Lattice}}
{\it Leading authors: H.-W. Lin, J. H. Weber, P. Nadolsky \\ \vspace{6pt}}

\subsection{Strong coupling calculations}

Precision determinations of PDFs go hand-in-hand with determinations of other QCD parameters: the coupling strength $\alpha_s$ and heavy-quark masses. 
Traditionally or phenomenologically, $\alpha_s(m_Z)$ is obtained by comparing
experimental data involving a hard scale $\nu_h$ to a function
$O(\alpha_s(n\nu_h),n)$ calculated in truncated perturbative QCD (pQCD);
see Refs.~\cite{ParticleDataGroup:2020ssz, dEnterria:2022hzv} for recent reviews.
Nonperturbative lattice gauge theory calculations, anchored to
low-energy QCD by tuning the bare quark masses, provide numerical
results for a wide range of quantities that may be used in place of
experimental data.
The basis of lattice QCD (LQCD) is a regularization of the path integral on a discretized Euclidean
space-time lattice with lattice spacing $a$, implicitly defined as a function of the bare
gauge coupling $g_0$, which permits stochastic evaluation via Markov-chain Monte-Carlo simulations.
Any LQCD predictions are dimensionless ratios, e.g. of dimensional
quantities in units of the spacing $a$. $a$ needs to be fixed in a somewhat
arbitrary \emph{scale setting} procedure that dictates the minimal errors for all
dimensional quantities.
LQCD is systematic, systematically improvable, and permits removing the
regulator (taking a continuum limit $g_0 \to 0$).
Moreover, just like pQCD, LQCD is not restricted to the physical world;
unphysical quantities are fair game, too.
Hence, high-precision LQCD calculations of various $O(\nu_h)$ now play a
major role in determining $\alpha_s(m_Z)$.
Functions $O(\alpha_s(n\nu_h),n)$ entail unknown truncation errors,
which usually dominate the error budget at scales $\nu_h \ll m_Z$.
As truncation errors can only be estimated, e.g., by varying the scale (the number
$n \sim 1$), one should use as many quantities with unrelated truncation errors as possible.
$\alpha_s(n\nu_h)$ is finally connected to $\alpha_s(m_Z)$ by
perturbative running and decoupling.

LQCD calculations suffer from a window problem.
The hierarchy $\Lambda_\text{QCD} \ll \nu_h \ll \sfrac{1}{a}$ is
mandated when comparing to $O(\alpha_s(n\nu_h),n)$; otherwise, (if too small) there
may be substantial nonperturbative effects, and the truncation introduces
large uncertainties, or (if too large) the hard scale is poorly resolved on the
lattice, making continuum extrapolation challenging.
On top of this, lattice simulations should also maintain the hierarchy
$\sfrac{1}{L} \ll m_\pi \ll \Lambda_\text{QCD}$ for a reliable
connection to low-energy QCD.
Unsatisfying realizations of the latter hierarchy are usually subleading
in the error budgets.
Topological freezing (incorrect sampling of the QCD vacuum's topological
sectors), which occurs on fine lattices, seems to have no significant impact on
short-distance quantities~\cite{Weber:2018bam} used in determining $\alpha_s(m_Z)$.
To achieve the sub-$5\%$ precision in the QCD Lambda parameter, with an aim to reach
$1\%$ accuracy in $\alpha_s(m_Z)$ in the next decade, electroweak
or isospin breaking effects can still be safely neglected.

There is a rich trove of literature on lattice determinations of the strong
coupling constant. For modern reviews, see Refs.~\cite{DallaBrida:2020pag,
Komijani:2020kst, dEnterria:2022hzv}.
There have been substantial  efforts to
formulate standardized quality criteria for lattice determinations of
hadronic quantities and $\alpha_s$.
The Flavor Lattice Averaging Group (FLAG) report~\cite{Aoki:2013ldr, Aoki:2016frl,
FlavourLatticeAveragingGroup:2019iem, Aoki:2021kgd} is the most impactful;
FLAG reports its most recent lattice average
\begin{align}
 &\alpha_s(m_Z) = 0.1184(8)&&\text{(FLAG lattice average)~\cite{Aoki:2021kgd}}.
 \label{eq:FLAG average}
\end{align}
There is broad consensus in the LQCD community that such quality criteria, if
applicable, should be applied to phenomenological determinations, too.
In the following, we summarize the most important
conceptually different lattice methods.
Similar to different classes of phenomenological determinations, these methods
are thought to have unrelated truncation errors.
Spread between or within these methods is usually rather narrow. The error
in Eq.~\eqref{eq:FLAG average} is taken to be the smallest among those of the
individual methods instead of the much smaller naive error of a weighted average.

The \emph{step-scaling} method~\cite{Luscher:1991wu, Luscher:1993gh,
deDivitiis:1994yz, Jansen:1995ck} 
relies on the
\emph{Schr\"odinger functional} scheme~\cite{Symanzik:1981wd, Luscher:1985iu,
Luscher:1992an, Sint:1993un}
and allows calculation of
$\alpha_s(\nu_h=\sfrac{1}{L})$ at large energy scales while avoiding the
window problem through a finite-volume ($V=L^3$) approach.
Relevant pQCD expressions are known at N2LO, resp.
$\mathcal{O}(\alpha_s^3(\nu_h)$~\cite{Bode:1999sm}.
The most recent result~\cite{Bruno:2017gxd} for $\alpha_s(m_Z)$ is widely regarded as the most
reliable one and dominates the FLAG average.

Another lattice method uses short distance observables $O(\nu_h)$, such
as \emph{small Wilson loops}~\cite{Mason:2005zx, Davies:2008sw, Maltman:2008bx, McNeile:2010ji}.
The key difference to other lattice methods is that comparison to pQCD is performed
at a finite lattice spacing, which is inversely proportional to the relevant hard
scale $\nu_h=d_O/a$, with a coefficient $d_O \sim \pi$ depending on the observable.
Relevant pQCD expressions are known at N2LO, resp.
$\mathcal{O}(\alpha_s^3(\nu_h))$~\cite{Mason:2005zx}, with higher-order terms limiting precision.

A third lattice method uses the \emph{QCD static energy}. The static energy or the (singlet) free energy~\cite{Bazavov:2018wmo} can be obtained  using Wilson loops or gauge-fixed Wilson and Polyakov line correlators.
As reaching the perturbative regime at
$r \lesssim 0.15\text{ fm}$~\cite{Bazavov:2014soa} presently requires fine
lattices and distances $r \lesssim 5a$ affected by non-smooth
lattice artifacts, the continuum limit is still
under investigation~\cite{Aoki:2021kgd, Komijani:2020kst}.
No scheme change is required to obtain $\alpha_s(\nu_h=\sfrac{1}{r})$, and
continuum pQCD expressions are known at N3LL, i.e. up to
$\mathcal{O}(\alpha_s^{4+n}(\nu_h)\ln^n(\alpha(\nu_h))),
~0 \le n$~\cite{Appelquist:1977tw, Fischler:1977yf, Billoire:1979ih, Schroder:1998vy,
Brambilla:1999qa, Brambilla:2006wp, Brambilla:2009bi, Anzai:2009tm, Smirnov:2009fh,
Lee:2016cgz, Lee:2016lvq}.
When computing the singlet free energy at very short distances, the error budget
is dominated by statistics~\cite{Bazavov:2019qoo}.
Otherwise, uncertainties due to the resummation of ultra-soft logs $\alpha_s^{4+n}(\nu_h)\ln^n(\alpha(\nu_h))$~\cite{
Pineda:2000gza, Brambilla:2004jw, Brambilla:2009bi} and scale variation
generate the lion's share of the error budget~\cite{Bazavov:2019qoo,
Takaura:2018lpw, Takaura:2018vcy, Ayala:2020odx}.

A fourth lattice method (similar to quarkonium sum rules~\cite{Dehnadi:2015fra, Boito:2020lyp}),
uses \emph{heavy-quark two-point correlators}.
In this method the valence heavy-quark mass serves as a hard scale $\nu_h = x m_h$ that can be varied across charm- and bottom-quark regions.
The moments $G_n$ are finite for $n\ge 4$ and known up to N3LO, resp.
$\mathcal{O}(\alpha_s^3(\nu_h))$, for $N_f$ massless and one massive
flavor~\cite{Sturm:2008eb,Kiyo:2009gb,Maier:2009fz}.
The large bare quark mass $am_{h0}$ necessitates improved quark actions,
usually HISQ~\cite{HPQCD:2008kxl,McNeile:2010ji,Chakraborty:2014aca,Maezawa:2016vgv,Petreczky:2019ozv,Petreczky:2020tky} or domain-wall fermions~\cite{Nakayama:2016atf}.
Reduced moments, e.g. $R_4=G_4^{\text{QCD}}/G_4^{\text{0}}$, cancel the tree-level
contribution and associated lattice artifacts~\cite{HPQCD:2008kxl}.
The continuum limit turned out to be challenging, in particular for $R_4$ at $m_h \gtrsim 2m_c$.
Results at $m_h = m_c$ are consistent up to known deficiencies \cite{HPQCD:2008kxl, McNeile:2010ji, Petreczky:2019ozv, Petreczky:2020tky,Nakayama:2016atf}. Recently, reliable results up to $m_h \le 4m_c$ became available~\cite{Petreczky:2020tky}.
The composition of the error budget for $\alpha_s(m_Z)$ varies with $m_h$.
At $m_h \ge 2 m_c$, statistical errors dominate, and nonperturbative contributions
are irrelevant, while at $m_h < 2 m_c$ truncation errors dominate.

A fifth, somewhat similar method uses \emph{light-quark two-point correlators} or
the hadronic vacuum polarization, computed via OPE in the isospin limit for Euclidean momenta $Q^2=-q^2>0$,
where the hard scale is $\nu_h=\sqrt{Q^2}$.
This OPE is done in terms of the Adler function, whose leading coefficient is known at
N4LO, resp. $\mathcal{O}(\alpha_s^4(\nu_h))$~\cite{Chetyrkin:1979bj,Surguladze:1990tg,Gorishnii:1990vf,Baikov:2008jh}.
Further terms are due to nonperturbative contributions.
As the window problem is severe, these calculations are very challenging both in
momentum space~\cite{JLQCD:2008bwj,Shintani:2010ph,Hudspith:2018bpz} or in position
space~\cite{Cali:2020hrj}.

The sixth widely used method, pioneered in Refs.~\cite{Alles:1996ka, Boucaud:2001qz},
uses \emph{QCD vertex functions} in a fixed gauge.
Requiring nonperturbative dressing factors of only ghost and gluon two-point functions,
the ghost-gluon vertex in Landau gauge is particularly simple~\cite{Boucaud:2008gn}.
The respective hard scale is $\nu_h=\sqrt{q^2}$, with $q^\mu$ being the four-momentum
of one ghost and the gluon.
Nonperturbative contributions to the renormalized coupling $\alpha_T(\nu_h)$ in an
intermediate MOM Taylor scheme, known at N4LO, i.e., \ $\mathcal{O}(\alpha_s^4(\nu_h))$~\cite{Blossier:2010ky}, are suppressed for large $\nu_h$,
while fundamental $n$-point functions themselves and the conversion to $\overline{\text{MS}}$ are
only known at N3LO, resp. $\mathcal{O}(\alpha_s^3(\nu_h))$~\cite{Chetyrkin:2000dq}.
Neither calculations with twisted-mass Wilson fermions~\cite{Blossier:2010ky,Blossier:2011tf,Blossier:2012ef,Blossier:2013ioa} nor with domain-wall fermions~\cite{Zafeiropoulos:2019flq} pass FLAG quality criteria.

A novel approach is the \emph{decoupling method}~\cite{DallaBrida:2019mqg, dEnterria:2022hzv}, in which $N_f$ massive quark flavors, with a large common mass $M$ serving as the hard scale $\nu_h=M$, are simultaneously decoupled to connect the running coupling to $N_f=0$. The decoupling relation is known at N4LO, resp. $\mathcal{O}(\alpha_s^4(\nu_h))$.
By matching the theories with different $N_f$ at a scale $M$ that in principle can be arbitrarily high, one expects the truncation errors to be small, and volume effects to be practically irrelevant.

A final, not yet widely used method relies on the \emph{eigenvalues of the Dirac operator}, which are known at N3LO, resp. $\mathcal{O}(\alpha_s^3(\nu_h))$~\cite{Chetyrkin:1994ex, Kneur:2015dda}.
While the resulting $\alpha_s$ turns out a bit high, its large reported errors~\cite{Nakayama:2018ubk} overlap with the FLAG average.

\subsection{Lattice calculations of PDFs}
Since the first proposal of Large-Momentum Effective Theory (LaMET, also called the ``quasi-PDF'' method)~\cite{Ji:2013dva,Ji:2014gla}, there has been rapid progress in computing the dependence of PDFs on the momentum fraction $x$ on the lattice.
LaMET evaluates equal-time spatial correlators, whose Fourier transforms, called "the quasi-PDFs", become the lightcone PDFs in the limit of an infinite hadron momentum.
For large but finite momenta accessible on a realistic lattice, LaMET relates quasi-PDFs to physical ones  in the $\overline{\text{MS}}$ scheme through a factorization theorem, the proof of which was developed in Refs.~\cite{Ma:2017pxb,Izubuchi:2018srq,Liu:2019urm}.
After the first lattice $x$-dependent PDF calculation~\cite{Lin:2014zya}, many other calculations were done.
Alternative approaches to lightcone PDFs in LQCD are ``operator product expansion (OPE) without OPE''~\cite{Aglietti:1998ur,Martinelli:1998hz,Dawson:1997ic,Capitani:1998fe,Capitani:1999fm,Chambers:2017dov,QCDSF-UKQCD-CSSM:2020tbz,Horsley:2020ltc},
``auxiliary heavy/light quark''~\cite{Detmold:2005gg,Detmold:2018kwu,Detmold:2020lev,Braun:2007wv},
``hadronic tensor''~\cite{Liu:1993cv,Liu:1998um,Liu:1999ak,Liu:2016djw,Liu:2017lpe,Liu:2020okp},
``good lattice cross sections''~\cite{Ma:2017pxb,Bali:2017gfr,Bali:2018spj,Sufian:2019bol,Sufian:2020vzb} and the pseudo-PDF approach~\cite{Radyushkin:2017cyf}.
For recent reviews on these topics, we refer readers to Refs.~\cite{Lin:2017snn,Cichy:2018mum,Zhao:2020vll,Ji:2020ect,Ji:2020byp,Constantinou:2020hdm}.

\subsubsection{Nucleon PDFs}
\begin{figure}[tb]
\subfloat[]{
\includegraphics[width=.47\textwidth]{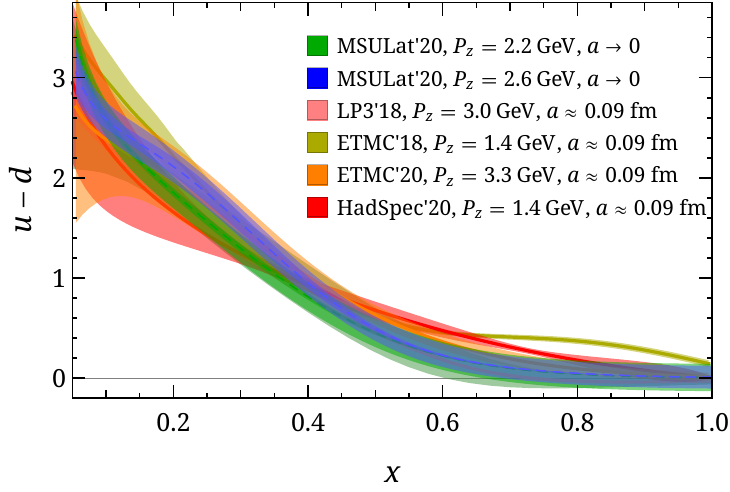}}
\subfloat[]{
\includegraphics[width=0.48\textwidth]{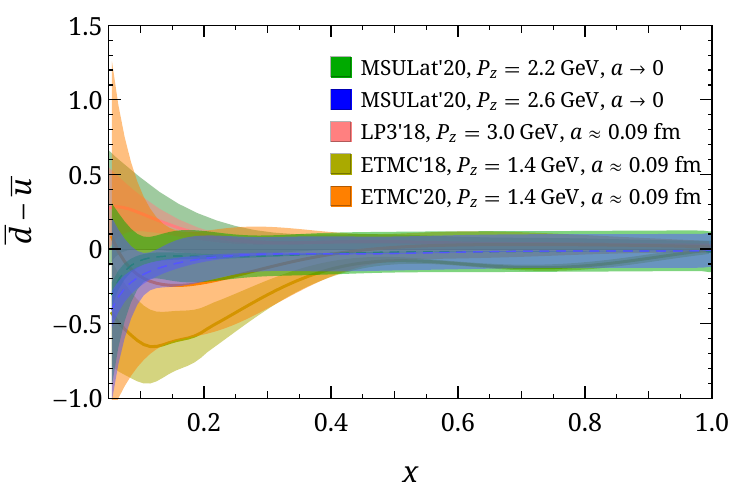}}\\
\subfloat[]{
\includegraphics[width=0.47\textwidth]{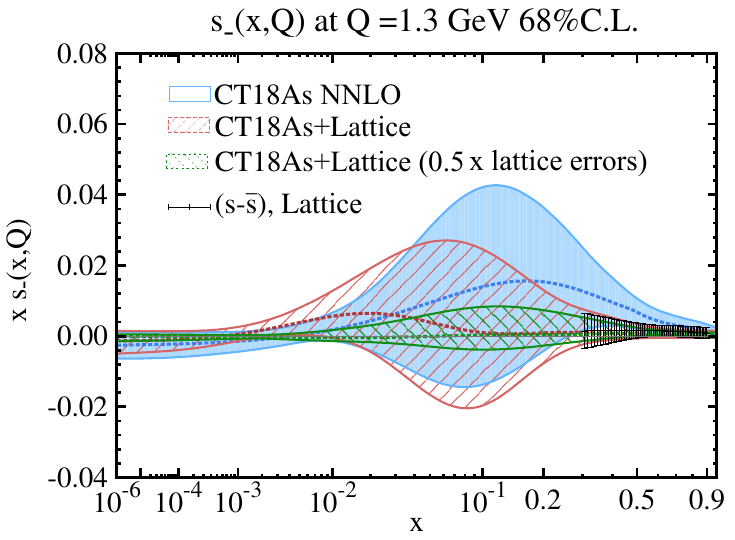}}
\subfloat[]{
\includegraphics[width=.48\textwidth]{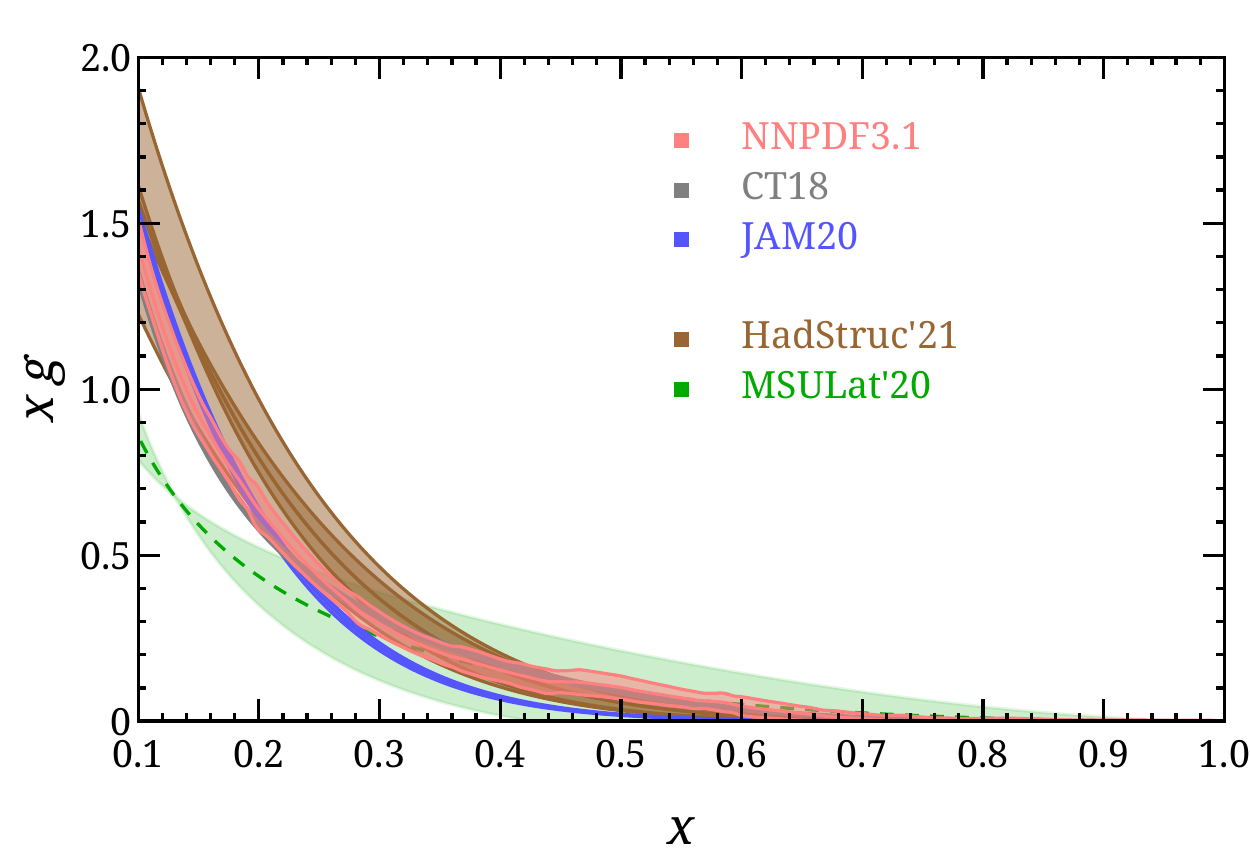}}
\caption{
Unpolarized nucleon PDFs. The isovector quark (a) and antiquark (b) PDFs from a lattice calculation ``MSULat'20''~\cite{Lin:2020fsj} in the physical-continuum limit, and from single-lattice-spacing calculations at (or extrapolated to) physical pion mass  using the LaMET ( ``LP3'18''~\cite{Chen:2018xof} and ``ETMC'18''~\cite{Alexandrou:2018pbm}),  and pseudo-PDF  (``ETMC'20''~\cite{Bhat:2020ktg} and ``HadSpec'20''\cite{Joo:2020spy}) methods.
(c) Impact of the constraints from LQCD (black dashed area) on the difference between the strange quark and antiquark PDFs in a recent CT18As N2LO fit \cite{Hou:2022sdf}. The red (green) error bands are obtained with the current (reduced by 50\%) LQCD errors.
(d) The  gluon PDF, $xg(x,Q)$, at $Q=2$~GeV,
obtained from fits to the lattice data at pion masses $M_\pi=135$ (extrapolated), 310 and 690~MeV by MSULat~\cite{Fan:2020cpa}, and at $M_\pi=380$~MeV by HadStruc21~\cite{HadStruc:2021wmh}, compared with the NLO JAM20 \cite{Moffat:2021dji} and N2LO CT18 \cite{Hou:2019efy} and NNPDF3.1 \cite{Ball:2017nwa} gluon PDFs.
}
\label{fig:nucleonPDF}
\end{figure}

\paragraph{Isovector quark PDFs.} The nucleon unpolarized isovector PDF combinations, $u(x)-d(x)$ and $\bar d(x)-\bar u (x)$, are perhaps  the most studied $x$-dependent structures, with 
multiple collaborations reporting either direct lattice calculations at physical pion mass or extrapolations to physical pion mass using quasi-PDF~\cite{Lin:2017ani,Alexandrou:2018pbm,Chen:2018xof} and pseudo-PDF methods~\cite{Bhat:2020ktg,Joo:2020spy}.
Reference~\cite{Lin:2020fsj} presents the first lattice-QCD calculation of the nucleon isovector unpolarized PDFs in the physical-continuum limit, using ensembles with multiple sea pion masses as low as around 135~MeV, three lattice spacings $a\in[0.06,0.12]$~fm, and multiple volumes with $M_\pi L$ ranging from 3.3 to 5.5. It performed
a simultaneous chiral-continuum extrapolation to obtain renormalized nucleon matrix elements in the RI/MOM scheme, using various Wilson-link displacements and four physical-continuum matrix elements.
Figures~\ref{fig:nucleonPDF}(a) and \ref{fig:nucleonPDF}(b) show the results of the lattice calculations for the isovector combinations using at least one near-physical pion mass.
The predictions have different systematics, some taken into account, some not.
Overall, they are in a reasonable agreement after scaling up the systematic uncertainties.
However, the antiquark isovector combination in Fig.~\ref{fig:nucleonPDF}(b) and the small-$x$ region suffer from large systematic uncertainties that can only be removed when using a large value of $P_z$~\cite{Chen:2018xof,Lin:2018qky}, as predicted previously~\cite{Lin:2017ani}.
Increasing the boost momentum of the lattice calculations will be critical for extending the impact of future lattice PDF calculations at both large and small $x$. Using large boost momenta in the nucleon, on the other hand, may enlarge the statistical errors beyond the manageable level even in high-statistics measurements.

\paragraph{Strangeness and charm PDFs.}
The strange and charm parton distributions were computed in Ref.~\cite{Zhang:2020dkn} by evaluating light ($M_\pi\approx310$~MeV) and strange ($M_\pi\approx690$~MeV) nucleon two-point correlators in the LaMET approach,  with 344,064 (57,344) measurements in total, and allowing extrapolation to physical pion mass.
The renormalized real matrix elements, quantifying the difference $q(x)-\bar q(x)$ at $x>0.3$, turned out to be zero within the statistical errors for both $q=s$ and $c$. This result supports the assumption of a good symmetry between $s$ (or $c$) quarks and antiquarks at large $x$ adopted in some phenomenological PDFs. The imaginary matrix elements quantify the sum of the quark and antiquark distribution, $q(x)+\bar q(x)$, and the respective strange contribution is about a factor of 5 or larger than the charm one.
They are consistently smaller than those from CT18 and NNPDF3.1, possibly due to the missing contributions from the mixing with gluon matrix elements in the renormalization.
The later work by ETMC~\cite{Alexandrou:2020uyt}, which calculated both light and strange lattice matrix elements at 260-MeV pion mass, extracted individual (anti)quark sea PDFs with the mixing in the quark and gluon sectors neglected.
Future lattice calculations, expected to include the gluon mixing, will be crucial for predicting the sea quark and antiquark PDFs.

Such lattice predictions can already provide constraints on poorly known PDF combinations at large momentum fractions. As an illustration, Ref.~\cite{Hou:2022sdf} replaced the default assumption $s(x,Q_0)=\bar s(x,Q_0)$ of the CT18 family of N2LO PDFs \cite{Hou:2019gfw} by allowing a small $s_-(x,Q_0) \equiv s(x,Q_0) -\bar s(x,Q_0)$ asymmetry at the initial scale $Q_0 = 1.3$~GeV according to the strategy from  Ref.~\cite{Lai:2007dq} and using updated experimental data and PDF functional forms. The resulting N2LO fit produced CT18As PDFs, i.e.  the alternative PDF set CT18A ensemble \cite{Hou:2019gfw} that includes the ATLAS $\sqrt{s} = 7$~TeV $W$, $Z$ combined cross section measurements \cite{ATLAS:2016nqi} and is augmented by allowing a non-zero $s_-(x,Q_0)$. 
The impact of the current lattice data points, available at $x>0.3$, on CT18As PDFs at $Q_0$ is shown in Fig.~\ref{fig:nucleonPDF}(c).  Compared to the CT18As error band, the uncertainty in lattice data points is quite small, so that including the lattice data in the CT18As\_Lat fit greatly reduced the $s_-$ uncertainty in the large-$x$ region. The degree of reduction in the uncertainty at lower $x$ depends on the chosen parametrization form for $s_-(x,Q_0)$. 
The projection for further feasible reduction of this uncertainty is illustrated by a PDF ensemble labeled ``CT18As\_HELat'',  which is obtained by including the same lattice data with the errors reduced by a half. 

\paragraph{Gluon PDF.} 

Since gluon quantities are much noisier than quark disconnected loops, LQCD calculations with very high statistics are necessary. An exploratory study applying LQCD to the gluon PDF \cite{Fan:2018dxu}, using overlap valence fermions on gauge ensembles with 2+1 flavors of domain-wall fermions at $M_\pi^\text{sea}=330$~MeV, calculated the gluon operators 
for all spacetime lattice sites at such high statistics. 207,872 measurements were taken of the two-point functions with valence quarks at the light sea and strange masses.
The coordinate-space gluon quasi-PDF matrix element ratios were
compared to the ones based on the gluon PDF from two NLO global fits: the \textsc{PDF4LHC15} combination~\cite{Butterworth:2015oua} and \textsc{CT14}~\cite{Dulat:2015mca}.
Up to perturbative matching and power corrections at $O(1/P_z^2)$, the lattice results were compatible with global fits within the statistical uncertainty at large $z$.

There have been more recent developments in improving the operators for gluon-PDF lattice calculations~\cite{Balitsky:2019krf,Wang:2019tgg,Zhang:2018diq}, which facilitates taking the continuum limit for the gluon PDFs in the future.
Figure~\ref{fig:nucleonPDF}(d) shows two nucleon gluon PDFs calculations, by MSULat~\cite{Fan:2020cpa} and HadStruc~\cite{HadStruc:2021wmh} (via SDF method).
MSULat~\cite{Fan:2020cpa} performed an exploratory study using about 30k nucleon measurements with pion masses 310 and 690~MeV, and extrapolating to the physical pion mass.
HadStruc~\cite{HadStruc:2021wmh} used multiple nucleon interpolating fields to solve the generalized eigenvalue problem in order to extract the gluonic matrix elements at 358-MeV pion mass.

\paragraph{Helicity and transversity.}
Early exploratory works have shown great promise for predicting quantitatively the helicity and transversity quark and antiquark distributions~\cite{Chen:2016utp}.
There have been two attempts,  by
ETMC~\cite{Alexandrou:2018pbm} and LP$^3$~\cite{Lin:2018qky}, to improve the helicity PDFs by removing the heavy pion-mass systematic.
Transversity computations have been dominated by the quasi-PDF method carried out by LP$^3$~\cite{Liu:2018hxv} and ETMC~\cite{Alexandrou:2018eet}, who reported transversity results at the physical pion mass in 2018.
Recently, HadStruc Collaboration reported results from the pseudo-PDF approach with a lattice spacing $a = 0.094$~fm and 358-MeV pion mass~\cite{HadStruc:2021qdf}.
Excited-state systematics in the above works are carefully studied using multiple source-sink separations. Some results include the errors coming from lattice-spacing and finite-volume effects, varying the renormalization scale, the choice of $z P_z$ in the Fourier transform, or approximations made in the matching formula. 

\begin{figure}[p]
\subfloat[]{
\includegraphics[width=.40\textwidth]{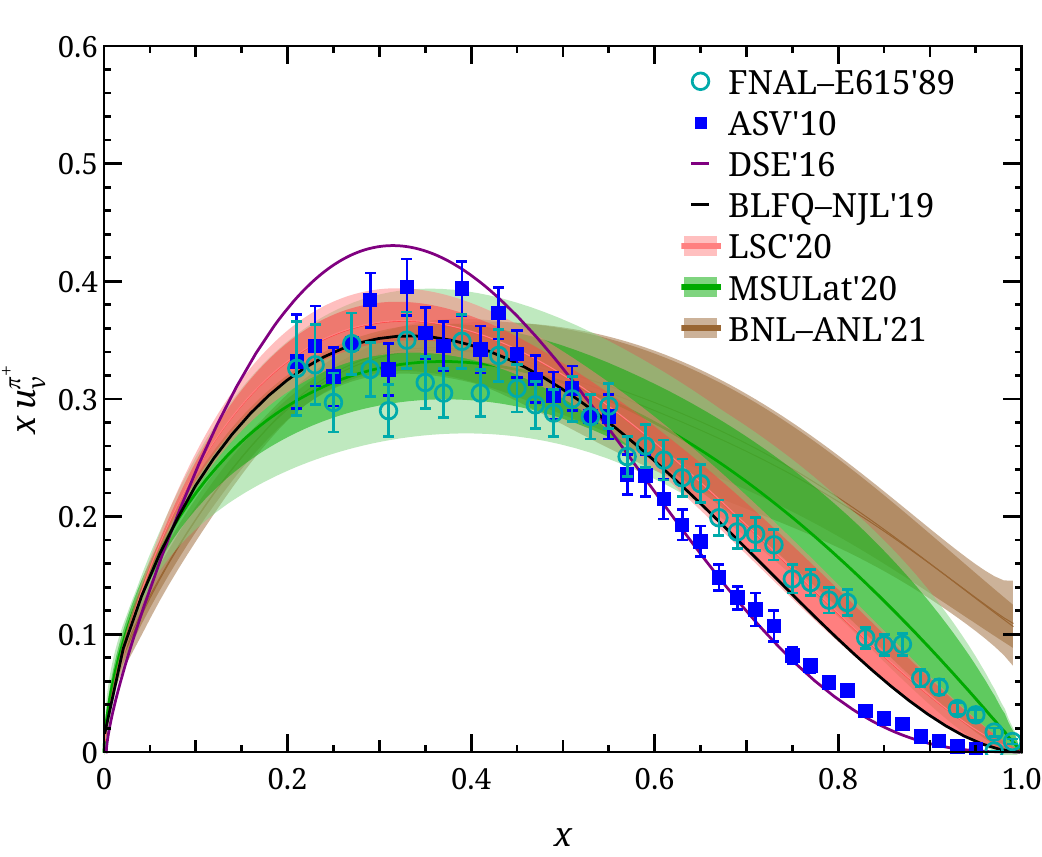}}
\subfloat[]{
\includegraphics[width=.48\textwidth]{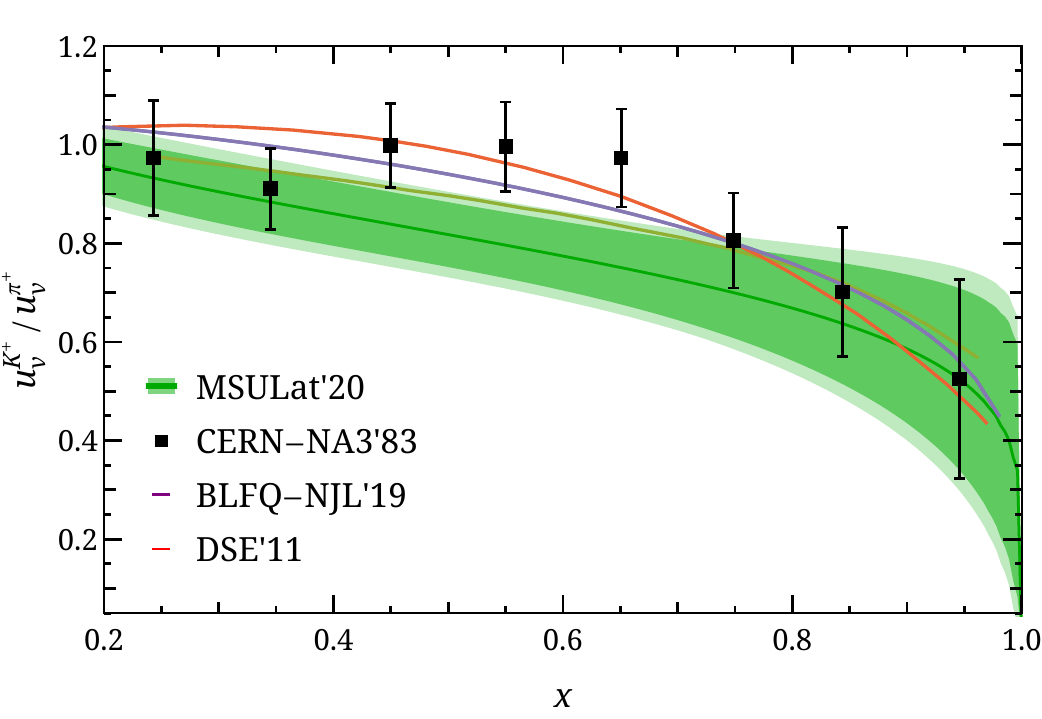}}\\
\vspace*{-10pt}\subfloat[]{
\includegraphics[width=.48\textwidth]{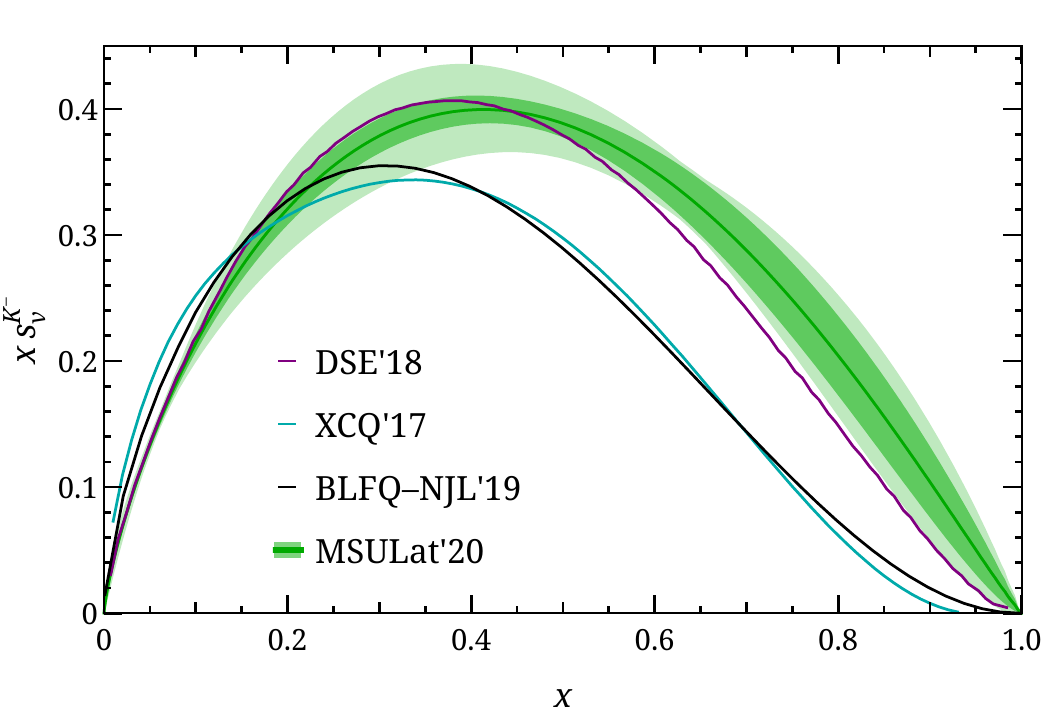}}\\
\vspace*{-10pt}\subfloat[]{
\includegraphics[width=.48\textwidth]{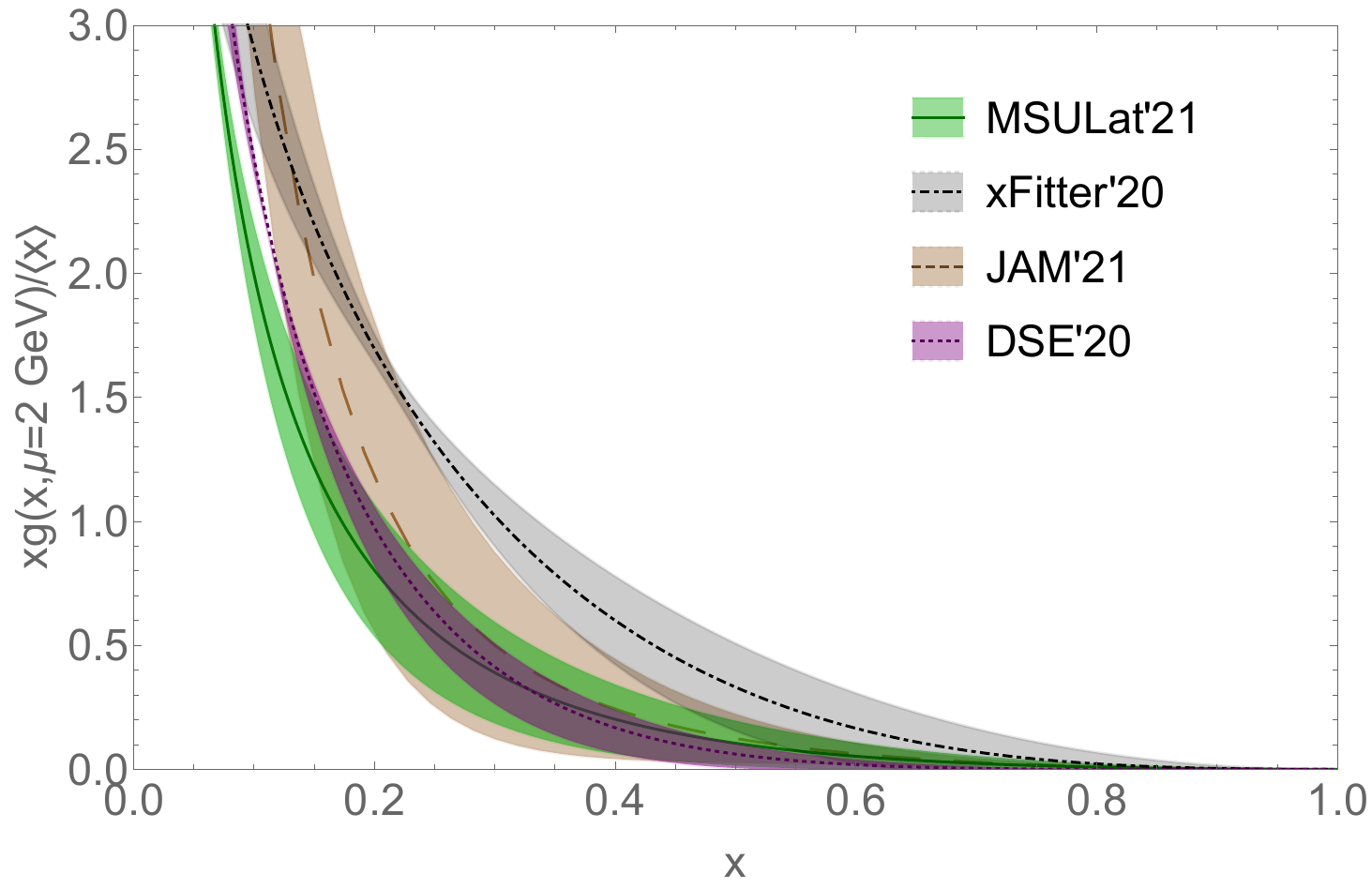}}
\subfloat[]{
\includegraphics[width=0.48\textwidth]{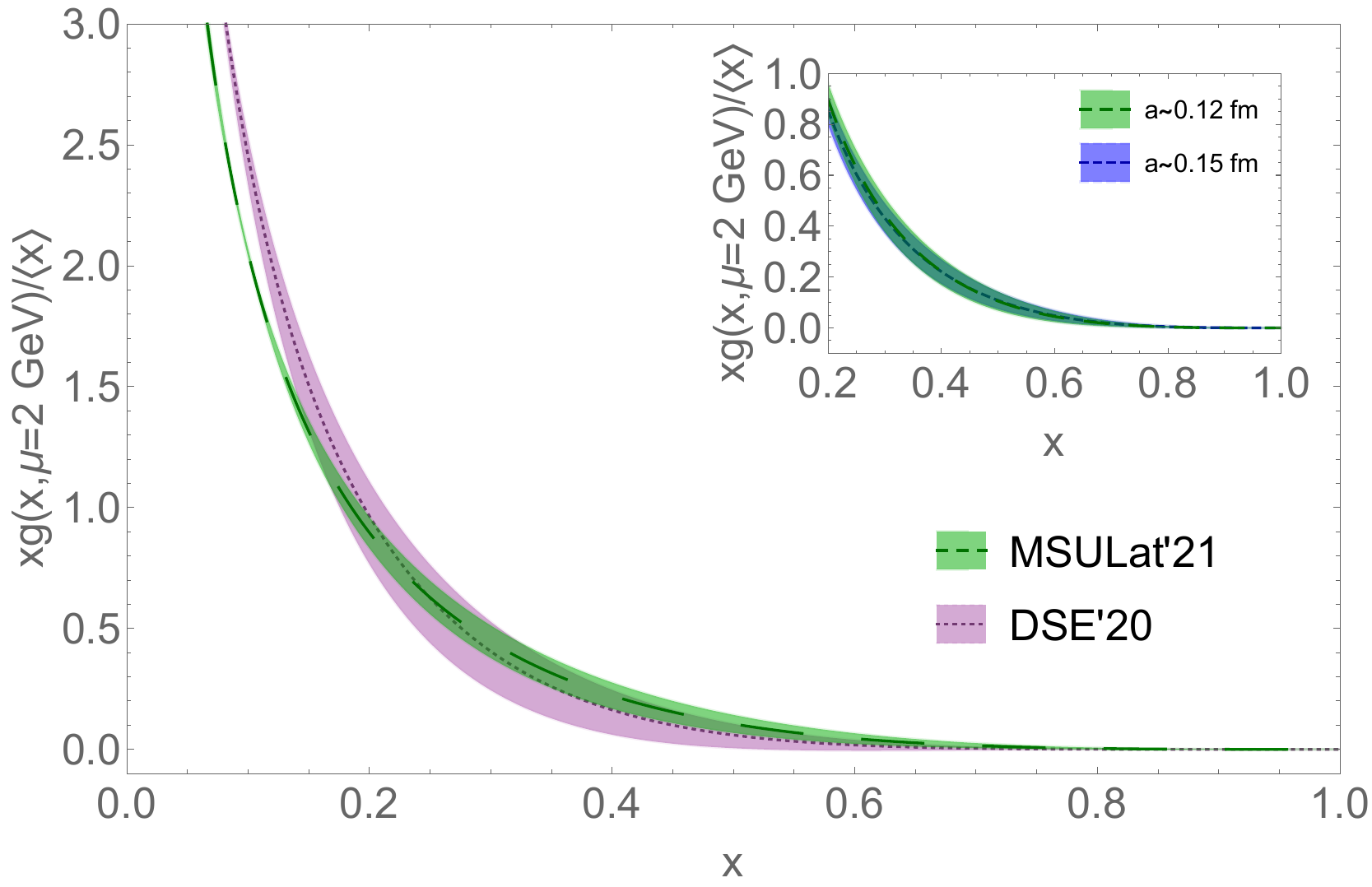}}
\caption{Meson PDFs. (a) Extraction of the valence quark distribution of the pion from the FNAL-E615 Drell-Yan data (open circles \cite{Conway:1989fs}), compared with a threshold resummation calculation ``ASV'10''  (closed circles \cite{Aicher:2010cb}), Dyson-Schwinger equation formalism \cite{Chen:2016sno}, and lattice calculations with the LCS (JLab and W\&M group~\cite{Sufian:2020vzb}) and LaMET (BNL~\cite{Gao:2021dbh} and MSULat~\cite{Lin:2020ssv})
methods.
(b) The ratio of the light-quark valence distribution of a kaon to that of a pion. (c) $x \overline{s}_v^K(x,\mu)$ at a scale of $\mu^2=27\text{ GeV}^2$, from lattice calculation by MSULat~\cite{Lin:2020ssv}, along with results from relevant experiments and other calculations.
The inner bands indicate statistical errors with the full range of $zP_z$ data, while the outer bands include errors from using different data choices and fit forms.
(d) 
The pion gluon PDF $xg(x, \mu)/\langle x \rangle_g$ obtained from a fit to the lattice data from a 220-MeV ensemble~\cite{Fan:2021bcr}.
(e)
The kaon gluon PDF $xg(x, \mu)/\langle x \rangle_g$ obtained from fits to the lattice data on ensembles with lattice spacing $a\approx\{0.12,0.15\}$~fm (inset plot), pion masses $M_\pi\approx310$~MeV at $a\approx 0.12$~fm \cite{Salas-Chavira:2021wui}, compared with the kaon gluon PDF from DSE'20 at $\mu=2$~GeV.
\label{fig:mesonPDFs}}
\end{figure}

\subsubsection{Meson PDFs}
\paragraph{Quark PDFs.} In one of its early lattice-QCD calculations of the pion and kaon valence-quark distribution functions~\cite{Zhang:2018nsy,Lin:2020ssv}, MSULat employed multiple pion masses (with the lightest one around 220~MeV), two lattice spacings $a=0.06$ and 0.12~fm, and $(M_\pi)_\text{min} L \approx 5.5$. 
The chiral-continuum extrapolation was performed to obtain the renormalized matrix elements at physical pion mass, using a simple ansatz to combine the data from 220, 310 and 690~MeV: $h^R_{i}(P_z, z, M_\pi) = c_{0,i} + c_{1,i} M_\pi^2 + c_{a} a^2$ with $i=K, \pi$.
Mixed actions, with light and strange quark masses tuned to reproduce the lightest sea light and strange pseudoscalar meson masses, can suffer from additional systematics at $O(a^2)$;
such artifacts are accounted for by the $c_{a}$ coefficients, and all the $c_{a}$ values were found to be consistent with zero.
The JLab and W\&M group reported another lattice study of the pion valence-quark distribution~\cite{Sufian:2020vzb}, using the current-current correlator method (also called ``lattice cross sections'', LCS) and
extrapolating to physical pion mass and continuum limit ($a=0.09$ and 0.12~fm, $(M_\pi)_\text{min}=278$~MeV). 
Most lattice calculations of PDFs in both SDF and LaMET methods employed NLO matching, or equivalently the NLO Wilson coefficients~\cite{Xiong:2013bka,Ma:2014jla,Ji:2017rah,Ji:2020ect}.
N2LO matching exists~\cite{Chen:2020ody,Li:2020xml} and has been employed by the BNL lattice group~\cite{Gao:2021dbh} in recent lattice calculations of the valence pion PDF with  very fine lattice spacings, 0.03 and 0.04~fm, at 310-MeV pion mass.

The behavior of meson valence PDFs in the limit $x\to 1$ draws attention as an interesting test of nonperturbative models of hadron structure. Quark counting rules ~\cite{Ezawa:1974wm,Farrar:1975yb,Berger:1979du,Soper:1976jc} and a number of nonperturbative approaches, such as the Dyson-Schwinger equations (DSE) \cite{Chen:2016sno,Ding:2019lwe}, predict the PDF falloff $f(x,Q) \propto (1-x)^\beta$ with $\beta \gtrsim 2$ for mesons and $\beta \gtrsim 3$ for baryons. Observing this primordial behavior requires one to go to $x \gtrsim 0.9$ \cite{Courtoy:2021xcc}, the region where higher-twist terms and threshold resummation \cite{Aicher:2010cb,Barry:2021osv} affect experimental observables. Instead of attempting to derive the nominal $(1-x)$ power $\beta$ of the parametrization, which is not unique because of the functional mimicry \cite{Courtoy:2020fex}, it is more informative to analyze an effective exponent $\beta_{\rm{eff}}(x,Q) \equiv\ln(f(x,Q))/\ln(1-x)$ \cite{Ball:2016spl,Courtoy:2020fex}, which reduces to $\beta$ in the extreme $x\to 1$ limit. 
The effective exponent can be computed both in phenomenological and nonperturbative approaches, including LQCD. Given the simpler structure of mesons and their smaller masses, interpretation of the $x\to 1$ PDF asymptotics is cleaner for mesons than for baryons.

As an illustration, Fig.~\ref{fig:mesonPDFs} shows several determinations of the pion valence momentum distribution, $x u_v^{\pi^+}(x, \mu)$, evolved to a scale $\mu^2 = 27\text{ GeV}^2$. A derivation of $x u_v^{\pi^+}(x, \mu)$ from an original analysis of the pion-tungsten Drell-Yan data by the FNAL-E615 experiment \cite{Conway:1989fs} is shown as full circles.
Superimposed is the nominal MSULat's parametrization \cite{Lin:2020ssv}, computed at a physical pion mass and approaching $x=1$ as $(1-x)^{1.01}$, with the shape consistent with the FNAL-E615 analysis. An LCS calculation from \cite{Sufian:2020vzb} shows a similar behavior. 
BNL's computation \cite{Gao:2021dbh} reliably determines $x u_v^{\pi^+}(x, \mu)$ for $x$ up to 0.8, where it is consistent with the other lattice results. On the other hand, at $x>0.6$, there is still substantial uncertainty. Here the DSE formalism \cite{Chen:2016sno} predicts a faster fall-off, at least as $(1-x)^2$ as $x \to 1$.\footnote{The nominal lowest $(1-x)$ power alone does not determine the shape at $x$ below 1, as it is correlated with higher powers in the parametrization \cite{Courtoy:2020fex}.} Re-analyses of the FNAL-E615 experiment data that include NLL threshold resummation corrections predict either a faster falloff, as for the shown `ASV'10'' extraction \cite{Aicher:2010cb}, or a slower falloff \cite{Barry:2021osv}, depending on the treatment of power-suppressed corrections. 

Figure~\ref{fig:mesonPDFs}(b) shows the ratios $u_v^{K^+}/u_v^{\pi^+}$ of the light-quark distributions in the kaon to the ones in the pion.
When comparing the LQCD result with the experimental determination of the valence quark distribution via the Drell-Yan process by NA3 Collaboration, good agreement is found between the LQCD results and the data.
The LQCD result approaches $0.4$ as $x \to 1$ and agrees nicely with other analyses, such as the constituent quark model, the DSE approach (``DSE'11''),
and basis light-front quantization with color-singlet Nambu--Jona-Lasinio interactions (``BLFQ-NJL'19'').

The LQCD prediction for $x s_v^{K}$ is also shown in Fig.~\ref{fig:mesonPDFs}(c), with the lowest three moments of $s_v^{K}$ being $0.261(8)_\text{stat}(8)_\text{syst}$, $0.120(7)_\text{stat}(9)_\text{syst}$, $0.069(6)_\text{stat}(8)_\text{syst}$, respectively. These results can be compared against the QCD-model estimates
from the chiral constituent-quark model (0.24, 0.096, 0.049) and DSE~\cite{Chen:2016sno} (0.36, 0.17, 0.092).

\paragraph{Gluon PDFs.} 
The gluon quasi-PDFs in the pion were studied for the first time in Ref.~\cite{Fan:2018dxu}, and features similar to those observed for the proton were revealed.
Figures~\ref{fig:mesonPDFs}(d) and \ref{fig:mesonPDFs}(e) show the gluon PDFs of the pion (2 lattice spacings: 0.09 and 0.12~fm with $M_\pi \approx 220$, 310 and 690~MeV)~\cite{Fan:2021bcr} (see also \cite{Fan:2021rvx}), as well as the first kaon gluon PDF~\cite{Salas-Chavira:2021wui} (right) at 310-MeV pion mass.

\subsubsection{Other lightcone quantities
\label{sec:LatticeLightconeQuantities}
}

There has been recent progress in determination of $x$-dependent  meson distribution amplitudes (DAs)~ \cite{Zhang:2017bzy,Zhang:2017zfe,Bali:2017gfr,Bali:2018spj,RQCD:2019osh,Hua:2020gnw,Zhang:2020gaj,Detmold:2021qln,Juliano:2021hys,LatticeParton:2022zqc,Gao:2022vyh}. 
DAs are important universal quantities to describe exclusive processes at large momentum transfers $Q^2 \gg \Lambda^2_\text{QCD}$ using factorization theorems. Some well-known examples of such processes, which are relevant to measuring fundamental parameters of the Standard Model, include $B \to \pi l \nu$, $\eta l \nu$ giving the CKM matrix element $ |V_{ub}|$, $B \to D \pi$ used for tagging, and $B \to \pi \pi$, $K \pi$, $K\bar{K}$, $\pi \eta$, etc., which are important channels for measuring CP violation (see e.g. Ref.~\cite{Stewart:2003gt}). The lattice studies also help us to 
understand the flavor SU(3) symmetry breaking among light flavors before attributing the effects to enhancement of higher-order amplitudes or even new physics.

New experiments and facilities will explore the three-dimensional  structure of hadrons described by the transverse-momentum-dependent (TMD) PDFs (discussed in Sec.~\ref{sec:TMDs}) and generalized parton distributions (GPD's). TMD PDFs depend on the parton's transverse momentum $k_T$, in addition to the longitudinal momentum fraction $x$. They are nonperturbative inputs for processes that follow TMD factorization, such as the Drell-Yan process and SIDIS.
Early lattice studies computed selected TMD functions at heavier-than-physical pion masses ranging down to about 300 MeV~\cite{Musch:2010ka,Musch:2011er,Engelhardt:2015xja,Yoon:2015ocs,Yoon:2017qzo}. Recently, there were first extraction of the Collins-Soper kernel, soft function and wavefunctions for TMDs~\cite{Shanahan:2020zxr,LatticeParton:2020uhz,Schlemmer:2021aij,Li:2021wvl,Zhang:2020dbb,Shanahan:2021tst}.
Like for the PDF calculations, lattice precision calculations of TMDs will require large hadron momentum to suppress the power corrections at ${\cal O}(1/(P_zb_T)^2)$.

GPDs are hybrid momentum and coordinate-space distributions of partons that bridge the standard nucleon structure observables, form factors and inelastic cross sections. 
Only recently have there been a few lattice calculations made for the pion GPDs with the pion mass of 310 MeV \cite{Chen:2019lcm}, and for nucleon GPDs with the pion masses of 260 MeV~\cite{Alexandrou:2020zbe} and 139 MeV~\cite{Lin:2020rxa,Lin:2021brq}.
These calculations require an increase in computational resources by at least an order of magnitude  relatively to PDF calculations due to the additional boost momenta required. For the best determinations of GPDs, the lattice results can be combined with experimental measurements in a global analysis. 

\subsection{Outlook and challenges
\label{sec:LatticeOutlook}
}
A Snowmass whitepaper~\cite{Constantinou:2022yye} provides more details on the rapid advances in LQCD calculations of PDFs and other QCD functions and has more complete references to relevant work.
Experimental exploration of the three-dimensional structure at the Jefferson Lab, EIC, other facilities will match the ongoing theoretical advancements that open doors to many previously unattainable predictions, from the $x$ dependence of collinear nucleon PDFs to TMDs~\cite{Musch:2010ka,Musch:2011er,Engelhardt:2015xja,Yoon:2016dyh,Yoon:2017qzo} and related functions~\cite{Shanahan:2020zxr,Zhang:2020dbb,Schlemmer:2021aij,Li:2021wvl,Shanahan:2021tst}, GPDs~\cite{Chen:2019lcm, Alexandrou:2020zbe,Lin:2020rxa,Lin:2021brq}, and higher-twist terms -- progress that was not envisioned as possible during the 2013 Snowmass study.

There remain challenges to be overcome in the lattice calculations, such as reducing the noise-to-signal ratio, extrapolating to the physical pion mass, and increasing hadronic boosts to suppress systematic uncertainties.
Computational resources place significant limitations on the achievable precision, as sufficiently large and fine lattices are necessary to suppress finite-size and higher-twist contaminating contributions.
New ideas can bypass these limitations.
With sufficient support, lattice QCD can fill in the gaps where the experiments are difficult or not yet available, improve the precision of global fits, and provide better SM inputs to aid new-physics searches across several HEP frontiers.
\section{Nuclear and meson PDFs \label{sec:nuc_PDFs}}
{\it Leading authors: T. J. Hobbs, E. R. Nocera, F. I. Olness}

%
The QCD theory of the strong interactions is among the most complex
and enigmatic, displaying both confinement of the quarks and gluons (at large distance scales) and asymptotic freedom (at short distance scales). Intriguing nonperturbative and nonlinear collective effects may be pronounced in high-density hadronic matter, and even more so in the extended nuclear medium. Even the structural details of stable nucleons and nuclei are not fully derivable from {\it ab initio} QCD theory.
On the other hand, it is reasonable to expect that the collinear factorization formalism operating with nuclear PDFs (nPDFs) provides a pathway to describe some of these nuclear phenomena within a perturbative framework. Similarly to the nucleon PDFs, the nuclear PDFs --  to the extent they obey regular, universal dependence on atomic weight $A$ and electric charge $Z$ -- in principle can be determined from a global analysis of high-energy scattering data on nuclear targets. For a concise summary of the common frameworks employed in the heavy-ion global analyses see, e.g., Sec. 5 of \cite{Kovarik:2019xvh}. 

Progress in nuclear PDF  analyses ~\cite{Kovarik:2019xvh,Ethier:2020way,AbdulKhalek:2020yuc,Eskola:2016oht,Kusina:2020lyz} was made rapidly in recent years due, in part, to new 
measurements by both fixed-target and collider (RHIC, LHC) experiments, see {\it e.g.}~\cite{Kovarik:2015cma,Helenius:2021tof,Khalek:2022zqe,Eskola:2021nhw} and references therein.
%
As compared to the proton PDFs, the nuclear PDFs have an extra dimension to explore, provided by the $A$ quantum number.  The nuclear $A$ dimension represents both a challenge and an opportunity. 
It is a challenge because the total size of the nuclear PDF data set is  
mostly comparable to the proton data set, but it has the nuclear $A$ as an extra degree of freedom.
It is an opportunity because  the freedom of the nuclear $A$ dimension allows us to compare a variety of different nuclei, as we look for patterns that 
may provide clues to a deeper understanding of QCD. 

\subsection{Connections between nuclear and nucleon PDF analyses}

Studies of nPDFs leverage techniques from the proton PDF analysis, and {\it vice versa}.
As we explore different $A$ values, we can move from the well-known limit of the proton ($A{=}1$) 
up to the very heavy gold and lead nuclei. As such,
free-nucleon global analyses generally serve as a constraining
boundary of the $A$ depenedence at $A\!=\!1$. Reciprocally, if corrections from the nuclear medium are reasonably known, the data on nuclear targets can, and is, employed for proton and meson PDF determinations. Data from fixed-target DIS and Drell-Yan experiments on deuteron and heavy targets helps to distinguish sea quark and antiquark flavors in proton PDF fits. 

For example, neutrino--nucleon ($\nu A$) DIS structure functions $\{F_{2,3}^{\nu},F_{2,3}^{\bar{\nu}}\}$ are a key data set in the nucleon fits. They are derived from DIS on an iron target that is required due to the small neutrino cross section. Therefore, the nuclear correction ratios (e.g., $F_2^{\rm Fe}/F_2^{p}$) are indispensable for translating the nuclear results to the proton PDFs.
As the precision of proton PDFs is steadily improving, 
it becomes critical to reduce the comparatively large uncertainties of the nuclear correction ratios. The progress in this area requires new measurements as well as improved theoretical analyses to pin down the flavor dependence of nuclear correction factors. This is an area where new approaches from machine learning,  artificial intelligence, and lattice QCD may be proven fruitful~\cite{Constantinou:2020hdm}. 

\subsection{Exploring nuclear $A$ dependence}  
As mentioned above, the nuclear $A$ dimension represents an opportunity to explore 
a data set that is comparable and more diverse than the measurements limited to only the proton. 
These nuclear PDF fits do typically  use a smooth parameterization in the $A$ value, 
and hence make the implicit assumption that the nPDFs vary smoothly in this dimension. 
While this may be a reasonable assumption for the heavier nuclei, 
this can be problematic for light nuclei, such as deuterium, where it may be more challenging to incorporate few-body bound state effects into a smooth parametrization. For instance, corrections associated with the deuteron can be sizable~\cite{Accardi:2016qay,Accardi:2021ysh}. 

Dynamics inside light nuclei can be complementary approached using lattice-QCD calculations. Preliminary analyses of light nuclei are in progress~\cite{Lin:2017snn,Constantinou:2020hdm,Constantinou:2022yye}, building upon improved PDF moment calculations and the quasi-PDF  methods that have proven beneficial  for the proton analysis.
For example, the NPLQCD collaboration computed the first moment of the unpolarized gluon
distribution for the deuteron and ${}^3$He using a higher-than-physical quark
mass; these investigations can serve  as a starting point for future developments~\cite{Winter:2017bfs}.
Although the lattice studies are limited to very light nuclei, within the nCTEQ parameterization
it has been observed that some of the $A$-dependent parameters evolve quickly at low A values~\cite{Kovarik:2015cma}. 
In particular, deuteron corrections have been studied extensively, and these have been determined to 
be important in fitting nuclear data~\cite{Accardi:2016qay}, especially since much of the nuclear structure function data are expressed 
as ratios of the form $F_2^A/F_2^D$. 
Thus, even additional insights on the first few nuclei may help us to improve our
description of the nPDFs in the low-$A$ region. 

\subsection{Extreme  kinematics} 
In contrast to the nucleon PDFs that have support on $x< 1$, nuclear PDFs also 
extend to $x$ values beyond unity ($x > 1$).
This is a region where target mass corrections are expected to be important. 
In the opposite limit of $x\ll 1$, we can explore the resummation of 
$\ln(1/x)$ contributions in the BFKL framework. 
Additionally, parton saturation and recombination effects are 
expected to grow at $x\to 0$ with a $A^{1/3}$ enhancement
for heavy nuclei. 
Of course, if the gluon saturation regime is reached as expected at some value of $x$, then collinear factorization must break down, and the concept of a PDF is not useful. While structure functions are physical observables, the concept of PDFs relies on collinear factorization. This becomes an even bigger issue for heavy nuclei, where these growing small-$x$ "higher-twist" effects are further enhanced by $A^{1/3}$. In this kinematic regime, all higher-point correlator functions are of the same order as the two-point functions, i.e. the PDFs. 

Finally, we can extend analyses into the low-$Q^2$ region,
where the increase of $\alpha_S(Q)$ pushes us into the
nonperturbative regime. 
Preliminary investigations have examined the effects of relaxing the
typical $Q^2$ and $W^2$
cuts in PDF fits~\cite{Harland-Lang:2016yfn,Segarra:2020gtj}.  
These works suggest that the characteristic $x$ dependence of nuclear structure-function
ratios persists into the resonance region at low-$W$ values and could be
a manifestation of the quark-hadron duality phenomenon.  If correct,
this may permit a description of nuclear structure functions in terms
of partonic degrees of freedom, even in kinematic regions of large $x$ where
resonance excitations dominate.

\subsection{Collective properties of QCD} 
Another important aspect of  nuclear studies is the observation of collective effects:
jet quenching in nuclei-nuclei collisions~\cite{CMS:2011iwn,Angerami:2012pja},
long-range correlations (the ridge effect) in both proton-proton~\cite{CMS:2010ifv}
and proton-lead~\cite{CMS:2012qk} collisions, 
quark-gluon plasma (GQP)~\cite{BRAHMS:2004adc}, 
color glass condensate (CGC)~\cite{CMS:2013jlh},
nuclear saturation~\cite{STAR:2006dgg},
cold nuclear matter effects~\cite{ALICE:2015sru},
as well as others.
In heavy-ion collisions, a superposition of the hot and cold nuclear matter effects is expected, and a quantitative evaluation of the latter is an important prerequisite for a detailed understanding of the former.
 The large value of $\alpha_s$ at soft momenta renders traditional small-coupling perturbation theory unavailable in the infrared, and consequently many collective phenomena in nuclei are nonperturbative.
Nevertheless, certain aspects of this dynamics still require nPDFs. 
Concerted application of the PQCD formalism for protons, nuclei and
mesons, together with advances in simulations of extended medium, could
deepen our understanding of such phenomena. 

\subsection{Outlook}
Looking forward, a wealth of new measurements with heavy ions are expected at JLab, RHIC, the LHC/HL-LHC, the
future EIC and neutrino experiments, as briefly reviewed in Sec. 7 of \cite{Begel:2022kwp}. A number of hard scattering processes (vector boson, heavy-quark, jet production at the LHC, neutral- and charged-current DIS at the EIC) will be measured with high statistics in collisions of various ion species. NNLO calculations will be adopted for heavy-ion scattering. This combination of experimental and theoretical
efforts will undoubtedly deepen our understanding of underlying
nuclear structure and dynamics, casting light onto the topics discussed in this Section.
\section{Transverse-momentum dependent distributions \label{sec:TMDs}}
{\it Leading authors: V. Bertone, C. Bissolotti, F. G. Celiberto, G. Schnell, and G. Vita \\ \vspace{6pt}}

\newcommand{\cI}{\mathcal{I}}
\newcommand{\cO}{\mathcal{O}}
\newcommand{\df}{\mathrm{d}}
\newcommand{\lqcd}{\Lambda_\mathrm{QCD}}
\newcommand{\GeV}{\,\mathrm{GeV}}
\newcommand{\qt}{{\vec q}_T}
\newcommand{\nn}{\nonumber}

Collinear factorization and the ensuing collinear parton distribution functions (PDFs) have proven to be powerful tools for the study of high-energy collisions involving hadrons. Nonetheless, the use of collinear factorization is limited to observables characterized by a single hard momentum scale and cannot be applied to observables that depend on two or more widely separate momentum scales. An example of observable that breaks collinear factorization is the transverse-momentum ($q_T$) distribution of the lepton pair in the Drell-Yan (DY) process at small values of $q_T$. In this regime, the presence of large logarithms of $q_T$ in the perturbative calculation of the hard cross sections spoils the perturbative convergence, effectively invalidating collinear factorization. An appropriate description of DY at low $q_T$ is instead achieved through \textit{transverse-momentum-dependent} (TMD) factorization~\cite{Collins:2011zzd} that has the ability to resum the large logarithms of $q_T$ to all orders in perturbation theory, thus producing sensible results at low values of $q_T$.

A ``byproduct'' of TMD factorization is the introduction of TMD distributions (TMDs). 
TMD parton distribution functions (TMD PDFs) and fragmentation functions (TMD FFs) 
can be regarded as generalizations of collinear PDFs and FFs in that they provide information on the transverse momentum distribution $k_T$ of partons within initial- and final-state hadrons.
As a consequence, TMD PDFs encode more information on the structure of hadrons than PDFs. Their knowledge has the potential to shed light on the origin of mass, spin, lifetime, and other key properties of hadrons.
Precise studies of TMD FFs, feasible at the Electron-Ion Collider (EIC)~\cite{Accardi:2012qut,AbdulKhalek:2021gbh,Khalek:2022bzd} and new-generation lepton-lepton machines~\cite{AlexanderAryshev:2022pkx}, will extend our knowledge of final-state QCD radiation and hadronization.

 \subsection{Quark TMD PDFs}
 
 When accounting for partonic transverse momentum, the interplay between hadron and parton polarizations gives rise to a much richer partonic structure of hadrons. It turns out that, for a spin-1/2 hadron, there exist eight independent leading-twist TMD PDFs~\cite{Mulders:1995dh}. A second important feature of TMDs is that, as opposed to collinear distributions, they break naive universality; in other words, they may depend on the process under consideration. This breaking of universality can be traced back to the presence of a Wilson line in the operator definition of a TMD PDF necessary to guarantee gauge invariance. The Wilson line depends on the integration path, which in turn is determined by the process in which the TMD PDF is participating. In the case of quark TMD PDFs, there are only two possible Wilson-line configurations relevant to phenomenological applications, often referred to as future-pointing $[+]$ and past-pointing $[-]$ staple links. For example, the $[-]$ configuration is to be used in DY, while the $[+]$ configuration enters in semi-inclusive deep inelastic scattering (SIDIS). Switching from one Wilson-line configuration to the other introduces a sign change for time-reversal-odd (T-odd) TMD PDFs, while time-reversal-even (T-even) TMD PDFs remain unaffected. This is the origin of the now well-known Sivers effect~\cite{Collins:2002kn}, predicting that the Sivers TMD PDFs change their sign depending on whether they enter the DY or SIDIS cross section.
 
The copious independent TMD distributions are less accessible experimentally, as compared to collinear PDFs. The consequence is that our present quantitative knowledge of TMD PDFs is generally far less accurate than that of PDFs. In fact, many of the eight leading-twist quark TMD PDFs are largely unknown, with much of the recent effort being put into the study of the unpolarized quark TMD PDFs, $f_1^q$, and the Sivers quark TMD PDFs, $f_{1T}^{\perp q}$.

Over many years, the relevance of the unpolarized TMD PDFs, $f_1^q$, has been fully recognized for questions related to the precision determination of the electroweak parameters of the Standard Model (SM) \cite{Ladinsky:1993zn,Balazs:1997xd,Landry:2002ix,Bozzi:2008bb}. Intense activities around these distributions have led to fairly accurate determinations. Beyond the pioneering determinations of $f_1^q$, such as those of Refs.~\cite{Ladinsky:1993zn,Landry:2002ix,Konychev:2005iy}, the recent years have seen  steady acceleration triggered by the data delivered by the LHC experiments and by important theoretical advances. See Secs.~\ref{sec:matching}-\ref{sec:TMDexperimentalprospects} for a more detailed discussion.

A second category of TMD PDFs that has seen raising interest in the past few years is the Sivers distribution, $f_{1T}^{\perp q}$. Recent determinations of Refs.~\cite{Bacchetta:2020gko,Echevarria:2020hpy,Bury:2021sue} have brought the precision of these distributions to an unprecedented level. A point worth mentioning is that the collinear dynamics of the quark Sivers distributions is driven by the twist-3 quark Qiu--Sterman (QS) distributions~\cite{Qiu:1998ia}. At present, our knowledge of the QS distributions is very limited, which left the door open to different approaches to the determination of the quark Sivers TMD PDFs --- the topic of an ongoing debate.

With many studies focusing on TMD PDFs for light quark flavors, TMD factorization for $c$ and $b$ quarks is interesting in its own right as a three-scale QCD factrorization problem \cite{Nadolsky:2002jr,Pietrulewicz:2017gxc} with phenomenological implications for EW precision physics \cite{Berge:2005rv}. Perturbative $c,b$ masses suppress very soft QCD radiation e.g. in the final-state dead-cone effect \cite{Dokshitzer:1991fd,ALICE:2021aqk,Llorente:2014bha,Maltoni:2016ays,Cunqueiro:2018jbh}. At the same time they introduce power-suppressed terms that modify soft factorization compared to the massless case \cite{Caola:2020xup,Gaggero:2022hmv}. As the most common collinear PDFs are given in the general-mass variable flavor number schemes, TMD calculations must consistently include heavy-quark masses in the same schemes both in the LHC \cite{Berge:2005rv,Bozzi:2015zja,Belyaev:2005bs,Cao:2019uor} and DIS \cite{Nadolsky:2002jr} processes. These investigations must be extended to N2LO and N3LO. 

 \subsection{Gluon TMD PDFs}

 The complete list of leading-twist gluon TMD PDFs for a spin-$1/2$ target was first given in Ref.~\cite{Mulders:2000sh} (see also Refs.~\cite{Meissner:2007rx,Lorce:2013pza}),
 where the polarization states of both of the parent nucleon and the struck gluon were accounted for. 
 Gluon TMD PDFs for a spin-$1$ target were listed in Ref.~\cite{Boer:2016xqr}, and this led to the emergence of 11 new distributions on top of the ones arising in the spin-$1/2$ case.
 The two gluon TMD functions that survive after the integration over transverse momentum are the distribution of unpolarized gluons inside an unpolarized nucleon, $f_1^g$, and of circularly-polarized gluons inside a longitudinally-polarized nucleon, $g_1^g$.
 They represent the TMD counterparts of the unpolarized and helicity gluon PDFs in the collinear regime. 

 Just as it happens in the quark case, the gluon TMD PDFs are sensitive to resummation of transverse-momentum logarithms. The logarithms constitute the perturbative contribution of the TMD PDFs (see, \emph{e.g.}, Refs.~\cite{Balazs:2000wv,Bozzi:2003jy,Bozzi:2005wk,Catani:2010pd,deFlorian:2011xf,Echevarria:2015uaa} and references therein). While our knowledge about the transverse-momentum resummation is quite solid, the nonperturbative component of gluon TMD PDFs, relevant to understanding the dynamics of intrinsic motion of partons inside nucleons, is poorly known, as it is strongly suppressed at electroweak scales. 

 Similarly to quark TMD PDFs, different classes of reactions probe distinct gauge-link structures for gluon TMD PDFs, each of them being given in term of a combination of two or more staple links. This leads to a more diversified kind of \emph{modified universality}.
 Two major gluon structures exist: the $f\text{-type}$ and the $d\text{-type}$ ones, also known in the context of small-$x$ studies as Weisz\"acker--Williams and dipole links, respectively~\cite{Kharzeev:2003wz,Dominguez:2011wm}.
 The antisymmetric $f_{abc}$ QCD color structure enters the analytic expression of the $f$-type T-odd gluon-TMD correlator, while the symmetric $d_{abc}$ structure is part of the $d$-type T-odd one. 
 The $f$-type gluon TMD PDFs depend on $[\pm,\pm]$ gauge-link combinations. 
 The $[+,+]$ structure is probed in reactions where the gluon interacts with a color-singlet initial particle (\emph{e.g.}, a photon in a DIS process) and two colored objects (\emph{e.g.}, two jets) are emitted in the final state. 
 The $[-,-]$ structure emerges in processes where a gluon interacts with another gluon (color-octet state), and a color-singlet state (\emph{e.g.}, a Higgs boson) is tagged in the final state.
 TMD factorization holds for all these reactions, and the following modified-universality relations for $f$-type distributions arise from time-reversal invariance (T-symmetry):
 \begin{align} \label{eq:mod_univ_f} \nonumber
  f_1^{g\,[+,+]} &= f_1^{g\,[-,-]} \; && \text{(T-even),} \\
 f_{1T}^{g\,\perp[+,+]} &= -f_{1T}^{g\,\perp[-,-]}  && \text{(T-odd)}.
 \end{align}
 Here the unpolarized gluon TMD PDF, $f_1^{g}$, is a representative of all the T-even distributions, while the Sivers gluon TMD PDF, $f_{1T}^{g}$, stands for all the T-odd functions.
 The $d$-type gluon TMD PDFs depend on $[\pm,\mp]$ gauge-link combinations and appear when a gluon interacts with a colored initial particle, and a colored final-state system is produced (\emph{e.g.}, when a photon is emitted together with a jet in proton-proton collisions).
 The $d$-type modified-universality relations are analogous to the $f$-type ones, given in Eq.~\eqref{eq:mod_univ_f}.
 In this case, TMD factorization has not been proven and might be affected by issues connected with color entanglements~\cite{Rogers:2013zha}. More intricate gauge-link structures are involved in processes where multiple color states are present in both the initial and final state~\cite{Bomhof:2006dp}. Here TMD factorization runs into even deeper issues.

 Experimental information on the nonperturbative part of the gluon TMD PDFs is  limited. [In electroweak-scale processes like $gg\to H^0$ \cite{Balazs:2000wv,Bozzi:2005wk,deFlorian:2011xf} or $gg\to \gamma\gamma$ production \cite{Nadolsky:2007ba,Balazs:1997hv, Balazs:1999yf, Balazs:2007hr}, the gluon TMD PDF is dominated by perturbative resummed contributions and is practically insensitive to nonperturbative dynamics.] Attempts at phenomenological analyses of the unpolarized gluon TMD PDF have been presented in Refs.~\cite{denDunnen:2014kjo,Lansberg:2017dzg,Bacchetta:2018ivt,Gutierrez-Reyes:2019rug,Scarpa:2019fol}. Experimental and phenomenological studies of the intrinsic motion of gluons in transversely-polarized protons quantified by the Sivers function can be found in Refs.~\cite{Adolph:2017pgv, DAlesio:2017rzj,DAlesio:2018rnv,DAlesio:2019qpk}.
 Thanks to its connection with the QCD Odderon, the gluon Sivers TMD PDF can also be studied in unpolarized electron-proton collisions~\cite{Boussarie:2019vmk}.

 In the \emph{high-energy factorization} regime, where gluons are extracted from nucleons with a small longitudinal fraction $x$ and a large transverse momentum, a relation can be established~\cite{Dominguez:2011wm} between the unpolarized and linearly-polarized gluon distributions, $f^g_1$ and $h^{\perp g}_1$, and the \emph{unintegrated gluon distribution} (UGD), whose evolution is controlled by the Balitsky--Fadin--Kuraev--Lipatov (BFKL) equation~\cite{Fadin:1975cb,Kuraev:1976ge,Kuraev:1977fs,Balitsky:1978ic} (see Refs.~\cite{Hentschinski:2012kr,Besse:2013muy,Bolognino:2018rhb,Bolognino:2018mlw,Bolognino:2019bko,Bolognino:2019pba,Celiberto:2019slj,Brzeminski:2016lwh,Garcia:2019tne,Celiberto:2018muu} for recent applications).
 A connection between the high-energy and the TMD factorization formalisms was recently highlighted in Refs.~\cite{Nefedov:2021vvy,Hentschinski:2021lsh}.
 The impact of embodying gluon-TMD inputs within high-energy factorization was recently assessed for vector-meson leptoproduction processes at the EIC~\cite{Bolognino:2021niq,Bolognino:2021gjm,Bolognino:2022uty,Bolognino:2022ndh}.

 The distribution of linearly polarized gluons in an unpolarized nucleon, $h_1^{\perp g}$, plays a crucial role in the dynamics of spin correlations in collisions of unpolarized hadrons~\cite{Nadolsky:2007ba,Catani:2010pd,Boer:2010zf,Sun:2011iw,Boer:2011kf,Pisano:2013cya,Dunnen:2014eta,Lansberg:2017tlc}. They are collectively known as the Boer--Mulders effect. Part of it is generated at large transverse momenta within perturbative QCD via the transverse-momentum resummation, and it represents the perturbative part of $h_1^{\perp g}$.
 A genuine perturbative-QCD treatment would miss, however, the polarization effect generated by the \emph{intrinsic} motion of gluons, which has a  nonperturbative nature and cannot be caught by the resummation, but needs to be quantified via fits on global data that will be collected at new-generation colliders~\cite{AbdulKhalek:2021gbh,Chapon:2020heu,Arbuzov:2020cqg,Abazov:2021hku,Anchordoqui:2021ghd,Feng:2022inv}.

 With the aim of bridging the gap between theory and experiment, phenomenology-suited models are needed to perform exploratory studies of gluon TMD PDFs.
 A recent calculation of all the unpolarized and polarized T-even gluon-TMD densities at twist-2 was done via an enhanced spectator-model approach~\cite{Bacchetta:2020vty} (see also Refs.~\cite{Bacchetta:2021oht,Celiberto:2021zww,Celiberto:2022fam}, and Refs.~\cite{Bacchetta:2008af,Bacchetta:2010si} for similar results in the quark case), where proton remnants after gluon emission are treated as a single on-shell effective fermion. Preliminary calculations of leading-twist T-odd functions were presented in Refs.~\cite{Bacchetta:2021lvw,Bacchetta:2021twk,Bacchetta:2022esb,Bacchetta:2022crh}.

\begin{figure}[htbp!]
\centering
\includegraphics[scale=0.300]{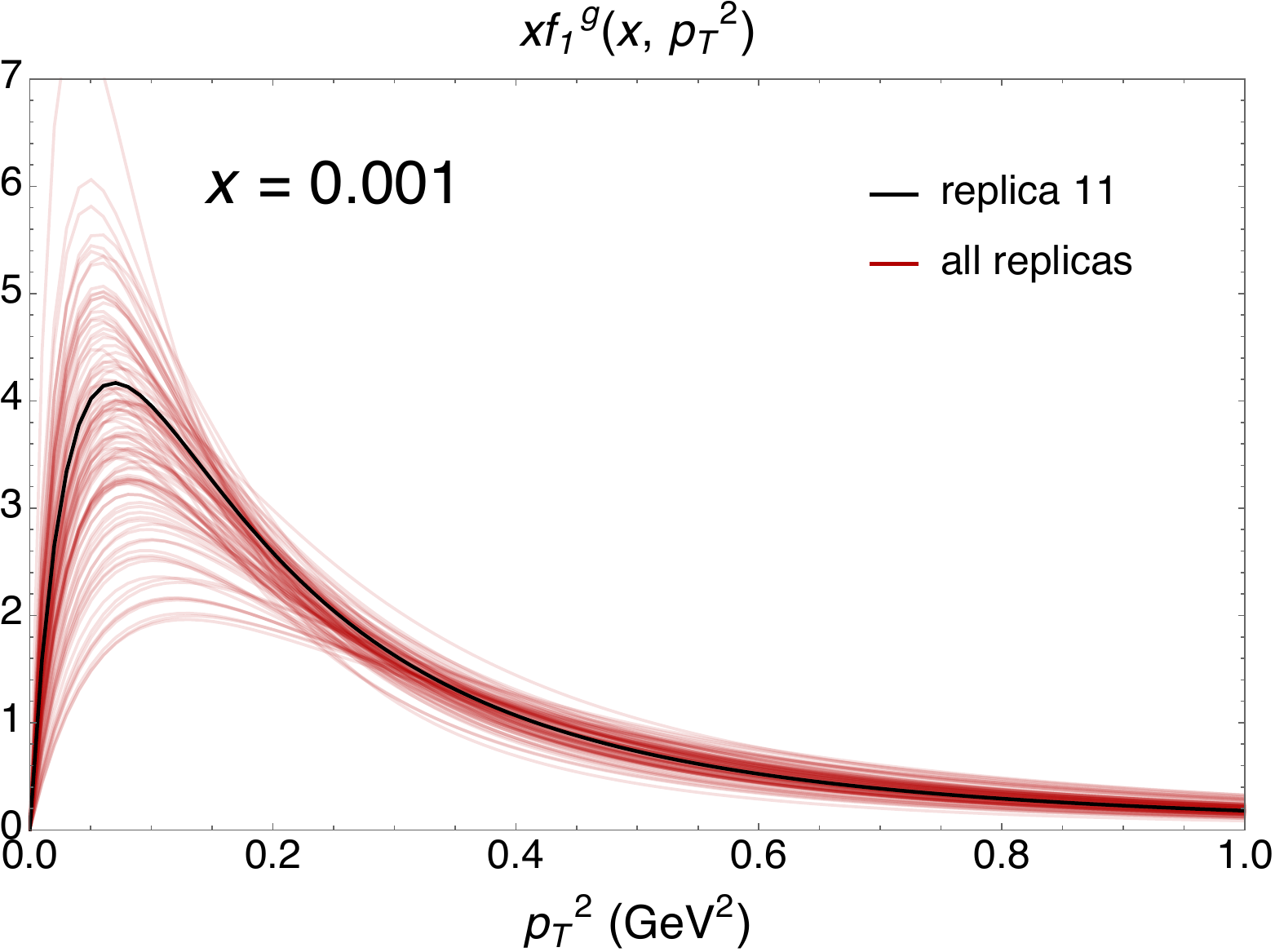} \hspace{0.50cm}
\includegraphics[scale=0.310]{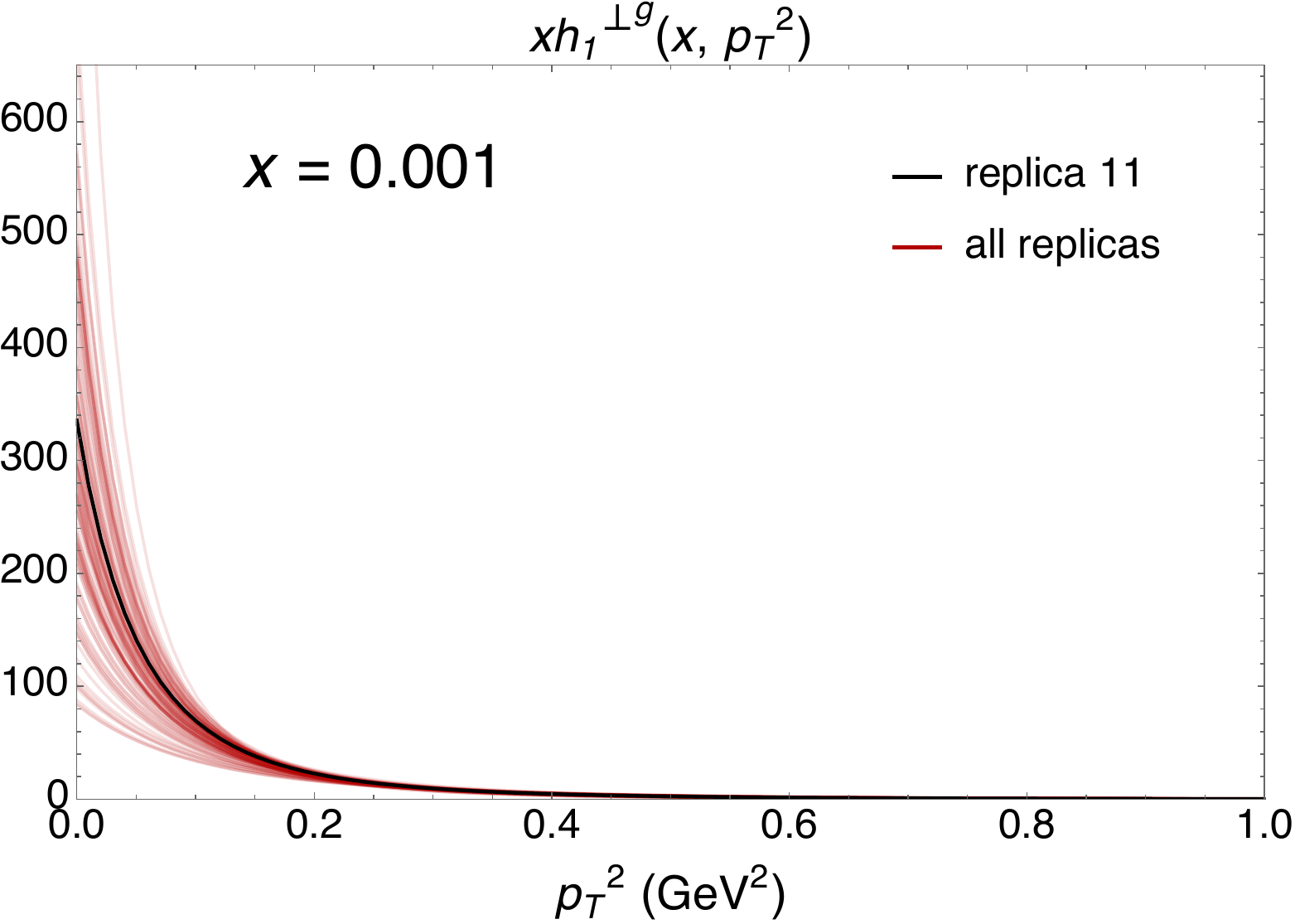} 
\caption{Unpolarized (left) and linearly-polarized (right) gluon TMD PDFs as functions of $\boldsymbol{p}_T^2$, for $x=10^{-3}$ and at the initial scale $Q_0 = 1.64$ GeV. Figures adapted from Ref.~\cite{Bacchetta:2020vty}.}
\label{fig:SM_gluon_TMDs}
\end{figure}

 Taking advantage of the link between TMD and collinear factorization, a consistent procedure was set up in Ref.~\cite{Bacchetta:2020vty} to simultaneously fit the unpolarized and helicity gluon TMD densities to the corresponding collinear PDFs obtained from \nnpdf \cite{Ball:2017otu,Nocera:2014gqa} at the initial scale $Q_0 = 1.64$ GeV.
 Predictions for the unpolarized and the linearly-polarized gluon TMD PDFs are presented in Fig.~\ref{fig:SM_gluon_TMDs} as functions of the transverse momentum squared, $\boldsymbol{p}_T^2$, for $x=10^{-3}$ and at the initial scale $Q_0$, namely without switching TMD evolution on. Thus, initial-scale results precisely refer to the nonperturbative part of our TMD densities. Predictions are given as a set of 100 replicas, which are statistically equivalent and reproduce well the unpolarized and helicity collinear PDFs. Each red line in plots represents a single replica, with the black line corresponding to the most representative one (n. 11). 
 
 We note that each TMD exhibits a peculiar trend both in $x$ and $\boldsymbol{p}_T^2$. The unpolarized TMD clearly shows a non-Gaussian pattern in $\boldsymbol{p}_T^2$, and goes to a small but non-vanishing value for $\boldsymbol{p}_T^2 \to 0$. The linearly-polarized gluon TMD is large at small $\boldsymbol{p}_T^2$ and decreases very fast. Both of them are increasingly large at small $x$, and their ratio is constant in the asymptotic limit $x \to 0$. This is in line with the BFKL behavior of the small-$x$ UGD, which predicts an ``equal number" of unpolarized and the linearly-polarized gluons up to higher-twist effects. This is a contact point between the TMD and the BFKL approach that could be explored via studies on processes featuring a \emph{natural stability} of the high-energy resummation~\cite{Celiberto:2020wpk,Bolognino:2019yls,Celiberto:2020tmb,Bolognino:2021mrc,Celiberto:2020rxb,Celiberto:2021dzy,Celiberto:2021fdp,Celiberto:2022dyf,Celiberto:2022rfj,Celiberto:2022zdg,Celiberto:2022keu,Celiberto:2022gji,Hentschinski:2020tbi,Celiberto:2022fgx,Hentschinski:2022xnd}.
 Furthermore, even if all replicas reproduce similar collinear PDFs, they predict very different results for the TMD PDFs in Fig.~\ref{fig:SM_gluon_TMDs}. Forthcoming data on gluon TMD PDFs are expected to exclude many replicas and constrain parameters not yet so well constrained by collinear PDFs.

\subsection{TMD evolution and matching to collinear PDFs}\label{sec:matching}
TMD PDFs arise in the context of 3D description of the internal dynamics of the proton. 
However, we also know that integration over transverse momenta produces cross sections dependent of simply collinear PDFs. 
It is therefore interesting to understand how this transition happens, and how we can improve theoretical predictions for TMD-sensitive observables across the large spectrum of transverse momenta probed by current and future experiments.
We start by noticing that, for TMD observables at colliders, there are typically three scales that characterize the physics at play: the scale of the hard scattering $Q$, the scale of transverse momenta measured for the observable of interest $q_T$, and $\lqcd$. 
For values of the transverse momentum that are perturbative, i.e. for $q_T \gg \lqcd $, it is possible to define an Operator Product Expansion (OPE), which matches TMD PDFs onto collinear PDFs up to corrections of $\cO(\lqcd/q_T)$. Schematically the OPE takes the form \cite{Collins:1981uw,Collins:1984kg}
\begin{align} \label{eq:beam_OPE_schematic}
 f^{\rm TMD}_i(z, \qt,\mu,\nu) = \sum_j \int_z^1\!\frac{\df z'}{z'} \, \cI_{ij}(z',\qt,\mu,\nu) f_j\Bigl(\frac{z}{z'},\mu \Bigr) \times \bigl[1 + \cO(q_T/\lqcd)\bigr]
\,,\end{align}
where $\cI_{ij}(z,\qt,\mu,\nu)$ is a perturbative matching kernel, $f_j\bigl(x,\mu \bigr)$ is the standard collinear PDF for flavor $j$. A rapidity scale $\nu$ reflects the presence of rapidity divergences in the renormalization of TMDs, which can take a variety of forms and notations depending on the renormalization procedure and scheme employed \cite{Collins:1981uk,Ji:2004wu,Beneke:2003pa, Chiu:2007yn, Becher:2011dz,Chiu:2011qc, Chiu:2012ir,Chiu:2009yx, GarciaEchevarria:2011rb,Li:2016axz,Ebert:2018gsn}. For an overview of different schemes for TMD definitions and rapidity regularization see, for example, App.B of ref. \cite{Ebert:2019okf}.

Throughout the years, significant progress has been made in the calculation of the matching kernels up to N2LO both for the quark \cite{Catani:2012qa,Gehrmann:2012ze,Gehrmann:2014yya,Echevarria:2016scs,Luo:2019hmp} and gluon \cite{Catani:2011kr,Gehrmann:2014yya,Echevarria:2016scs,Luo:2019bmw} cases.
Recently, their calculation has been pushed to N3LO \cite{Luo:2019szz,Ebert:2020yqt}.
It is important to note that, given the complexity of these analytic calculations, achieving such level of accuracy from the perturbative side required significant innovation in the way such objects are calculated. New methods for performing multiloop computation in the context of effective field theory have been developed, such as generalized integration-by-parts identites for the treatment of rapidity regulators \cite{Li:2016axz,Luo:2019szz} and a framework for the collinear expansion of analytic cross sections \cite{Ebert:2020lxs}. 
Originally developed for the calculation of the TMD matching kernels, these new tools have been further applied to obtain a variety of N3LO quantities, such as $N$-jettiness beam functions, time-like splitting functions, TMD FFs, and energy-energy correlators \cite{Ebert:2020unb,Chen:2020uvt,Luo:2020epw,Ebert:2020qef,Ebert:2020sfi}.

The evolution of the TMD PDFs is dictated by a coupled system of differential equations \cite{Collins:1981uk,Collins:1981va,Chiu:2012ir,Li:2016axz}, which becomes multiplicative in transverse position ($b_T$) space. Using the $b_T$ as the conjugate variable of $q_T$, the RGEs take the form 
\begin{align} \label{eq:RGEs}
 \mu \frac{\df}{\df\mu}{\tilde f_i(x,b_T,\mu,\nu/\omega)} &
 = \tilde\gamma_{\mu}^i(\mu,\nu/\omega)\, \tilde f_i(x,b_T,\mu,\nu/\omega)
\,,\nn\\
 \nu \frac{\df}{\df\nu}{\tilde f_i(x,b_T,\mu,\nu/\omega)} &
 = -\frac{1}{2}\tilde\gamma_\nu^i(b_T,\mu)\, \tilde f_i(x,b_T,\mu,\nu/\omega)
\,,\end{align}
where $\tilde\gamma_\mu^i(\mu,\nu/\omega)$ is related to the collinear and threshold anomalous dimensions,  and $\tilde\gamma_\nu^i(b_T,\mu)$ is the so called rapidity anomalous dimension \cite{Chiu:2012ir}, which is closely related to the Collins-Soper kernel \cite{Collins:1981uk,Collins:1981va} and has been obtained at N3LO in \cite{Li:2016ctv}.
\begin{figure*}
 \centering
 \includegraphics[width=0.49\textwidth]{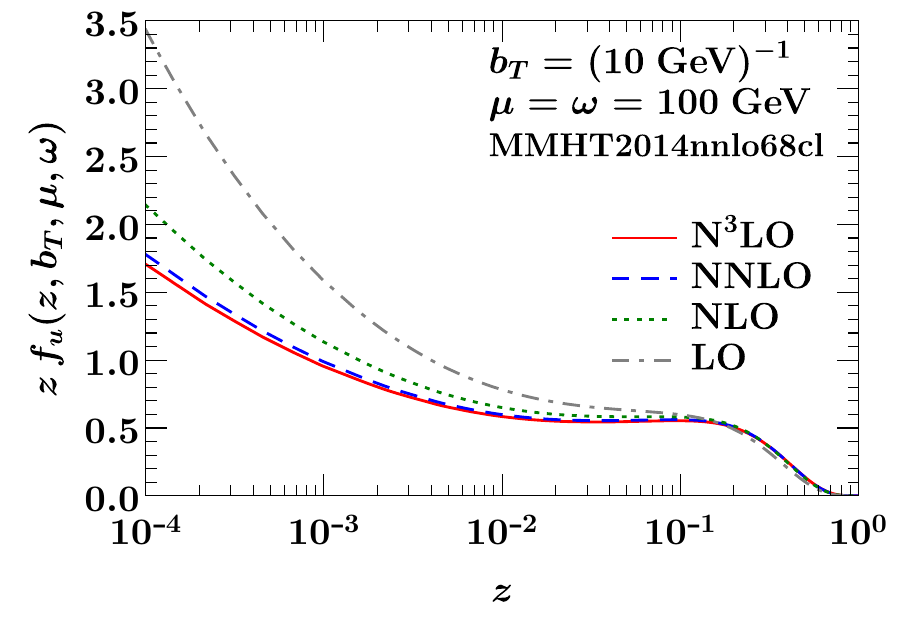}
 \hfill
 \includegraphics[width=0.49\textwidth]{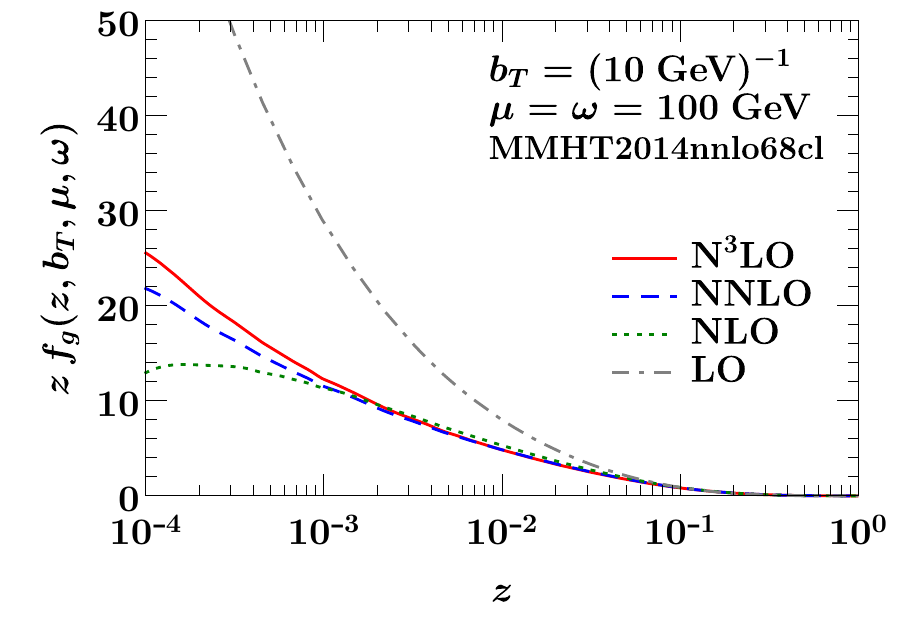}
 \caption{The $u$-quark TMD PDF (left) and the gluon TMD PDF (right) as a function of $z$ for fixed $b_T = (10~\GeV)^{-1}$ and $\mu=\omega=100~\GeV$. We show the result at LO (which is equivalent to the PDF since the matching kernel is trivial at this order), NLO, N2LO, and N3LO. Plots taken from \cite{Ebert:2020yqt}.}
 \label{fig:TMDPDF_convergence}
\end{figure*}

Progress in resummation accuracy is crucial for precise phenomenology.  
Determination of the complete singular analytic structure for various TMD observables at N3LO, obtained via the calculation of the N3LO TMD beam functions \cite{Ebert:2020yqt,Luo:2019szz} and FFs \cite{Luo:2020epw,Ebert:2020qef}, enabled the push of the TMD resummation accuracy to N3LL$^\prime$. This was first applied to the description of the energy-energy correlator in the back-to-back limit \cite{Ebert:2020sfi}, an event shape in electron-positron colliders, and then to transverse momentum distributions and fiducial cross sections at the LHC \cite{Billis:2021ecs,Neumann:2021zkb}. 
In both cases, the perturbative uncertainties decrease significantly as the resummation accuracy increases to N3LL$^\prime$. 
Precise control of perturbative uncertainties on resummed cross sections, thanks to calculation of anomalous dimensions and boundary functions up to three loops and beyond, will be even more relevant over a huge kinematic region accessible at future colliders.

 \subsection{Status of unpolarized TMD extractions}\label{sec:TMDdeterminations}

DY and SIDIS processes are the crucial sources of information on the functional form of TMDs. For these processes, factorization theorems allow us to write the cross section in term of convolutions of TMDs.
 In the so--called \textit{TMD factorization region}, where $q_T \ll Q$, the DY cross section is proportional to a convolution of two TMD PDFs, and the SIDIS cross section can be expressed in terms of a convolution of one TMD PDF and one TMD FF.
 TMDs are partially computable by means of well--established perturbative methods that take into account soft and collinear radiation to all orders.
 However, calculations based on perturbative QCD become unreliable for values of transverse positions $b_T > 0.5\mbox{ GeV}^{-1}$. In this regime, nonperturbative components have to be constrained through fits to experimental data.
 
 Recent works directly performed extractions of TMDs from DY data~\cite{DAlesio:2014mrz,Scimemi:2017etj,Bertone:2019nxa}, SIDIS data~\cite{Signori:2013mda,Anselmino:2013lza,BermudezMartinez:2018fsv}, or both~\cite{Echevarria:2014xaa,Su:2014wpa,Bacchetta:2017gcc,Scimemi:2019cmh}.
 At the present time, the best known quark TMD is the unpolarized TMD PDF $f_1(x, k_\perp)$, whose latest extractions reach the state-of-the-art perturbative accuracy of N3LL~\cite{Bacchetta:2019sam}.
 \begin{figure}[h!]
  \begin{centering}
    \includegraphics[width=0.49\textwidth]{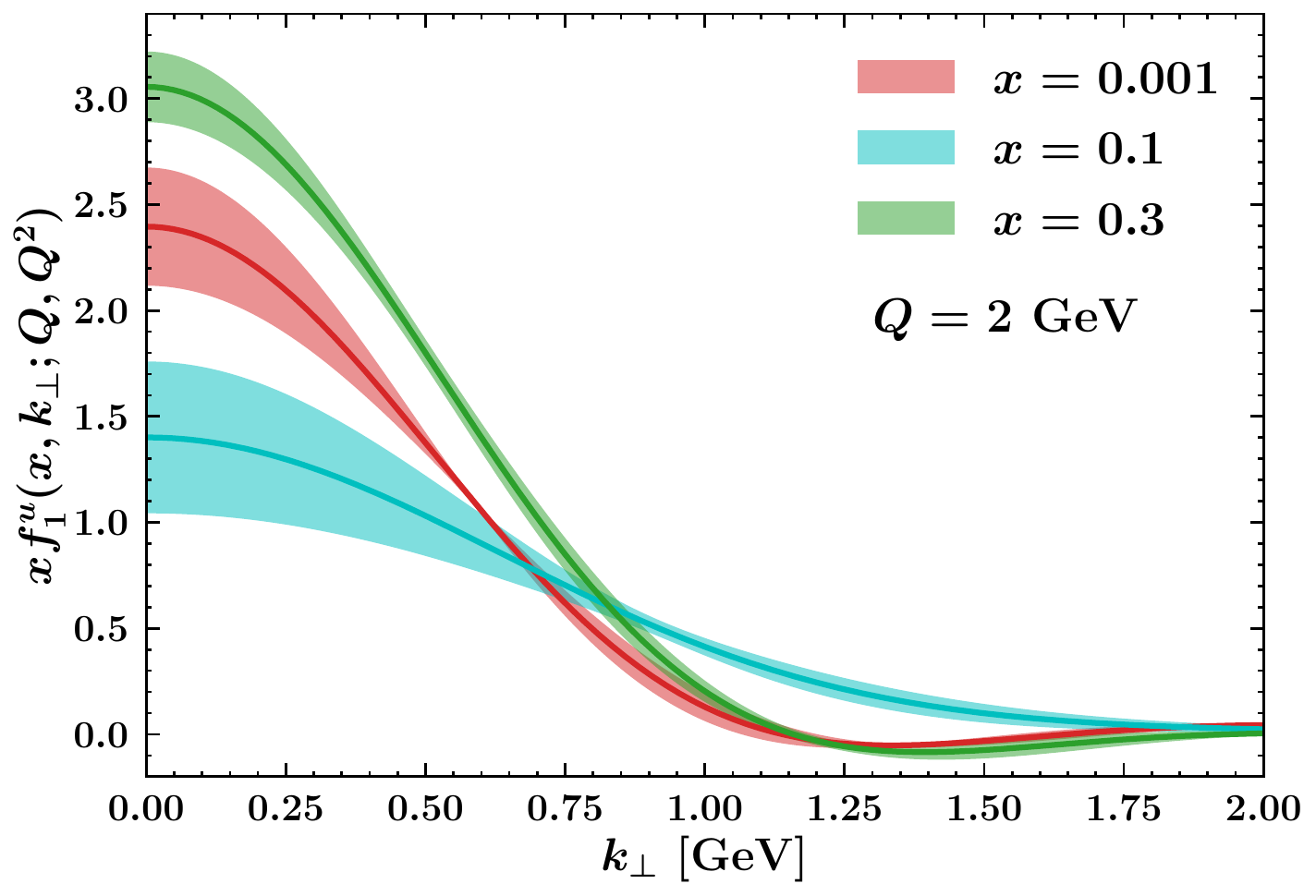}
    \includegraphics[width=0.49\textwidth]{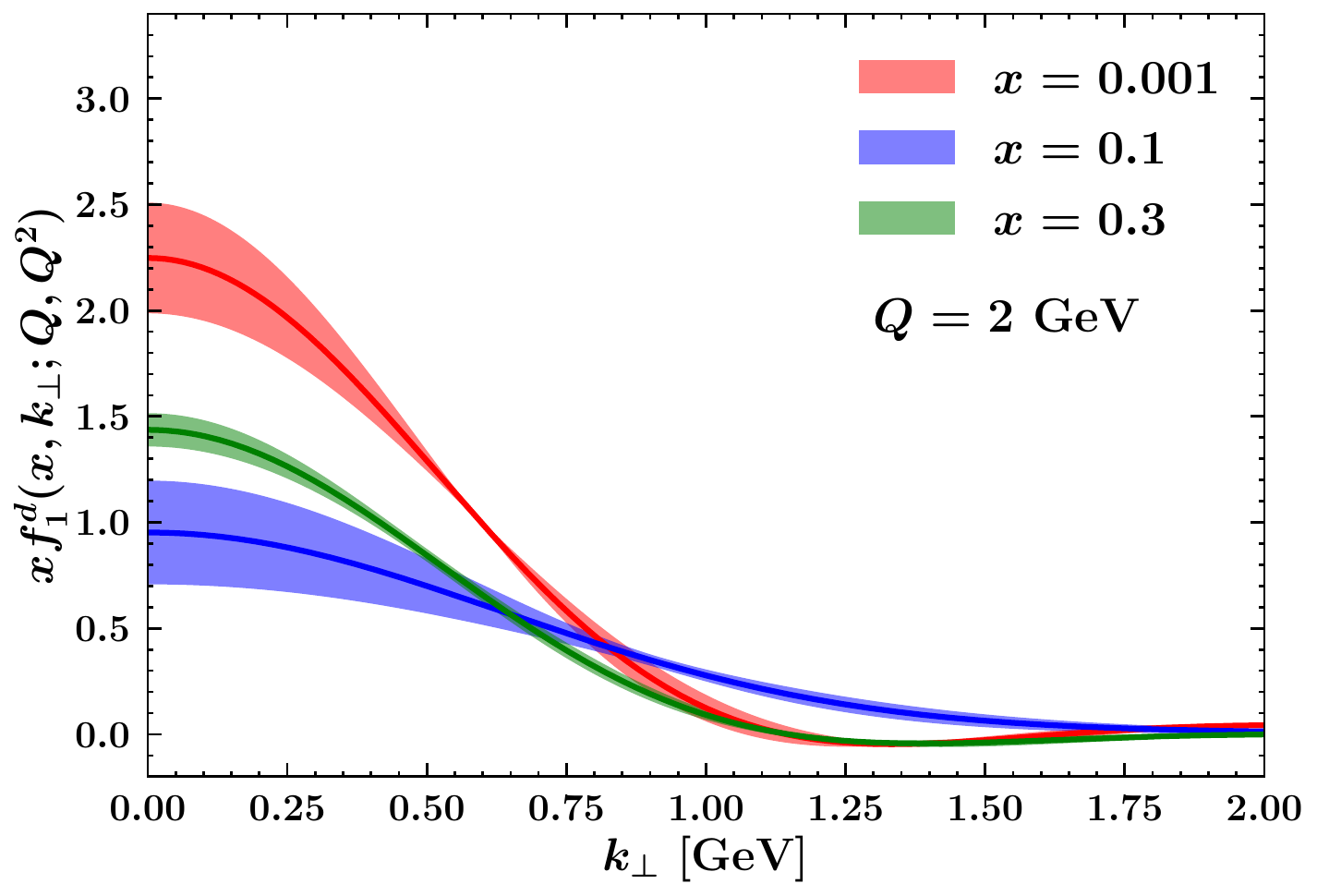}
    \caption{The unpolarized TMD PDFs of the up and down quarks at $Q = 2$~GeV
      as a function of the partonic transverse momentum $k_\perp$ for different
      values of $x$. The bands indicate $1\sigma$
      uncertainties.\label{fig:TMDs}}
  \end{centering}
\end{figure}
 In Fig.~\ref{fig:TMDs}, we show the results of the $f_1(x, k_{\perp})$ extraction performed in Ref.~\cite{Bacchetta:2019sam}, and we compare the unpolarized TMD PDFs for the up and down quark at $Q = 2$~GeV for $x = 10^{-3}$, 0.1, and 0.3.
 
 \subsection{Experimental prospects}\label{sec:TMDexperimentalprospects}
 
 While integrated PDFs have been extensively studied, the knowledge of TMDs is still very limited, especially concerning their flavor dependence. In addition to $e^+e^-$ hadroproduction, SIDIS, and DY -- the processes traditionally used to assess the TMDs -- recently, hadron+jet production has been put forth as a complementary channel, in particular to study TMD FFs. Current data for all those processes cover a much smaller kinematic space than for collinear PDFs (cf. fig.~5 of Ref.~\cite{Scimemi:2019cmh} vs. fig.~2.1 of Ref.~\cite{Ball:2021leu}) or FFs. In particular, the region in $x$ below 0.02 and below the scale of $Z$ production, which in the case of collinear PDFs is predominantly covered by collider DIS data, currently only has HERA-1 data on SIDIS transverse energy flows and particle multiplicities, examined in the first phenomenological analysis of SIDIS resummation in \cite{Nadolsky:1999kb, Nadolsky:2000ky} based on the formalism from \cite{Meng:1991da,Meng:1995yn}.
 
 Looking at the future of TMD physics, SIDIS at the EIC will be of paramount importance. The EIC a unique accelerator will be colliding polarized electrons (and potentially positrons) with polarized protons and light nuclei at various energies and with unprecedented luminosity for a lepton-hadron collider. Together this will yield the precision and spin degrees of freedom necessary to pursue an ambitious physics program of exploration of the spin structure and  acquisition of multi-dimensional tomographic images of protons and nuclei (see the ``Hadronic Tomography at the EIC and the Energy Frontier" Snowmass White Paper~\cite{Khalek:2022bzd}).
 The huge impact that the EIC will have on TMD extractions has been discussed e.g. in Sec.~7.2 of the EIC Yellow Report~\cite{AbdulKhalek:2021gbh}, where impact studies based on pseudodata coming from {\tt PYTHIA} simulations~\cite{Sjostrand:2006za} have been performed. 
 Impact studies carried out by fitting simultaneously both existing data and pseudodata showed a significant reduction (up a to factor of $\sim4$ in the kinematic regions not covered by present data) of the uncertainty bands for the unpolarized quark TMD PDFs.
 Moreover, a reduction of the uncertainties of a factor of $\sim10$ is foreseen in the determination of the nonperturbative part of the evolution for the unpolarized TMD PDFs $f_1(x, k_{\perp})$ and $D_1(z, P_{\perp})$.
 
On the low-energy side, the Jefferson Lab 12 GeV program will continue to contribute to TMD mapping with orders-of-magnitude higher luminosity and a wide range of polarization and target configurations (see, e.g., Ref.~\cite{Arrington:2021alx}). CLAS12~\cite{Burkert:2020akg}, the SBS, and the future SoLID~\cite{Chen:2014psa} experiments will explore the valence region with unprecedented precision. These data will also provide crucial input in evolution studies of TMDs. Currently, ideas are put forward to expand the kinematic reach by increasing the beam energy to 24 GeV (cf. Appendix~C of Ref.~\cite{Arrington:2021alx}).

The LHC will continue to provide crucial data especially on the high-energy end. Its importance lies also in the different processes used to study TMDs, allowing for tackling questions of factorization and universality. 
So far, the LHC is perceived as a machine for only unpolarized TMD studies. Installing polarized  targets at the LHC would permit to also enter the domain of polarized TMDs, most notably the Sivers function that is expected to change sign when probed in DY vs. SIDIS. Indeed, such ideas have been put forward~\cite{Hadjidakis:2018ifr} and are extensively pursued within the \hyperlink{https://pbc.web.cern.ch}{Physics Beyond Collider Study Group} at CERN~\cite{Jaeckel:2020dxj}. Injecting polarized nucleons into a storage cell internal to the LHC ring in front of the LHCb detector~\cite{Steffens:2015kvp}, similar to what was done for the HERMES experiment at HERA~\cite{HERMES:2004vsf}, is currently the most promising avenue. Even the use of unpolarized gases in such fixed-target setup, as already foreseen for the LHC run 3~\cite{Bursche:2018orf}, opens up unique opportunities of studying nucleon TMD PDFs at large scales and very large $x$~\cite{Hadjidakis:2018ifr,QCDWorkingGroup:2019dyv}. Dedicated DY studies of TMD PDFs are also foreseen by \hyperlink{https://spinquest.fnal.gov}{SpinQuest} at Fermilab and are part of the remaining program at RHIC~\cite{Aschenauer:2016our}. All these activities, using lepton-hadron as well as hadron-hadron reactions, will help tremendously to expand the currently limited kinematic reach of spin-dependent TMD studies (see, e.g., Fig.~12 of \cite{QCDWorkingGroup:2019dyv}). 
Last but not least, extraction of TMD FFs will crucially profit from the advent of Belle II~\cite{Belle-II:2018jsg}, building upon previous unique and complementary TMD measurements in $e^+e^-$ hadroproduction at Belle \cite{Belle:2005dmx,Belle:2008fdv,Belle:2018ttu,Belle:2019nve,Belle:2019ywy}, but with 50 times higher luminosity.
\section{Computing needs and computing tools}
\label{sec:computing}

\subsection{The \lhapdf library and other user interfaces for PDFs}
\textit{Leading authors: A. Buckley\\ \vspace{6pt}}

\lhapdf \cite{LHAPDFWebsite} is the community standard resource to provide PDF parametrizations across high-energy experiments and phenomenological computations. Its current incarnation (version 6, since 2013) contains more than 1150 PDF sets encoded in a uniform data format and interpolated with standard algorithms. While these have generally met or exceeded the required precision for MC calculations, the CPU expense of PDF interpolation is a non-negligible, sometimes even dominant, factor in many modern computations. For N3LO calculations, the default local-bicubic interpolation in $\log{x}$--$\log{Q^2}$ space has been found insufficiently stable~\cite{Dulat:2017prg,Nagar:2019gij,Diehl:2021gvs}.

Work in 2020-21 succeeded in both reducing the CPU cost (by optimization of intrinsic  \lhapdf routines as well as the interface to MC generators) and developing smoother Lagrange-based interpolators for stability in high-precision calculations. The latter, as well as support for GPU workflows (cf.~Python-oriented tools like \texttt{PdfFlow}~\cite{Carrazza:2020qwu}) and more general error-set combination rules, will shortly appear in upcoming \lhapdf releases. 

Longer-term requirements on PDFs, especially from precision hadron-collider studies, $\gamma N$ and $\gamma\gamma$ scattering,  and lepton-hadron physics at the EIC, will require extension of the current nucleon-specific \lhapdf machinery and an interface to support also resolved-photon and transverse-momentum dependent (TMD) PDFs. These extensions, while motivated by distinct physics processes, share the common feature of requiring interpolation in more than two variables: as the standard 2-variable $x$--$Q^2$ interpolation is implemented as composition of 1D interpolator functions, the extensions will be implemented by recursive strategies for higher-order composition. This generalization may also be a useful opportunity to agree on community-standard interfaces for PDF querying, to allow better interoperation of \lhapdf6 with PDF-fitting toolkits such as \texttt{ApfelGrid}~\cite{Bertone:2013vaa} and \xfitter~\cite{Alekhin:2014irh}.

\subsection{Public PDF fitting codes}

For their final use, most PDF sets are made publicly available via the \lhapdf interface described in the previous section. However, until recently only the outcomes of the global PDF fits, namely the \lhapdf interpolation grid files, were released, while the PDF fitting codes themselves 
remained private. This implied that results were not reproducible by external parties. Another limitation of private PDF codes is that benchmarking studies, such as those described in Sec.~\ref{sec:benchmarking}, become more labyrinthine due to the challenge of disentangling the various components that determine the final outcome.

\xfitter~\cite{Alekhin:2014irh} is the first open-source code to perform global fits of PDFs and related nonperturbative functions. The successor of the HERAPDF fitting code, it provides various features essential for performing global QCD analyses. The \nnpdf code~\cite{NNPDF:2021uiq} was also recently made available and offers complementary functionalities as compared to those in \xfitter: machine-learning tools for the NN PDF parametrization and automated determination of the minimization algorithm~\cite{Carrazza:2019mzf}, routines to estimate the robustness of PDF analysis via closure tests~\cite{DelDebbio:2021whr}, and an extensive suite of statistical validation and plotting tools. In what follows we describe the two available public frameworks in more detail. 

\subsubsection{\xfitter: an  open-source QCD analysis framework}
{\it Leading authors: F. Giuli, F. I. Olness \\ \vspace{6pt}}

\xfitter~\cite{Alekhin:2014irh} is an open-source software package (available at \cite{xFitterWebsite}) that provides a framework for the determination of the PDFs, FFs, and related functions. 
\xfitter version 2.2.0 has recently been released and offers an expanded set of tools and options. It incorporates experimental data from a wide range of experiments including fixed-target, Tevatron, HERA, and LHC data sets. \xfitter can analyze these data using a variety of theoretical calculations up to N2LO in perturbation theory, as well as including numerous methodological options for carrying out PDF fits and plotting tools to visualize the results. While primarily supporting collinear factorization, \xfitter also provides tools for fitting dipole models and TMD distributions.

First and foremost, \xfitter provides a flexible open-source framework for performing PDF fits to data. \xfitter can also automatically generate comparison plots of data vs.\ theory. There are a variety of options for the definition of the $\chi^2$ function and the treatment of experimental uncertainties. Examples are presented in Ref.~\cite{Alekhin:2014irh}.

\xfitter is able to perform PDF profiling and reweighting studies. The reweighting method allows \xfitter to update the probability distribution of a PDF uncertainty set to reflect the influence of new data. For PDF profiling, \xfitter compares data and MC predictions based on the $\chi^2$-minimization and then constrains the individual PDF eigenvector sets taking into account the data uncertainties. We note that the reduction in the PDF uncertainty with both methods depends on the tolerance convention \cite[Sec. 4L in][]{Kovarik:2019xvh}, which must be consistent with that adopted in the profiled PDF ensemble. Default profiling in \xfitter assumes a constant $\Delta \chi^2=1$ tolerance, consistently with HERAPDF/ATLASPDF ensembles and not with CT or MSHT ones. For the latter, Hessian profiling can be consistently performed by the \texttt{ePump} package \cite{Schmidt:2018hvu,Hou:2019gfw}, which implements both global and dynamic tolerance prescriptions. Profiling with too aggressive tolerance generally overestimates the constraints from the profiled experiment on a global PDF set, as demonstrated on the example of profiling ATLAS 7 TeV $W,Z$ production measurements in Appendix F of \cite{Hou:2019efy}.

The \xfitter package can quantify the impact of both existing and future  colliders, in the latter case by fitting or profiling pseudodata for a proposed experiment (e.g. the LHeC or EIC). For example, Ref.~\cite{Abdolmaleki:2019acd} used charged-current DIS charm production pseudodata for the LHeC to constrain the strange PDF. Additionally, it has been shown that measurements of lepton angular distributions can be used to improve the accuracy of theoretical predictions for Higgs boson production cross sections at the LHC~\cite{Amoroso:2020fjw}. The high-statistics determinations of the longitudinally polarized angular coefficient at the LHC Run III and high-luminosity HL-LHC improve the PDF systematic uncertainties of the Higgs boson cross section predictions by 50\% over a broad range of Higgs boson rapidities. Moreover, the complementarity of the lepton-charge and forward-backward asymmetries in DY processes has been studied, and the impact in reducing PDF uncertainties in observables relevant to both SM and BSM physics has been assessed~\cite{Fiaschi:2021okg}.

\xfitter can also study the impact of the $\ln(1/x)$-resummation corrections to the DGLAP splitting functions using DIS coefficient functions from the public code HELL~\cite{Bonvini:2016wki,Bonvini:2017ogt}; these effects are illustrated in Ref.~\cite{Abdolmaleki:2018jln}. In a related study~\cite{Bonvini:2019wxf}, \xfitter employed a more flexible PDF parametrization to achieve a better description of the combined inclusive HERA I+II data, especially at low $x$.

Another  feature of \xfitter is the ability to handle both pole and $\overline{MS}$ heavy-quark masses. While the pole mass is more closely connected to the kinematic features seen in experiments, the $\overline{MS}$ mass has an advantage of better perturbative convergence. As an example, \xfitter was used to perform a high-precision determination of the $\overline{MS}$ charm mass from combined HERA DIS data~\cite{Bertone:2016ywq}.

Finally, while many PDF analyses are now extended out to N2LO QCD, the NLO QED effects may be comparable in some observables. For example, inclusion of QED radiation in the parton evolution breaks the quark charge symmetry, as the up and down quarks have different couplings to the photon. \xfitter includes NLO QED effects, and this is illustrated in Ref.~\cite{Giuli:2017oii}, which determines the photon PDF in a N2LO QCD + NLO QED analysis.

\subsubsection{\nnpdf: an open-source machine learning framework for global analyses of PDFs}
{\it Leading author: M. Ubiali \\ \vspace{6pt}}
\label{sec:nnpdffittingcode}

Along with the recent release of the NNPDF4.0 PDF set~\cite{Ball:2021leu}, in a companion paper~\cite{NNPDF:2021uiq}, the public release of the complete
software framework underlying the NNPDF4.0 global determination was presented, together with user-friendly examples and  extensive documentation.\footnote{The \nnpdf code can be downloaded from a repository at \url{https://github.com/NNPDF/}. The \nnpdf code documentation webpage is located at \url{https://docs.nnpdf.science/}.}
In addition to the \nnpdf fitting code itself, the public repository includes the
original and filtered experimental data, the fast NLO interpolation grids relevant for the computation of hadronic observables, and, whenever available, the bin-by-bin N2LO QCD and NLO electroweak K-factors for all processes entering the fit. Furthermore, the code comes accompanied by a battery of plotting, statistical, and diagnosis tools providing the user with an extensive characterization of the PDF fit output.

These statistical analysis and plotting tools are provided by the \validphys toolkit, which is at the heart of the \nnpdf code base, bridging together the other components and providing basic data structures, compatibility interfaces, I/O operations and algorithms. The \validphys code is, in turn, built on top of \reportengine~\cite{zahari_kassabov_2019_2571601}, a user-friendly data analysis framework providing a declarative interface to specify the required analysis with a minimal amount of information in the form of a run card. 

The availability of the \nnpdf open-source code enables users to perform new PDF analyses based on the NNPDF methodology and modifications thereof.
Some examples of potential applications include assessing the impact of new measurements in the global fit; producing variants based on reduced data sets, carrying out PDF determinations with different theory settings, such as different values of $\alpha_s$, heavy quark masses, electroweak parameters; estimating the impact on the PDFs of theoretical constraints and calculations e.g. from
nonperturbative QCD models~\cite{Ball:2016spl} or lattice calculations~\cite{Lin:2017snn,Cichy:2019ebf}; and quantifying the role of theoretical uncertainties from missing higher orders to nuclear effects. One could also deploy the \nnpdf code as a toolbox to pin down the possible effects of beyond the Standard Model physics at the LHC, such as Effective Field Theory corrections in high-$p_T$ tails~\cite{Carrazza:2019sec,Greljo:2021kvv} or modified DGLAP evolution from new BSM light degrees of freedom~\cite{Berger:2010rj}. Furthermore, while the current version of the \nnpdf code focuses on unpolarized parton distributions, its modular and flexible infrastructure makes it amenable to determination of closely related nonperturbative collinear QCD quantities such as polarized PDFs, nuclear PDFs, fragmentation functions, or even the parton distributions of mesons like pions and kaons. 

\subsection{Fast interfaces for pQCD computation}

\newcommand{\nnlojet}{\texttt{NNLOJET}\xspace}

\newcommand{\rd}{\ensuremath{\mathrm{d}}}

\newcommand{\Pe}{{\ensuremath{\mathrm{e}}}\xspace}
\newcommand{\Pp}{{\ensuremath{\mathrm{p}}}\xspace}
\newcommand{\PZ}{{\ensuremath{\mathrm{Z}}}\xspace}
\newcommand{\PW}{{\ensuremath{\mathrm{W}}}\xspace}
\newcommand{\jet}{\text{jet}\xspace}

\newcommand{\murf}{\ensuremath{\mu_{\rm R/F}}\xspace}
\newcommand{\mup}{\ensuremath{\mu_{0}}\xspace}

\newcommand{\as}{\ensuremath{\alpha_{\mathrm{s}}}\xspace}
\newcommand{\asmz}{\ensuremath{\as(M_{\PZ})}\xspace}
\newcommand{\asmur}{\ensuremath{\as(\mur)}\xspace}
\newcommand{\tilmu}{\ensuremath{\tilde{\mu}}\xspace}

\newcommand{\chisq}{\ensuremath{\chi^{2}}\xspace}
\newcommand{\ndf}{\ensuremath{n_\mathrm{dof}}\xspace}
\newcommand{\ptjet}{\ensuremath{p_\mathrm{T}^\mathrm{jet}}\xspace}
\newcommand{\etalab}{\ensuremath{\eta_\mathrm{lab}^\mathrm{jet}}\xspace}
\newcommand{\GeVsq}{\ensuremath{\,\mathrm{GeV}^2}\xspace}
\newcommand{\pt}{\ensuremath{p_\mathrm{T}}\xspace}
\newcommand{\Qsq}{\ensuremath{Q^{2}}\xspace}


Modern calculations of higher-order corrections in perturbative QCD are computationally very
demanding, with typical calculations at N2LO taking of order
$\mathcal{O}(10^5)$ CPU hours due to the complicated singularity structure of the
real-emission amplitudes and delicate numerical cancellations they
entail.
%
%
The data for relevant LHC cross sections  are becoming increasingly precise, 
and so N2LO predictions must be repeated in comparisons to such data thousands of times 
using varied values for \as, PDFs, and renormalization or factorization scales, $\mu_R$ or $\mu_F$.
It is therefore computationally prohibitive to run the full calculation at 
N2LO for each phase space point as required in such an analysis.

Storing the perturbative
coefficients on a grid, before their combination with the parton luminosity and \as, allows the convolution with arbitrary PDFs to be performed later, with miniscule computational cost. In this approach, variations of \asmz, $\mu_R$, and $\mu_F$ are also possible.
The grid technique, as introduced in Ref.~\cite{Adloff:2000tq}, is implemented independently in the 
\appl~\cite{Carli:2005,Carli:2010rw}
and \fastnlo~\cite{Kluge:2006xs,Britzger:2012bs}
packages.
The technique works by using interpolation functions to
distribute each single weight from the integration over the momentum fractions $x$ and QCD
scales in the convolution. The \applfast project~\cite{Britzger:2019kkb}
implements an interface of \appl and \fastnlo with the \nnlojet program~\cite{Gehrmann:2018szu}. These programs and their \applfast interface are described in Section~\ref{sec:applfast}. In Section~\ref{sec:pineapple} we describe the \pineappl interface, which allows the inclusion of NLO EW corrections.

\subsubsection{The \applfast project}
\label{sec:applfast}

{\it Leading authors: D. Britzger, C. Gwenlan, A. Huss, J. Pires, K. Rabbertz, M. R. Sutton \\ \vspace{6pt}}

The grid technique works by accurately interpolating the full behaviour of any 
function $f(x)$ from the function's values at discrete nodes in  
$a\equiv x^{[0]} < x^{[1]} < \ldots < x^{[N]}\equiv b$ that partition
the interval $[x_{\rm min},x_{\rm max}]$.
Interpolating polynomials, $E_i(x)$, of degree $n$  are used for each node $i$, such that 
$f(x)$ can be approximated by
\begin{align}
  f(x) &\simeq \sum_{i=0}^{N} f^{[i]} \; E_i(x) \quad \text{with \ $f^{[i]} \equiv f(x^{[i]})$} .
\end{align}
To elevate the accuracy of the interpolation with equally spaced nodes, a
variable transformation $x \longmapsto  y(x)$ increases the 
density of nodes in regions where $f(x)$ varies more rapidly.
The corresponding interpolation functions are denoted by $E^y_i(x)$.
The integration can then be approximated by a sum over 
the nodes $i$,
\begin{equation}
  \int_a^b \rd x \; f(x) \; g(x) 
  \simeq
  \sum_{i=0}^{N} f^{[i]} \; g_{[i]}, 
{\rm \ \ where \ \ } 
  g_{[i]} \equiv \int_a^b \rd x \;  E_i(x) \; g(x), 
  \label{eq:grid}
\end{equation}
and the time-consuming computation of the Monte Carlo integral in~Eq.\eqref{eq:grid} is 
performed once and for all to produce a grid $g_{[i]}$ ($i=0,\ldots,N$) for subsequent use. The integral in 
Eq.~\eqref{eq:grid} can then 
be {\em a posteriori} approximated for different functions $f(x)$ using only the summation over the
$N$ grid nodes.

For DIS processes, the parton densities $f_a(x,\muf)$ can be included directly using 
the grid technique. In this case, a two-dimensional grid in the two independent variables $x$ and $\muf$ is 
constructed.
As described in detail elsewhere~\cite{Britzger:2019kkb}, for any value of $x$ and $\mu$, both the PDFs and 
the running of the strong coupling can then be represented by a sum over the interpolation nodes, $i$ and $j$, %
\begin{align}
  \as(\mu) \; f_a(x,\mu) &\simeq
  \sum_{i,j} \as^{[j]} \; f^{[i,j]}_{a} \; E^y_i(x) \; E^\tau_j(\mu)
  \, ,
\end{align}
with $\mur=\muf\equiv\mu$ for simplicity. The index $a$ represents the different partons in the 
cross section, and the calculation includes an implicit sum over these partons. 
In practice, the parton summations often reduce to simple factors and sums over the up-type and 
down-type quarks, and the gluons.  
The computationally expensive convolution with
the PDFs in Eq.~\eqref{eq:grid}, for each order  in the calculation, $\as^p$, can thus be approximated by a summation,
\begin{eqnarray}
  \sigma = \sum_p
  \int \rd x \left(\frac{\as(\mu)}{2\pi}\right)^{k+p} f_a(x,\mu) \; \hat{\sigma}^{(p)}_a(x,\mu)
  \ &\approx& \
  \sum_p \sum_{i,j} \biggl(\frac{\as^{[j]}}{2\pi}\biggr)^{k+p} f^{[i,j]}_{a} \; \hat{\sigma}^{(p)}_{a[i,j]},
  \,\nonumber\\  {\rm \ where \ \ }
   \hat{\sigma}^{(p)}_{a[i,j]} = 
\sum_{m=1}^{M_p}
  E^y_i(x_m)
  \; E^\tau_j(\mu_{m}) & &\hspace{-4mm}
  \; w^{(p)}_{a;m}
  \; \hat{\sigma}^{(p)}_{a;m},
  \label{eq:grid_gen:dis}
\end{eqnarray}
and where the sum over $i$ and $j$ runs over the grid nodes  $x_i$ and $\mu_j$. 
In the product, one interpolation variable is 
needed per independent variable, such that, with one scale and one momentum fraction, 
only two variables are needed, and a separate grid is required for each parton contribution.
To include $\mu_{R},$ $\mu_F$ variations, the summation over the grids 
$\hat{\sigma}^{(p)}_{a;m}$ is modified by adding relevant terms with logarithms of the two scales. 
The full expression can be seen in Eq. (16) from ~\cite{Britzger:2019kkb}. A comprehensive study of the N2LO predictions, as well as an application in PDF and $\alpha_s$ determinations was presented further in Ref.~\cite{H1:2017bml}. Grids for inclusive jet and dijet production at HERA in N2LO are available at~\cite{ploughshare}.

\begin{figure}[bt]
\includegraphics[height=0.335\textwidth]{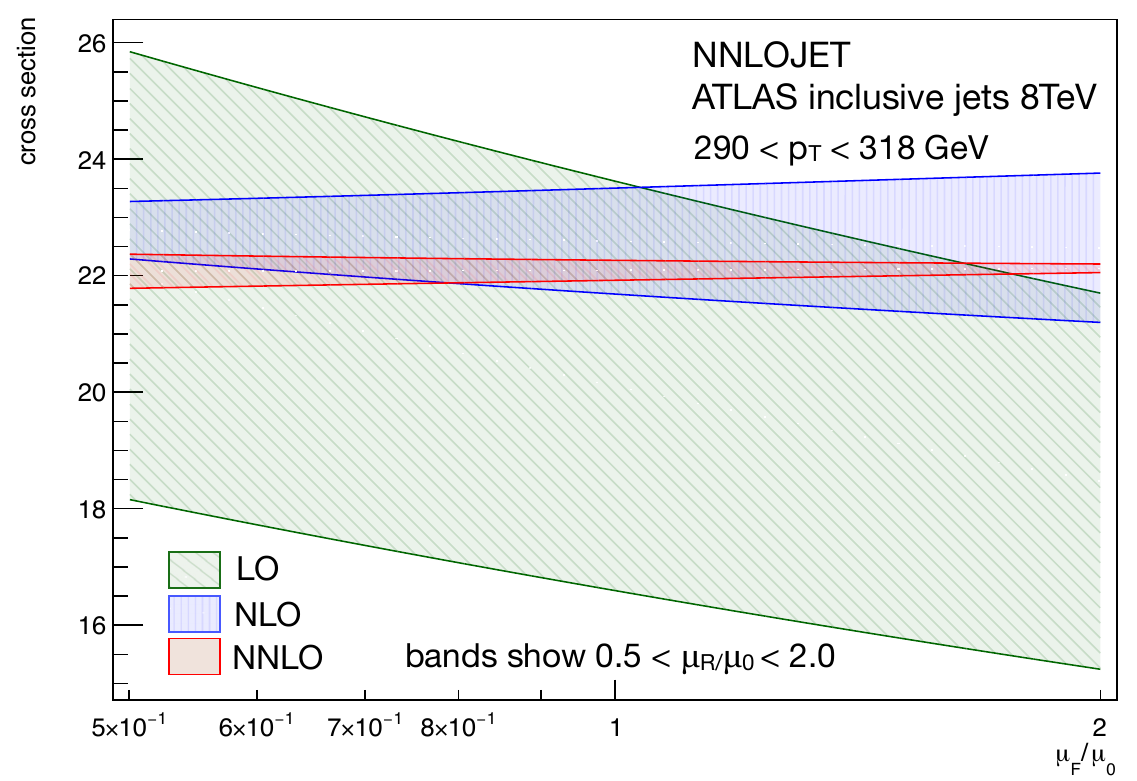}
\includegraphics[height=0.335\textwidth]{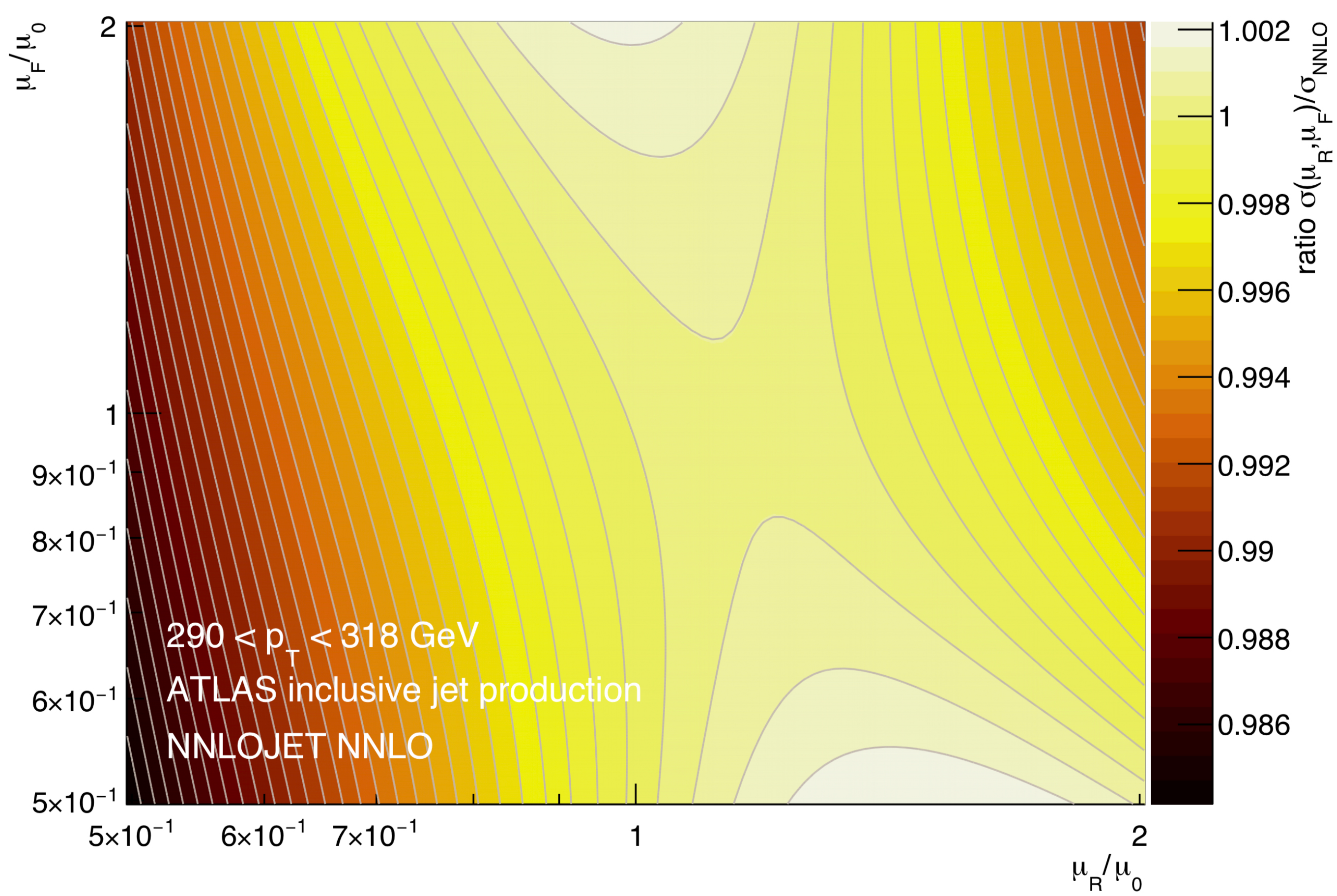}
\caption{Detailed scale variation analysis for the ATLAS inclusive jet production cross section
at 8 TeV in the range \mbox{290 $< \ptjet< 318$~GeV}. Left: the error bands correspond to the renormalization scale variation $0.5 \leq \mu_R/\mu_{0}$ at the indicated QCD orders, with the factorization scale factor $\mu_F/\mu_{0}$ specified on the horizontal axis. Right:
relative variations of the N2LO cross section as a function of the full two-dimensional variation in both the renormalization and factorization scales.}
\label{fig:allscales}
\end{figure}

For cross section predictions for hadron--hadron collisions, the convolution 
over the underlying partonic hard scattering includes a separate PDF for each target 
hadron and so requires an additional interpolation for the momentum fraction from the second hadron, $x_2$, resulting in 
the overall interpolation function 
\begin{equation}
  \as(\mu) \; f_a(x_1,\mu)  f_b(x_2,\mu) \simeq
  \sum_{i,j,k} \as^{[k]} \; f^{[i,k]}_{a} f^{[j,k]}_{b} \; E^y_i(x_1)  E^y_j(x_2) \; E^\tau_k(\mu)
  \, ,
  \label{eq:kernels}
\end{equation}
with  $\mur=\muf\equiv\mu$ again for simplicity, and the transformations $x_i\to y_i$, $\mu^2\to\tau$ performed. 
The summation over $i$, $j$, and $k$ represents the summation over the nodes for $x_1$, $x_2$, and $\mu$ respectively.

A separate grid is needed for each parton contribution, so for $pp$ collisions, 121 grids (excluding top as a parton) 
would be needed for  summations over  partons $a$ and $b$. 
This  would make the grid extremely large and potentially prohibitive for any practical application, in which many 
grids need to be stored in memory. To reduce the number of contributions in the grids,  symmetries within the hard 
subprocesses should be exploited to 
produce a smaller set of unique weighted parton luminosities $F_\lambda(x_1,x_2,\mu)$ relevant for the process.
The summation over the full set of parton flavor combinations, $a$ and $b$, is then replaced by a single 
summation over a significantly smaller set of parton luminosities:
\begin{eqnarray}
  \sigma \simeq
  \sum_{n} \sum_{i,j,k} \biggl(\frac{\as^{[k]}}{2\pi}\biggr)^{p+n} F^{[i,j,k]}_{\lambda} \; \hat{\sigma}^{(n)}_{\lambda[i,j,k]}
  \, .
\end{eqnarray}
For example, for jet production in hadron--hadron collisions, 
the number of separate grids is reduced from  121 down to 13.
\nnlojet automatically performs this reduction via mapping of the separate parts of
the calculation onto a smaller number of unique parton luminosities. 
The final grid is then obtained by accumulating the weights according to
\begin{eqnarray}
  \hat{\sigma}^{(n)}_{\lambda[i,j,k]} = &\xrightarrow{\text{MC}}
  \sum_{m=1}^{M_n}
  E^y_i(x_{1;m})
  E^y_j(x_{2;m})
  E^\tau_k(\mu_{m})
  \; w^{(n)}_{\lambda;m}
  \; \rd\hat{\sigma}^{(n)}_{\lambda;m}
  \, ,
  \label{eq:grid_gen}
\end{eqnarray}
where now the terms $w^{(n)}_{\lambda;m}$ correspond to the weights $w^{(n)}_{ab;m}$ associated with the individual $\lambda$.
As in the case of the DIS cross section, the full grid convolution including scale variations is significantly more 
complex and includes terms that are logarithmic in the scales.  

The grid technique speeds up phenomenological studies that would be impossible otherwise.
In order to facilitate analyses of various jet production cross sections at the LHC up to N2LO in QCD,  the authors are undertaking a
campaign of high-precision grid production using \nnlojet.
For such calculations, high-statistics running --- typically hundreds of thousands of CPU hours --- is required to produce a stable cross section at high orders. Typically the grids are able to reproduce the reference cross section to within 0.1\%.
It is expected that a number of grids for LHC jet cross sections from both 
ATLAS and CMS will be made available to the wider community on the \texttt{ploughshare} web site~\cite{ploughshare} in the near future.

As an example of the tremendous speedup feasible with \applfast, Fig.~\ref{fig:allscales} illustrates the $\mu_R$ and $\mu_F$ dependence of the LO, NLO and N2LO cross sections 
for ATLAS inclusive jet production~\cite{Aaboud:2017dvo} in the range \mbox{290 $< \ptjet < 318$~GeV}. 
The left panel illustrates the $\mu_F$ dependence,  
with the bands quantifying the $\mu_R$ scale uncertainty at the indicated orders. 
The right panel illustrates, with high granularity, the changes in the N2LO cross sections under simultaneous variations of both the renormalization 
and factorization scales.

The full range of the cross section variation at N2LO 
in the shown range is approximately 
1.6\% in total (much smaller 
than the 11\% variation seen at NLO), with a saddle point located close to the nominal cross section. The high resolution on both  $\mu_R$ and $\mu_F$ is indispensable in this case for establishing the saddle-like dependence of the cross section.
As in even more influential applications in fits of \as and PDFs, such a detailed exploration of the behavior of the N2LO cross section is only possible using a fast grid.

\subsubsection{The \pineappl{} interface}
\label{sec:pineapple}
{\it Leading authors: A. Candido, F. Hekhorn, J. Cruz-Martinez, C. Schwan \\ \vspace{6pt}}
  
\pineappl{}~\cite{Carrazza:2020gss,christopher_schwan_2022_3890291} is the newest addition to the family of interpolation grid codes and was developed to support arbitrary coupling orders in $\alpha_\mathrm{s}$ and $\alpha$.
In particular, this includes NLO EW corrections, but also mixed corrections like N2LO QCD--EW corrections, which are not supported by other interpolation grid libraries.
Support for these corrections is needed to fit PDFs with these additional corrections~\cite{Schwan:2021txc}.

\begin{figure}[tb]
\includegraphics[scale=0.53]{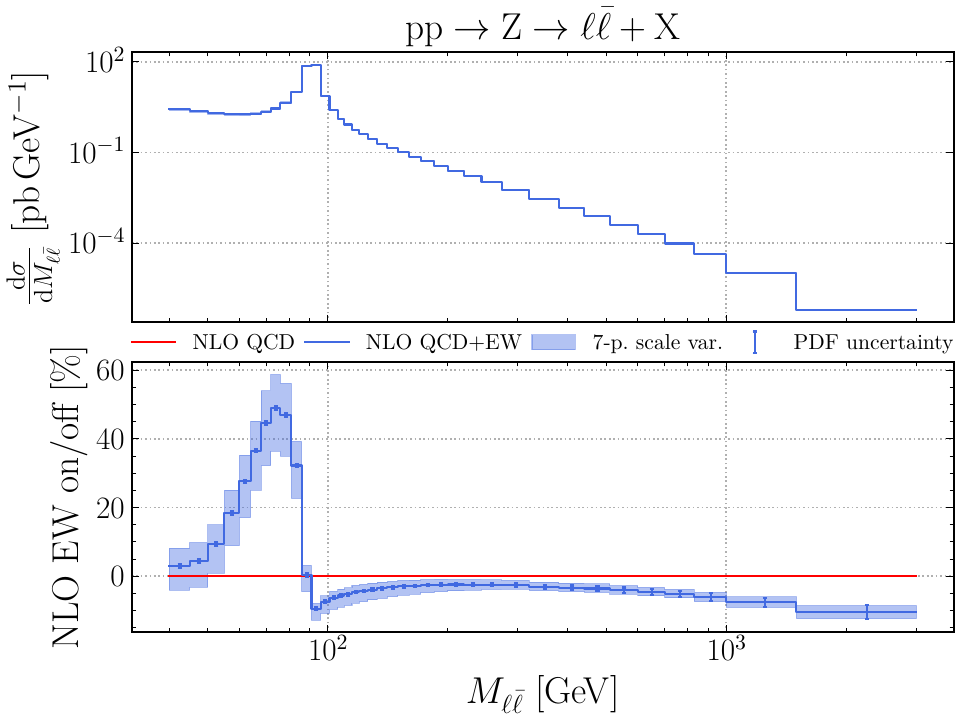}
\includegraphics[scale=0.53]{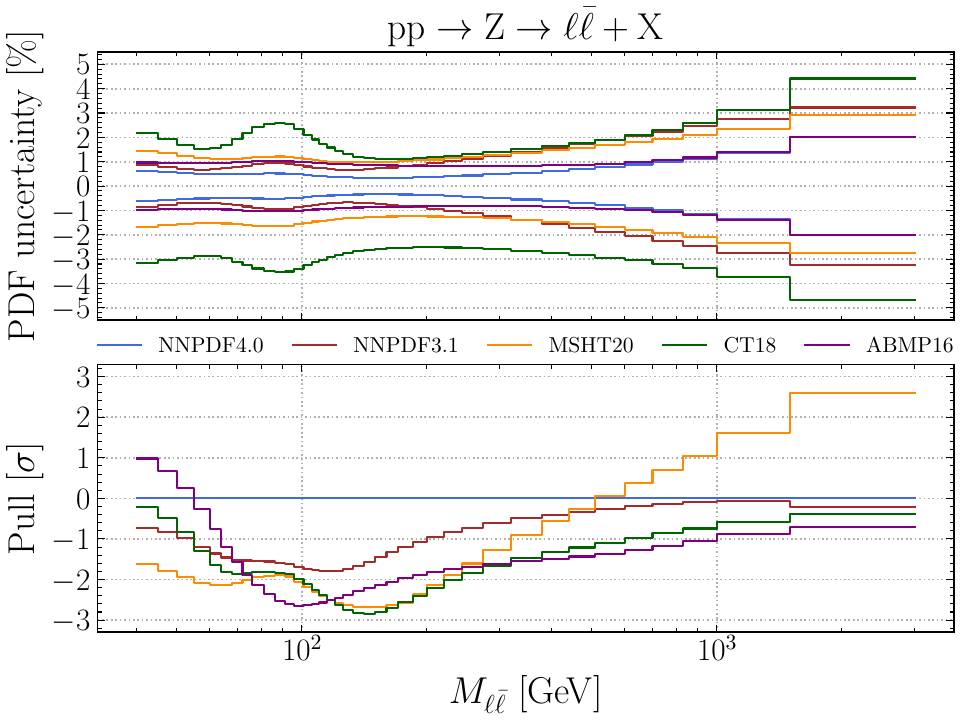}
\caption{NLO QCD+EW corrections for DY lepton-pair production at the LHC at 14~TeV, showing absolute predictions (top left) and relative size of the EW corrections (bottom left), PDF uncertainties (top right) and pulls (bottom right) for different PDF sets.
See Ref.~\cite{Ball:2021leu}, Sec.~9.2 and 9.3, for full information; all plots have been generated with \pineappl{}'s command-line interface.}
\label{fig:pineappl-plots}
\end{figure}
\pineappl{} is interfaced with \madgraph{}~\cite{Frederix:2018nkq} and \yadism{}~\cite{yadism,candido_alessandro_2021_5795955}, which we use to produce interpolation grids for hadron--hadron and hadron--lepton collider processes, respectively.
Although \pineappl{} is written in Rust, interfaces in C, C++, Fortran and Python are also provided, so it can be easily integrated into any Monte Carlo (MC) generator.
\begin{figure}[tb]
    \centering
    \includegraphics[width=0.9\textwidth]{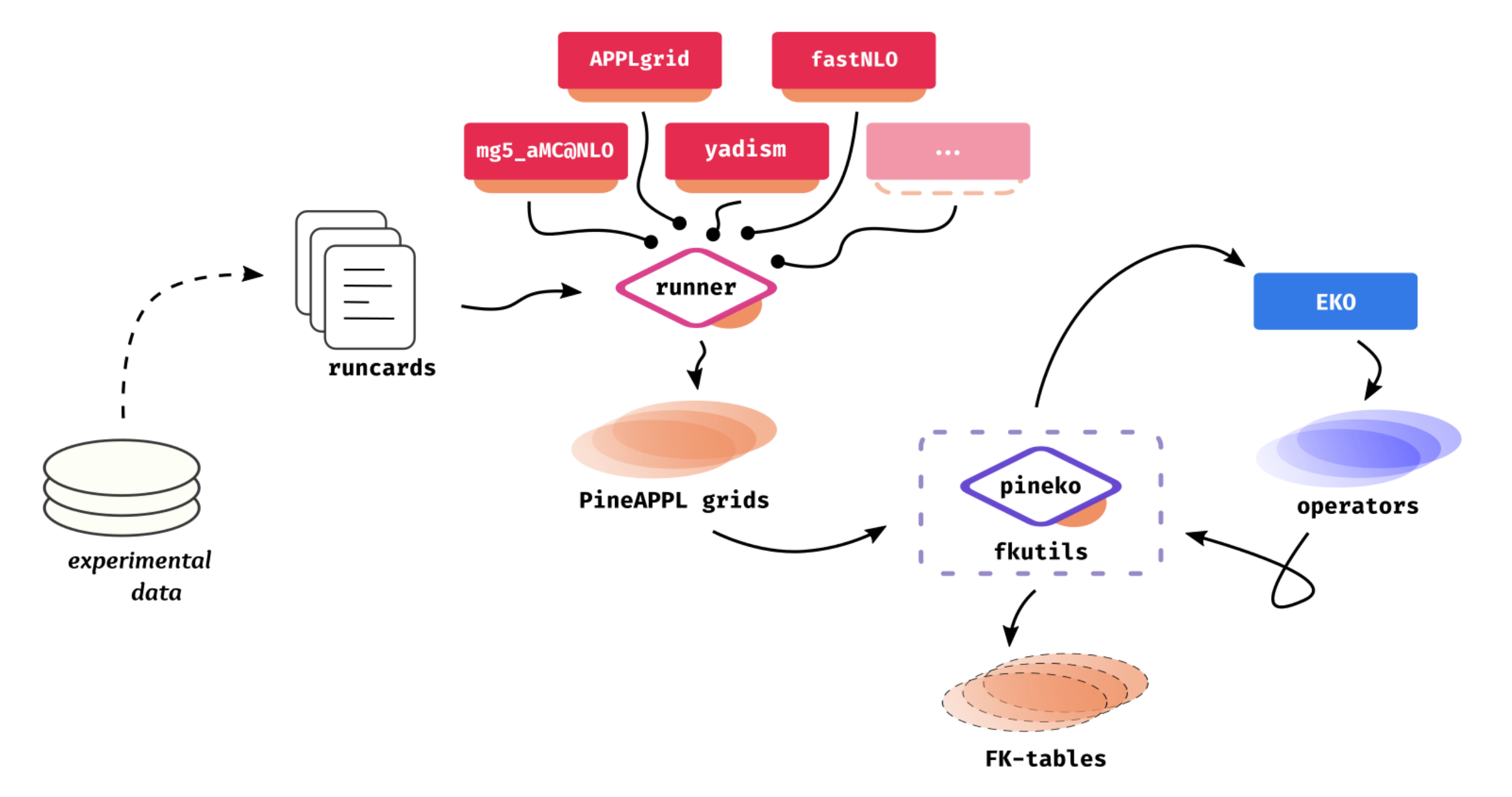}
    \caption{Updated pipeline for NNPDF's theory predictions. \pineappl{} interpolation grids are produced by a runner tool that either converts existing \appl{} or \fastnlo{} tables or runs a chosen MC event generator (\madgraph, \yadism, ...) using run cards to input the parameters for a selected experimental observable (phase-space cuts, scale choices, binning of events, ...).. Afterwards the interpolation grids are queried to generate suitable evolution operators with \eko, which are then utilized to produce the desired FK tables. All orange insets attached to the programs' boxes represent usage of the \pineappl interface.}
    \label{fig:fkstack}
\end{figure}
Interfaces to more MCs, including those with N2LO precision, are being worked on.
Existing interpolation grids from \appl{} and \fastnlo{} can be converted into the \pineappl{} format using one of the supplied programs.
Finally, \pineappl{} comes with a command-line interface, which allows easy convolutions of grids with PDFs, and even more operations, such as: plot predictions and pulls (see Fig.~\ref{fig:pineappl-plots}), list the sizes of all partonic channels, show differences between two grids, show the size of the different coupling orders, calculate PDF uncertainties, and calculate the pull between two PDF sets.
\pineappl{} will be used in an updated version of the \nnpdf fitting code \cite{NNPDF:2021uiq}, for which a part that will be updated is shown in Fig.~\ref{fig:fkstack}.

The interpolation grids generated by MCs are not directly used by the \nnpdf fitting code, but instead they are first converted into so-called \texttt{Fast Kernel (FK)} tables~\cite{Ball:2008by,Ball:2010de,DelDebbio:2007ee}. Using DGLAP equations, the grids at factorization scale values set by the process are evolved to a (typically smaller) single scale, at which the PDFs are fitted. This procedure reduces the evaluation of theory predictions down to a simple linear algebra operation, which can be implemented efficiently and easily parallelized.

At the technical level these operations are shown in Fig.~\ref{fig:fkstack}.
First, a \pineappl{} grid must be generated, either by converting existing \appl and \fastnlo{} tables, or by running programs that the compute the c, for example \yadism{}.
In the latter case, run cards must be written to specify how the process is calculated for a given experimental measurement.
Next, the \pineappl{} grid is evolved into an \texttt{FK} table.
This is performed by \texttt{pineko}, which instructs \eko~\cite{Candido:2022tld,candido_alessandro_2022_5896965} to generate the evolution kernel operators (EKO) and subsequently uses the operators to perform the evolution.
The program \texttt{fkutils} integrates this process for all the processes in NNPDF and finally provides the FK tables to the fitting code.

\section{Benchmarking and combination of global PDF analyses: PDF4LHC21 recommendation}
\label{sec:benchmarking}

\noindent
{\it Leading authors: A. M. Cooper-Sarkar, A. Courtoy, T. Cridge, J. Rojo, K. Xie \\ \vspace{6pt}}

The  highly challenging endeavor of precise and accurate determination of the proton's PDFs~\cite{Gao:2017yyd,Kovarik:2019xvh} tackles interconnected issues associated with limitations of fixed-order theory calculations, internal or external
inconsistencies of the experimental measurements, ill-defined correlation models,
choice of techniques for PDF error estimate
and propagation, choice of the PDF parametrization, implementation of
theoretical constraints on the PDF shape like positivity,
integrability, counting rules, or Regge asymptotics, implementation of heavy-quark contributions, and the choice of SM parameters.
In-depth  understanding of the differences and similarities between global
PDF determinations can be achieved through 
dedicated benchmark exercises involving close collaboration of
the  PDF-fitting groups among themselves and with the experimental groups who published the fitted data.

To advance progress in our understanding of the proton structure,
the PDF4LHC Working Group was established in 2008~\cite{DeRoeck:2009zz} with the mission of coordinating scientific discussions and collaborative projects within the PDF theory and LHC experimental communities.
The first PDF4LHC benchmarking exercise was performed in 2010~\cite{Alekhin:2011sk}, resulting in
an initial set of recommendations~\cite{Botje:2011sn} for PDF usage at Run I of the LHC.
Subsequently, several dedicated studies and benchmark exercises
were carried out~\cite{Ball:2012wy,Rojo:2015acz,Andersen:2016qtm,Andersen:2014efa}.
Then in 2015, following a year-long study, the PDF4LHC15 combined sets were released~\cite{Butterworth:2015oua}
together with an updated set of recommendations for PDF usage and uncertainty estimate at the LHC Run II.
PDF4LHC15 was based on the combination of the CT14~\cite{Dulat:2015mca}, MMHT2014~\cite{Harland-Lang:2014zoa},
and NNPDF3.0~\cite{NNPDF:2014otw} global analyses and  was made possible thanks to
development of techniques for transformation of
Hessian PDF sets into their MC representation~\cite{Watt:2012tq} and vice versa~\cite{Carrazza:2015aoa,Gao:2013bia,Carrazza:2016htc},
and for compression of MC replica sets~\cite{Carrazza:2015hva}.

Since the release of PDF4LHC15, several developments took place in topics of direct relevance for global
PDF determinations.
First of all, the availability of a large number of new data sets from the LHC, which provide significant constraints
on the proton PDFs in a wide kinematic range and for many complementary flavor 
combinations.
Second, the completion of crucial N2LO QCD calculations~\cite{Heinrich:2020ybq} for processes such as inclusive jet~\cite{Currie:2016bfm} and dijet~\cite{Currie:2017eqf} production, direct photon production~\cite{Campbell:2016lzl}, differential top-quark pair production~\cite{Czakon:2015owf}, and
charged-current deep-inelastic scattering with heavy-quark mass effects~\cite{Gao:2017kkx},
of key relevance for global PDF fits~\cite{AbdulKhalek:2020jut,Faura:2020oom,Czakon:2016olj,Campbell:2018wfu,Czakon:2019yrx,Harland-Lang:2017ytb}.
Third, steady progress in the developments of novel fitting methodologies,
such as improved parameterization strategies and machine learning techniques.
An update of the PDF4LHC15 combination was both timely and relevant, especially taking into account the upcoming
restart of data-taking at the LHC during its Run III and subsequently of its high-luminosity
era~\cite{Cepeda:2019klc,Azzi:2019yne}.

This state of affairs has motivated a recent
PDF4LHC21 study~\cite{Ball:2022hsh} based on the combination of three updated
global PDF analyses, CT18~\cite{Hou:2019efy}, MSHT2020~\cite{Bailey:2020ooq}, and
NNPDF3.1~\cite{NNPDF:2017mvq},
and the subsequent assessment of its implications for the phenomenology program of the LHC Run III.
As a requisite for this new combination, comprehensive benchmark comparisons \cite{Cridge:2021qjj} were carried out,
aiming to better pinpoint the origin of the differences between the three global PDF fits either in terms of the input data, theory settings, or fitting methodology.
Special attention has also been paid in these benchmarking exercises to the role played by the assumptions underlying
the experimental correlation models
in the interpretation of high-precision LHC measurements,
which are often limited
by systematic uncertainties~\cite{Harland-Lang:2017ytb,Bailey:2019yze,Harland-Lang:2017ytb,Thorne:2019mpt,Boughezal:2017nla,AbdulKhalek:2020jut,Amat:2019upj,Hou:2019gfw,Aad:2015mbv,ATLAS:2018owm,Amoroso:2020lgh}.
The PDF4LHC21 study also
benefited from the lessons provided by the PDF analyses published by
ATLAS~\cite{ATLAS:2021vod,ATLAS:2021qnl} and CMS~\cite{CMS:2021yzl}.
These benchmarking studies in the context of
the PDF4LHC21 combination demonstrated that many differences
observed between the three global PDF sets can be explained 
by genuinely valid choices related to the input data set,
theory settings, and fitting methodology adopted by the PDF-fitting groups.

\begin{figure}[t]
    \includegraphics[width=0.94\textwidth]{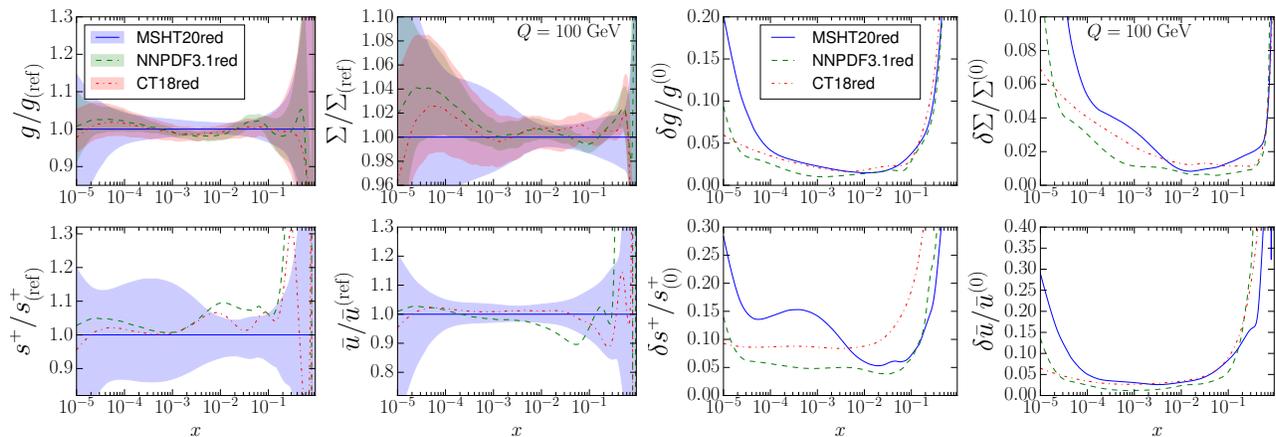}
    \caption{Left: comparison between the reduced data set PDFs from the CT18, MSHT20, and
      NNPDF3.1 groups at $Q=100$ GeV (normalized to the central value of MSHT20).
      Right: same for the corresponding one-sigma PDF uncertainties.
    }
    \label{fig:pdf_benchmarking_fig1}
\end{figure}

One of the main ingredients of the PDF4LHC21 benchmarking study
was production and comparisons of variants of
the CT18, MSHT20, and NNPDF3.1 fits based on a reduced,
identical data set, for which one has striven to homogenize, as much as possible, the settings of the underling theory calculations \cite{Cridge:2021qjj}.
Figure~\ref{fig:pdf_benchmarking_fig1} compares the reduced-data set PDFs from the CT18, MSHT20, and
      NNPDF3.1 groups at $Q=100$ GeV, normalized to the central value of MSHT20,
      as well as the corresponding one-sigma PDF uncertainties.
Good agreement between the three reduced fits is found, and in particular
their agreement is improved as compared to the corresponding global fits based on
the baseline data set from each group.
This good agreement is clearly visible e.g. for the gluon and the total quark singlet PDFs
across the whole range of $x$.
Some differences observed in the baseline fits also persist in the reduced fits, such as in the magnitude of the PDF uncertainties. This observation
indicates that the methodological choices adopted by each group, for example due to the parametrization form, tolerance, or fitting methodology, remain significant even
when fitting the same data set and  can be, in some cases, as
large or even larger than the PDF uncertainties associated with the input fitted data.

\begin{figure}[tb]
 \includegraphics[width=0.32\textwidth]{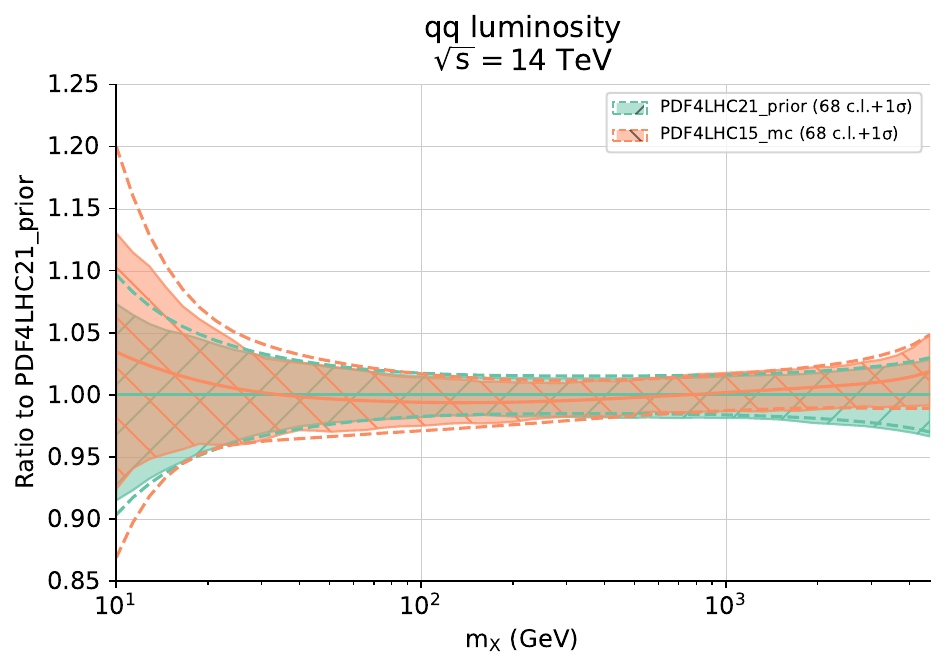}
\includegraphics[width=0.32\textwidth]{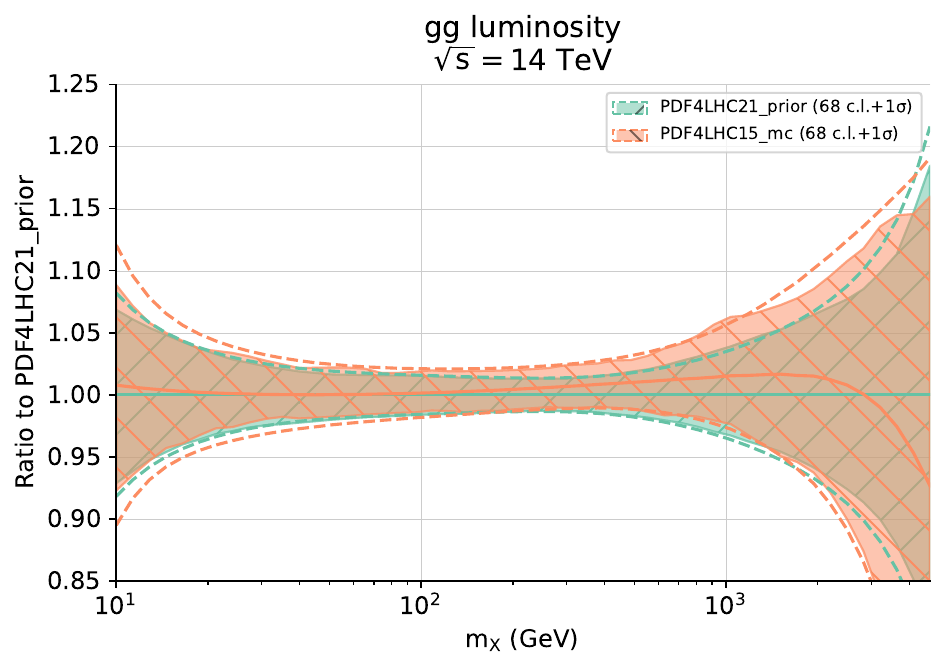}
\includegraphics[width=0.32\textwidth]{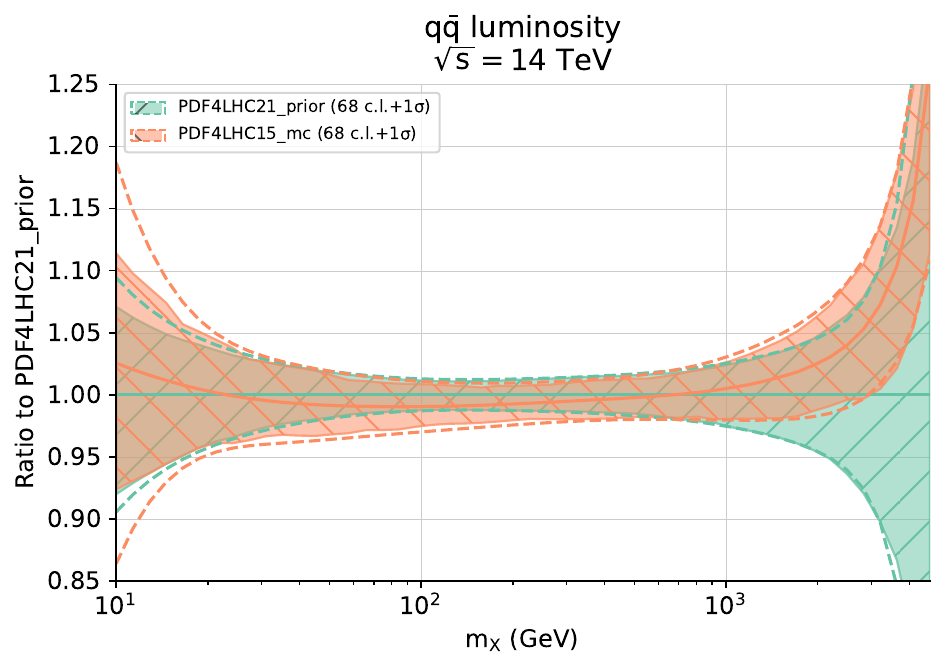}
\includegraphics[width=0.32\textwidth]{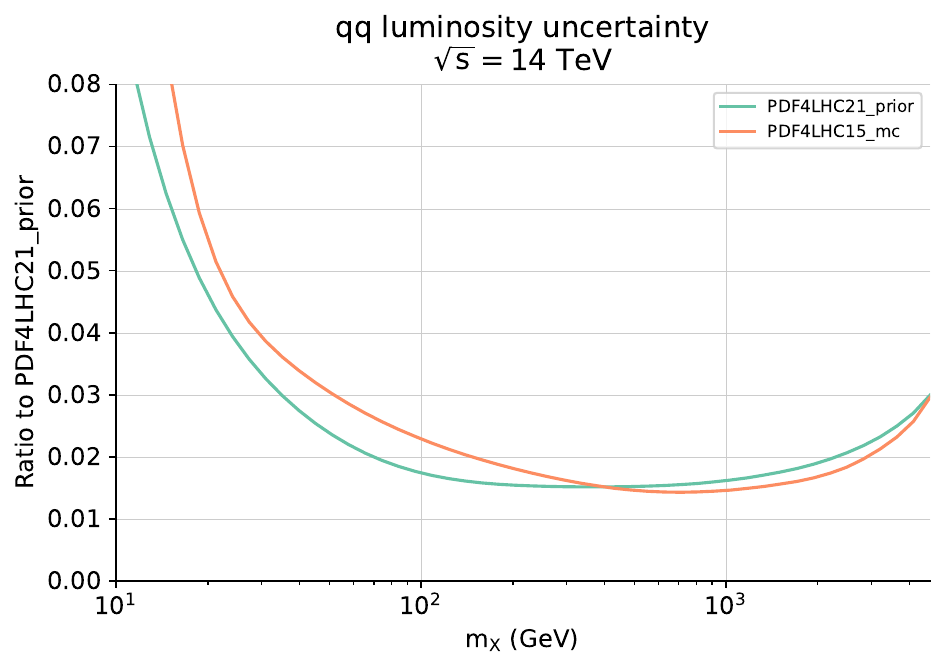}
\includegraphics[width=0.32\textwidth]{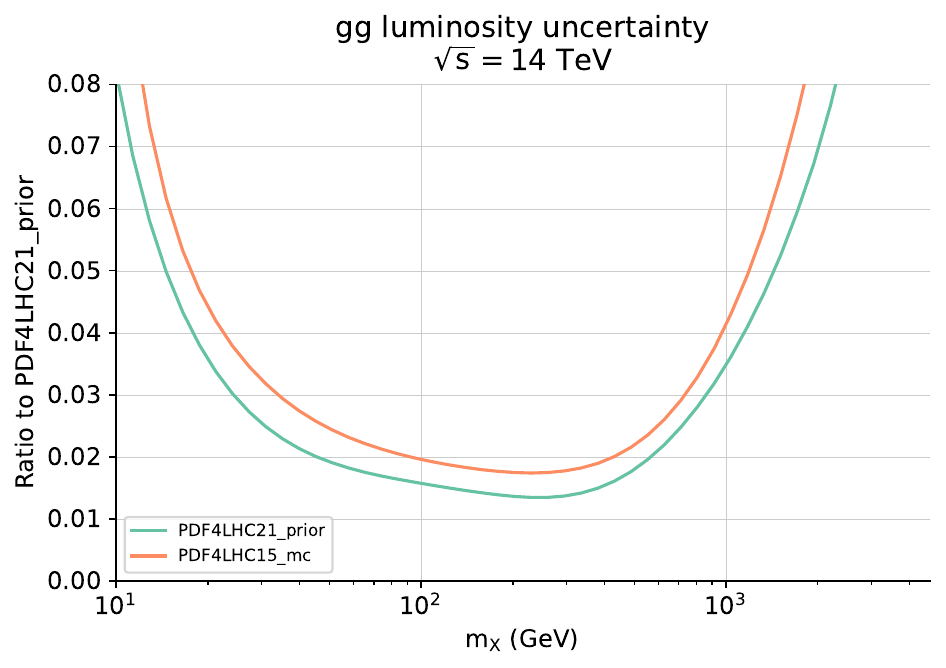}
\includegraphics[width=0.32\textwidth]{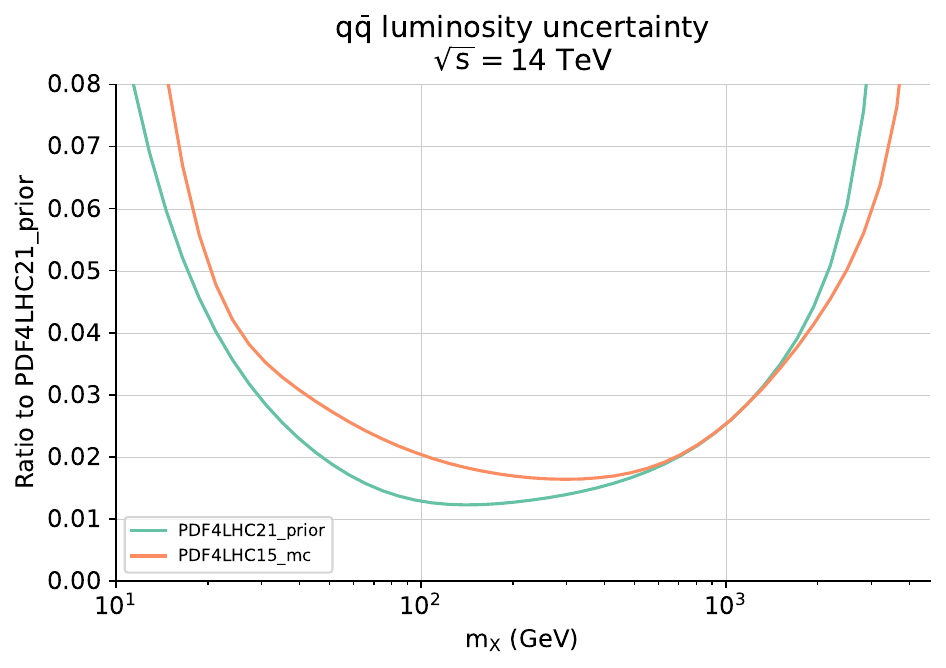}
\caption{Comparison between the partonic luminosities at the LHC 14 TeV 
    of PDF4LHC21 (baseline combination with $N_{\rm rep}=900$ replicas, labeled as ``prior") and of PDF4LHC15
    (compressed set with $N_{\rm rep}=100$ replicas).
    The upper panels display the ratio of the central
    value of PDF4LHC21, and the lower panels the relative
    $1\sigma$ PDF uncertainty in each case.
}
\label{fig:pdf4lhc21}
\end{figure}

Having established that the differences between CT18, MSHT20, and NNPDF3.1
are mostly associated with the choices
related to methodology and data set, the three global
fits were combined by taking $N_{\rm rep}=300$ MC replicas from each group, to form the PDF4LHC21 baseline set with $N_{\rm rep}=900$ replicas in total.
Figure~\ref{fig:pdf4lhc21} displays the
comparison between the partonic luminosities at 14 TeV of the resulting PDF4LHC21 baseline set  and PDF4LHC15 (in this case, the compressed MC variant with $N_{\rm rep}=100$
replicas).
We display the quark-quark, gluon-gluon, and quark-antiquark
luminosities at the LHC 14 TeV, normalized to the central value
of PDF4LHC21 in the upper panels, and for their $1\sigma$ relative
uncertainty in the lower panels.
   Despite the various changes that the three constituent sets have undergone from
the previous to the current combination, PDF4LHC21 not only agrees
within uncertainties with PDF4LHC15 in the kinematic range relevant
for the LHC, but also exhibits a moderate reduction of the PDF uncertainties
in the gluon sector and for the quark luminosities
in the invariant mass region $m_X\le 1$ TeV.
Hence, theory predictions based on PDF4LHC21 will have
reduced PDF uncertainties for several precision LHC observables.

\begin{figure}[!t]
\centering
\includegraphics[width=0.32\textwidth]{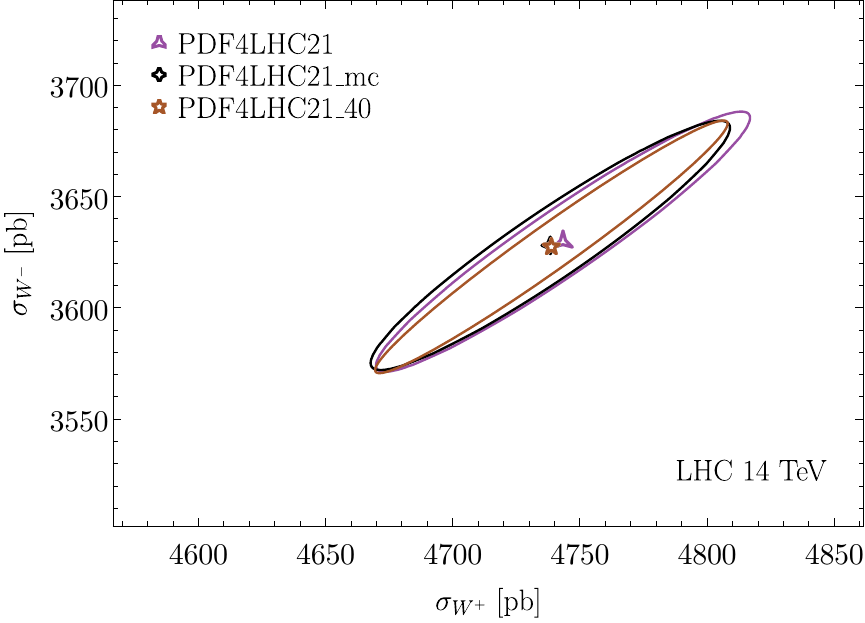}
\includegraphics[width=0.32\textwidth]{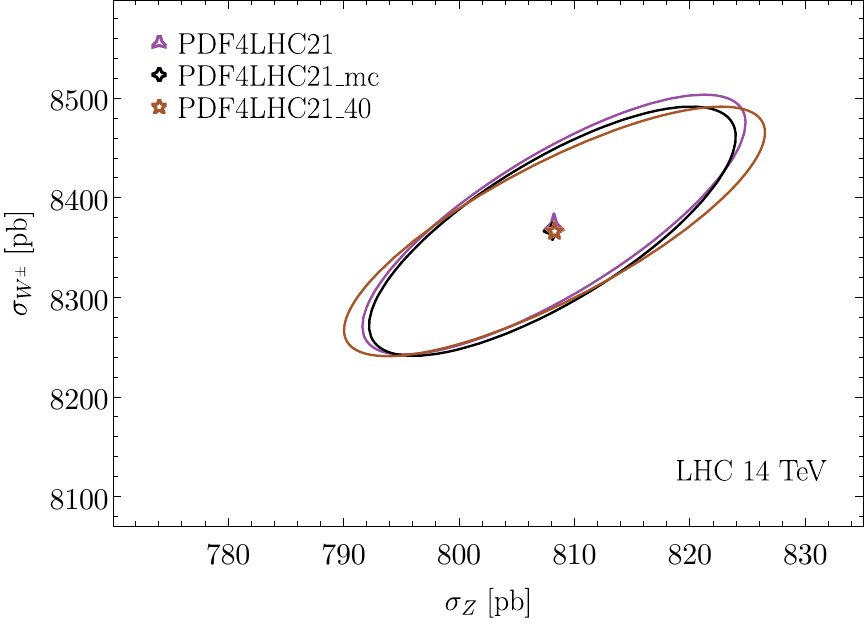}
\includegraphics[width=0.32\textwidth]{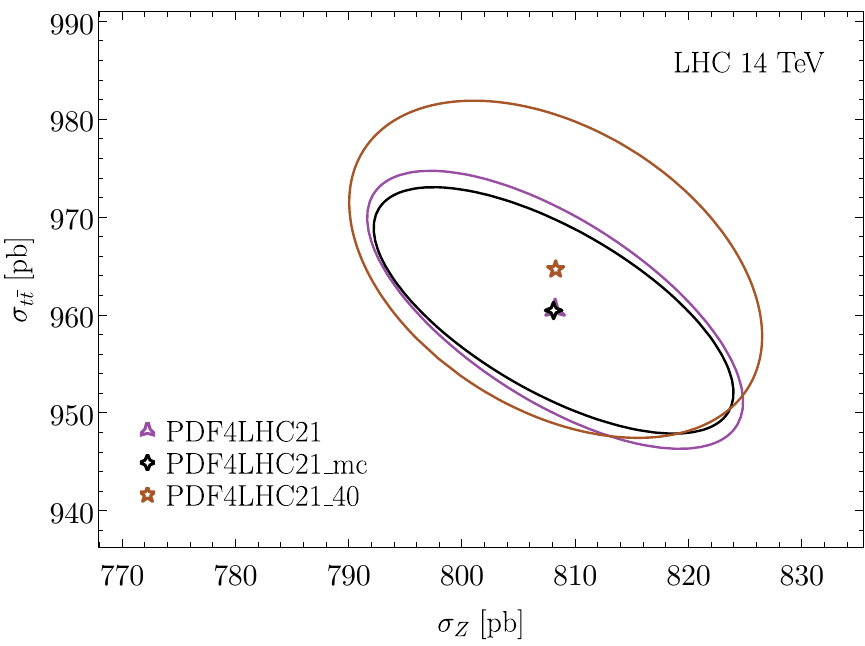}
\includegraphics[width=0.32\textwidth]{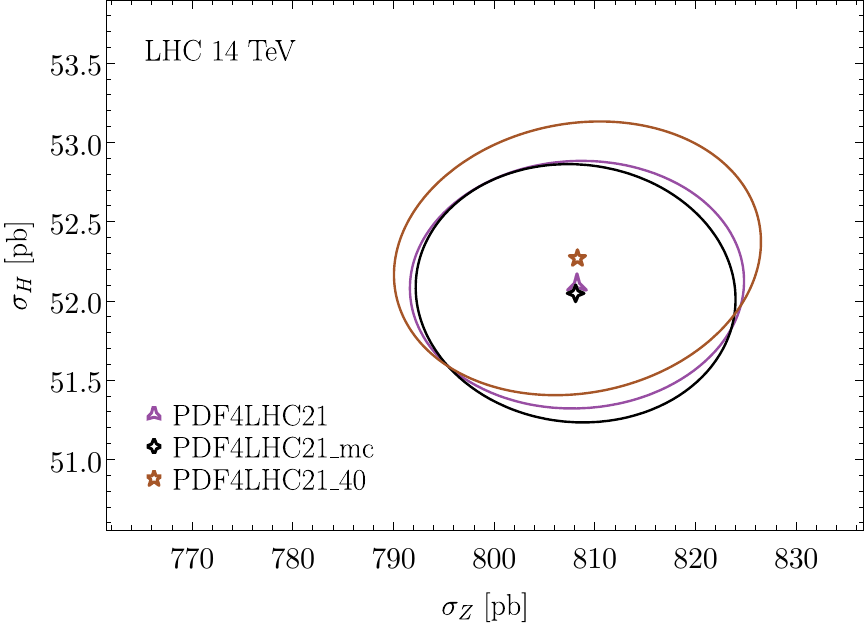}
\includegraphics[width=0.32\textwidth]{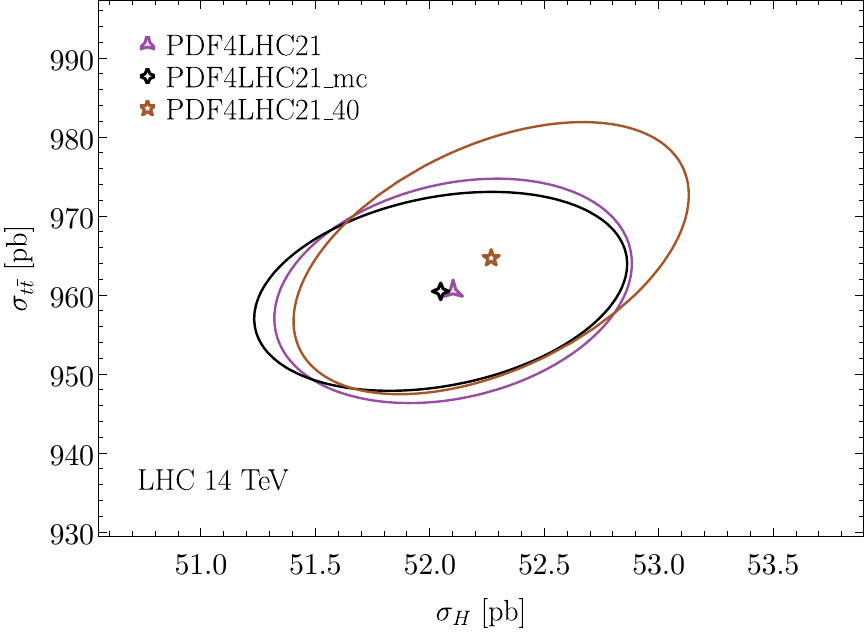}
\includegraphics[width=0.32\textwidth]{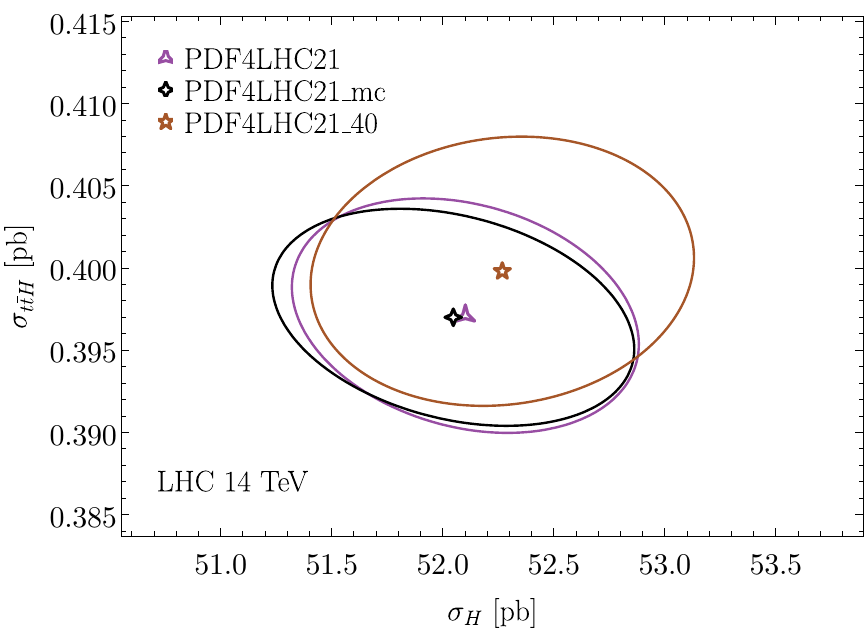}
\caption{Comparison between the predictions using the baseline PDF4LHC21 baseline set 
($N_{\rm rep}=900$ replicas) and those of its Hessian ($N_{\rm eig}=40$)
and compressed MC ($N_{\rm rep}=100$) representations for the $1\sigma$ correlation ellipses for pairs of inclusive cross sections among the $W^\pm,Z,t\bar{t},H,t\bar{t}H$ production processes at the LHC 14 TeV.}
\label{fig:CMC2META}
\end{figure}

As was the case of the previous combination~\cite{Butterworth:2015oua},
the $N_{\rm rep}=900$ replicas of the PDF4LHC21 baseline are 
reduced down to a more manageable number of error PDFs for applications
at the LHC.
We have considered two techniques to obtain a Hessian representation
of PDF4LHC21, namely the  \texttt{META-PDF} approach~\cite{Gao:2013bia}
and the \texttt{mc2hessian} algorithm~\cite{Carrazza:2015aoa,Carrazza:2016htc}.
The \texttt{META-PDF} method is based on constructing
a common meta-parametrization of the replicas that constitute the baseline
using Bernstein polynomials.
All input replicas end up having associated the same parametric form,
each with different numerical parameters of the Bernstein polynomials.
Then, dimensionality reduction is performed in the space of meta-parameters  by Principal Component Analysis (PCA).
Within this method, it is also possible to impose the positive-definiteness of the
central member of the  PDF ensemble.
The basic idea of \texttt{mc2hessian} is to use the MC replicas of the prior themselves
to  construct a Hessian representation with the replicas' linear expansion basis,
and then to determine the numerical coefficients of the expansion to ensure
that the mean, variance, and correlations of the baseline distribution are
reproduced  based
on the combination of PCA and
Singular Value Decomposition (SVD). 
After the comparison of two methods, the final deliverable Hessian set, 
\texttt{PDF4LHC21\_40}, was chosen to be the reduced set obtained through 
the updated \texttt{META-PDF} technique with 40 eigenvector sets and a feature ensuring positive-definite central PDFs of the resulting Hessian set.

The second public error ensemble of the PDF4LHC21 distribution, based on a reduced Monte Carlo representation, is also constructed by means of the replica compression
algorithm~\cite{Carrazza:2015hva,Carrazza:2021hny}, whose goal is to extract the subset 
of the replicas  that most faithfully reproduces the statistical properties of the prior distribution.
The compression
methodology relies on two main ingredients: a proper definition of a distance metric
that quantifies the distinguishability between the baseline and the compressed
probability distributions,
and an appropriate minimization algorithm that explores the space of possible combinations
of PDF replicas which leads to such a minima. The final deliverable MC set, called \texttt{PDF4LHC21\_mc}, contains 100 compressed MC replicas.

Figure~\ref{fig:CMC2META} compares the predictions for the $1\sigma$ correlation ellipses at
the LHC at $\sqrt{s}=14~{\rm TeV}$ for representative inclusive cross sections between
the PDF4LHC21 baseline combination and its Hessian and compressed MC representations, \texttt{PDF4LHC21\_40} and \texttt{PDF4LHC21\_mc}.
We have considered production of $W^\pm$ and $Z$ gauge bosons, top-quark pairs, Higgs bosons in gluon fusion, and $t\bar{t}$ paris associated with a Higgs boson. 
The $W^\pm/Z$ cross sections correspond to the fiducial volume measured at ATLAS 13 TeV~\cite{ATLAS:2016fij}, while others refer to the full phase space.
One finds generally good agreement between the baseline and its compressed MC and Hessian reduced sets. The small shift in central value in the Hessian set as compared to the baseline is related to imposing positivity of the central PDF in the former, with the difference contained within the uncertainties  of the baseline set.
Extensive comparisons for other LHC observables at the inclusive and differential level confirm that PDF4LHC21 is compatible with PDF4LHC15, while exhibiting a modest reduction of PDF uncertainties, and that, furthermore, the compressed MC and Hessian reduced sets provide an adequate, user-friendly representation of the baseline combination.
Additional studies of the phenomenological implications of PDF4LHC21 are reported in~\cite{Ball:2022hsh}.

There are several directions in which the PDF4LHC21 studies could be expanded.
To begin with, one could extend the analysis of PDF fits based on a common reduced data set
by adding other measurements,
since an even wider ``reduced'' data set could further highlight which differences observed between PDF groups can be traced back to the underlying methodological choices.
One could also consider investigations of why the PDF uncertainties between
various groups differ even when a similar input data set is considered, such as the one recently pursued in \cite{Courtoy:2022ocu}.
Also, the PDF4LHC21 combination will have to be eventually updated once new releases
from the various PDF fitting collaborations are presented.
Furthermore, future combinations will also have to account for not only the PDF contribution to the total
uncertainty, but also other sources such as MHOUs which will be strongly correlated between the groups,
as well as combinations between PDF sets including QED corrections and the photon
PDF~\cite{Bertone:2017bme,Ball:2013hta,xFitterDevelopersTeam:2017fxf,Harland-Lang:2019pla,Schmidt:2015zda}.


\section{Conclusion: precision PDFs in the United States
\label{sec:Conclusions}}
Among several groups (ABM, CTEQ-TEA, HERAPDF, MMHT, NNPDF) working on the determination of general-purpose N2LO PDFs, one group (CTEQ-TEA \cite{CTEQTEAWebsite,Hou:2019efy,Dulat:2015mca,Hou:2016nqm,Hou:2016sho,Hobbs:2019gob,Wang:2018heo,Hou:2019gfw}) is currently based in the US. Each general-purpose global analysis of PDFs is a major undertaking, involving significant investment in development, testing, and tuning of theoretical and computational frameworks. Recall that it took more than ten years from the publication of N2LO DGLAP equations \cite{Moch:2004pa,Vogt:2004mw} to the release of N2LO PDF parametrizations with benchmarked accuracy \cite{Butterworth:2015oua}.
Further advancements require support for the critical mass of the personnel with the specialized expertise. These advancements greatly benefit from the collaborations between experimentalists and theorists, and from international collaborations. 

Since the Electron-Ion Collider can provide powerful new constraints on large-$x$ PDFs \cite{AbdulKhalek:2021gbh}, it makes sense to forge novel collaborations between the HEP and nuclear physics communities in the US.
Unique data with high sensitivity to a wide range of PDF phenomena may
be also collected with the LHeC experiment \cite{LHeC:2020van} at CERN in the 2030s. The US
nuclear and particle physics community should feel encouraged to support
CERN in its efforts to realize that unique experiment, and subsequently
benefit from the open collaboration and access to these data. Looking even further into the future, the Muon-Ion Collider \cite{Acosta:2022ejc} in the US may become a factory of precision measurements of the hadron structure. 

The precision physics frontier at the HL-LHC and EIC opens up new fascinating opportunities and challenges in the field of PDF determination. The HL-LHC projections are very encouraging, with a foreseen reduction of PDF uncertainties by factor 2-3. However, reaching this accuracy target requires coordinated advancements in experimental measurements, theoretical computations, and global analysis methodology. In particular, to be able to reduce the PDF uncertainties, the precision experiments that probe the PDFs should strive to reach better agreement among themselves than has been possible until now. We believe that, to reach such agreement, it is critical that new experiments and theory calculations implement consistent error control at all stages, from experimental measurements to the distribution of final PDFs. Efforts in this direction should go hand-in-hand with, and be as adequately supported as the investment into 
new conceptual advancements, such as the PDFs with electroweak constituents \cite{Manohar:2017eqh, Han:2020uid, Buonocore:2020nai}, as well as 
computations of new radiative contributions, such as those associated with N3LO QCD and NLO EW terms. As important is to continue development of the robust methodology for the global fits, including advanced statistical tests of the goodness of fit, methods for estimating theoretical uncertainties, novel statistical inference techniques inspired by the large-scale data science and artificial intelligence, and a new generation of computer programs for the global fits and fast multi-loop computations.

\acknowledgments

This work was supported by the U.S. Department of Energy under Contracts DE-AC02-76SF00515, DE-AC02-06CH11357,
DE-FG02-91ER40684, DE-SC0010129, DE-SC0007914;
by the U.S. National Science Foundation under Grants No. PHY-1820760, PHY-2112025, PHY-2013791, PHY 1653405;
by the Deutsche Forschungsgemeinschaft (DFG, German Research Foundation) Research Unit FOR 292, project 40824754, and project 417533893/GRK2575 “Rethinking Quantum Field Theory”; by the European Research Council under the European Union's Horizon 2020 research and innovation Programme
(grant agreements 740006, 824093, 950246);
by the UK Royal Society grant DH150088;
by the UK Science and Technology Facilities Council (STFC) grants  ST/P000274/1, ST/L000377/1,  ST/P000630/1, ST/T000600/1,
 ST/T000694/1, ST/T000856/1, ST/T000864/1.

N. Armesto acknowledges financial support by Xunta de Galicia (Centro singular de investigaci\'on de Galicia accreditation 2019-2022); the "Mar\'\i{}a de Maeztu" Units of Excellence program MDM2016-0692 and the Spanish Research State Agency under project PID2020-119632GB-I00; European Union ERDF; the European Research Council under project ERC-2018-ADG-835105 YoctoLHC; MSCA RISE 823947 "Heavy ion collisions: collectivity and precision in saturation physics" (HIEIC).
F. G. Celiberto acknowledges support from the INFN/NINPHA project and thanks the Universit\`a degli Studi di Pavia for the warm hospitality.
A. M. Cooper-Sarkar wishes to thank the Leverhulme Trust. 
A. Courtoy is supported by the UNAM Grant No. DGAPA-PAPIIT IN111222 and CONACyT Ciencia de Frontera
2019 No. 51244 (FORDECYT-PRONACES).
M. Hentschinski acknowledges Support by Consejo Nacional de Ciencia y Tecnolog{\'\i}a grant number A1 S-43940 (CONACYT-SEP Ciencias B{\'a}sicas).
The work of H.-W. Lin is partially supported by the US National Science Foundation under grant PHY 1653405 ``CAREER: Constraining Parton Distribution Functions for New-Physics Searches'' and  by the  Research  Corporation  for  Science  Advancement through the Cottrell Scholar Award. 
B. Malaescu gratefully acknowledges the continuous support from LPNHE, CNRS/IN2P3, Sorbonne Universit\'e and Universit\'e de Paris.
J. Rojo is partially supported by the Dutch Science Council (NWO).
A. Si\'odmok is supported by the National Science Centre, Poland, (Grant no. 2019/34/E/ST2/00457: A. Si\'odmok and J. Whitehead).
The work of A. Si\'odmok was also funded by the Priority Research Area Digiworld under the program Excellence Initiative -- Research University at the Jagiellonian University in Cracow.
K. Xie was supported by the U.S. Department of Energy under grant No. DE-SC0007914, the U.S. National Science Foundation under Grant No. PHY-2112829, and also in part by the PITT PACC. 
C.-P.~Yuan is also grateful for the support from the Wu-Ki Tung endowed chair in particle physics. Bei Zhou was supported by the Simons Foundation.

\clearpage\newpage


\end{document}